\documentclass{scrreprt} % <= Druckversion: "scrbook" / Bildschirmversion: "scrreprt"
\usepackage[english,biber]{osm-thesis} % <= Sprache der Arbeit ("ngerman"/"english"), Biblatex-Backend ("bibtex"/"biber")
\usepackage{multicol}
\usepackage{longtable} 

\usepackage[frozencache=true,cachedir=minted-cache]{minted2}
\setminted{
  fontsize=\small,
  tabsize=4
}
\usepackage{xcolor}
\definecolor{LightGray}{gray}{0.95}
\makeatletter
\AddToHook{begindocument/before}{\@ifpackageloaded{minted}{\removefromtoclist[float]{lol}}{}}
\makeatother

\usepackage{pdflscape}

\usepackage{array}
\newcolumntype{P}[1]{>{\raggedright\arraybackslash}p{#1}}
\usepackage{svg}

\usepackage{listings}
\usepackage[margin=1in]{geometry}
\usepackage{courier}
\lstdefinelanguage{MyAsm}{
    sensitive=true,
    morecomment=[l]{;},
    morestring=[b]"
}

\usepackage[capitalise]{cleveref}
\usepackage[disable]{todonotes}

\newcommand{\thesisauthor}{Moritz Steffin and Jiska Classen}
\newcommand{\thesistitle}{Modern iOS Security Features -- A Deep Dive into SPTM, TXM, and Exclaves}

\usepackage[type={CC},modifier={by-sa},version={4.0},lang=English]{doclicense}

\usepackage{subcaption}

\newcommand{\thesishandindate}{\today}%
\usepackage{setspace}

\expandafter\def\expandafter\UrlBreaks\expandafter{%
  \UrlBreaks
  \do\a\do\s\do\:\do\t\do\y\do\n\do\t\do\a\do\D\do\-\do\_\do\.\do\u
}

\makeglossaries

\newacronym{SoC}{SoC}{System-on-a-Chip}
\newacronym{SPRR}{SPRR}{Shadow Permissions Remapping Register}
\newacronym{APRR}{APRR}{Access Permissions Remapping Register}
\newacronym{SPTM}{SPTM}{Secure Page Table Monitor}
\newacronym{GL}{GL}{Guarded Level}
\newacronym{EL}{EL}{Exception Level}
\newacronym{GXF}{GXF}{Guarded Execution Feature}
\newacronym{PPL}{PPL}{Page Protection Layer}
\newacronym{HVC}{HVC}{Hypervisor Call}
\newacronym{SVC}{SVC}{Supervisor Call}
\newacronym{ELR}{ELR}{Exception Link Register}
\newacronym{SPSR}{SPSR}{Saved Program Status Register}
\newacronym{TXM}{TXM}{Trusted Execution Monitor}
\newacronym{SPRRUPERMEL0}{SPRR\_UPERM\_EL0}{SPRR User Permission Configuration Register (EL0)}
\newacronym{TPIDR}{TPIDR}{Software Thread ID Register}
\newacronym{FPSR}{FPSR}{Floating Point Status Register}
\newacronym{FPCR}{FPCR}{Floating Point Control Register}
\newacronym{FTE}{FTE}{Frame Table Entry} 
\newacronym{VBAR}{VBAR}{Vector Base Address Register}
\newacronym{IOMMU}{IOMMU}{Input/Output Memory Management Unit}
\newacronym{RPC}{RPC}{Remote Procedure Call}
\newacronym{IPC}{IPC}{Inter-Process Communication}
\newacronym{HCREL2}{HCR\_EL2}{Hypervisor Configuration Register EL2}
\newacronym{SK}{SK}{Secure Kernel}
\newacronym{WFE}{WFE}{Wait-for-Event}
\newacronym{PMAP}{PMAP}{Physical Map}
\newacronym{PTE}{PTE}{Page Table Entry}
\newacronym{ANE}{ANE}{Apple Neural Engine}
\newacronym{SIL}{SIL}{Secure Indicator Light}
\newacronym{EIC}{EIC}{Exclave Indicator Controller}
\newacronym{ESR}{ESR}{Exception Syndrome Register}
\newacronym{AP}{AP}{Access Permission}
\newacronym{UXN}{UXN}{Unprivileged Execute Never}
\newacronym{PXN}{PXN}{Privileged Execute Never}
\newacronym{SRD}{SRD}{Security Research Device}
\newacronym{DCP}{DCP}{Display Coprocessor}
\newacronym{XNU}{XNU}{X is Not Unix}
\newacronym{FPR}{FPR}{Fast Permission Restrictions}
\newacronym{DMG}{DMG}{Apple Disk Image}
\newacronym{POSIX}{POSIX}{Portable Operating System Interface}
\newacronym{DYLD}{DYLD}{Dynamic Linker}
\newacronym{SCID}{SCID}{Scheduling Context ID}
\newacronym{CTID}{CTID}{Compact Thread ID}
\newacronym{XPC}{XPC}{Cross-Process Communication}
\newacronym{TGE}{TGE}{Trap General Exception}
\newacronym{PAPT}{PAPT}{Physical Aperture Table}
\newacronym{NSA}{NSA}{National Security Agency}
\newacronym{SDK}{SDK}{Software Development Kit}

\begin{document}
\crefname{figure}{Figure}{Figures}
\crefname{table}{Table}{Tables}
\crefname{lstlisting}{Listing}{Listings}
\Crefname{lstlisting}{Listing}{Listings}
\crefname{enumi}{Item}{items}
\Crefname{enumi}{Item}{Items}
\crefname{equation}{Equation}{Equations}
\crefname{chapter}{Chapter}{Chapters}
\crefname{section}{Section}{Sections}
\crefname{subsection}{Subsection}{Subsections}
	% Einband
	%\pagenumbering{alph}
	%\ifisbook\include{content/coverpage}\fi
	%\ifisbook\cleardoubleemptypage\fi

	% (Haupt-)Titelseite, Abstract, ggf. Danksagung & Inhaltsverzeichnis
	\pagenumbering{roman}
	\begin{titlepage}
	\setcounter{page}{1}
	\centering

	% TITLE PAGE

	\raisebox{-0.5\height}{\includegraphics[width=5.5cm]{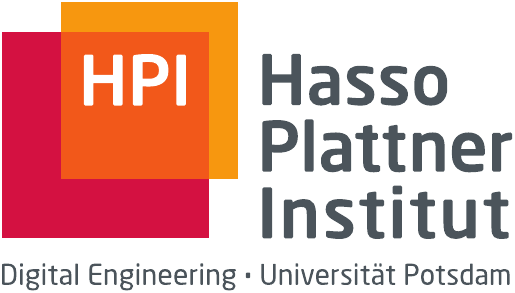}}
	\hspace*{.2\textwidth}
	\raisebox{-0.5\height}{\includegraphics[width=4cm]{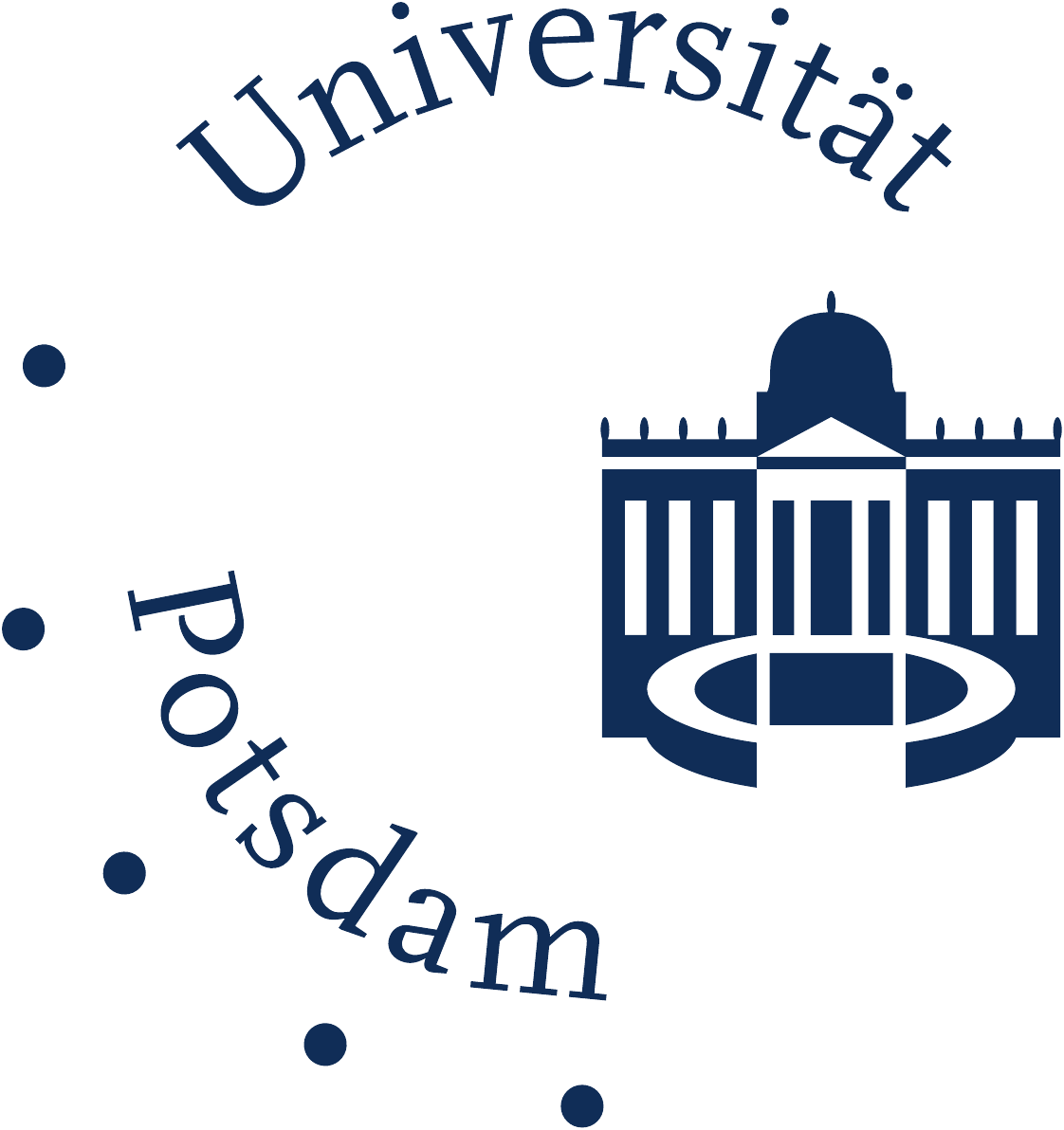}}

	\vfill
	{\usekomafont{title}\thesistitle\par}\par
	\vspace*{\baselineskip}
	{\smallskip\usekomafont{author}\thesisauthor\par}\par

	\vfill\vfill
	{\usekomafont{date}Based on the M.Sc. thesis by Moritz Steffin.\par}\par

	\clearpage
	% BACK TITLE PAGE

	\begingroup
	\ifcsvoid{doclicenseThis}{\phantom{}}{%
		\begin{minipage}[t]{\textwidth}
			\noindent%
			\begin{flushleft}
				\small
				This work is licensed under a \doclicenseLongType license:\\
				\doclicenseIcon\doclicenseLongNameRef.\\
				This does not apply to quoted content from other authors and works based on other permissions.\\
				To view a copy of this license, visit \url{\doclicenseURL}
			\end{flushleft}
		\end{minipage}}%
	\par\vfill
	
	\endgroup

\end{titlepage}
	% => Wenn die Arbeit auf Deutsch verfasst wurde, verlangt das Studienreferat KEINEN englischen Abstract

%  englischer Abstract
\null\vfil
\begin{otherlanguage}{english}
\begin{center}\textsf{\textbf{\abstractname}}\end{center}

The \gls{XNU} kernel is the basis of Apple's operating systems.
Although labeled as a hybrid kernel, it is found to generally operate in a monolithic manner by defining a single privileged trust zone in which all system functionality resides.
This has security implications, as a kernel compromise has immediate and significant effects on the entire system. 
Over the past few years, Apple has taken steps towards a more compartmentalized kernel architecture and a more microkernel-like design.
Whilst initially limited to moving key sensitive components (page table manipulation, cryptographic functionality) to kernel-separated scopes, with the introduction of the \gls{SPTM} in 2023, this movement towards compartmentalization has intensified.

To date, there has been no scientific discussion of \gls{SPTM} and related security mechanisms. Therefore, the understanding of the system and the underlying security mechanisms is minimal. 
In this paper, we provide a comprehensive analysis of new security mechanisms and their interplay, and create the first conclusive writeup considering all current mitigations.

\gls{SPTM} is be the highest-privileged system component running in \gls{GL} 2 using the \gls{GXF}. The \gls{GXF} implements horizontal guarded exception levels in addition to the standard exception levels, and is used to protect sensitive system activity from unrestricted access by XNU.

\gls{SPTM} acts as the sole authority regarding memory retyping. Our analysis reveals that, through \gls{SPTM} domains based on frame retyping and memory mapping rule sets, \gls{SPTM} introduces domains of trust into the system, effectively gapping different functionalities from one another. Gapped functionality includes, among others, the \gls{TXM}, responsible for code signing and entitlement verification. We further demonstrate how this introduction lays the groundwork for the most recent security feature of Exclaves, and conduct an in-depth analysis of its communication mechanisms. We discover multifold ways of communication, most notably \texttt{xnuproxy} as a secure world request handler, and the Tightbeam \gls{IPC} framework. Finally, we provide a top-level overview of components running in \gls{GL} and their interactions.

The architecture changes are found to increase system security, with key and sensitive components being moved out of XNU's direct reach. This also provides additional security guarantees in the event of a kernel compromise, which is no longer an immediate threat at the highest trust level.

\end{otherlanguage}
\vfil\null % example
    
    %\include{content/dedication}
	%\ifisbook\cleardoubleemptypage\fi\include{content/dedication}
	\tableofcontents
	\cleardoublepage

    \setstretch{0.95}

	% Textteil
	\pagenumbering{arabic}
{
\raggedright
\printglossaries
}
\listoffigures        % generates the Table of Figures
\listoftables 
\listoflistings
    \chapter{Introduction}
\section{Motivation}
Apple's \gls{XNU} kernel builds the base of the Darwin operating system. 
Upon this operating system, a multitude of Apple proprietary operating systems, such as macOS and iOS, are built.
According to Apple, the XNU kernel is a hybrid kernel~\cite{XNU:2025}. It combines the \emph{Mach microkernel}, built at Carnegie Mellon University, with the monolithic FreeBSD kernel~\cite{FreeBSD:2015}. 

Operating systems based on monolithic kernels run ``every basic system service'' in the privileged kernel space. This includes services such as process management, interrupt handling, drivers, and \gls{IPC}.
In contrast to that, microkernel-based operating systems provide a very basic set of functionality in the kernel space, including ``basic process communication and I/O control'', whilst placing other system services in user space~\cite{roch2004monolithic}. While microkernels are generally considered superior in terms of system security due to their reduced attack surface compared to components running at the highest privilege levels, they are typically regarded as inferior in terms of performance.

Despite its labeling as a hybrid kernel, the implementation of the XNU kernel is monolithic mainly, as it ``places all system functions within the same privileged scope''~\cite{OnAppleExclaves:2025}.
This implementation comes with the known security flaws of monolithic kernels.
Especially the lack of kernel address space separation poses a non-negligible security risk.
With Apple's chosen implementation, faults and weaknesses in any system call or driver can be exploited to ``maliciously manipulate kernel code'', posing a risk of a full and privileged kernel compromise, as the kernel is the most privileged part of the system ~\cite{monolithicKernelSecurity}.

With the release of the SPTM and the TXM for A15/M2 or newer \gls{SoC} in 2023, Apple is addressing these issues by beginning to compartmentalize its kernel into different privilege spaces.
This aims to ``protect page tables for both user and kernel processes against modifications, even when the attackers have kernel write capabilities [...]''. As advertised on the Apple Support pages, these systems are intended to replace the previous \gls{PPL}, which we will investigate. They aim to ``provide[] a smaller attack surface that does not rely on trust of the kernel''~\cite{appleOSSecurity:2024}. This was further advanced with the introduction of Exclaves, a security feature designed to isolate sensitive operations away from XNU.

Publicly available information regarding the actual implementation and implications of these features is scarce.
To date, Apple has not issued any official releases on this matter, apart from the previously cited minor publications in the context of operating system integrity guides.
Therefore, to infer a potential change in security and the underlying threat model addressed, a deep analysis of the underlying architecture is of high interest.
In doing so, we aim to provide an introduction to more intensive research into modern iOS security and memory system mechanisms.

\section{Research Goals \& Contributions}
Our analysis aims to provide an initial understanding of the architectural design of the SPTM-based system and memory security. 
We shed light on the underlying security assumptions and their impact on the security of real-world systems.
Public analysis of modern iOS security features, such as SPTM, TXM, and Exclaves, and their underlying mechanisms, is minimal.
Most sources focus on individual aspects or are limited in their applicability~\cite{df-f-1:2023, df-f-2:2025, df-f-3:2025}.
While working on this topic, more detailed blog posts were published~\cite{OnAppleExclaves:2025,RandomAugustine-2:2025,DisassemblingAppleExclaves:2025}, but various open questions remain.
Furthermore, official information from Apple is extremely limited. Our contributions are as follows:
\begin{itemize}
    \item Providing the most comprehensive analysis to date of components running in GXF, their interaction with each other, and interaction with XNU, in light of the recent restructuring of \gls{GXF} components with the release of Exclaves. 
    \item Reverse engineering of SPTM with a focus on its interaction with its clients and the discovery of its underlying rule set regarding memory frame retyping and memory mapping, as well as the derived security implications, despite the lack of any prior official information regarding its inner workings.
    \item First comprehensive public discussion of Exclave communication mechanisms with a focus on Tightbeam and xnuproxy. Discovery of Tightbeam as an Exclaves secure world \gls{IPC} framework.
    \item To our knowledge, we have performed the first practical experiment in directly interacting with conclaves from XNU user space.
\end{itemize}

\section{Outline}
Our work aims to provide a view into individual components and their interplay in the realization of memory and system security. We will begin by giving a brief top-down overview of the system in question (see \cref{CHAP:Overview}) and then proceed to a more detailed analysis of its specific components.
Our analysis of SPTM will be the core part of this work (see \cref{CHAP:SPTM}), with a focus on the underlying security mechanisms and implications.
Furthermore, we provide insights into Exclaves, including their underlying memory safety assumptions, and focus on communication mechanisms (see \cref{exclaves}).
As Exclaves run in the GXF, we examine the Secure Kernel (see \cref{secureKernel}), a central component of GXF, and provide a brief overview of TXM.
Our focus in this analysis will be the cross-component communication and memory security based on SPTM.

 % example
	\chapter{Fundamentals and Background}
\section{AArch64 Fundamentals}
The chips in use in Apple's hardware are based on the AArch64 architecture by ARM.
While Apple employs a variety of proprietary, Apple-specific customizations, the architectural fundamentals still apply and build the base for the respective firmware.
As the security mitigations examined in this paper primarily focus on granular privilege scoping and memory access privileges, two key components of the architecture are AArch64 exception levels and AArch64 memory addressing.

\subsection{AArch64 Exception Levels \& Exceptions}
\begin{figure}[H]
    \centering
\includegraphics[width=200pt]{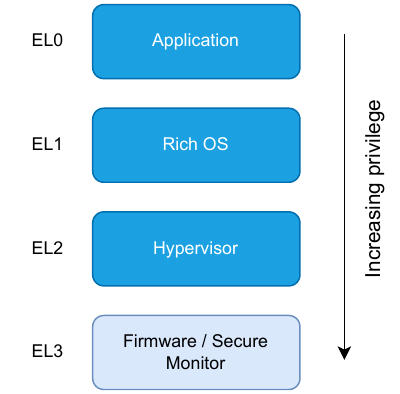}
\caption[AArch64 exception levels.]{Exception levels on AArch64 with the common usage model applied, adapted from~\cite{ARM-exceptionLevels:2024}.}
    \label{lab:ExceptionLevels}
\end{figure}

Exception levels are the AArch64 realization of privilege levels. They are usually referred to as \texttt{EL\_X}, whereby \texttt{X} is the exception level in question. Higher exception levels are associated with higher privileges. The AArch64 architecture has four native exception levels, as shown in \cref{lab:ExceptionLevels}. 

The usage of the above exception levels is not strictly defined by the architecture, but rather by the common usage model.
Switching between privilege levels typically occurs when taking an exception or returning from an exception.
However, it may also change during the debug state, upon exiting the debug state, and upon processor reset~\cite{ARM-exceptionLevels:2024}. 

An exception in the sense of the ARM architecture model is ``any condition that requires the core to halt normal execution and instead execute a dedicated software routine known as an exception handler [...]''~\cite{ARM-exceptionHandling:2024}. An example of this is an EL0 application making \glspl{SVC} to invoke system functionality at a higher privilege level~\cite{ARM_SVC:2025}.

The architecture utilizes a variety of key system registers to facilitate exception handling. These include a register holding information on the reason for an exception (\gls{ESR}), a register holding the address of the instruction causing the exception (\gls{ELR}), and a register that sets the exception base address for exception handling in a specific exception level (\gls{VBAR})~\cite{ARM-typesOfPrivilege:2024}.
We will have a more in-depth look at these specific registers and their usage in \cref{CHAP:SPTM}.

\subsection{AArch64 Memory Access}
\label{memoryAccess}
Memory access in the AArch64 architecture is realized via a virtual address space. 
Instead of directly operating on physical memory addresses, applications are provided with virtual addresses by the operating system.
This allows for easy sandboxing by giving each application own address space, or for abstraction on the underlying hardware~\cite{ARM-virtualMemory:2024}.

The operating system performs the mapping between physical and virtual memory through so-called translation tables.
At its essence, they hold entries for every virtual address, with a mapping to the corresponding physical address, along with various flags and attributes.
In practice, this lookup table system is typically multilevel, with indirection from translation tables to further higher-level translation tables.
However, the key concept of memory mapping remains the same~\cite{ARM-multilevelTranslation:2024}. Additionally, mapping is usually not done on individual memory addresses, but instead on predefined-sized memory chunks. 
A memory chunk in the \emph{virtual} address space is referred to as a memory \emph{page}, while a memory chunk in the \emph{physical} address space is referred to as a memory \emph{frame}.

\begin{figure}[H]
    \centering
\includegraphics[width=400pt]{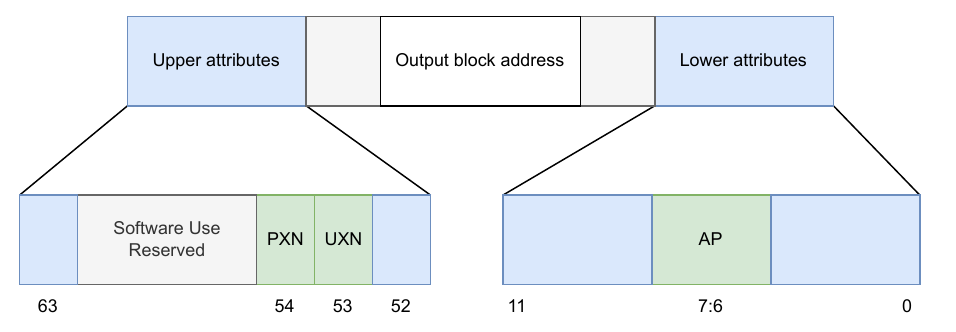}
    \caption[Simplified ARMv8-A translation table block entry.]{Simplified ARMv8-A translation table block entry, adapted~\cite{ARM-v8AddressTranslations:2024}.}
    \label{lab:blockenty}
\end{figure}

\cref{lab:blockenty} shows a simplified level one translation table block entry layout for the ARMv8-A architecture (on which the AArch64 execution state is based). Whilst different kinds of entries and entries on other translation table levels may differ, the general concept remains the same.

The block entry displays the architecture's default memory access permission mechanism, as represented by the so-called \gls{AP} bits, at bit positions 6 and 7. They encode the memory access permission, with the interpretation being dependent on the context of privileged (EL1+) and unprivileged (EL0) access attempt. The underlying encoding is illustrated in \cref{permissionTab}.

\begin{table}[h!]
\centering
\begin{tabular}{|c|c|c|}
\hline
\textbf{Access Permission Bits} & \textbf{Unprivileged (EL0)} & \textbf{Privileged (EL1/2/3)} \\
\hline
00 & No access     & Read/Write   \\
01 & Read/Write    & Read/Write   \\
10 & No access     & Read-only    \\
11 & Read-only     & Read-only    \\
\hline
\end{tabular}
\caption[Access permissions for different exception levels.]{Access permissions for different exception levels~\cite{ARM-permissions:2024}.}
\label{permissionTab}
\end{table}

As the \gls{AP} bits only encode access (i.e., read/write) permissions, an additional mechanism is required for execution permissions.
This is implemented via the \gls{UXN} and \gls{PXN} bits at bit position 54 and 53, respectively.
If the relevant bit is set, the memory mapped by the specific \gls{PTE} is marked as non-executable from both the unprivileged/privileged exception levels.
Note that there are additional mechanisms that modify how the exact permission is interpreted, depending on factors such as the processor state at the time of access~\cite{ARM-permissions:2024}. 

\section{Exception Levels Usage for iOS}
Apple operating systems are built upon the ARMv8 architecture. Therefore, the implementation of privilege levels is done via ARMv8 exception levels. 
In Apple's interpretation of the common usage model, user-space code (i.e., applications, most daemons) runs on EL0.
Apple provides no reliable information as to which exception level the kernel uses, but we find a multitude of both EL1 and EL2 register accesses in the kernel binary.
This indicates that Apple deviates from the standard of running a hypervisor in EL2, but instead runs the kernel there. This is nothing out of the ordinary per se, and can be achieved with techniques like ARM64 Virtualization Host Extensions~\cite{SPRRandGXF:2021}. Since 2019, iOS abandoned the concept of running a secure monitor in EL3~\cite{P0-PAC:2024}.

\section{Related Mitigations}
The introduction of Exclaves and the deployment of SPTM is not the first novel approach towards compartmentalizing the XNU kernel, but rather the latest release in a line of multiple previous mitigations released over the past few years, aimed at enhancing kernel security.
To provide a basis for the most recent mitigations, this section provides a brief background on a selection of these.

\subsection{Fast Permission Restrictions}
\gls{FPR} are a hardware primitive used for the fast and efficient remapping of memory permissions, without the need for computationally heavy page table walks.
For example, it can help to ensure secure just-in-time compilation as it can quickly switch between writable and executable memory.
This fundamental ability is the core capability for other mitigations that have been released throughout the past years. 

\gls{FPR} are realized through special APRR registers~\cite{appleOSSecurity:2024}. They have been introduced with the 2017 Apple A11 Bionic \gls{SoC}. Whilst not officially confirmed, APRR is assumed to denote \gls{APRR}.
As this exact implementation is no longer in use with current devices, we consider it out of scope.
It is, however, comparable to the implementation of the SPRR registers, which is looked at in \cref{subs:SPRR}. 

\subsection{Page Protection Layer}
Introduced in 2017, Apple designed \gls{PPL} to deny user-space code modification after successful code signature verification.
\gls{PPL} was introduced as the new sole mean to alter specific protected pages, especially those containing page tables. \gls{PPL} was realized by employing the previously mentioned \gls{FPR} and, more specifically, the proprietary APRR register~\cite{appleOSSecurity:2024}. PPL offers a specific trampoline code (\texttt{\_\_PPLTRAMP} segment), which, after invoking it, employs the APRR register to efficiently remap large amounts of page table permissions, and vice versa on exit~\cite{casaDePaPeL:2019}. The actual protection is therefore achieved by mapping sensitive memory (in this case, mainly page tables) and the PPL code itself as \texttt{r--} (read-only) to the kernel.
The PPL trampoline uses the APRR register to flip the permissions of the page tables to \texttt{rw-} (read-write) and the actual PPL code as \texttt{r-x} (read-execute). This PPL-trampoline is the only entry point to PPL, and therefore significantly reduces the attack surface with respect to page table manipulation~\cite{SPRRandGXF:2021}.
\gls{SPTM} has replaced the PPL mitigation on A15 / M2 or newer SoCs~\cite{appleOSSecurity:2024}.

\subsection{Guarded Execution Feature}
\label{GXF}
The GXF, presumably released in 2021, introduces lateral exception levels next to the architecture's default exception levels. These levels are called \acrfullpl{GL}.
Entering these GLs is achieved via the Apple-proprietary \texttt{GENTER} instruction (opcode: \texttt{0x00201420}), whilst exiting is done via \texttt{GEXIT} respectively (opcode: \texttt{0x00201420}). Comparable to standard AArch64 exception levels, the GLs have relevant system registers (such as \gls{TPIDR}, \gls{VBAR}, \gls{ESR}, and \gls{ELR}), those being another Apple-proprietary architectural addition~\cite{SPRR_Asahi:2024}. The lateral levels introduce a different set of permissions compared to default exception levels, allowing for a more granular distribution of privilege. This is implemented via the so-called \glspl{SPRR}.

\subsection{Shadow Permission Remap Register}
\label{subs:SPRR}
The \glspl{SPRR} are the tool used to implement the \gls{GXF}.
They reimplement the usage of the \acrfull{AP} bits and the \gls{UXN}/\gls{PXN} bits in page table entries looked at previously by adding a layer of indirection to them (initially introduced via \gls{APRR}). Instead of directly encoding the memory access/execute permissions into the relevant bitfields, they are used to generate an index for a lookup into a system register~\cite{SPRRandGXF:2021, SPRR_Asahi:2024}.
The bitwise \gls{SPRR} index calculation based on the permission fields of a \acrfull{PTE} is shown below:

\begin{figure}[H]
    \centering
\includegraphics[scale=0.9]{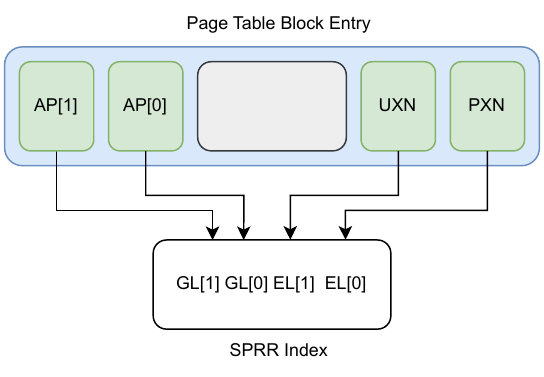}
   \caption[SPRR index calculation.]{SPRR index calculation based on the previously seen arguments AP, UXN, and PXN in a page table block entry, adapted~\cite{SPRR_Asahi:2024}.}
    \label{fig:SPRRINDEX}
\end{figure}

\begin{figure}[H]
    \centering
\includegraphics[scale=0.9]{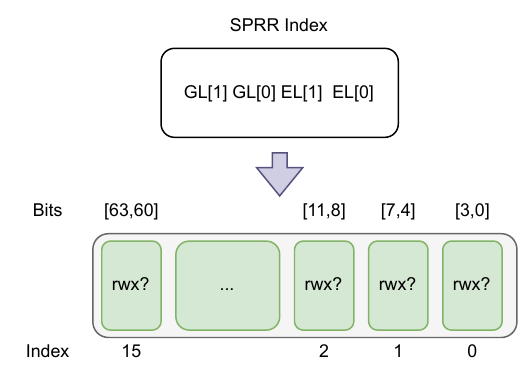}
    \hspace{-5cm}\caption[SPRR permission lookup.]{Permission lookup scheme in SPRR system register based on the SPRR index.}
    \label{fig:SPRRREgister}
\end{figure}

The retrieved index is then used to access privilege level-specific system registers, in which the actual permissions are encoded.
The concept of this index retrieval is shown in \cref{fig:SPRRREgister}.

In 2021, Sven Peter reverse engineered the permissions mapped behind the usage of SPRR on an M1~\cite{SPRRandGXF:2021}. The following table is based directly on his findings. The exact permissions mapped to each index have likely changed since then, but we deem the table informative enough to understand the general workings of SPRR permissions. The table also includes found index usage in Peters's reverse engineering.

\begin{table}[h!]
\centering
\begin{tabular}{|c|c|c|c|l|}
\hline
\textbf{Index} & \textbf{SPRR EL0} & \textbf{SPRR EL2} & \textbf{SPRR GL2} & \textbf{Usage} \\
\hline
1  & \texttt{---} & \texttt{r--} & \texttt{rw-} & Page Tables \\
3  & \texttt{---} & \texttt{rw-} & \texttt{rw-} & Kernel Data \\
5  & \texttt{rw-} & \texttt{r-x} & \texttt{r--} & Userland MAP\_JIT \\
7  & \texttt{rw-} & \texttt{rw-} & \texttt{rw-} & Userland Data \\
8  & \texttt{---} & \texttt{r--} & \texttt{r-x} & PPL Code \\
10 & \texttt{---} & \texttt{r-x} & \texttt{r-x} & Kernel Code \\
11 & \texttt{---} & \texttt{r--} & \texttt{r--} & Kernel Read-only Data \\
13 & \texttt{r-x} & \texttt{r--} & \texttt{---} & Userland Code \\
15 & \texttt{r--} & \texttt{r--} & \texttt{---} & Userland Read-only Data \\
\hline
\end{tabular}
\caption[SPRR index to permissions mapping and usage.]{SPRR index to permissions mapping and usage~\cite{SPRRandGXF:2021}.}
\label{tab:permissions}
\end{table}

It is noteworthy that at the time of initial release, all remaining indices showed no access permissions on any privilege level. This indicates that there was still room for a more granular implementation of fast-mapped permissions. 

\paragraph*{Implications for Security}

As shown in \cref{tab:permissions}, the \gls{SPRR}-based permissions significantly restrict the possibility of altering page tables.
Whilst read access is still possible from EL2, page table writes are completely encapsulated in GL2.
Page tables, and as such, a highly vulnerable part of the operating system, are protected even from privileged kernel code, requiring the use of very specific \texttt{GENTER}/\texttt{GEXIT} functionality to write access them.

\section{Previous Work}
The Apple security mechanisms analyzed in this work have, to this date, not yet been analyzed in any scientific paper.
This is partially due to their recency, but also a result of Apple's lack of public communication regarding their releases and a near-complete lack of official sources on their inner workings.
There are few public sources on SPTM, TXM, and Exclaves. 

Dataflow Forensics has provided a fairly comprehensive write-up about SPTM, TXM, and their XNU interaction in the form of three blog posts spanning from August 2023 to February 2025.
The blog posts provide an entry point into understanding SPTM, and especially shed light on communication between XNU and SPTM, as well as SPTM and TXM.
As SPTM and components running in Apple's GXF are subject to constant change, especially since they are relatively new, the posts do not, in parts, reflect the current state of things.
Nonetheless, we build our analysis of SPTM upon them and the concepts they display.
They will be referenced repeatedly throughout this work for core aspects of SPTM understanding.

On the question of Apple Exclaves, the only original public discussion is found in three key blog posts by Random Augustine~\cite{OnAppleExclaves:2025, RandomAugustine-2:2025, DisassemblingAppleExclaves:2025}. They discuss the general concept of Apple Exclaves and provide a high-level analysis, primarily based on XNU public sources. We will reference this work in our analysis of Exclaves, as it provides a fairly in-depth look into Exclave thread scheduling, a concept we have excluded from our analysis.
 % example
    \chapter{Setup}
The following analysis has primarily been conducted on specific hardware and firmware specifications.
As the topic of \gls{SPTM}, \gls{TXM}, and Exclaves is a highly recent one, these implementations are naturally subject to change. 
Therefore, the findings presented in this paper may no longer apply to newer hardware and firmware.
Nonetheless, in favor of practicability, we have chosen and decided on specific versions and have not taken newer releases into account for most of this paper.
This poses no issue, as any analysis and reverse engineering of still actively developed software, firmware, and hardware is always just a momentary snapshot of the current state.
The following section will shortly list the used hardware and firmware specifications.

\section{Firmware}
Regarding firmware, the analysis was primarily conducted on the iOS 18.4 (22E240) firmware for the iPhone 16.
The firmware was obtained via \url{https://appledb.dev}. The release correlates with the kernel version \emph{xnu-11417.102.9}.

Significant parts of the following analysis are based on Apple's open source publications of kernel sources~\cite{XNU:2025}.
The primary open source release used was \emph{xnu-10063}~\cite{XNU_10063:2024}, due to its availability at the time of our research.
All references to XNU open-source code without an explicit version specification shall refer to this version. We furthermore declare that all path specifications provided throughout this work, if not specified otherwise, shall refer to said XNU open source code. 

\section{Hardware}
Regarding the dynamic analysis of the new security mitigations, multiple hardware devices were utilized.
\begin{enumerate}
    \item iPad Pro 11-inch (M4) running iPadOS 17.5.1, used for initial log analysis via the macOS Console app.
    \item iPhone 16 \gls{SRD} running iOS 18.5, used for dynamic reverse engineering on recent iOS versions.
    \item iPhone 8 running iOS Version 16.7.10, jailbroken via the palera1n jailbreak version v2.1-beta~\cite{palera1n:2025}. This device was used for dynamic reverse engineering in the context of security analysis comparison on the \emph{Recording Indicator/Secure Indicator Light}.
\end{enumerate}

\subsection*{Security Research Device}
The \textit{Apple Security Research Device Program}\footnote{Apple Security Research Device Program: \url{https://security.apple.com/research-device/}, last accessed: 14.09.2025} allows specific individuals with a strong and proven background in security research on Apple and other modern operating systems to obtain so-called \gls{SRD} from Apple. These devices ``allow[ ] you to perform iOS security research without having to bypass its security features''~\cite{AppleSecurityResearchProgram:2025}.
Such a research device enables the analysis of even the most recent firmware versions without waiting for the corresponding jailbreaks to be released.
The capabilities on these devices are extensive, as they support shell access, allow for setting individual entitlements, and even support kernel customization.

The \gls{SRD} used for analysis during this paper was provided by Apple to Dr. Jiska Classen. In compliance with Apple's Security Research Device guidelines, all analysis on the device itself was performed by Dr. Jiska Classen, the authorized Apple user. 

\section{Tooling}
As this paper was mainly focused on static reverse engineering, we employed a variety of such tools.
\begin{description}
    \item[Ghidra] Ghidra is a software reverse engineering framework developed by the \gls{NSA}~\cite{ghidra:2025}. It has been the primary tool used in our analysis.
    \item[ipsw] A command-line research tool for iOS and macOS, and has been used in our analysis to dissect Apple IPSW files~\cite{ipsw:2025}.
    \item[disarm] A command-line binary analyzer we additionally employed for reverse engineering~\cite{disarm:2025}.
    \item[symbolicator] A tool used for symbolication of XNU binaries~\cite{symbolicator:2025}.
    \item[lldb] The debugger from the LLVM project, used for dynamic tracing of calls on the \gls{SRD}~\cite{LLDB2024}.
\end{description}

\section{Binary Naming Convention}
To allow for easier reference to binaries analyzed in this work, we will define the following naming convention for binaries at paths in the firmware:
\begin{itemize}
    \item SPTM binary: \\ \path{22E240__iPhone17,3/Firmware/sptm.t8140.release}
    \item TXM binary: \\ \path{22E240__iPhone17,3/Firmware/Firmware/txm.iphoneos.release}
    \item Exclavecore binary: \\ \path{22E240__iPhone17,3/Firmware/image4/exclavecore_bundle.t8140.} \\ \path{RELEASE}
    \item Secure Kernel binary: \\ \path{22E240__iPhone17,3/Firmware/image4/SYSTEM/kernel}
    \item \gls{XNU} kernelcache binary: \\ \path{22E240__iPhone17,3/kernelcache.release.iPhone17,3}
\end{itemize}

\section{Symbol Enumeration}
During our reverse engineering for this work, we symbolicated a variety of different functions and structures in all the binaries we examined. We include a table with symbol-to-address mappings for the various binaries analyzed in \cref{symbolTable}, which ensures reproducibility of our analysis.

\section{Apple-Proprietary Registers}
For the realization of features like the \gls{GXF}, Apple uses its own custom and proprietary registers. As these specifications are usually not officially released, some reverse engineering tools may not yet have incorporated these.
To facilitate an understanding of the underlying functionality, we have utilized releases of these Apple-specific system registers from previous reverse engineering work. 
As these are unofficial, there is no claim to correctness and completeness.
These sources include a recent leak as well as Asahi Linux's tools~\cite{appleRegisters:2025, appleRegistersLeak:2025}
For the sake of clarity, we will use the definitions of Asahi Linux's \texttt{m1n1} tool, as they form the basis of our reverse engineering work.
 % example

	\chapter{Overview on Architectural Design}
\label{CHAP:Overview}
\begin{figure}[H]
\centering
\includegraphics[width=\textwidth]{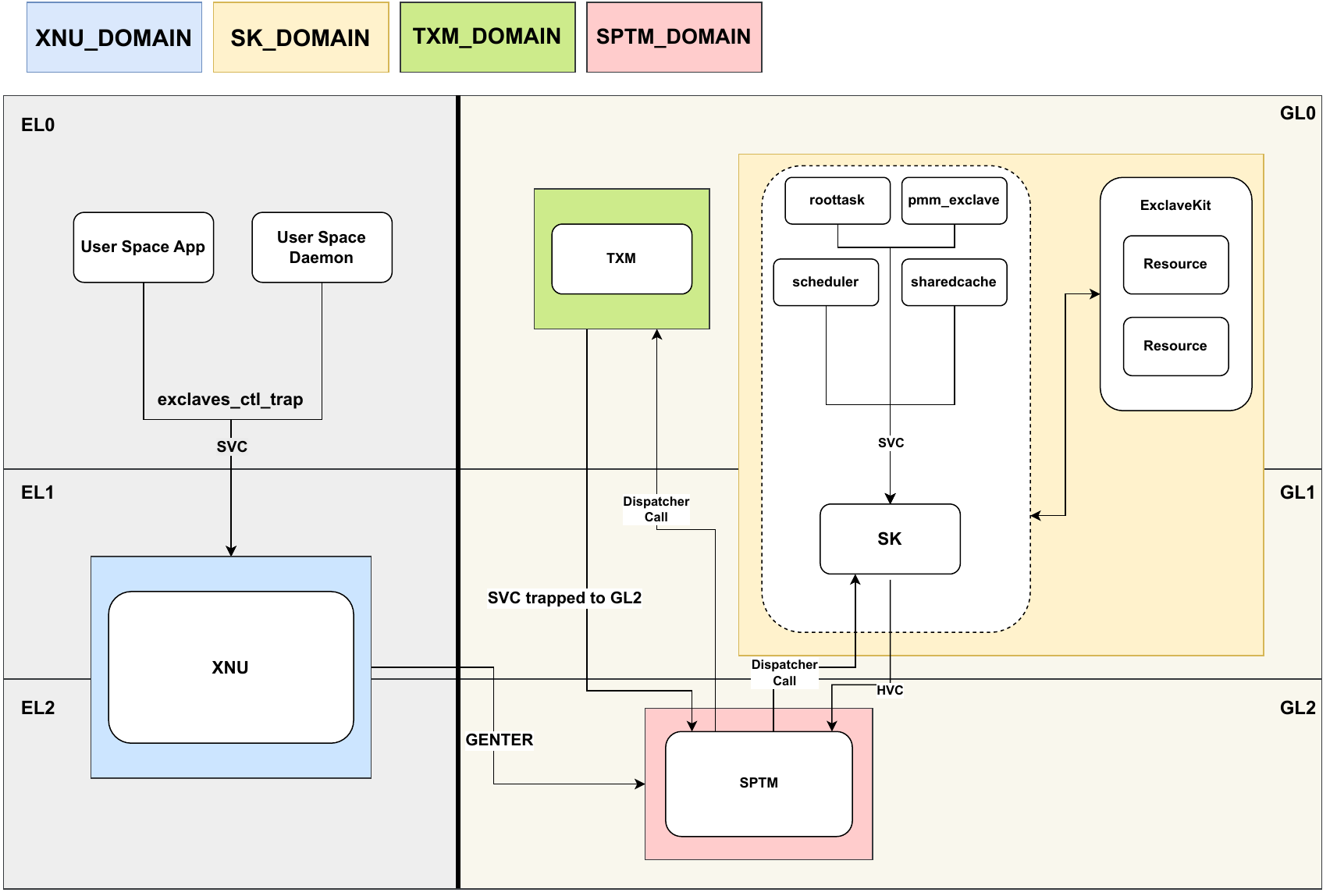}
\caption[Full system overview across both ELs and GLs.]{Overview of the system analyzed by us, with focus on communication paths between individual components across ELs and GLs.}
\label{Overview}
\end{figure}

This chapter provides a brief overview of the system's architectural design, as reverse-engineered by us. We will shortly introduce the underlying usage of \gls{GXF} in Apple operating systems. In doing so, we provide the fundamentals for the upcoming in-depth analysis of system components and their interactions. 
An overview of the system, our assumptions, and findings based on this analysis are shown \cref{Overview}.

\section*{System Layout}
The novel system is based on the \acrfull{GXF} we have previously introduced in \cref{GXF}.
We find that all components (except the \gls{XNU} kernel) analyzed by us are running in \acrfullpl{GL}, for which we know different access and execution permissions are realized via \gls{SPRR} (see \cref{subs:SPRR}).

The new \gls{GXF} central management and dispatching component is \acrfull{SPTM}.
XNU can call into SPTM via the proprietary \texttt{GENTER} instruction, and based on a provided dispatch target, SPTM either handles the request itself or dispatches the call into other guarded-level components.
\gls{SPTM} replaces the \gls{PPL} in performing page table manipulation. It has further advanced its responsibilities regarding memory safety, introducing the concept of SPTM domains.
These domains act as a source of trust, with limited memory mapping interaction across domains.

Regarding other protected-level components, \acrfull{TXM} is running in GL0 and is responsible for, among other things, code signing and entitlement validation.
It can call into SPTM via \gls{SVC} instructions.

With Exclaves, the guarded levels are extended with a complex and extensive system.
Exclaves are found to be access-scoped groupings of sensitive resources, which require special privileges to call into, and reduce XNU's ability to interact with security-relevant operations. 
The Exclave ecosystem introduced the \gls{SK}, which handles \gls{SPTM} requests towards Exclaves and serves as a GL1 microkernel.
The \gls{SK} can call into SPTM via \gls{HVC} instructions.
Furthermore, a variety of Exclave-related GL0 binaries have been introduced, which appear to handle, among other things, Exclave scheduling and calling. We discover Tightbeam as a newly introduced \gls{IPC} mechanism for calling into Exclaves.

 % example
    \chapter{Secure Page Table Monitor}
\label{CHAP:SPTM}
\section{SPTM Fundamentals}
\gls{SPTM} is a \gls{GXF} component running in GL2\footnote{We know this from GL2 register calls.}. GL2 is the highest guarded level observed so far, and therefore the most privileged one across the entire system. \gls{SPTM} is the sole component detected so far running in that privilege level, and is therefore the most privileged component. This puts it in a unique position, making it responsible for the most security-relevant operations.

We will demonstrate that \gls{SPTM} is invoked from \gls{XNU} via \texttt{GENTER}, with this call entering the GXF.
\gls{TXM}, running in GL0, may call directly into \gls{SPTM} via \glspl{SVC}.
This is achieved through a special hypervisor configuration for \gls{SPTM} trapping exceptions, targeting GL1 to GL2.
GL1 components (the Secure Kernel) may call into \gls{SPTM} directly via \glspl{HVC}. 

We find \gls{SPTM} to be the key component responsible for performing memory mappings, which it restricts based on a specific ruleset.
These restrictions are demonstrated through \gls{SPTM} frame types and \gls{SPTM} domains, which scope components into privilege spaces concerning memory mapping. 
This compartmentalization within privileged kernel code appears to be the first step towards a more microkernel-oriented XNU architecture, which is further supported by our finding of seL4 microkernel components used to realize Exclaves (see \cref{exclaves}). 

\section{SPTM Setup}
We know \gls{SPTM} to be running in GL2 and therefore the \gls{GXF}. As previously looked at, GXF is entered via the \texttt{GENTER} instruction.
Among other GXF configurations, the GXF entry point, that is, the address to be jumped to when entering GXF, is defined via a system register called \texttt{GXF\_ENTRY\_ELx}~\cite{SPRRandGXF:2021}.
Looking for these registers in \gls{SPTM} reveals two \gls{GXF} setup functions. The two functions are shown in \cref{lst:early_GXF_setup} and \ref{lst:late_GXF_setup}.

\begin{listing}[H]
    \begin{minted}[linenos, breaklines, bgcolor=LightGray, frame=lines]{c}
void gxf_setup_early(void)

{
  GXF_CONFIG_EL1 = 1;
  GXF_PABENTRY_EL1 = 0xfffffff027097934;
  GXF_PABENTRY_EL1 = 0xfffffff02708b88c;
  InstructionSynchronizationBarrier();
  GENTER();
}
    \end{minted}
    \captionof{lstlisting}[SPTM \texttt{gxf\_setup\_early} function.]{\texttt{gxf\_setup\_early} function in SPTM performing GXF setup after initial SPTM entry. The source code was disassembled by Ghidra.}
    \label{lst:early_GXF_setup}
\end{listing}

\begin{listing}[H]
    \begin{minted}[linenos, breaklines, bgcolor=LightGray, frame=lines]{c}
undefined8 gxf_setup_late(void)

{
  undefined8 uVar1;
  
  spsel = 1;
  spsel = 0;

  GXF_PABENTRY_EL1 = 0xfffffff027097934);
  GXF_ENTRY_EL1 = 0xfffffff02708053c;
  VBAR_GL1 = 0xfffffff027084000;
  
  InstructionSynchronizationBarrier();
  
  uVar1 = 0x2f;
  if (DAT_fffffff027079618 == '\0') {
    GXF_CONFIG_EL12 = 0;
    GXF_ENTRY_EL12 = 0;
    GXF_PABENTRY_EL12 = 0;
    uVar1 = 0x6f;
  }
  GXF_CONFIG_EL1 = uVar;
  InstructionSynchronizationBarrier();
  return uVar1;
}    
    \end{minted}
    \captionof{lstlisting}[SPTM \texttt{gxf\_setup\_late function}.]{\texttt{gxf\_setup\_late} function in SPTM performing GXF setup, not part of the initial SPTM entry process. The source code was disassembled by Ghidra.}
    \label{lst:late_GXF_setup}
\end{listing}

The \texttt{GXF\_ENTRY} registers store the \gls{GXF} entry vector for the specified exception level, and \texttt{GXF\_PABENTRY} registers store the respective \gls{GXF} abort vector~\cite{appleRegisters:2025}. We assume that \texttt{GXF\_CONFIG} stores general \gls{GXF} configuration for the specific exception level, but no detailed information is available on this register. 

\paragraph*{GXF Entry Point -- Early Setup}

The core setup in the \texttt{gxf\_setup\_early} function sets the \texttt{GXF\_entry} register. The address stored is the entry vector into \gls{GXF}; therefore, the address that is jumped to on \texttt{GENTER} invocations.
The address points to a function we named \texttt{GXF\_early\_entry}\todo{table}, which performs context setting.
Most notably, it sets the EL1 \gls{SPRR} kernel permission configuration register \texttt{SPRR\_PPERM\_EL1}, which encodes access permissions for EL1 kernel components.
This might ensure specific permissions for kernel components during guarded-level execution.
It further sets the EL1 \gls{SPRR} configuration register (\texttt{SPRR\_CONFIG\_EL1}) to \texttt{0xff}, enabling \gls{SPRR}, and locking down its configuration and both user and kernel space permissions, preventing further alterations~\cite{SPRRandGXF:2021, appleRegisters:2025}.
The key setup part regarding further request handling is setting the GL1 \gls{VBAR}, which we can assume works similarly to the EL1 equivalent known to point to a base address, offset from which exception handlers are registered~\cite{VBAR_EL1:2025, TakingAnException:2025}.

Looking at exception handlers at the defined offsets~\cite{TakingAnException:2025}, we find a pattern of reading exception information based on the calling context, and storing it in memory structures.
There appears to be no further request handling performed. We assume this to be a very early request handling setup, which handles incoming requests by storing them, but does not actually act on them. This is in line with this GXF entry point being set nearly directly after the \gls{SPTM} \texttt{entry} point.

\paragraph*{GXF Entry Point -- Late Setup}

As the early setup GXF entry point did not reveal much regarding actual \gls{SPTM} request handling, we assume that \texttt{GXF\_late\_setup} performs the actual request handling setup.
Different from the previous \texttt{GXF\_setup\_early} function, it only sets the \texttt{GXF\_PABENTRY} and \texttt{GXF\_ENTRY} registers.
It additionally directly sets the \texttt{VBAR\_GL1} register, storing an address based on which we find a variety of actual exception handlers.
We will examine the request handler in more depth in \cref{SPTM:svc_handling}.
The address stored in the \texttt{GXF\_ENTRY\_EL1} register points to a function we denote as \texttt{GXF\_entry\_point}, which handles \texttt{GENTER} invocations and will be examined in \cref{GENTERHandling}.
We assume \texttt{gxf\_setup\_late} to be executed in later setup stages of SPTM, when \gls{SPTM} is actually prepared to begin handling requests, and can therefore set up the required handlers. 

\section{Calling into SPTM}
\label{lab:callingIntoSPTM}
As SPTM, in its essence, provides services to clients. These clients (among them \gls{XNU} and \gls{TXM}) must be able to invoke such services.
As previously determined, XNU interacts with the GXF and \gls{SPTM} via the \texttt{GENTER} instruction, as it has to switch from the highest \acrfull{EL} to a \acrfull{GL}.
Note that this is not the case for other \gls{SPTM} clients already running in a \gls{GL}, which instead use an \gls{SVC} or \gls{HVC}.
When XNU calls into \gls{SPTM}, \texttt{GENTER} is parameterized via the \texttt{x16} register.
An exemplary code fragment that shows this parameterization and call into \gls{SPTM} is shown below.
\begin{listing}[H]
\begin{minted}[linenos, breaklines, bgcolor=LightGray, frame=lines]{asm}
ff00ab30080 10 00 e0 f2     movk       x16,#0x0, LSL #48
ff00ab30084 10 00 c0 f2     movk       x16,#0x0, LSL #32
ff00ab30088 10 00 a0 f2     movk       x16,#0x0, LSL #16
ff00ab3008c f0 01 80 f2     movk       x16,#0xf
ff00ab30090 20 14 20 00     GENTER       
\end{minted}
\captionof{lstlisting}[Exemplary parameterized \texttt{GENTER} call from XNU towards SPTM.]{Parameterized \texttt{GENTER} call from XNU, with \texttt{x16} register setup followed by \texttt{GENTER} invocation.}
\label{lst:x16_param}
\end{listing}    

This is a recurring pattern that can be found repeatedly in \gls{SPTM} clients. Previous research indicates that the parameter in the \texttt{x16} register denotes a so-called ``subsystem'' together with a calling code~\cite{df-f-2:2025}.
This assumption does not hold entirely and is instead corrected by a later release of \gls{SPTM} headers directly from Apple (see \cref{ALookIntoSPTMHeaders}). However, it remains a valid working hypothesis.

\begin{figure}[H]
    \centering
\includegraphics[scale=0.75]{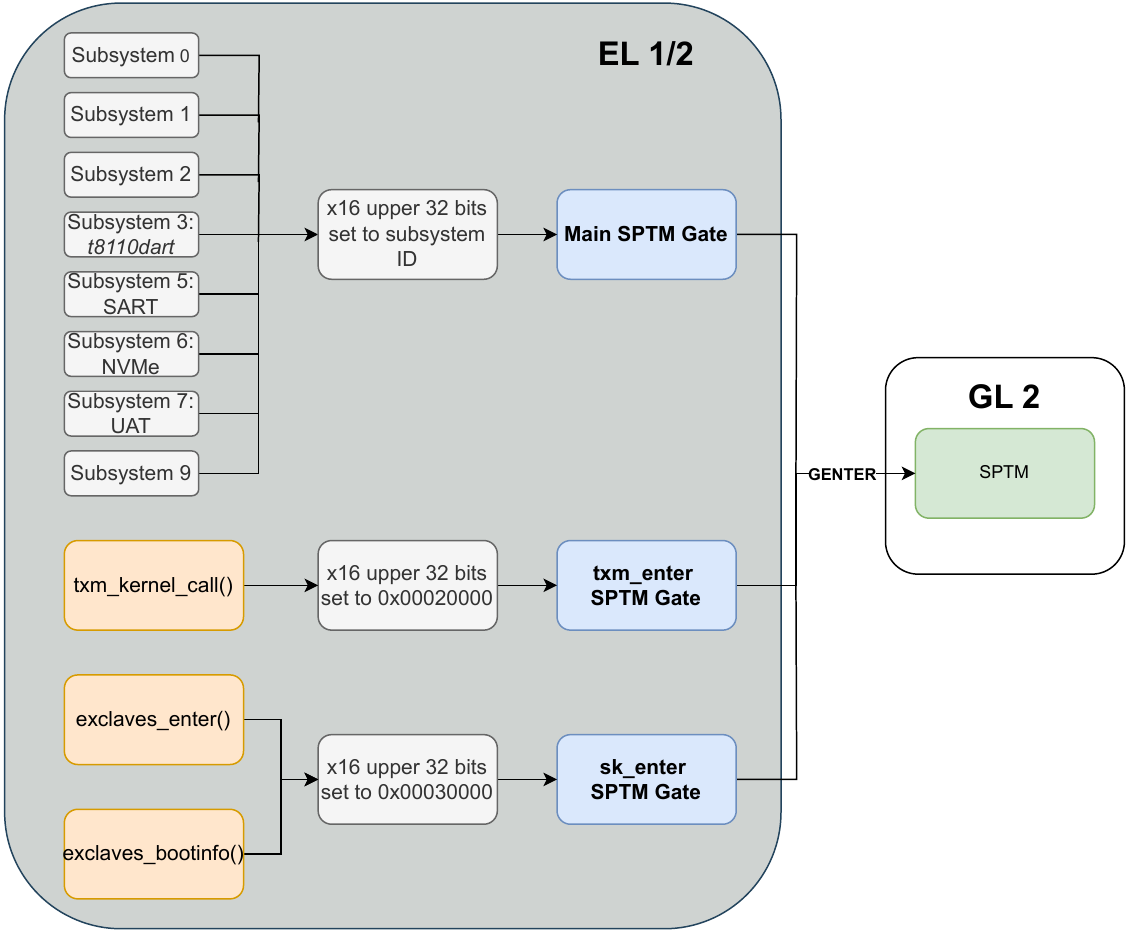}
\caption[XNU calls into SPTM.]{Depiction of XNU call structure into \gls{SPTM} -- \gls{SPTM} is called through three specific gates. The main \gls{SPTM} gate is used by what appears to be a multitude of different subsystems with calls parameterized via the subsystem ID. Two special \gls{SPTM} gates are only invoked from individual functions and employ different \texttt{x16} parameterization schemes.}
\label{fig:callsFromXNU}
\end{figure}

The following sections will list calls to \gls{SPTM} from various clients and examine their characteristics. An overview of all detected XNU calls into \gls{SPTM} is shown in~\cref{fig:callsFromXNU}.

\subsection{Calls from XNU}
\label{lab:sptmCallsFromXNU}
XNU makes various calls to different subsystems for \gls{SPTM}.
It does so by invoking \texttt{GENTER} from different ``SPTM gates'' depending on the calling context, which we will look at and list the relevant parameterized calls. To do this, XNU uses multiple different functions to invoke \texttt{GENTER} in different contexts.

\paragraph*{Main SPTM Gate}

We find a function invoking \texttt{GENTER}\todo{0xfffffff00ab3510c table} in XNU, which we labeled \texttt{GENTER\_main\_gate}.
Based on its frequent usage (77  direct calls), we infer that this is the \textbf{main \gls{SPTM} caller}.
The function calls a function that we consider to be a minor \texttt{GENTER} setup function (\texttt{genter\_setup}\todo{table}), which verifies that the current privilege level is not yet guarded. After this, \texttt{GENTER} is called directly.

However, we are more interested in the calling functions, as we find that these set the \texttt{x16} parameter via the scheme listed in~\cref{lst:x16_param}. Enumerating the \texttt{x16} parameters used shows the following calling values from XNU forwarded to the aforementioned default \gls{SPTM} entry point depicted in \cref{XNU_calls_table}.
We are furthermore able to partially map specific subsystems to names by matching symbolized functions that call into the specific subsystems \gls{SPTM} calls.
\begin{table}[h!]
\center
\footnotesize
\begin{tabular}{|l|l|l|l|l|}
\hline
Subsystem ID & Subsystem Name & \#Calls & Call Values & \begin{tabular}[c]{@{}l@{}}Invoking Functions - Excerpt\end{tabular} \\ \hline
0 & - & 22 & 0x0-0xe, 0x10-0x15, 0x21 &  \\ \hline
3 & t8110dart & 17 & \begin{tabular}[c]{@{}l@{}}0x300000000\\ - 0x300000010\end{tabular} & \texttt{\begin{tabular}[c]{@{}l@{}}\_t8110dart\_iovmfree,\\ table\_t8110dart\_map, \\\_t8110dart\_ioctl\end{tabular}} \\ \hline
5 & SART & 3 & \begin{tabular}[c]{@{}l@{}}0x500000000   \\ - 0x500000002\end{tabular} & \texttt{\_sart\_map, \_sart\_unmap} \\ \hline
6 & NVME & 8 & \begin{tabular}[c]{@{}l@{}}0x600000000 \\ - 0x600000007\end{tabular} & \texttt{\_nvme\_ioctl, \_nvme\_init} \\ \hline
7 & UAT & 13 & \begin{tabular}[c]{@{}l@{}}0x700000000\\ - 0x70000000c\end{tabular} & \texttt{\begin{tabular}[c]{@{}l@{}}\_uat\_ioctl, \\ \_uat\_get\_ptep,\\ \_uat\_init\end{tabular}} \\ \hline
9 & CPUTRACE & 13 & \begin{tabular}[c]{@{}l@{}}0x900000000\\ - 0x90000000c\end{tabular} &  \\ \hline
\end{tabular}
\caption[XNU SPTM calls parameter values.]{Call values for \texttt{GENTER} calls into SPTM parameterized via \texttt{x16}. Subsystems are named, if possible, by the given context and calling functions.}
\label{XNU_calls_table}
\end{table}

\begin{figure}[H]
    \centering
\includegraphics[width=125pt]{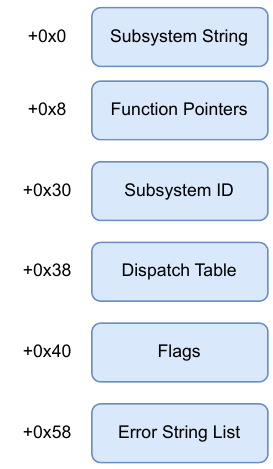}
    \caption[SPTM subsystem descriptor layout.]{Layout of a subsystem descriptor as found in the SPTM binary.}
    \label{subsystemDescriptor}
\end{figure}

These subsystem name assignments can be validated using the approach of Dataflow Forensics~\cite{df-f-2:2025} by examining the \gls{SPTM} binary. We find structures that we shall denote as subsystem descriptors. They are identifiable by the usage of the subsystem name string.
Their structure is shown below in \cref{subsystemDescriptor}. 

The descriptor contains the subsystem name, a number of function pointers, the subsystem ID, a pointer to a list of functions that we assume to be a dispatch table, various options and flags, and a list of error strings.
We can confirm our previously identified subsystems through this, and find a fourth subsystem called \texttt{CPUTRACE}, which we find to be subsystem number 9.
The subsystem descriptors are detectable via their use of the name string.
We will show their actual use in \cref{subs:registering}.\todo{gucken ob das alle haben oder nur iphone 16 / m4 mac}

\paragraph*{txm\_enter -- SPTM Gate}

We find another \texttt{GENTER}-calling function in XNU.
Unlike the main \gls{SPTM} gate, parameterization occurs directly before the call, not in another calling function.
This specific \texttt{GENTER} call is dynamically parameterized.
The upper 32 bits are statically set to \texttt{0x00020000}, while the lower 32 bits are dynamically set based on a call parameter.
We assume this \gls{SPTM} gate to be \texttt{txm\_genter}\todo{Tabelle}. We can infer this based on the only calling function to it, which we, in turn, assume to be \texttt{txm\_kernel\_call}\todo{table}, based on debug string usage in \path{bsd/kern/txm.c} in the XNU open-source code. Based on the \texttt{GENTER} call in the binary and the usage in the aforementioned file, we see the lower 32 bits of the \texttt{x16} register are parameterized via a selector parameter, which we assume indicates a specific function to invoke. We will enumerate these function selectors and further look into them in \cref{txm}.

\paragraph*{sk\_enter -- SPTM Gate}

We find a third \texttt{GENTER} call in XNU.
Comparable to the \texttt{txm\_enter} call, parameterization also occurs right before the \texttt{GENTER} call and is also dynamic based on an argument regarding the lower 32 bits of \texttt{x16}.
The upper bits of \texttt{x16} are statically set to \texttt{0x00030000}. There are two invocations of this \gls{SPTM} gate found in the XNU binary. 
Based on the debug string usage in \texttt{osfmk/kern/exclaves.c} and comparison of control flow, we assume the invoking functions to be \texttt{exclaves\_enter}\todo{table} and \texttt{exclaves\_bootinfo}\todo{table} respectively.
Using this knowledge, we can deduce that this specific \gls{SPTM} gate corresponds to the \texttt{sk\_enter} function.
We can further infer the potential values with which \texttt{sk\_enter} can be called.
The endpoint, which is stored in the lower 32 bits of \texttt{x16}, is either \texttt{RINGGATE\_EP\_ENTER} or \texttt{RINGGATE\_EP\_INFO}, which we can assume to be values 0 and 1, respectively.
In summary, we identify the following \gls{SPTM} calls from this gate:
\begin{description}
    \item[Secure Kernel Call \#0] \texttt{x16} value \texttt{0x3000000000000}, corresponds to\newline \path{exclaves_enter}
    \item[Secure Kernel Call \#1] \texttt{x16} value \texttt{0x3000000000001}, corresponds to\newline \path{exclaves_bootinfo}
\end{description}

\subsection{Calls from the Secure Kernel}
\label{SK_SPTM_CALLS}
As already mentioned, the \texttt{GENTER} call serves as a gate into the \gls{SPTM} via the GXF. As will be shown later, the \acrfull{SK} already runs in guarded mode. As it is even running at a privileged guarded level (GL1), it does not invoke \gls{SPTM} via \glspl{SVC}, but instead via \glspl{HVC}. Nonetheless, we can still find similar \texttt{x16} parameterization.
This section will only briefly list \gls{SPTM} calls from the \gls{SK} without further delving into their workings.
We will look at these in \cref{secureKernel}.

The secure kernel issues \glspl{HVC} from two functions. We denote them as \path{main_sptm_gate}\todo{0xffffff8000001784} and \path{special_sptm_gate}\todo{0xffffff800000178c}. 

\paragraph*{Main Secure Kernel SPTM Gate}

We identify this as the regular \gls{SPTM} caller from the \gls{SK}.
\texttt{x16} is parameterized comparable to the earlier subsystem calls from XNU. More precisely, it is set as follows:
\begin{description}
    \item[Subsystem 2] 5 different parameters, with the following \texttt{x16} values: \texttt{0x200000000} -- \texttt{0x200000004}
    \item[Subsystem 4] 2 different parameters, with the following \texttt{x16} values: \texttt{0x400000000} -- \texttt{0x400000001}
\end{description}

Based on the fact that the \gls{SK} seems to be the only caller using these subsystems, we assume these to be dedicated \gls{SK} subsystems. This is in line with previous reverse engineering of the \gls{SK}~\cite{DisassemblingAppleExclaves:2025}.

\paragraph*{Special Secure Kernel SPTM Gate}

This special gate is surrounded by a function that zeroes general-purpose registers \texttt{x2} to \texttt{x15}, and \texttt{x17} to \texttt{x18}. It also completely zeroes the NEON vectors~\cite{ARM_Vector:2025}. There are three invocations to this:
\begin{itemize}
    \item Call with \texttt{x16 = 0xff00000000}
    \item Call with \texttt{x16 = 0xfe00000000}, in this case \texttt{x1} is also zeroed
    \item Call with \texttt{x16 = 0xfd00000000}, in this case \texttt{x1} is also zeroed
\end{itemize}

Without examining the \gls{SPTM} handling side of these calls, we cannot yet deduce any clear meaning from them. However, based on the aforementioned calling preparation, we can assume this is not normal calling behavior, but probably handles unexpected errors or similar issues.

\subsection{Calls from TXM}
\label{CallsFroMTXM}
\gls{SPTM} \texttt{x16} parameterization is also found in \gls{TXM}, preceding SVC executions. This appears to be in line with \gls{TXM} running in GL0, which we will examine more in-depth in \cref{txm}.
We find a single used \gls{SPTM} gate in \gls{TXM}, which we denote \texttt{core\_SPTM\_gate}\todo{table}. It is called with the following parameters:
\begin{itemize}
    \item \textbf{Subsystem 1:} 4 different parameters, with the following \texttt{x16} values: \newline \texttt{0x10000001} -- \texttt{0x10000004}
    \item Call with \texttt{x16 = 0xff00000000}
    \item Call with \texttt{x16 = 0xfe00000000}
    \item Call with \texttt{x16 = 0xfd00000000}
\end{itemize}

Similar ``special'' calls outside of the standard subsystem numeration are found, as in the previous paragraph on \gls{SK} \gls{SPTM} gates. This supports the theory that these are special error-handling calls. 

\subsection{Calling SPTM -- A Look into SPTM Headers}
\label{ALookIntoSPTMHeaders}

\begin{figure}[H]
    \centering
\includegraphics[scale=0.75]{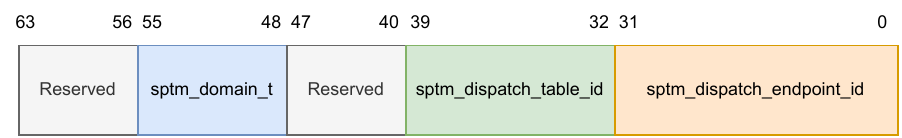}
    \caption[SPTM \texttt{sptm\_disaptch\_target\_t} type.]{\texttt{sptm\_dispatch\_target\_t}, adapted from \path{sptm_common.h}.}
    \label{fig:callingConvention}
\end{figure}

This preliminary analysis provided us with a fairly extensive view of interaction with \gls{SPTM}.
We can further this by incorporating header file releases by Apple, which were published with the \texttt{MacOSX15.5.sdk} release in April 2025~\cite{apple2025sequoia155}.
The \gls{SDK} contains three \gls{SPTM}-specific header files, namely \path{debug_headers.h}, \path{sptm_common.h} and \path{sptm_xnu.h} located at \path{System/Library/Frameworks/Kernel.framework/Versions/A/platform/sptm}.

\path{sptm_common.h} provides us with key additional and ``official'' information on \gls{SPTM} calling.
The following provides insight into the parameterization and generally supports previous results.
In \cref{fig:callingConvention}, we see a type description for the \texttt{sptm\_dispatch\_target\_t} type, which we assume is the data type used in all \gls{SPTM} calls and is stored in register \texttt{x16} for cross-component communication.

Regarding the \texttt{sptm\_domain\_t}, \path{sptm_dispatch_table_id} and \path{sptm_dispatch_endpoint_id} types used in \cref{fig:callingConvention}, we can glean significant information from the relevant header files.

\paragraph*{sptm\_domain\_t}

Based on the \path{stpm_common.h} header, we find  five \gls{SPTM} domains, namely \texttt{SPTM\_DOMAIN}, \texttt{XNU\_DOMAIN}, \texttt{TXM\_DOMAIN}, \texttt{SK\_DOMAIN} and \texttt{XNU\_HIB\_DOMAIN} (see \cref{app:domains}).
We assume \texttt{XNU\_HIB\_DOMAIN} to be the XNU hibernation domain\footnote{XNU hibernation is tasked with ``suspending the entire system state to RAM'', see \path{doc/lifecycle/hibernation.md} in XNU open-source code.}. 

Through the previous reverse engineering, we identified the relevant bits (48-55) as being set in only three specific calls from XNU.
The special cases are invocations of \texttt{exclaves\_enter}, \texttt{exclaves\_bootinfo}, and \texttt{txm\_enter} from XNU (see \cref{lab:sptmCallsFromXNU}).
For the first two, these specific domain bits were set to \texttt{0x3}, corresponds to the \texttt{SK\_DOMAIN}.
This aligns with our assumption that calls to Exclaves are managed by Secure Kernel as an intermediary between Exclaves and SPTM/XNU.

For the \gls{TXM} call, the domain bit was set to \texttt{0x2}, which corresponds to the \texttt{TXM\_DOMAIN}, and is in line with our assumption that this calls into \gls{TXM}.  All other calls detected so far are made with the domain entry set to \texttt{0x0}, indicating that the called domain is \texttt{SPTM\_DOMAIN}, which we assume to invoke general \gls{SPTM} services.

Although there is no public information available on the concept of \gls{SPTM} domains in Apple's operating systems, we can make some assumptions based on the analysis of the header file. It appears that domains scope execution context and separate different parts of the operating system. They are not equivalent to privilege levels, but instead provide a new level of differentiation. We assume that \gls{SPTM} domains are introduced to scope different security-relevant components from each other, thereby limiting access to each other's resources and reducing the risk of full system compromise. We will look into this scoping based on domains in \cref{pagemap}.

\paragraph*{sptm\_dispatch\_table\_id\_t}

What we have previously labeled as subsystems, as indicated by previous research, are in fact IDs of dispatch tables (see \cref{app:dispatchTables}). They are listed in \path{sptm_xnu.h}.  This is in line with what we determined via reverse engineering on the \gls{SPTM} dispatching process and will be further shown in \cref{subs:registering}. 

This does, however, confirm our previous ``subsystem'' mapping, which we now know are dispatch table IDs. As determined, dispatch table ID 3 corresponds to \texttt{T8110dart}, table ID 5 to \texttt{SART}, table ID 6 to \texttt{NVMe} and table ID 7 to \texttt{UAT}.
Interestingly, table ID 9 for the \texttt{CPUTRACE} system cannot be confirmed by this, as ID 9 is found to be \texttt{SPTM\_DISPATCH\_TABLE\_RESERVED}. This might be an indication that this dispatch table is still under development or not supported by all devices.

We further assume that dispatch tables are assigned to specific called domains, with the listing seen in \cref{app:dispatchTables} being \texttt{SPTM\_DOMAIN} specific. This is indicated by the usage of these dispatch tables in calls with the domain parameter set to \texttt{SPTM\_DOMAIN}. We will find this assumption validated in our analysis of dispatch table registration (see \cref{subs:registering}).

\paragraph*{sptm\_dispatch\_endpoint\_id\_t}

The lower 32 bits of the parameter seem to encode the specific function/endpoint call. This aligns with what has been discovered so far. We will later use the now-available mapping for what we assume to be the \texttt{XNU\_BOOTSTRAP} dispatch table (\cref{app:functionIDs}) to infer \gls{SPTM} usage and functionality. We can assume these endpoint IDs to be defined for specific dispatch table IDs in specific called domains.

\subsection{SPTM Dispatch Tables -- Registration}
\label{subs:registering}
\begin{figure}[H]
    \centering
\includegraphics[scale=0.75]{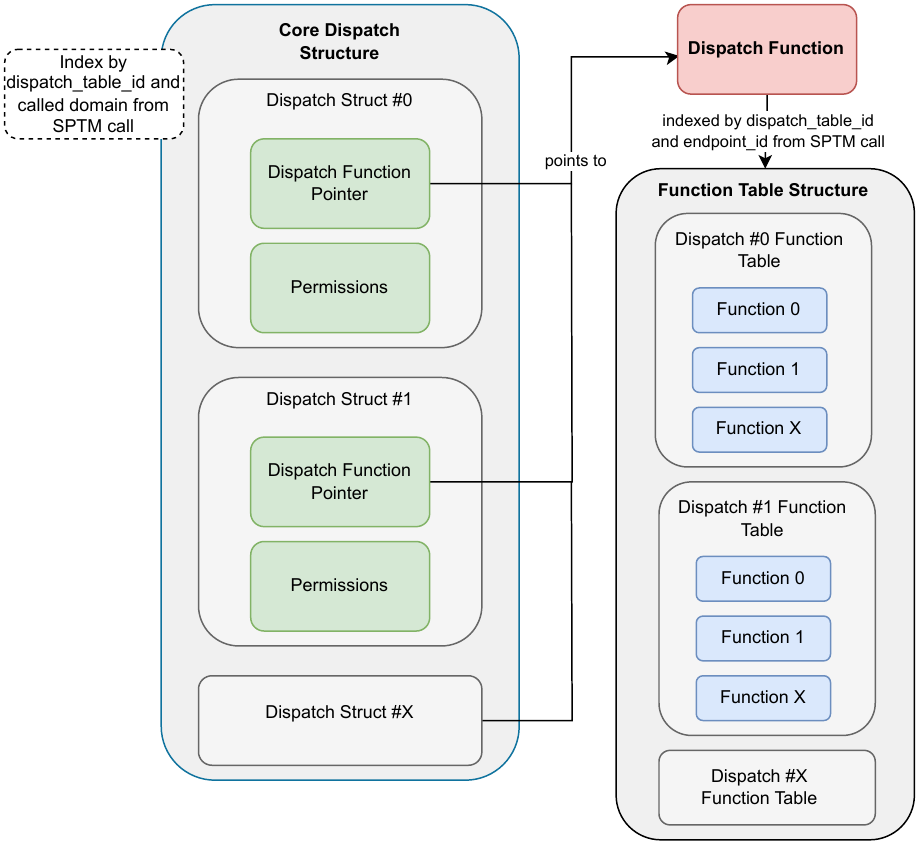}
\caption[SPTM dispatch structure.]{Function dispatching structure found in the SPTM binary. The \emph{Core Dispatch Structure} holds multiple \emph{Dispatch Structs}, and is indexed by the called domain and the dispatch table ID extracted from the SPTM call. Each entry contains a \emph{Dispatch Function Pointer} and \emph{Permissions}, with the permissions being used to validate the call to it, and the dispatch function pointer containing a pointer to a function handling the actual function dispatch. The \emph{Dispatch Function} accesses the corresponding \emph{Dispatch Function Table} based on the SPTM call endpoint value, and indexes the specific function based on the SPTM call dispatch table.}
\label{fig:dispatchStructure}
\end{figure}

\gls{SPTM} is invoked in a variety of different ways. It offers a custom portfolio of available functions for different calling contexts. These different contexts are realized by domains and dispatch tables, both of which are part of the \gls{SPTM} parameterization, as shown in \cref{fig:callingConvention}. \gls{SPTM} supports a variety of different dispatch tables (\cref{app:dispatchTables}). The general dispatch structure is illustrated in \cref{fig:dispatchStructure}.

Significant parts of the setup and registration of the dispatch logic are performed at runtime and are therefore not fully recoverable from a static analysis of the \gls{SPTM} binary; however, key parts for request handling are still detectable.
We find a structure we deem the \emph{Function Table Structure}\todo{tableref}, which holds four pointers to dispatch function tables in the binary. These four table pointers are the \texttt{XNU\_BOOTSTRAP}\todo{table}, \texttt{TXM\_BOOTSTRAP}\todo{table}, \texttt{SK\_BOOTSTRAP}\todo{table} and \texttt{HIB}\todo{table} table respectively. These names were inferred by index matching the table entries with the \gls{SPTM} headers (see \cref{app:dispatchTables}), and from the calling context.

With regards to the \emph{Core Dispatch Structure}, we presume a pointer to it to reside at a specific point in memory, which we call \emph{Core Dispatch Structure Pointer}\todo{table}. It is, however, not initialized in the binary, and we presume it gets set up at runtime. Its location is clear via its usage in \texttt{sptm\_register\_dispatch\_table}\todo{table}. This function name is known from the debug string usage within its body.
We further know the signature of this function, as it is included in \texttt{sptm\_common.h}. The signature is as follows:
\begin{flushleft}
\texttt{void sptm\_register\_dispatch\_table(} \\
\quad \texttt{sptm\_dispatch\_table\_id\_t table\_id,} \\
\quad \texttt{sptm\_vaddr\_t dispatch\_entry,} \\
\quad \texttt{uint64\_t permissions);}
\end{flushleft}
From this signature and the function body, we can verify that the \texttt{dispatch\_entry}, which we have denoted \texttt{dispatch\_function\_pointer}, and the specific access permissions are indeed written to the specific \emph{Dispatch Struct}. During population, the \emph{Core Dispatch Structure} is indexed and populated not only based on the dispatch table ID, but also on what we presume to be the caller domain, stored in the thread-identifying information. This makes sense as calling into different domains should result in different dispatch tables.

\paragraph*{Registering for XNU, TXM, SK, and HIB}

For these four dispatch ``systems'', the \emph{Dispatch Function Table} pointer is statically set up in the \emph{Function Table Structure}. Therefore, they are directly set up with regards to their permission and \emph{Dispatch Function} via calls to \texttt{sptm\_register\_dispatch\_tables} from a function we presume to be \texttt{init\_xnu\_ro\_data}\todo{table} based on debug string usage. All except the \texttt{HIB} table are registered unconditionally, whilst for the \texttt{HIB} table, a flag in memory is checked. This might be based on system support for hibernation, which is not provided for all Apple devices supporting \gls{SPTM} (MacBooks vs. iPhones).
From the invocations, we can infer the permissions set for each table. 
\begin{listing}[H]
    \begin{minted}[linenos, breaklines, bgcolor=LightGray, frame=lines]{c}
sptm_register_dispatch_table(0, sptm_dispatch, 2);
sptm_register_dispatch_table(1, sptm_dispatch, 4);
sptm_register_dispatch_table(2, sptm_dispatch, 8);
if (hib_flag == '\x01') {
    sptm_register_dispatch_table(10, sptm_dispatch, 0x10);
}    
    \end{minted}
    \captionof{lstlisting}[SPTM dispatch table registration calls from the \texttt{init\_xnu\_ro\_data} function.]{SPTM dispatch table registration from \texttt{init\_xnu\_ro\_data}. The source code was disassembled by Ghidra.}
    \label{lst:sptm_dispatch}
\end{listing}
    
We can extract the specific bit set in the permission flag for each call and compare it to the corresponding domain ID (see \cref{Tab:SPTMDomainIDs}). Additionally, we have further mapped the dispatch table ID to the corresponding table name (see \cref{Tab:SPTMDispatchTableIDs}). 
\begin{table}[h!]
\small
\centering
\begin{tabular}{|c|l|c|c|l|}
\hline
\textbf{Table ID} & \textbf{Matching Table Name}              & \textbf{Permission} & \textbf{Permission Bit Set} & \textbf{Matching Domain}    \\
\hline
0  & \texttt{XNU\_BOOTSTRAP}     & 0x2   & 1   & \texttt{XNU\_DOMAIN}     \\
1  & \texttt{TXM\_BOOTSTRAP}     & 0x4   & 2   & \texttt{TXM\_DOMAIN}     \\
2  & \texttt{SK\_BOOTSTRAP}      & 0x8   & 3 & \texttt{SK\_DOMAIN}      \\
10 & \texttt{HIB}               & 0x10 & 4 & \texttt{XNU\_HIB\_DOMAIN} \\
\hline
\end{tabular}
\caption[SPTM dispatch table registration and corresponding permissions from \texttt{init\_xnu\-ro\_data} function.]{SPTM dispatch table registrations and permissions from \texttt{init\_xnu\_ro\_data}.}
\label{tab:sptm_dispatch_tables}
\end{table}

We will later examine the exact usage of these permissions during \gls{SPTM} call invocations, but it appears logical that only corresponding domains should have access to their specific dispatching logic.

\paragraph*{Registering for Input/Output Memory Management Units (IOMMUs)}

\gls{SPTM} not only serves as the access permissions handler for XNU, \gls{TXM}, and \gls{SK}, but also for various different IOMMUs. Therefore, there is a dedicated function for bootstrapping IOMMUs\todo{glossary entry} in \gls{SPTM}, which we have labeled \texttt{IOMMU\_bootstrap}\todo{table}, inferred by panic string usage.
It is called from a large \gls{IOMMU} setup function. This calling function invoked \texttt{IOMMU\_bootstrap} with five different integer parameters. We can map these integers to the corresponding \gls{IOMMU} ID (see  \cref{Tab:IOMMUIdentifiers}). This confirms that the bootstrap function is called for SART, NVME, UAT, and DART\_T8020. It is further called for \gls{IOMMU} ID 7, which we do not find in the respective header file. This might indicate that support for the \gls{IOMMU} has been removed from \gls{SPTM}. 

At a high level, the bootstrap function validates the \gls{IOMMU} and reserves memory frames for it. The reserved memory frames are typed \texttt{SPTM\_IOMMU\_BOOTSTRAP}. The number of frames to reserve seems to be dependent on an \gls{IOMMU}-specific global structure, which \gls{SPTM} accesses based on the \gls{IOMMU} ID.

The actual dispatch table registration for IOMMUs is performed via a call to \texttt{register\_iommu}\todo{table}. This call is parameterized with the \gls{IOMMU} ID, the corresponding dispatch table ID and dispatch table, and permissions that are set based on the \gls{IOMMU} ID. Permissions are set to 0x2 for all but the NVME registration. In that case, the permissions are set to 0x12. A listing of the permissions set can be seen in \cref{tab:IOMMUPerms}.

\begin{table}[h!]
\small
\centering
\renewcommand{\arraystretch}{1.3} % optional for nicer vertical spacing
\begin{tabular}{|c|l|c|c|p{3cm}|}
\hline
\textbf{IOMMU ID} & \textbf{IOMMU name} & \textbf{Permission} & \textbf{Permission Bits Set} & \textbf{Matching Domain} \\
\hline
1  & \texttt{SART}         & 0x2  & 1    & \texttt{XNU\_DOMAIN} \\
\hline
2  & \texttt{NVME}         & 0x12 & 2,3  & \texttt{TXM\_DOMAIN, SK\_DOMAIN} \\
\hline
3  & \texttt{UAT}          & 0x2  & 1    & \texttt{XNU\_DOMAIN} \\
\hline
5  & \texttt{DART\_T8110}  & 0x2  & 1    & \texttt{XNU\_DOMAIN} \\
\hline
7  & \texttt{?}            & 0x2  & 1    & \texttt{XNU\_DOMAIN} \\
\hline
\end{tabular}
\caption[SPTM IOMMU IDs and corresponding permissions.]{IOMMU ID and corresponding permissions set in the dispatch table registration.}
\label{tab:IOMMUPerms}
\end{table}

It appears that XNU is allowed to call into the dispatching tables for all but the NVME \gls{IOMMU}, which seems to be only valid for \gls{TXM} and \gls{SK}. The implications of this are still unclear.

This \texttt{register\_iommu} function invokes the previously looked at \path{sptm_register_dispatch_table} function to place the relevant dispatching information into the \texttt{Core Dispatch Structure}, and additionally sets the \texttt{Function Table Structure} for the specific \gls{IOMMU}. A disassembled and labeled version of the \texttt{IOMMU\_bootstrap} function can be found in~\cref{IOMMU_bootstrap}.

\section{SPTM Request Handling}
\label{SPTMRequestHandling}
As \gls{SPTM} can be invoked from a variety of different contexts, it must employ multiple means of handling incoming requests. Regardless of the request's origin, \gls{SPTM} employs an extensive internal dispatching regime, with which it tracks the execution state and allows future request invocations. We will examine the internal dispatching in \cref{InternalDispatching} before analyzing the actual handling of calls invoked from both \gls{EL} and \gls{GL} components.

\subsection{Internal Dispatching}
\label{InternalDispatching}
\gls{SPTM} clients can call into it in different ways: \texttt{GENTER}, \gls{SVC}, and \gls{HVC}. Most of these calls are, in one way or another, forwarded to a core request handling functionality within SPTM. Before examining the handling of special invocations from other components, we will dissect this core function, which we denote \texttt{CORE\_SPTM\_FUNCTION}\todo{table}.

\begin{figure}[H]
    \centering
\includegraphics[scale=0.75]{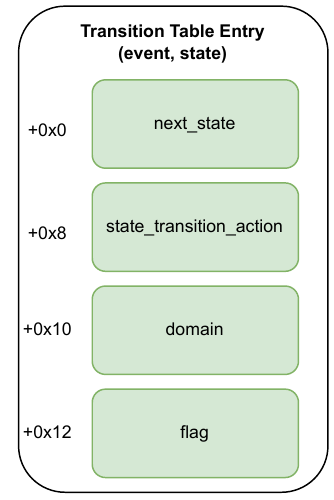}
 \caption[Transition table entry used by \texttt{SPTM\_CORE\_FUNCTION}.]{A transition table entry as retrieved by the \texttt{SPTM\_CORE\_FUNCTION}, indexed by the \texttt{event} parameter and \texttt{state} retrieved from thread identifying information.}
    \label{fig:TransitionTableStruct}
\end{figure}

Based on context, the function is parameterized with an \texttt{event\_type}, and a 64-bit call parameter, which we assume to be of type \texttt{sptm\_dispatch\_target\_t} (see \cref{fig:callingConvention}). The function does some initial verification on the \texttt{event\_id}\footnote{Provides us with the knowledge that the \texttt{event\_id} must be at most 14.} and the state stored in the current thread identifying information (TPIDR register)\footnote{Provides us with the knowledge that the state must be at most 22.}. If these checks are passed, an entry from a table is retrieved from an in-memory transition table. This retrieval is based on the \texttt{event\_id} and current state. As a first layer of implemented security, the \texttt{next\_state} field is checked from the transition table entry and verified to be within bounds. In a similar fashion, a so-called \texttt{state\_transition\_action} (name inferred from panic string usage) is retrieved and verified to be set in the table entry. Any failure on these conditions indicates a forbidden transition attempt, leading to an immediate panic. Furthermore, a flag is retrieved from the entry, and a fourth field denoting an \gls{SPTM} domain is written to the TPIDR caller domain field. An assumed transition table entry struct based on the fields accessed so far is shown in \cref{fig:TransitionTableStruct}. 

An overview of the values retrieved from the transition structure is presented in \cref{app:stateTransitions}. If the low bit is not set in the aforementioned flag, the \path{dispatch_entry_point} is set to be empty. This might correspond to transitions that require no further dispatching. Otherwise, it is calculated based on the \path{sptm_dispatch_target_t} parameter. The calculation is shown below.
\begin{figure}[H]
    \centering
\includegraphics[scale=0.75]{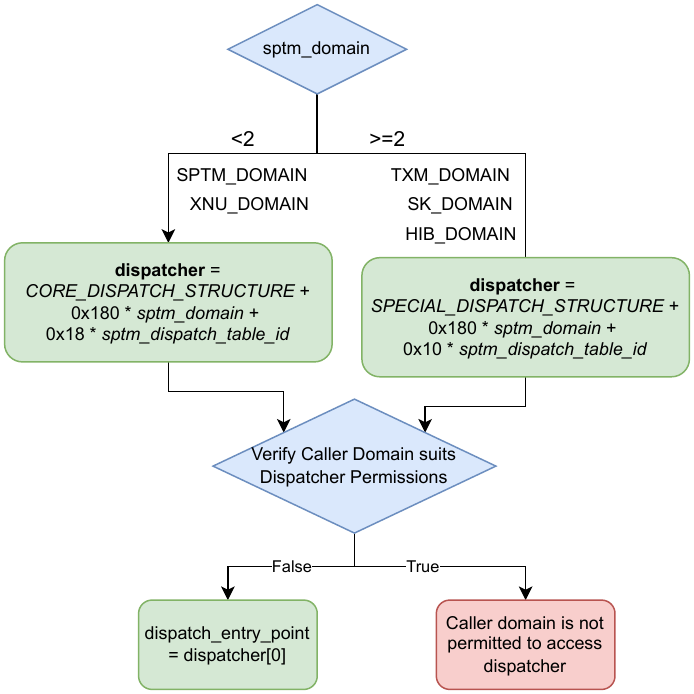}
\caption[\texttt{dispatch\_entry\_point} calculation in \texttt{SPTM\_CORE\_FUNCTION}.]{Calculation of the \texttt{dispatch\_entry\_point in} in \texttt{SPTM\_CORE\_FUNCTION}. Depending on the \texttt{sptm\_domain\_t} value from the input \texttt{sptm\_dispatch\_target\_t} value, a dispatcher is retrieved from memory. }
\label{fig:calculationDispatch}
\end{figure}

Based on the domain specified in the parameter, a different dispatcher is retrieved. If the domain is \texttt{XNU\_DOMAIN} or \texttt{SPTM\_DOMAIN}, we retrieve it from the \texttt{CORE\_DISPATCHER\_STRUCTURE}, for which we dissected the population in \cref{subs:registering}. Otherwise, we retrieve it from a specific memory address we have denoted \path{SPECIAL_DISPATCH_STRUCTURE}. Our analysis of \gls{TXM} and \gls{SK} will show that they register their respective dispatch tables at that location.

Seeing this differentiation, we can look for specific \gls{SPTM} calls in \cref{lab:sptmCallsFromXNU} as a short interjection. Interestingly, we observe that for nearly all calls, the \texttt{sptm\_domain} field is not set, hence corresponding to the \texttt{SPTM\_DOMAIN} value. We find two exceptions in the \texttt{txm\_enter} \gls{SPTM} gate and the \texttt{sk\_enter} \gls{SPTM} gate, which are performed with the respective domain set in the request. Logically, we can assume that for these calls, \gls{SPTM} acts only as an intermediary, and ``forwards'' them to the actual target.

As can be seen in \cref{fig:calculationDispatch}, the function further validates that the permissions set for the dispatcher retrieved are in line with the caller domain that has been set based on the domain flag in the transition structure. This ensures that only intended caller domains can perform specific state transactions.

After a variety of further conditional checks, the actual state transition action is performed, parameterized with the \texttt{sptm\_dispatch\_target\_t} parameter, and the dispatch entry point calculated. With this, the actual request handling is performed, which will be looked at in \cref{GENTERHandling} and \cref{SPTM:svc_handling}.

\paragraph*{Internal Dispatch State Machine}
For state/event combinations, we find some, but not nearly all of the transition actions implemented. From this, we can deduce that there is a relatively rigid internal state dispatching logic that has to be adhered to.

\begin{figure}[H]
    \centering
\includegraphics[width=0.9\textwidth]{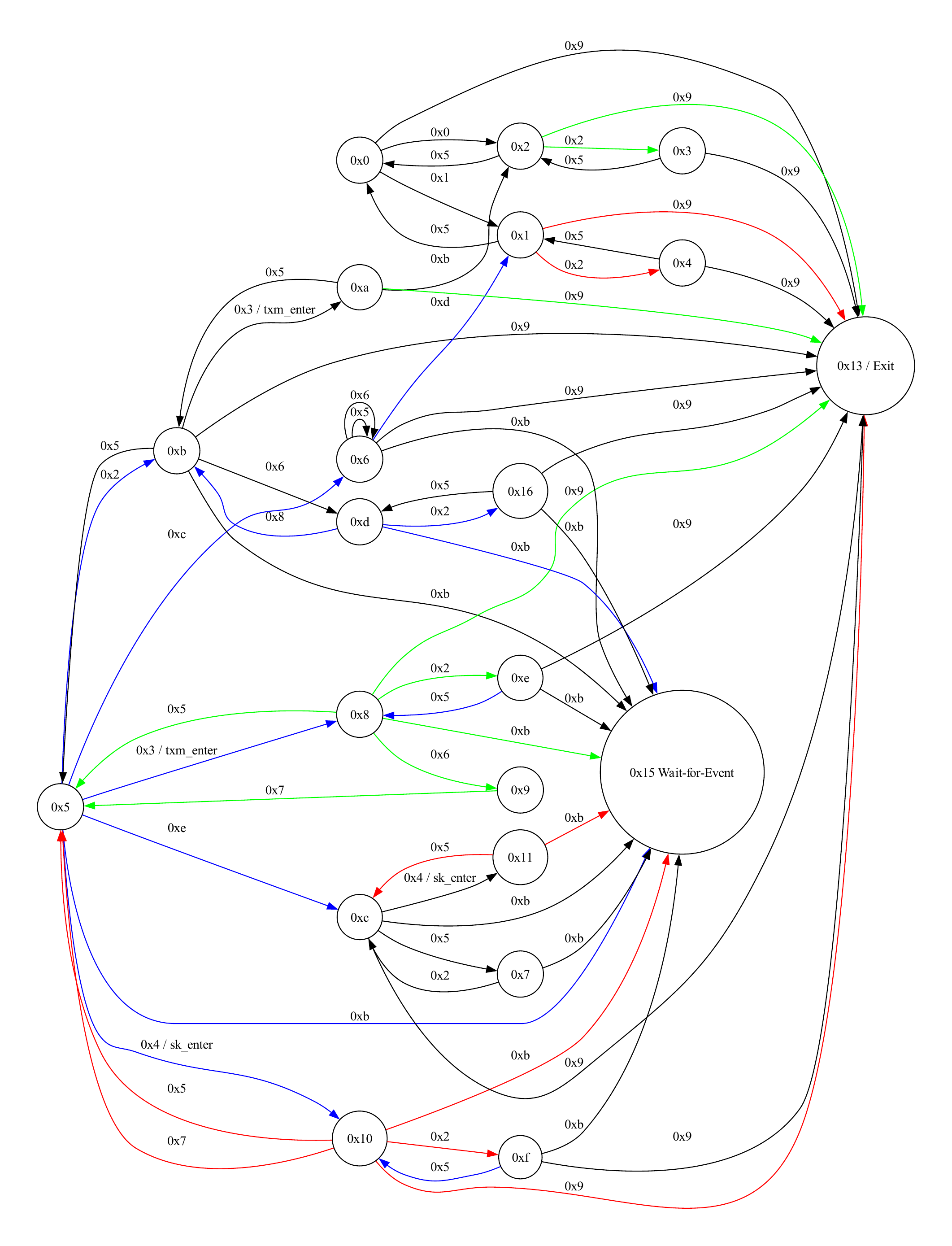}
 \caption[SPTM-internal state graph.]{SPTM-internal state graph. Vertices show states, and edges show events. The graph is colored based on the domain flag set in the transition structure entry and is shown in \cref{app:stateTransitions}. \textcolor{red}{\texttt{SK\_DOMAIN}} is colored red, \texttt{\textbf{SPTM\_DOMAIN}} is colored black, \textcolor{green}{\texttt{TXM\_DOMAIN}} is colored green, and \textcolor{blue}{\texttt{XNU\_DOMAIN}} is colored blue. States and events have partially been labeled based on our reverse engineering.}
    \label{fig:SPTMTransitionGraph}
\end{figure}

The complexity of the state transition actions implemented is highly diverse. Some are fairly simple, involving only system register writes, while others require dispatching and appear to trigger additional state transitions after implementing functionality. An enumeration of valid transitions can be found in \cref{App:StateTransitionTable}. The table lists the transition action label, the next state entry, the domain entry (which gets written to the TPIDR caller domain field), the flag entry, and the corresponding domain name for the respective domain ID for valid state event combinations. Valid combinations are those that have a transition action set.

\cref{fig:SPTMTransitionGraph} presents a visual representation of valid transitions from \cref{App:StateTransitionTable}. Vertices are state IDs, and edges are event IDs. We can partially infer the internal \gls{SPTM} states from this graph. It is directly observable that both state \texttt{0x13} and \texttt{0x15} act as sinks in this state machine, being transitioned into, but with no valid outgoing transitions.

Towards state \texttt{0x13} we find transitions through event \texttt{0x9} from the following states: \texttt{0x0}, \texttt{0x1}, \texttt{0x2}, \texttt{0x3}, \texttt{0x4}, \texttt{0x6}, \texttt{0x8}, \texttt{0xa}, \texttt{0xb}, \texttt{0xc}, \texttt{0xe}, \texttt{0xf}, \texttt{0x10}, \texttt{0x16}. For states \texttt{0x0}, \texttt{0x2}, \texttt{0x3}, \texttt{0x4} we find a call to a low-level logger function \path{low_level_logger}\todo{table ref}. This function appears to write a string to a hardware buffer. The string written is \emph{[PANIC DURING BOOTSTRAP]}. The same happens for state \texttt{0x1}, but for the string \emph{[SK BOOTSTRAP PANIC]}. After these writes, both transitions switch into a \emph{Wait-for-Event}-loop. From all other states, event \texttt{0x9} leads towards a \texttt{GEXIT} instruction, with system state manipulation performed before. To summarize, state \texttt{0x13} appears to be an internal state transition that occurs when exiting SPTM. These exits are performed via event ID \texttt{0x9}. Exits can either be a direct result of a panic or due to other unknown reasons.

With regards to state \texttt{0x15}, we find transitions towards it from the following states: \texttt{0x5}, \texttt{0x6}, \texttt{0x7}, \texttt{0x8}, \texttt{0xb}, \texttt{0xc}, \texttt{0xd}, \texttt{0xe}, \texttt{0xf}, \texttt{0x10}, \texttt{0x11}, \texttt{0x16}. All these transitions invoke the same function, which zeroes a currently unknown system register with encoding (\texttt{0x0}, \texttt{0x3}, \texttt{0x3}, \texttt{0xb}, \texttt{0x4}), and then enters a \emph{Wait-for-Event}-loop. This might indicate that state 0x13 is a ``successfully finished'' state for \gls{SPTM} internal tasks that do not require returning to a caller.

We have further labeled edges we determined to be exclusively used for entering \gls{SK} and \gls{TXM} accordingly. Apart from that, we have not performed a deeper analysis of this internal state machine, and leave this as future work.

%%%%%%%%%%%%% TODO bis hier zweite runde nochmal gelesen

\subsection{GENTER}
\label{GENTERHandling}
We have previously determined that the \gls{SPTM} \texttt{GENTER} handler is set via a register write to the \texttt{GXF\_ENTRY\_EL1} register, which points to the handler. The address points to a function we denoted \texttt{GXF\_entry\_point}. The function performs large context-setting operations and conditionally branches based on the \texttt{ESR\_GL1}~\cite{ESR_EL1:2025}. The register stores information on exceptions taken to GL1. The exact working of this handling is unclear at present, as we do not expect the function to be called in a normal exception handling path, but rather only on \gls{GXF} entry via \texttt{GENTER}. The key function call here is, however, a branch to \texttt{genter\_dispatch\_entry}, parameterized with the received \texttt{x16} \texttt{sptm\_dispatch\_target\_t} value. The full function can be found in \cref{app:genterDispatchEntry}. 

\begin{figure}[H]
    \centering
\includegraphics[scale=0.75]{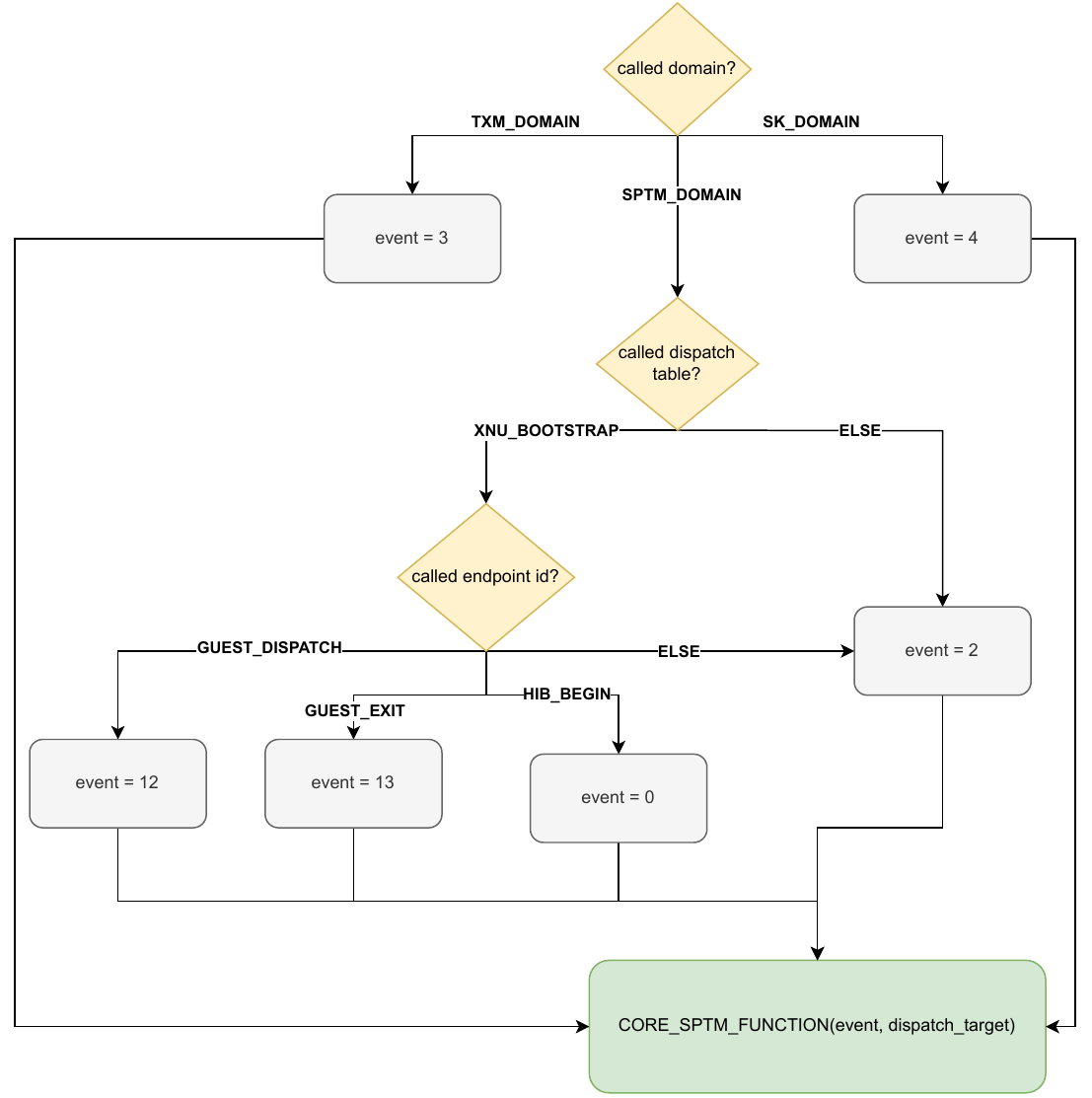}
\caption[\texttt{genter\_dispatch\_entry} control flow for internal event calculation.]{Control flow of \texttt{genter\_dispatch\_entry} for retrieving the internal event code based on \texttt{GENTER} input parameter \texttt{dispatch\_target}.}
\label{fig:genter_entry-_graphics}
\end{figure}

\cref{fig:genter_entry-_graphics} shows a schematic  of the \texttt{genter\_dispatch\_entry} control flow. We find event 3 to be the internal event for calling into \gls{TXM}, and event 4 to be the internal event for calling into the Secure Kernel, respectively. For calls into the \gls{SPTM} domain, event 2 appears to be the default handling event, with special event codes being denoted to specific endpoint calls in the \texttt{XNU\_BOOTSTRAP\_TABLE}. The actual dispatch handling is performed via a call to \texttt{CORE\_SPTM\_FUNCTION}, which we described in \cref{InternalDispatching}.

\subsection{SVC/HVC Handling}
\label{SPTM:svc_handling}
We have previously looked at the request handling for \gls{SPTM} requests from the \gls{EL} client XNU. These were found to be performed via \texttt{GENTER} to switch context into \gls{GXF}. For \gls{SPTM} clients already running in \gls{GXF}, these calls are performed via \gls{SVC} or \gls{HVC} instructions.

\subsubsection{GL0 SVC Rerouting to SPTM}
We have found GL clients of \gls{SPTM} to call into it with \gls{SVC} or \gls{HVC} instructions, depending on their respective guarded level. Whilst it is standard for Secure Kernel to call into \gls{SPTM} via \glspl{HVC} from GL1, SVC instructions executed at GL0 would usually be routed to GL1, rather than to \gls{SPTM} in GL2. We find that such \glspl{SVC} are conditionally trapped towards \gls{SPTM} in GL2 by the usage of the \gls{HCREL2}. The register provides configuration for virtualization, and appears to also affect \gls{SPTM} in GL2. The relevant register field is the \glspl{TGE} bit at position 27. If set, in standard Arm exceptions routed to EL1 are rerouted to EL2~\cite{HCR_EL2:2025}. We can assume that for components running in GLs, this performs exception rerouting to GL2.

We have not reverse-engineered the conditional setting of this flag as of now. The exact inner working of the request handling logic is still unknown, considering we will show in \cref{secureKernel} that GL0 components actually directly call into Secure Kernel in GL1. Based on the clear \gls{SPTM} parameterization of \gls{SPTM} calls, we assume these calls are routed to SPTM. The exact handling mechanisms for allowing GL0 components to call into Secure Kernel at GL1 via SVCs and \gls{TXM} calling into \gls{SPTM} at GL2 via SVCs at the same time have yet to be discovered.

\subsubsection{Request Handling}
GL clients of \gls{SPTM} are found to call into it via \gls{SVC} and \gls{HVC} instructions. Based on the initial analysis by Dataflow Forensics~\cite{df-f-1:2023} and the aforementioned \gls{GXF} setup functions, it is possible to deduce information on the SVC/HVC handling by SPTM.

The function \texttt{gxf\_setup\_late}\todo{tableReference} includes a system register write to \texttt{VBAR\_GL1}. As reasonably assumed by Dataflow Forensics, this appears to work in a similar fashion to the Arm standard register \texttt{VBAR\_EL1}, and therefore contains the base address for various exception handlers~\cite{VBAR_EL1:2025}. More specific information on this can be found in the Arm documentation on exception handling~\cite{TakingAnException:2025}. We expect a synchronous exception handler for exceptions taken from lower levels at \texttt{VBAR\_GL1} + \texttt{0x400}. At the specified address, we find the branch to the actual handler we denoted \texttt{synchronous\_exception\_handler\_from\_lower}.

The handler implements complex functionality regarding exception call handling. Request handling is differentiated based on the aforementioned hypervisor configuration register. The full listing of the handler can be found in \cref{app_syncrhonous_from_lower}. We find the handler to implement functionality for both SVC and HVC handling.

\subsubsection{SVC Exception Handling}
\begin{figure}[H]
    \centering
\includegraphics[scale=0.85]{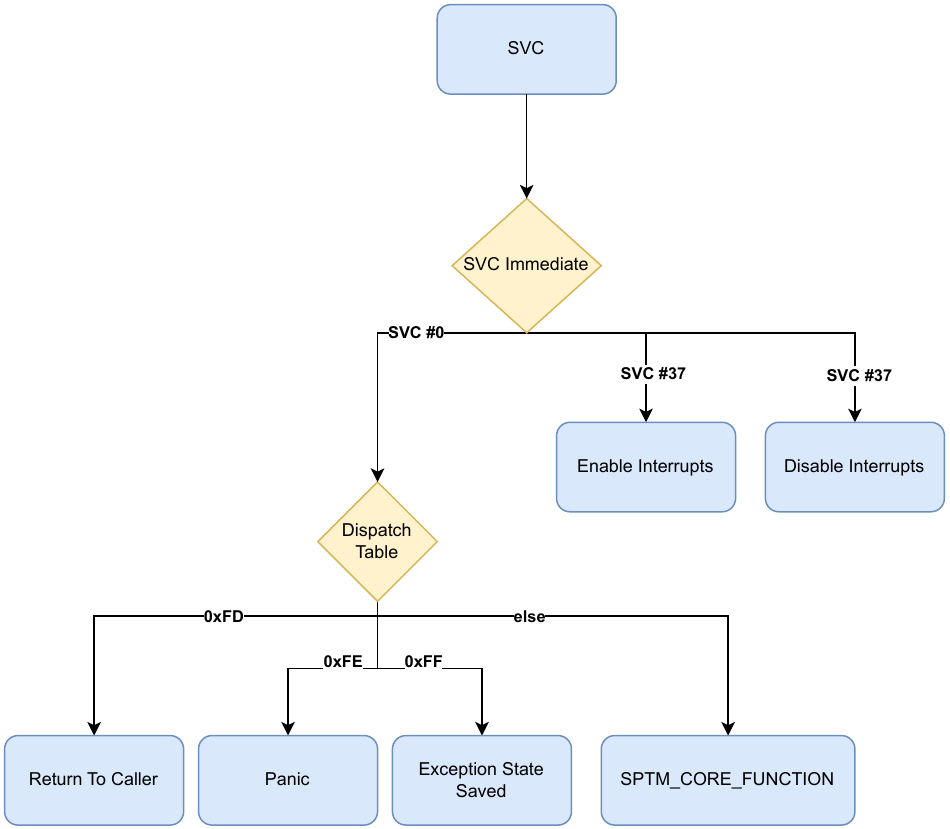}
\caption[SPTM SVC exception handling control flow.]{Conceptual SVC exception handling control flow of SPTM.}
\label{SVC0_PDF}
\end{figure}

A conceptual overview of \gls{SPTM} SVC handling can be seen in \cref{SVC0_PDF}. Furthermore, we show an excerpt on SVC handling in \texttt{synchronous\_exception\_from\_lower} in \cref{lst:sptm_SVC-handler}.

\paragraph*{SVC \#37 \& SVC \#38 -- Interrupts}

The workings of these two \glspl{SVC} have already been thoroughly reverse-engineered by Dataflow Forensics~\cite{df-f-1:2023}, and our analysis of a newer firmware version verifies their results; therefore, a comparable approach is employed in the current SPTM. \cref{lst:svc37_handler} shows the handler for an SVC \#37 call.
\begin{listing}[H]
\begin{minted}[linenos, breaklines, bgcolor=LightGray, frame=lines]{asm}
                   SVC_37_HANDLER                
fffffff027081fb4 68 fa 3e d5     mrs        x8,SPSR_GL1
fffffff027081fb8 09 38 80 d2     mov        x9,#0x1c0
fffffff027081fbc 08 01 29 8a     bic        x8,x8,x9
fffffff027081fc0 68 fa 1e d5     msr        SPSR_GL1,x8
fffffff027081fc4 e0 03 9f d6     eret
fffffff027081fc8 a0 d5 9b d2     mov        x0,#0xdead
\end{minted}
\captionof{lstlisting}[SPTM SVC \#37 handler.]{SVC \#37 handler in SPTM}
\label{lst:svc37_handler}
\end{listing}
The system register operated on is the \gls{SPSR} (GL1). The operation performed on it sets the A, I, and F bits of the register to zero. These bits, if set, mask SError exceptions, IRQ exceptions, and FIQ exceptions~\cite{SPSR:2025}. We therefore confirm that SVC \#37 enables interrupts. We assume this to be called when exiting critical code execution, which is usually protected from disruptions by disabling interrupts.
Analogue to this, SVC \# 38 disables all interrupts by setting the A, I, and F bits of the \gls{SPSR} (GL1). The respective handler can be seen in \cref{lst:svc38_handler}.

\begin{listing}[H]
\begin{minted}[linenos, breaklines, bgcolor=LightGray, frame=lines]{asm}
                   SVC_38_HANDLER  
fffffff027081fd4 68 fa 3e d5     mrs        x8,SPSR_GL1
fffffff027081fd8 08 09 7a b2     orr        x8,x8,#0x1c0
fffffff027081fdc 68 fa 1e d5     msr        SPSR_GL1,x8
fffffff027081fe0 e0 03 9f d6     eret
fffffff027081fe4 a0 d5 9b d2     mov        x0,#0xdead    
\end{minted}
    \captionof{lstlisting}[SPTM SVC \#38 handler.]{SVC \#38 handler in SPTM}
    \label{lst:svc38_handler}
\end{listing}

Both of these handlers end the exception handling via the \texttt{eret} instruction, returning to the caller of the SVC. Therefore, we return to a lower-privileged process here.
We find such calls solely performed by \gls{TXM}.

\paragraph*{SVC \#0 -- Dispatch Targets}

SVC \#0 is used as the main call to invoke \gls{SPTM} functionality from its GL0 clients. The handling differentiates based on the provided dispatch table ID in the \texttt{x16} \texttt{sptm\_dispatch\_target\_t} parameter. We know from \cref{Tab:SPTMDispatchTableIDs} that \texttt{0xFF}, \texttt{0xFE} and \texttt{0xFD} denote special dispatch table IDs that invoke control functions. For any other dispatch table ID, we find the call forwarded to \texttt{CORE\_SPTM\_HANDLER} after performing context preparation. 

\subsubsection*{HVC Exception Handling}
Comparable to SVC exception handling, HVC \#0 calls are handled depending on the set dispatch table ID. The same control functions as for SVC calls can be invoked via dispatch table ID values \texttt{0xFF}, \texttt{0xFE}, and \texttt{0xFD}. For any other value, the handler dispatches to \texttt{CORE\_SPTM\_FUNCTION} via another function call. There is no check for other immediate HVC values, so we assume GL1 clients of SPTM, namely Secure Kernel, do not either rely on interrupt disabling via \gls{SPTM} or direct this towards lower-level components. 

\section{SPTM Functions}
Whilst \gls{SPTM} performs a variety of different tasks for which it can be called into by its clients, frame retyping and page mapping appear to be the most central ones. Frame retyping enables \gls{SPTM} clients to modify the underlying \gls{SPTM} frame type of memory frames according to a predefined rule set. Regarding page mapping, \gls{SPTM} serves as the sole authority and component privileged to modify page table data. Memory mappings can be requested by its clients, with \gls{SPTM} performing them based on a rule set built upon the aforementioned \gls{SPTM} frame types. We will examine these functionalities in more detail in the following sections. 

\section{SPTM Frame Retyping -- Overview}
SPTM introduces so-called frame types. A conclusive list of all available \gls{SPTM} frame types can be seen in \cref{App:SPTMTypesNew}. Types are separated into SPTM-owned, XNU-owned, \gls{TXM}-owned, and \gls{SK}-owned memory types. As we will demonstrate in \cref{pagemap}, memory frame types are a crucial building block for memory mapping and system security. \gls{SPTM} performs frame retyping via the \texttt{retype} function, which we find registered in the bootstrap table for XNU, \gls{TXM}, and Secure Kernel, respectively. They all, therefore, appear to request frame retyping.

We find \gls{SPTM} frame types to realize memory frame ownership, with owners of a page being one of the known \gls{SPTM} domains. Ownership of memory frames via the \gls{SPTM} type is required to map memory into them via \texttt{sptm\_map\_page}, which is the sole function usable for page table alterations.

\subsection{SPTM Frame Retyping -- Calling from XNU}
We find XNU invocations to \gls{SPTM} via \texttt{sptm\_retype}, and can infer on the function signature based on the XNU open-source code. The signature as extracted from XNU can be seen below:

\begin{center}
\begin{lstlisting}[language=C]
void sptm_retype(pmap_paddr_t physical_address, sptm_frame_type_t previous_type, sptm_frame_type_t new_type, sptm_retype_params_t retype_params)
\end{lstlisting}
\end{center}

The function is parameterized with the physical address of the frame to retype, the current type, the new type, and retyping parameters. The fact that the current page type is provided suggests a list of allowed frame retyping transitions, which we will confirm through our analysis. As of now, we are uncertain why the current frame type must be provided by the caller. As \gls{SPTM} is tracking frame types, it is aware of the specified frame's type.

\subsection{SPTM Frame Retyping -- Fundamentals}
We find \gls{SPTM} to perform retyping on memory frames. Recalling \cref{memoryAccess}, a memory frame refers to a chunk of physical address space. It appears that \gls{SPTM} is tasked with tracking physical memory in a frame table. A frame table typically refers to a table containing a list of all physical memory frames, along with additional state information on them~\cite{wienand_bottomupcs_ch6s04}. 
Entries in SPTM's frame table are referred to as \glspl{FTE}. We find the underlying frame type of a frame stored in the relevant \gls{FTE}.

\section{SPTM Frame Retyping -- In-Depth}
Retyping of memory frames is the key task of SPTM. On the \gls{SPTM} side, it is performed via the \texttt{retype}\todo{table} function. We know it to be invoked from XNU with \texttt{x16} parameter value 0x1 by the \gls{SPTM} function ID mapping (see \cref{app:functionIDs}). As already determined, \texttt{sptm\_retype} is called on a physical address, with a current type, a new type, and ``sptm retype\_params'' (e.g. \path{osfmk/sptm/pmap.c}, line 6328-6341). This is also confirmed in the \texttt{sptm\_common.h} header. We further know \texttt{retype\_params} to store ``Type-specific information set at retype time'' (see \texttt{sptm\_common.h} header). Based on invocations observed in the XNU open-source code, these parameters are frequently left empty, indicating that they are not required for most type transitions.

\subsection{Handling Retyping Requests}
While we can infer on the inner workings of \gls{SPTM} \texttt{retype}, it is a highly complex function for which we have not fully reverse-engineered the entire control flow. Instead of a complete understanding of the underlying function, we are more interested in implemented security mechanisms, and we will put our focus on those. 

\subsubsection{Early Validation}
\begin{figure}[H]
    \centering
\includegraphics[scale=0.9]{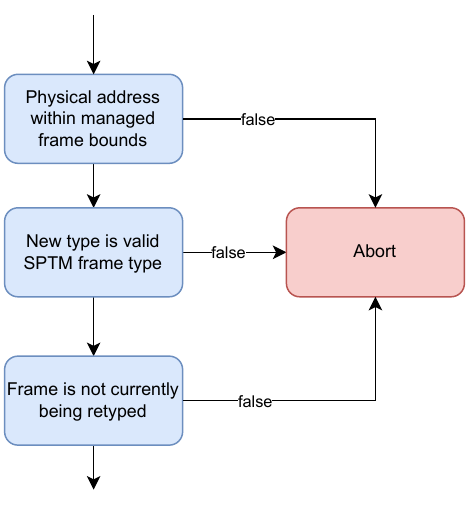}
    \caption[SPTM \texttt{retype} early validation.]{Early validation in \texttt{retype}.}
    \label{fig:RetypeEarlyValidation}
\end{figure}

On call, the function performs early validation, confirming the relevant preconditions for executing the function. It initially validates that the provided physical address is actually a managed frame. This is done by checking if it is within the globally stored relevant bounds. We can infer these bounds from error handling within another function\footnote{Error handling in \texttt{FUN\_fffffff0270b4118}.}. The next conditional check verifies that the new type is at most 62, which is in line with the \gls{SPTM} types found in the \gls{SPTM} headers (see \cref{App:SPTMTypesNew}). What follows is what we assume to be a \gls{FTE} retrieval, with the \gls{FTE} holding various flags and information on the frame it describes. An ``in use'' flag of this descriptor is checked and set, whilst panicking if already set\footnote{The meaning of this flag was inferable from later usage and debug strings, especially the ``Attempted to update PAPT permissions outside of a retype() operation, this is unsafe.'' string after flag usage at \texttt{0xfffffff0270b2aa8}.}. The early input validation is illustrated in \cref{fig:RetypeEarlyValidation}.

\subsubsection{Type-Specific Call Validity Checks}
\label{callerValidityCheck}
After initial verification passes, more type-specific call validity checks are performed. The \texttt{retype} function validates the caller domain with respect to the frame's current type. This check is only executed if the current frame type is not \texttt{SPTM\_UNTYPED}, which indicates that retypes to such frames are not limited based on the caller domain. For this check, a single byte is indexed from a global structure we denoted \texttt{AllowedCallerDomains}\todo{table}, based on the frame type. A full listing of the allowed caller domains for specific frame types can be found in \cref{app:AllowedCallerDomain}. This is largely in line with expectations, with each domain being able to retype memory frames typed to one of its respective \gls{SPTM} types. The exception to this seems to be the \texttt{XNU\_TAG\_STORAGE} type, for which the allowed domain is the \gls{SPTM} domain.

\begin{figure}[H]
    \centering
\includegraphics[width=175pt]{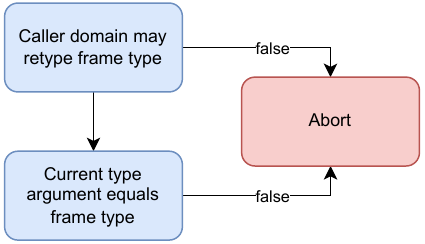}
    \caption[SPTM \texttt{retype} caller validity check.]{Caller validity check in \texttt{retype}.}
    \label{fig:CallerValidityCheck}
\end{figure}

A further key check here is comparing the function parameter \texttt{previous\_type} with the actual type in the \gls{FTE}. If they do not match, the retype fails. This mitigation prevents \gls{SPTM} retyping of frames in unexpected conditions, and also protects against tampering with the previous conditional check by faking the original frame type. The caller validity check is illustrated in \cref{fig:CallerValidityCheck}.

\subsubsection{Retyping Validity Checks}
After early validation and caller validity checks have been passed successfully, the first actual retyping validity check is performed. To achieve this, an entry is retrieved from a global table, which we denote as \texttt{SPTM\_Retype\_Global\_Table}, indexed based on the current type of the frame. The retrieved value is shifted by the new type argument, and it is checked if, after this shift, the low bit is set, indicating an allowed retype operation. We find this to be classic validity checking logic, where a bit set at a specific position in the table entry value indicates a valid retype to that type. The full list of allowed frame retype operations can be seen in \cref{App:allowedTransition}. The conditional check can be seen in \cref{lst:retypeAllowedCheck}.
\begin{listing}[H]
\begin{minted}[linenos, breaklines, bgcolor=LightGray, frame=lines]{c}
if ((*(ulong *)(&SPTM_Retype_Global_Table + CurrentFrameTypeOffset) >> (new_type & 0x3f) & 1) == 0
 ) {
    DEBUG();
} 
\end{minted}
\captionof{lstlisting}[SPTM retyping validity check with regards to allowed new frame types.]{Retyping validity check based on the current frame type, retrieving and checking all allowed types to retype to. The source code was disassembled by Ghidra.}
\label{lst:retypeAllowedCheck}
\end{listing}

As expected, \texttt{SPTM\_UNTYPED} is allowed to be retyped to any frame type. As we have previously seen in \cref{callerValidityCheck}, for frames of type \texttt{SPTM\_UNTYPED}, the caller domain is not considered, so we assume this to be the default frame type for which components running in other \gls{SPTM} domains can retype memory to their domain-specific types. 
\todo{Interpretation}

After this transition validity check, a type-specific \texttt{type\_out} function is retrieved from a structure we denote \texttt{RETYPE\_Typeout\_Structure}. The structure is indexed based on the current frame type. We find the existence of these \texttt{type\_out} functions supported by strings in the binary, where we see references to these functions for specific types.\footnote{An example for this is the \texttt{sk\_types\_retype\_out} string. We shall denote the corresponding function as \texttt{sk\_types\_retype\_out}\todo{label}. Looking at references to the function, we find them at \texttt{RETYPE\_Typeout\_Structure} offset by multiples of types 59/60/61, respectively. Verifying with \cref{App:SPTMTypesNew}, we find these values to in fact correspond to secure kernel types. Interestingly, the \texttt{SK\_IO} type does not have this function called as retype out.}Notably, for most entries this function is not defined. For frame types with typeout functions implemented, they are listed in  \cref{tab:type-function-table}. The \texttt{type\_out} functions generally implement further type-specific validity checking logic, which we find partially duplicates the retype validity checking performed in the retype function. Following this, a further entry is retrieved from a structure we denote \texttt{RETYPE\_Flag\_\-Structure}, indexed based on the current type. These type-specific flags are found in  \cref{app:AllowedCallerDomain}. The FTE is altered based on the flag. The exact workings of this are still unclear, but we assume it will alter frame parameters if necessary for the new frame type.

\begin{figure}[H]
    \centering
\includegraphics[scale=0.75]{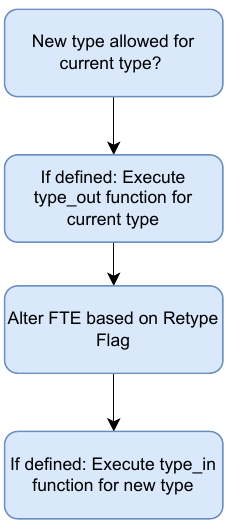}
\caption[SPTM \texttt{retype} advanced control flow.]{Control flow in SPTM \texttt{retype} after initial and caller validation passed.}
\label{controlFlowRetype}
\end{figure}

Comparable to the \texttt{retype\_out} operation performed before, a \texttt{retype\_in} operation is then performed based on the new type. If defined, the function is parameterized with the FTE, the new type, the retyping parameters, and an output pointer. The listing of defined functions can also be found in  \cref{tab:type-function-table}. The control flow to this point can be seen \cref{controlFlowRetype}.

\subsubsection{PAPT SPRR Updating}
Finally, we find a call to a method we denoted \texttt{PAPT\_permission\_update}\todo{table}. The name is inferred from assumed functionality and debug strings. Our understanding of this function call is limited. Based on our speculation, we assume it maps physical addresses back into virtual memory using the \gls{PAPT} and updates its SPRR index based on input parameters. The SPRR index to update is retrieved from the same global retyping structure we accessed before. We have denoted its base address as \texttt{SPRR\_Index\_From\_Type}.
The full listing of \gls{SPTM} types to \gls{SPRR} index mappings can be found in \cref{SPTM_to_SPRR}.

\subsubsection{Finalizing the Call}
The \texttt{retype} call is finalized by accessing and updating an in-memory \gls{SPTM} data structure, which we assume to be responsible for tracking and enforcing type and permission metadata for memory frames. As \gls{SPTM} appears to be the sole authority regarding memory frame management, it is logical to track the changes performed. There is currently no information available regarding these internal structures, and we have not yet reverse-engineered them. 

\subsection{SPTM Frame Retyping -- Security Implications}
Frame retyping via \gls{SPTM} employs a strict ruleset and a variety of conditional checks to ensure that only a specific, pre-allowed set of retypings can occur. Generally speaking, frame retyping is scoped to occur within the same \gls{SPTM} domain, with a few exceptions. The underlying security implications become clear when examining the second core \gls{SPTM} functionality, specifically page mapping.

\section{SPTM Page Mapping -- Overview}
\label{pagemap}
We have previously discovered that one of SPTM's capabilities is the mapping of memory pages. This was indicated by the existence of an endpoint ID \texttt{MAP\_PAGE} for the XNU bootstrap table (see \cref{table_typedef_tbendpoint}). We find multiple calls to \texttt{sptm\_map\_page} in \path{osfmk/arm64/sptm/pmap.c} in the XNU open-source code, which helps us infer information about the calling parameters. The function signature is shown below:
\begin{center}
\begin{lstlisting}[language=C]
sptm_return_t sptm_map_page(pmap_paddr_t ttep, vm_address_t va, pt_entry_t new_pte)
\end{lstlisting}
\end{center}

The function is usually called with the \texttt{ttep} parameter set to the \texttt{ttep} field of a \gls{PMAP}. A physical map is a per-process structure holding information on address translation. The \texttt{ttep} field of a pmap is the ``physical page of the root translation table'' (see \texttt{osfmk/arm64/stpm/pmap.h}), that is, the physical address of the initial translation table to consult when trying to translate a virtual memory address to a physical address. The second function parameter is a virtual address, and the final parameter of type \texttt{pt\_entry\_t} is a \gls{PTE}.

\subsection{Core Call to \texttt{sptm\_map\_page}} 

Whilst \texttt{sptm\_map\_page} is invoked at a variety of locations, we find the core one in the XNU open-source code from \texttt{pmap\_enter\_pte} (\path{osfmk/arm64/sptm/pmap/pmap.c}). To provide context for the \gls{SPTM} invocations, we trace back its callers to \texttt{pmap\_enter\_options} (\path{osfmk/arm64/sptm/pmap/pmap.c}). We know this function to be responsible for inserting translation table entries and introducing new virtual-to-physical address mappings~\cite{azad2020ppl}. The function calls an internal function that performs the actual page table entry creation, and then invokes \texttt{pmap\_enter\_pte} to update the created page table entry. As XNU is not entitled to write to page tables (recall \cref{subs:SPRR}), it needs to perform this update via the \texttt{sptm\_map\_page} invocation.

\begin{figure}[H]
    \centering
\includegraphics[scale=0.75]{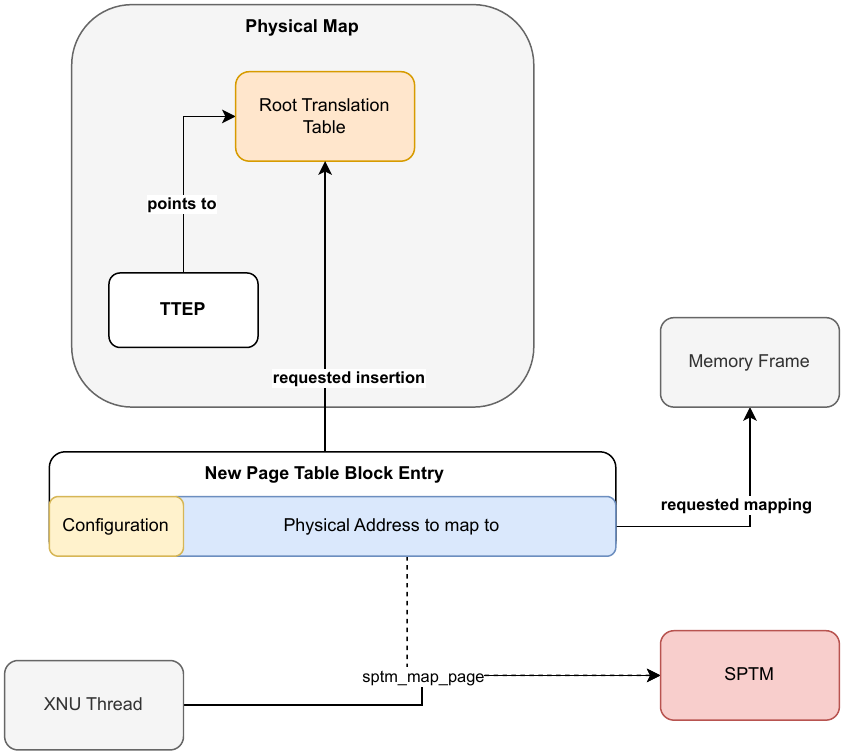}
\caption[\texttt{sptm\_map\_page} invocation context from XNU.]{Schematic of the process leading to an invocation of \texttt{sptm\_map\_page}. A XNU thread attempts to create a new page table entry and insert it into a physical map with a mapping to a targeted physical frame.}
\label{xnu_mapping_target}
\end{figure}

Recalling its function parameters, the \texttt{ttep} parameter in this case is the physical frame of the root translation table for the \texttt{pmap} into which XNU wishes to insert a page table entry. The \texttt{va} parameter holds the virtual address mapped by the newly created pointer, and the \texttt{new\_pte} parameter stores the value to store in the new page table entry. This \texttt{new\_pte} is a block entry, and therefore directly holds the output physical address to map the virtual address to. The logic around this function is shown in \cref{xnu_mapping_target}.

\subsection{\texttt{sptm\_map\_page} Security Mechanisms}
We know \texttt{sptm\_map\_page} to perform the actual insertion of the page table entry into the requested memory mapping. It is handled on the \gls{SPTM} side via \texttt{map\_page}. This function is highly complex and performs a variety of conditional checks. Inferring its exact inner workings is made difficult by the lack of symbols and the use of strings in the code. We do not aim to provide a comprehensive explanation of its inner workings and make no claim to completeness regarding security features, but instead attempt to highlight key mechanisms. We know from the previously examined \gls{SPTM} headers that \texttt{map\_page} corresponds to endpoint ID 2 concerning \gls{SPTM} calls to the \gls{XNU} bootstrap table. We can therefore easily find the \texttt{map\_page}\todo{table} function in the \gls{SPTM} binary. 

\paragraph*{Unparameterized Memory Frame Typecheck}

A key security check is the validation of the \gls{SPTM} type of the memory frame to map to via the new page table entry. The excerpt can be seen in \cref{lst:frameVerificationPageMap}.
\begin{listing}[H]
    \begin{minted}[linenos, breaklines, bgcolor=LightGray, frame=lines]{c}
  if ((0x4fffffd1c0fe177eU >> (frame_sptm_type & 0x3f) & 1) != 0) {
    DEBUG();
  }    
    \end{minted}
    \captionof{lstlisting}[SPTM \texttt{sptm\_map\_page} type validity check.]{SPTM type validity checking for frame to be mapped to via \texttt{sptm\_map\_page}. The source code was disassembled by Ghidra.}
    \label{lst:frameVerificationPageMap}
\end{listing}

A check is performed by right shifting the provided constant by the frame \gls{SPTM} type. Every bit set in the value seen in \cref{lst:frameVerificationPageMap} indicates a valid mapping to memory with the corresponding frame type. Integer to type mappings can be performed via the available list of all types (see \cref{App:SPTMTypesNew}). Valid page types that XNU is allowed to map can be seen below in \cref{allowdMappingsTab}.
\begin{table}[h!]
\centering
\begin{tabular}{|c|l|}
\hline
\textbf{Type ID} & \textbf{Type Name} \\
\hline
0   & \texttt{SPTM\_UNTYPED} \\ \hline
7   & \texttt{SPTM\_XNU\_CODE\_DBG\_RW} \\ \hline
11  & \texttt{XNU\_DEFAULT} \\ \hline
13  & \texttt{XNU\_RO\_DBG\_RW} \\ \hline
14  & \texttt{XNU\_USER\_EXEC} \\ \hline
15  & \texttt{XNU\_USER\_DEBUG} \\ \hline
16  & \texttt{XNU\_USER\_JIT} \\ \hline
24  & \texttt{XNU\_ROZONE} \\ \hline
25  & \texttt{XNU\_IO} \\ \hline
26  & \texttt{XNU\_PROTECTED\_IO} \\ \hline
27  & \texttt{XNU\_COMMPAGE\_RW} \\ \hline
28  & \texttt{XNU\_COMMPAGE\_RO} \\ \hline
29  & \texttt{XNU\_COMMPAGE\_RX} \\ \hline
33  & \texttt{XNU\_KERNEL\_RESTRICTED} \\ \hline
34  & \texttt{XNU\_RESERVED\_1} \\ \hline
35  & \texttt{XNU\_RESERVED\_2} \\ \hline
37  & \texttt{XNU\_RESTRICTED\_IO\_TELEMETRY} \\ \hline
60  & \texttt{SK\_SHARED\_RO} \\ \hline
61  & \texttt{SK\_SHARED\_RW} \\ \hline
63  & \texttt{UNKNOWN\_TYPE} \\ \hline
\end{tabular}
\caption[XNU mappable SPTM frame types.]{SPTM frame types that can be mapped to via a page table entry from XNU.}
\label{allowdMappingsTab}
\end{table}

As can be seen, XNU is only entitled to map a page to a small subsection of all \gls{SPTM} frame types. Notably, it is not allowed to create mappings to any \texttt{ROOT\_TABLE} or \texttt{PAGE\_TABLE} frame types, even for those belonging to the \texttt{XNU\_DOMAIN}. Such frame types are deemed security-relevant and can only be altered by SPTM. It is, however, able to map into \texttt{SK\_SHARED\_RO} and \texttt{SK\_SHARED\_RW} typed memory frames, which allows \gls{SK} to share data with XNU. More details regarding \gls{SK} will be provided in \cref{secureKernel}.

\paragraph*{Target Table Typechecking}
A second security mechanism implemented is a conditional check based on the \gls{SPTM} frame type of the table; the page table entry is supposed to be inserted. We know all possible types that the page table can be inserted into from its previous retrieval in the function \texttt{table\_acquire}. The valid \gls{SPTM} frame types for the table to insert the \gls{PTE} into are listed in \cref{validTableTypes}.
\begin{table}[h!]
\centering
\begin{tabular}{|c|l|}
\hline
\textbf{Type ID} & \textbf{Type Name} \\ \hline
8  & \texttt{SPTM\_KERNEL\_ROOT\_TABLE} \\ \hline
9  & \texttt{SPTM\_PAGE\_TABLE} \\ \hline
17 & \texttt{XNU\_USER\_ROOT\_TABLE} \\ \hline
18 & \texttt{XNU\_SHARED\_ROOT\_TABLE} \\ \hline
19 & \texttt{XNU\_PAGE\_TABLE} \\ \hline
20 & \texttt{XNU\_PAGE\_TABLE\_SHARED} \\ \hline
21 & \texttt{XNU\_PAGE\_TABLE\_ROZONE} \\ \hline
22 & \texttt{XNU\_PAGE\_TABLE\_COMMPAGE} \\ \hline
31 & \texttt{XNU\_STAGE2\_ROOT\_TABLE} \\ \hline
32 & \texttt{XNU\_STAGE2\_PAGE\_TABLE} \\ \hline
\end{tabular}
\caption[Valid table SPTM frame types to insert PTE into from XNU.]{Valid SPTM frame types for tables to insert PTE into from XNU.}
\label{validTableTypes}
\end{table}

The performed validation ensures that XNU can only map memory using page and root tables that belong to its domain, thereby creating a full memory separation between different \gls{SPTM} domains.

Whilst the previously shown \cref{lst:frameVerificationPageMap} performed an unconditional check for generally allowed types to map to, \gls{SPTM} further proceeds to lock down the ability to map to memory frames. This is done based on the \gls{SPTM} frame type of the table to which it is being inserted. Only specific frame types may be inserted into specific pages or root tables. This is achieved by retrieving a value from the global \gls{SPTM} retyping structure, indexed by the \gls{SPTM} type of the table to be inserted into. The check can be seen in \cref{lst:allowedTYpesToMap}.
\begin{listing}[H]
    \begin{minted}[linenos, breaklines, bgcolor=LightGray, frame=lines]{c}
if ((*(ulong *)(&REMAP_STRUCTURE + (ulong)table_sptm_type * 0x60) >> (frame_sptm_type & 0x3f) & 1
      ) == 0) {
        DEBUG();
  }    
    \end{minted}
    \captionof{lstlisting}[SPTM page mapping validity check in \texttt{map\_page}.]{SPTM check for mapping to a specific frame with frame type \texttt{frame\_sptm\_type} via a table of frame type \texttt{table\_sptm\_type}. The source code was disassembled by Ghidra.}
    \label{lst:allowedTYpesToMap}
\end{listing}

Based on the \texttt{table\_sptm\_type}, the global \texttt{RETYPE\_STRUCTURE} is indexed and a value retrieved that encodes valid \gls{SPTM} frame types to be mapped from the specific table type. The complete list of frame types that can be mapped from specific tables is included in \cref{appendixAllowedTableMaps}.
The results are, in general, as expected. Root tables can map to the corresponding page tables. \gls{SPTM} page tables can map to any other memory frame, regardless of its type, which appears logical given that \gls{SPTM} is the page table management component running at the highest privilege level.

\paragraph*{Further Validations}

We find the \texttt{map\_page} function to perform multiple additional validations based on the frame to map to, the frame type of the table to insert into, and other conditional checks. New \glspl{PTE} are validated, among other things, concerning their \gls{SPRR} index; we were not yet able to reverse engineer. Allowed SPRR indexes appear to be defined based on the \gls{SPTM} type of the table to insert into. We cannot provide a conclusive understanding of these mechanisms at this time, as we would require a current and up-to-date mapping of SPRR indices for this. However, we assume this to enforce specific access and execution permissions for specific frame types. Exemplary, dedicated read-only page types should not be made writable via SPRR indices.

We find the \texttt{map\_page} \gls{SPTM} function to perform a variety of further validity checks, with further underlying rule sets governing these mappings.

\section{SPTM Page Mapping -- Conclusion}
\gls{SPTM} is the sole entity capable of altering page table data and introducing new virtual-to-physical memory mappings. It handles memory-mapping requests on behalf of XNU, based on a rigid set of rules regarding allowed mappings. All XNU functions altering page table data do so via an invocation of SPTM. XNU can only map to memory frames with a significantly reduced subset of \gls{SPTM} types. This limits its ability to access security-relevant or isolated memory. \gls{SPTM} further ensures that SPRR indices set for page table entries to be inserted into page tables also adhere to a predetermined set of rules, prohibiting XNU from arbitrarily setting \gls{SPRR} indices and, consequently, access and execution rights to the memory it maps.

This restriction of memory-mapping ability via \gls{XNU} essentially introduces a separation of trust throughout the system. XNU is no longer able to access nearly all parts of the system; instead, specific vulnerable system components can be scoped away in different \gls{SPTM} domains, thereby limiting XNU's ability to access them. Such an architectural shift aims to protect against the dangers of a full kernel compromise in a monolithic kernel design, which typically implies a complete loss of control over all system functionality.  % example
    \chapter{Secure Kernel}
\label{secureKernel}
\section{Fundamentals}
The \acrfull{SK} is a \gls{GXF} component running in GL1. It is part of the \emph{Exclavecore} (see \cref{exclaves}). From our previous analysis of \gls{SPTM}, we know SK to have its own SPTM domain \texttt{SPTM\_DOMAIN\_SK}. There are four SK-specific SPTM frame types, namely \texttt{SK\_DEFAULT}, \texttt{SK\_SHARED\_RO}, \texttt{SK\_SHARED\_RW} and \texttt{SK\_IO}, which SK operates on.

We find entering the \texttt{SK\_DOMAIN} to be made possible via SPTM through two dispatch functions registered with SPTM by SK. These functions are invoked when SPTM clients (e.g., XNU) try to call into \texttt{SK\_DOMAIN}. Entering SK from XNU via \texttt{exclaves\_enter} ultimately resumes execution in GL0 via an \texttt{ERET} call. SK appears to handle requests from GL0 components. Core requests include SPTM frame retyping to XNU and SK shared frame types, indicating that GL0 clients share data with XNU via this mechanism. SK appears to be the privileged call dispatching and management component in \gls{GXF}, with SPTM acting as the overhead hypervisor for memory management.

\section{SPTM Calls}
We have previously discovered SPTM calls from SK in \cref{SK_SPTM_CALLS}.
Mapping them based on the known SPTM dispatch structure provides the results seen in \cref{SK_calls_table}. As all calls from SK are directed towards the \texttt{SPTM\_DOMAIN}, we redact this field for readability.

\begin{table}[h!]
\centering
\begin{tabular}{|l|l|l|}
\hline
\textbf{Dispatch Table} & \textbf{Endpoint ID} & \textbf{SPTM Function} \\ \hline
\texttt{SK\_BOOTSTRAP} & 0 & \texttt{register\_dispatch\_table} \\ \hline
\texttt{SK\_BOOTSTRAP} & 1 & \texttt{retype} \\ \hline
\texttt{SK\_BOOTSTRAP} & 2 & \texttt{get\_frame\_type} \\ \hline
\texttt{SK\_BOOTSTRAP} & 3 & Not statically provided \\ \hline
\texttt{T8110\_DART\_SK} & 0 & \texttt{sptm\_t8110\_dart\_map\_table} \\ \hline
\texttt{T8110\_DART\_SK} & 1 & \texttt{sptm\_t8110\_dart\_unmap\_table} \\ \hline
\texttt{RETURN\_TO\_CALLER} & - & Control Function \\ \hline
\texttt{PANIC} & - & Control Function \\ \hline
\texttt{EXCEPTION\_STATE\_SAVED} & - & Control Function \\ \hline
\end{tabular}
\caption[SPTM calls performed from SK.]{SPTM calls performed from SK. Based on the assumption that the \texttt{T8110\_DART\_SK} dispatch table is equal to the \texttt{T8110\_DART} dispatch table.}
\label{SK_calls_table}
\end{table}
As can be seen, SK performs dispatch table registrations, which we will look into more closely in \cref{SK_registration}. It can also retype physical frames and retrieve frame types using its \texttt{SK\_BOOTSTRAP} dispatch table in SPTM. Regarding the dispatch table, we were unable to find function endpoint 3 in SPTM. This might indicate it is set dynamically but it still could also be caused by issues in the decompiler.

With regards to the \texttt{T8110\_DART\_SK} dispatch table, \gls{SK} performs table mapping and unmapping via SPTM. \gls{SK} further utilizes the SPTM-provided control functions for panicking, saving exception states, and returning to callers.

Due to the complete lack of available symbols in the SK binary and other general information on its workings, and to stay within the scope of this work, we have decided to significantly limit our analysis of \gls{SK}'s inner workings to what we deem to be key aspects of it. This key aspect is frame retyping through SPTM, which is examined in \cref{frameRetypingSK}. Our goal is not to understand every individual SPTM function call, but to provide an idea of the underlying concepts and their impacts on system security.

\subsection*{SPTM Calls -- Frame Retyping}
\label{frameRetypingSK}

We find a variety of functions performing SPTM frame type remapping in \gls{SK}. A list of all invocations is shown in \cref{retypingSKCalls}.
\begin{table}[h!]
\centering
\begin{tabular}{|l|l|l|}
\hline
\textbf{Current Frame Type} & \textbf{New Frame Type} & \textbf{Invocation} \\ \hline
\texttt{SK\_DEFAULT} & \texttt{SK\_SHARED\_RO} & \texttt{retype\_to\_shared\_ro} \\ \hline
- & \texttt{SK\_SHARED\_RO} & \texttt{retype\_to\_shared1} \\ \hline
- & \texttt{SK\_SHARED\_RW} & \texttt{retype\_to\_shared1} \\ \hline
- & \texttt{SK\_DEFAULT} & \texttt{map\_to\_sk\_default1} \\ \hline
\texttt{SK\_DEFAULT} & \texttt{SK\_SHARED\_RO} & \texttt{retype\_to\_shared2} \\ \hline
\texttt{SK\_DEFAULT} & \texttt{SK\_SHARED\_RW} & \texttt{retype\_to\_shared2} \\ \hline
- & \texttt{XNU\_DEFAULT} & \texttt{retype\_to\_XNU\_default} \\ \hline
- & \texttt{SK\_DEFAULT} & \texttt{retype\_to\_sk\_default2} \\ \hline
\end{tabular}
\caption[SPTM frame retyping invocations from SK.]{Found invocations of SPTM frame retyping from SK with the corresponding function identifier. Unspecified types imply that they are dynamically retrieved from the frame to retype.}
\label{retypingSKCalls}
\end{table}

While unspecified current frame types indicate that the frame type is retrieved dynamically before the retyping operation is performed, it is essential to note that SK must adhere to the SPTM-provided rule set of allowed retyping (see \cref{App:allowedTransition}). 
We find SK to perform frame retyping mappings from \texttt{SK\_DEFAULT} to \texttt{SK\_SHARED\_RO} or \texttt{SK\_SHARED\_RW} on multiple occasions. An exemplary invocation of retyping can be seen in \cref{sk_map_to_shared}.

\begin{listing}[H]
    \begin{minted}[linenos, breaklines, bgcolor=LightGray, frame=lines]{c}
bool retype_to_shared2(undefined8 physical_address,int rw_ro_conditional)
{
  int current_frame_type;
  undefined4 new_type;
  
  current_frame_type = sptm_get_frame_type();
  if (current_frame_type == SK_DEFAULT) {
    new_type = SK_SHARED_RO;
    if (rw_ro_conditional != 0) {
      new_type = SK_SHARED_RW;
    }
    retype(physical_address,SK_DEFAULT,new_type,0);
  }
  return current_frame_type == SK_DEFAULT;
}    
\end{minted}
\captionof{lstlisting}[SK \texttt{retype\_to\_shared2} function.]{\texttt{retype\_to\_shared2} function in SK mapping a frame based on a provided physical address to \texttt{SK\_SHARED\_RW} or \texttt{SK\_SHARED\_RO} based on an input parameter. Only performs mapping if the underlying frame for the address is of type \texttt{SK\_DEFAULT}. The source code was disassembled by Ghidra.}
\label{sk_map_to_shared}
\end{listing}

Recalling \cref{allowdMappingsTab}, we know XNU can map pages into memory of type \texttt{SK\_SHARED\_RO} and \texttt{SK\_SHARED\_RW}, which leads us to determine that this function makes SK-owned physical memory available to XNU. \gls{SK} is responsible for and able to map memory in its domain, which is usually gapped from the kernel available to XNU if it is required for the task at hand. \todo{refExclaves}

\section{SVC Handling}
As SK is running in GL1, we find that it registers exception handlers for exceptions taken from GL0 components. 
As mentioned previously, multiple components running in \gls{GL} issue \glspl{SVC} to the secure kernel. The handling of such calls will be looked at in this section.

Similar to our analysis of \gls{SPTM} exception handlers, we can examine writes to the \texttt{VBAR\_GL1} register in SK and identify various writes to it. Next to the actual exception handler, we also find an early entry exception handler registration, which acts similarly to the SPTM one. The actual exception handler base is registered at \texttt{ExceptionHandlerBase}\todo{table}. The function performing the exception handling for synchronous exceptions from lower levels can be found at offset \texttt{+0x400}~\cite{VBAR_EL1:2025}. Following this path shows a wrapper method, and finally the actual handler \path{SVC_handler_from_lower_EL}\todo{table}.

We refrain from performing a full analysis on this function, but rather look at it conceptually. \gls{SK} handles \glspl{SVC} with immediate values from 0 to 5. For other values, it appears to perform a context restore and returns to the caller via \texttt{RestoreContextAndEret}\todo{table}. For SVC \#0 calls, we find the following implementation.

\begin{listing}[H]
    \begin{minted}[linenos, breaklines, breakanywhere, bgcolor=LightGray, frame=lines]{c}
    if (svc_imm == 0) {
      allinged_pointer_1 =
           (ulong *)(pointer_1 & 0b0000000000000000000000000000111111111111111111111111111111000000)
      ;
      if (allinged_pointer_1 != (ulong *)0x0) {
        (*(code *)(&FunctionTable)[(*allinged_pointer_1 >> 0x3a) * 0xb])
                  (allinged_pointer_1,pointer_2);
        goto PrepareContextAndEret;
      } 
\end{minted}
\captionof{lstlisting}[SK SVC dispatching logic for SVC \#0.]{Request dispatching in \texttt{SVC\_handler\_from\_lower\_EL} in SK, for SVC call with IMM value 0. A function is indexed in a function table based on data pointed to by a pointer provided by the SVC call. The source code was disassembled by Ghidra.}
\label{sk_svc0}
\end{listing}

\begin{figure}[H]
    \centering
\includegraphics[scale=0.65]{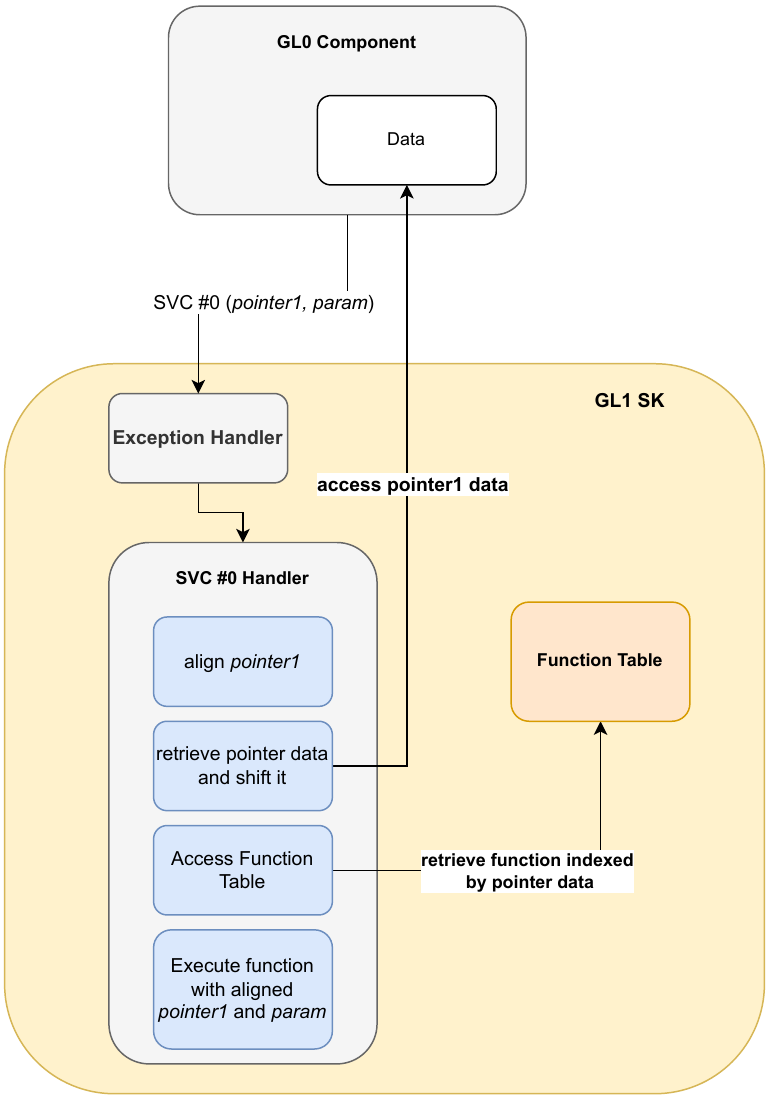}
\caption[SK SVC \#0 handling.]{Assumed SVC \#0 handling by SK. A client-provided pointer is aligned and used to access GL0 user space data. Based on the data retrieved, an internal function table is accessed and the function executed with two client-provided arguments.}
\label{fig:svc0SK}
\end{figure}

An input pointer provided from context at exception call is aligned by setting its lower 6 bits to zero. Furthermore, its uppermost 32\,bits are also set to zero. This ensures the pointer is within a specific range and aligned to a structure. The fact that the upper 32\,bits of the input pointer are not considered indicates that it is a userspace pointer from the GL0 component calling into SK\footnote{We can assume this due to the standard memory layout of ARM, where we find user space virtual addresses usually assigned to the lower memory regions~\cite{ARM-v8AddressTranslations:2024}.}. SK then retrieves the value pointed at by the updated pointer and performs a right shift operation on it. The result of said shift is used as an index into \texttt{FunctionTable}\todo{table}. Essentially, we assume SK retrieves user-provided data to index an internal function table, serving the client's request. The control flow of such a retrieval can be seen in \cref{fig:svc0SK}.

We are not provided with any table limits for the function table accessed. Looking through the table entries, we find function calls that call the previously discovered retyping function (\cref{retypingSKCalls}). From this, we know that GL0 clients of SK call into it for the retyping of memory frames they are using. Such retyping may be performed to make memory available to XNU (retyping to \texttt{SHARED} types, or to acquire new SK memory frames (retyping to \texttt{SK\_DEFAULT}).

The \gls{SVC} handling for different \texttt{IMM} values and the further functionality offered by SK to its GL0 clients has not yet been reverse-engineered by us and is left for future work. Based on the seen function table, SK appears to offer a wide range of functions; however, with the lack of any symbols, reverse engineering these would go beyond the scope of this work.

\subsection*{Dispatch Table Registration}
\label{SK_registration}
We have previously seen that XNU calls into the SK domain through STPM (see \cref{lab:sptmCallsFromXNU}). These calls from XNU functions \texttt{exclaves\_enter} and \texttt{exclaves\_bootinfo} are dispatched by SPTM to SK. However, dispatch tables for \texttt{SK\_DOMAIN} are found not to be statically included in SPTM, but are instead dynamically registered at runtime by SK. This is done via \texttt{sptm\_register\_dispatch\_table} calls, which are performed by SK (see \cref{SK_calls_table}). The two register calls performed can be seen in \cref{sk_register_tables}.

\begin{listing}[H]
\begin{minted}[linenos, breaklines, bgcolor=LightGray, frame=lines]{c}
sptm_register_dispatch_table(0,&SPTM_dispatch_function_0,2);
sptm_register_dispatch_table(1,&SPTM_dispatch_function_1,1);    
\end{minted}
\captionof{lstlisting}[Calls to \protect \path{sptm_register_dispatch_table} from SK.]{Calls to \texttt{sptm\_register\_dispatch\_table} from SK function \path{FUN_ffffff800000aecc}. SK registers two dispatch functions that allow calling into SK when calling the domain \texttt{SK\_DOMAIN}. The source code was disassembled by Ghidra.}
\label{sk_register_tables}
\end{listing}

\begin{figure}[H]
    \centering
\includegraphics[width=\textwidth]{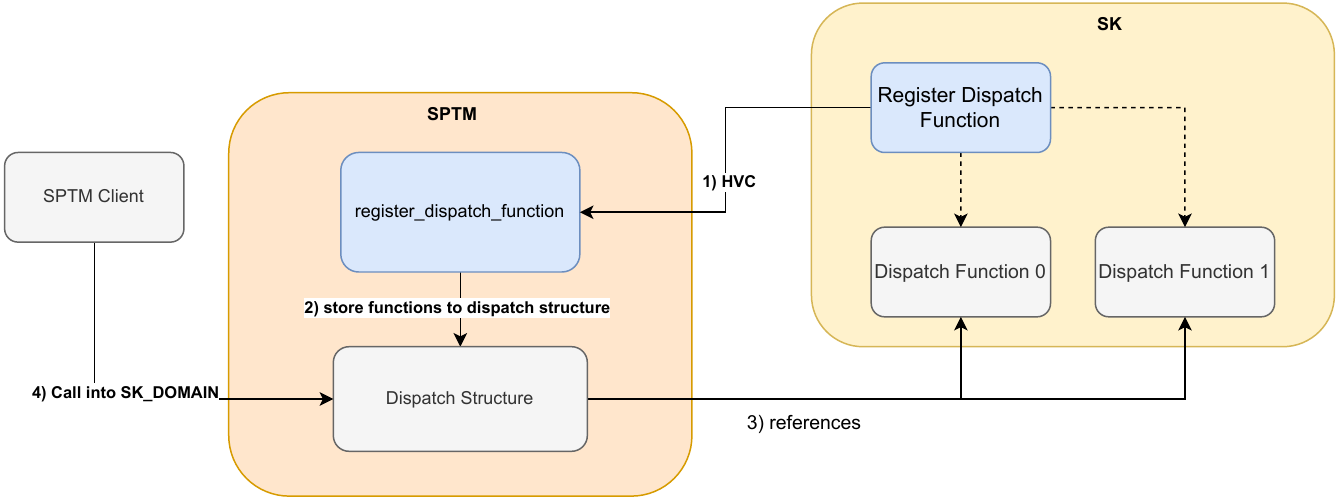}
\caption[SK dispatch function registration.]{Process of SK dispatch function registration in SPTM.}
\label{skDispatchRegistration}
\end{figure}

The function calling these registrations is called nested from the SK entry function, by which we know \gls{SK} makes itself available to SPTM soon after boot. 
The two functions SK registers as dispatch functions are \path{SPTM_dispatch_function_0} and \path{SPTM_dispatch_function_1}. As known from our previous reverse engineering of the SPTM dispatch table registration process (see \cref{subs:registering}), the third parameter of the registration call is the bitwise-encoded permission. A set bit in the value indicates that a domain is allowed to call into the corresponding dispatch table. We therefore know that \path{SPTM_dispatch_function_1} is only callable by \texttt{XNU\_DOMAIN}, and \path{SPTM_dispatch_function_2} is only callable by \texttt{SPTM\_DOMAIN}.
The dispatch registration performed by SK is shown in \cref{skDispatchRegistration}.

Looking at \texttt{SPTM\_dispatch\_function\_0}, we find a call to a function we named \path{SK_called_from_sptm_dispatch_function_0}. This function performs the actual dispatching via further method calls to \path{dispatch_0_handle_endpoints}. It checks whether the provided endpoint is either 0 or 1 (recall 0 for \texttt{EXCLAVES\_ENTER}, 1 for \texttt{EXCLAVES\_BOOTINFO}). For the bootinfo endpoint call, a return to the caller is performed via the specific SPTM call. For the Exclaves enter endpoint, we find a nested method call to \texttt{RestoreContextAndEret}.
\begin{listing}[H]
    \begin{minted}[linenos, breaklines, bgcolor=LightGray, frame=lines]{c}
undefined1  [16] RestoreContextAndEret(void)

{
  undefined8 *puVar1;
  undefined1 (*saved_registers) [16];
  
  puVar1 = (undefined8 *)TPIDR_GL1;
  saved_registers = (undefined1 (*) [16])*puVar1;
  if (*(long *)(saved_registers[0x13] + 8) == 0) {
    tpidr_el0 = *(undefined8 *)saved_registers[0x12];
    tpidrro_el0 = *(undefined8 *)(saved_registers[0x12] + 8);
    fpcr = *(undefined8 *)(saved_registers[0x11] + 8);
    fpsr = *(undefined8 *)saved_registers[0x11];
    SPSR_GL1 = *(undefined8 *) (saved_registers[0x10] + 8));
    sp_el0 = *(undefined8 *)saved_registers[0x10];
    ELR_GL1 = *(undefined8 *)(saved_registers[0xf] + 8);
    ExceptionReturn();
    return *saved_registers;
  }
  *(undefined8 *)(saved_registers[0x13] + 8) = 0;
  tpidr_el0 = *(undefined8 *)saved_registers[0x12];
  tpidrro_el0 = *(undefined8 *)(saved_registers[0x12] + 8);
  fpcr = *(undefined8 *)(saved_registers[0x11] + 8);
  fpsr = *(undefined8 *)saved_registers[0x11];
  SPSR_GL1 = *(undefined8 *)(saved_registers[0x10] + 8);
  sp_el0 = *(undefined8 *)saved_registers[0x10];
  ELR_GL1 = *(undefined8 *)(saved_registers[0xf] + 8));
  ExceptionReturn();
  return *saved_registers;
}}    
\end{minted}
\captionof{lstlisting}[SK \texttt{RestoreContextAndEret} function.]{\texttt{RestoredContextAndEret} function in SK. The source code was disassembled by Ghidra.}
\label{SK_ERET}
\end{listing}    

We find what appears to be a context restore and preparation. The GL1 \texttt{ELR} register is written to a value retrieved from the thread identifying information register (\texttt{TPIDR\_GL1}). Finally, both cases perform an \texttt{ExceptionReturn()} (\texttt{ERET}~\cite{ERET:2025}) call, which returns to the address previously set to the \texttt{ELR}.

We deduce that whilst SK acts as SPTM call taker, it delegates actual request handling to GL0 components it returns to via \texttt{ERET}. We cannot confirm where execution is continued after the instruction, but assume it to be \texttt{xnuproxy} (contained in the \texttt{sharedcache} binary; more on \texttt{xnuproxy} in \cref{xnuproxySECTION}).
 % example
    \chapter{Exclaves}
\label{exclaves}
Exclaves are found to contain specific subsets of sensitive resources, which are scoped in the \texttt{SK\_DOMAIN}, and by that naturally isolated from XNU. We find them to offer services that clients can call via exposed interfaces. Exclaves offer a variety of different communication mechanisms, including a dedicated \gls{IPC} framework called Tightbeam. They are used for further compartmentalization and division of trust away from XNU and towards a more microkernel-style architecture.
\todo{introduction to exclaves}

\section{Exclave System Components}
Exclaves encompass a range of components, which we will list briefly here.
\subsection{Exclavecore}
A large part of the Apple Exclave ecosystem is the \emph{Exclavecore}. It consists of the \gls{SK} found running in GL1, and a variety of other system binaries we determine to run in GL0. The GL0 binaries are the following:
\begin{itemize}
    \item ExclaveStackshotServer
    \item pmm\_exclave
    \item roottask
    \item scheduler
    \item sharedcache
\end{itemize}
Except for sharedcache in the context of our analysis of \texttt{xnuproxy} (see \cref{xnuproxySECTION}), we have not further analyzed any of these binaries so far. We do, however, find them all to perform \glspl{SVC}, which we assume to be handled by \gls{SK} as described in \cref{secureKernel}. GL0 components in the Exclavecore appear to be subject to significant changes between different firmware versions. We assume the reason is that Exclaves are a relatively new security mitigation still under constant development.

The Exclavecore bundle is located at \path{Firmware/image4/exclavecore.<version>.RELEASE.im4p} in the firmware IPSW file. It can be extracted using the \texttt{ipsw} tool with the command \texttt{ipsw extract -x *.ipsw}.

\subsection{ExclaveKit}
On devices supporting Exclaves, a new \gls{DMG} is found. We find this to contain the \texttt{ExclaveKit}. The \texttt{ExclaveKit} contains a variety of \texttt{Frameworks} and \texttt{PrivateFrameworks}. Frameworks are dynamic libraries, with \texttt{Frameworks} being usable from App Store Apps, whilst \texttt{PrivateFrameworks} are meant to be only used by Apple's apps~\cite{applewiki_frameworks_2025}. A complete listing of the frameworks discovered in the \texttt{ExclaveKit} DMG can be found in \cref{frameworkList}. We assume the \texttt{ExclaveKit} to store the actual Exclave's code. Among others, we find the private \path{Tightbeam.framework}, which will be a focus of our analysis in \cref{Tightbeam}.

\section{Exclave Memory Fundamentals}
\begin{figure}[H]
    \centering
\includegraphics[width=\textwidth]{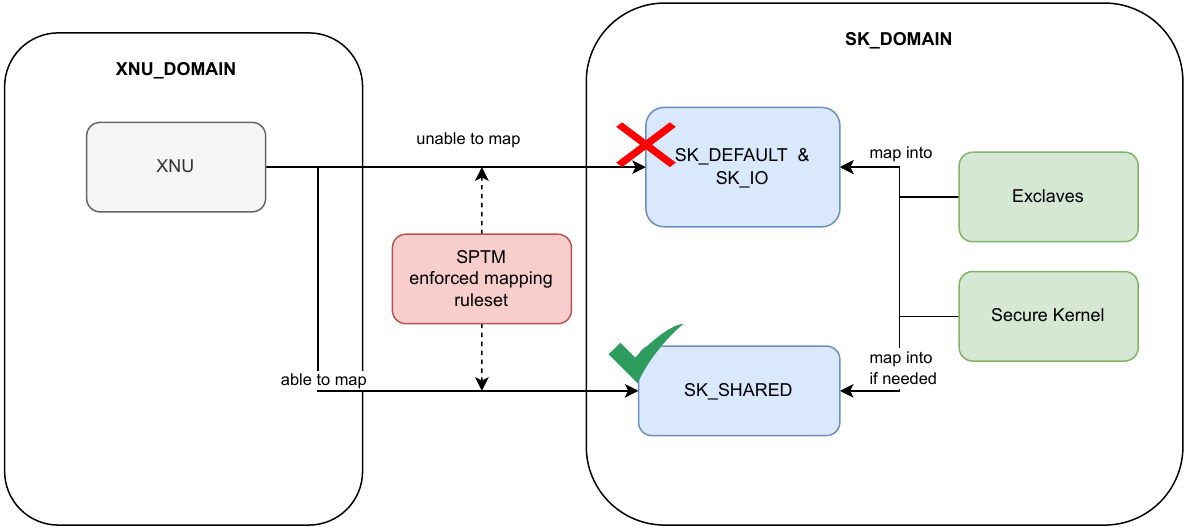}
    \caption[Memory mapping restrictions between XNU and the SK\_DOMAIN.]{Exclaves run in the \texttt{SK\_DOMAIN}, mapping memory into \texttt{SK\_DEFAULT} and \texttt{SK\_IO} type frames for execution separated from XNU. If the need arises for information exchange with XNU, Exclaves may also map into shared frame types.}
    \label{SK_Exclaves_mapping}
\end{figure}

Although it may appear premature, examining Exclave memory fundamentals provides important context for the upcoming analysis. \path{osfmk/arm/machine_routines.h} lists a \texttt{ml\_paddr\_is\_exclaves\_owned} function, which, for a provided physical address, verifies whether it is ``owned by the secure world''. From the underlying implementation we find this is the case for frame type \texttt{SK\_DEFAULT} and \texttt{SK\_IO}. This verifies that Exclaves are running in the secure world, which appears to denote the \texttt{SK\_DOMAIN}. Only the two named types are exclusively used by the secure world. \texttt{SK\_SHARED\_RW} and \texttt{SK\_SHARED\_RO} frames are not ``exclusively exclaves frames'', but instead can be shared with XNU (see \texttt{osfmk/arm64/machine-routines.c}, l. 2955). This confirms our previous analysis results in \cref{allowdMappingsTab}, which shows that XNU is allowed to map memory to frames of the named types, which is a requirement for memory access. This concept is depicted in \cref{SK_Exclaves_mapping}. \todo{caption}

\section{Resources}
\label{resources}
The open-source code parts of \gls{XNU} offer some key insights into resources within Exclaves (\texttt{osfmk/kern/exclaves\_resource.h}). According to comments in the source code, ``exclaves provide a fixed static set of resources available to XNU''. Types of resources include, among others, Conclave managers, services, sensors, and buffers. As noted in the comment, Exclave resources appear to be static at runtime, being enumerated once during boot for setup, and not being alterable after. This also entails that if resources gap specific capabilities from other parts of the XNU kernel, we can not alter this past boot and must rely on what was initially set up. Furthermore, if Exclave resources are gapped from XNU but made available, there must be mechanisms to call them, apart from standard memory mapping. We will analyze these in \cref{CallingIntoExclaves}. 

Resources have a specific name, a type, and an identifier, with the identifier being used for addressing between XNU and Exclaves. They are further scoped in so-called domains, which will be examined more closely in \cref{domain}. The \texttt{exclaves\_resource\_t} type definition as made available in \path{osfmk/kern/exclaves_resource.h} is shown in \cref{exclaveType}.
\begin{listing}[H]
    \begin{minted}[linenos, breaklines, bgcolor=LightGray, frame=lines]{c}
#define EXCLAVES_RESOURCE_NAME_MAX 128
typedef struct exclaves_resource {
	char                r_name[EXCLAVES_RESOURCE_NAME_MAX];
	xnuproxy_resourcetype_s r_type;
	uint64_t            r_id;
	_Atomic uint32_t    r_usecnt;
	ipc_port_t          r_port;
	lck_mtx_t           r_mutex;
	bool                r_active;
	bool                r_connected;

	union {
		conclave_resource_t     r_conclave;
		sensor_resource_t       r_sensor;
		exclaves_notification_t r_notification;
		shared_memory_resource_t r_shared_memory;
	};
} exclaves_resource_t;    
\end{minted}
\captionof{lstlisting}[\texttt{exclaves\_resource\_t} type definition.]{The \texttt{exclaves\_resource\_t} type as made available in \path{osfmk/kern/exclaves_resource.h}.}
\label{exclaveType}
\end{listing}

It stores general information about the resource, such as name, type, and identifier, and, more interestingly, type-specific additional information in the form of either a Conclave, notification, shared memory, or sensor resource. The specific type definitions can be found in \cref{typeDefinitionsExclaves}. We will analyze Conclaves in detail in \cref{Conclave}.
Regarding Exclave resource types, we find references to the following resource types in the XNU open-source code:
\begin{itemize}
    \item \texttt{XNUPROXY\_RESOURCE\_CONCLAVE\_MANAGER}
    \item \texttt{XNUPROXY\_RESOURCE\_NOTIFICATION}
    \item \texttt{XNUPROXY\_RESOURCE\_SERVICE}
    \item \texttt{XNUPROXY\_RESOURCE\_NAMED\_BUFFER}
    \item \texttt{XNUPROXY\_RESOURCE\_ARBITRATED\_AUDIO\_BUFFER}
    \item \texttt{XNUPROXY\_RESOURCE\_SENSOR}
    \item \texttt{XNUPROXY\_RESOURCE\_SHARED\_MEMORY}
    \item \texttt{XNUPROXY\_RESOURCE\_ARBITRATED\_AUDIO\_MEMORY}
\end{itemize}
We further find that Exclave resources appear to have a standard \texttt{ipc\_port\_t} indicating potential access via standard \gls{IPC} mechanisms. This will be discussed in \cref{CallingIntoExclaves}.

\section{Exclave Domains}
\label{domain}
The \gls{XNU} open-source code reveals a great deal about Exclaves and their structuring (\texttt{osfmk/kern/exclave\_resources.c}, l.90). We know that resources ``are scoped by what entities are allowed to access them''. This concept of a scope is what is denoted as \texttt{domains}. These domains are not interchangeable with \gls{SPTM} domains, but rather represent a distinct concept. There appears to be a two-level table scheme indexing Exclave resources for access. A root table indexing Exclave domains, with second-level tables storing actual Exclave resources.
\begin{figure}[H]
    \centering
\includegraphics[scale=0.75]{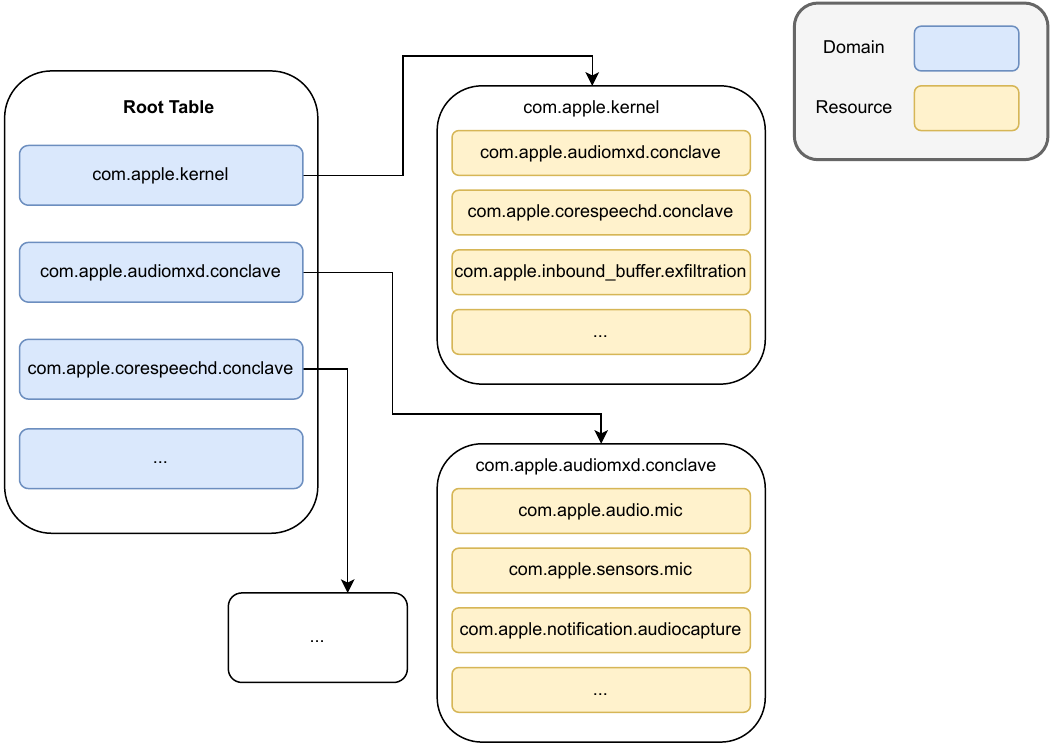}
    \caption[Exclaves table structure.]{Exclaves table structure with a root table holding domain references and a pointer to the actual domains with their respective Exclave resources, adapted~\cite{XNU_10063:2024}.}
    \label{fig:ExclaveTables}
\end{figure}

\subsection*{Resource Enumeration} 
\label{resourceEnumSection}If resources are predetermined at boot, we have to find them within the firmware. By searching for known resource name patterns like \texttt{com.apple.conclave.*} extracted from \path{osfmk/kern/exclaves_resource.c}, we can determine where in the firmware the resource list is stored. We find them in the \texttt{sharedcache} binary, a part of the \texttt{Exclavecore}, which will be looked at in more detail in our analysis of the underlying \texttt{xnuproxy} in \cref{xnuproxySECTION}. We can, however, already determine a \texttt{resource\_registration}\todo{table} function, which accesses a global structure containing resource and domain pairings. The extracted list of domains with corresponding resources can be found in \cref{exclaveResources}. The script used to extract these resources is provided in \cref{lst:resourceEnumScript}. A variety of resources are found via this, including sensors, notifications, inbound buffers, and generic services. It also appears that specific resources can be correlated with multiple domains at a time, as exemplified by the \path{com.apple.sensors.mic} resource. We also find resource duplication, even within the same domain space, which might hint towards resources with different permissions.

\paragraph*{Resource Discovery and Initialization}

Resources are discovered and initialized as part of the early Exclaves boot process. The discovery is performed by the \path{exclaves_resource_init} function in \path{osfmk/kern/exclaves_resource.c}. The function initializes the root table and then proceeds to iterate every resource via a call to \path{exclaves_xnu_proxy_send} with command \path{XNURPOXY_CMD_RESOURCE_INFO}. We will analyze calls to \texttt{xnuproxy} more thoroughly in \cref{xnuproxySECTION}, but can infer that it is responsible for resource management.

The \texttt{exclaves\_resource\_init} function proceeds by verifying that the resource domain exists in the root table and creates it if it does not. This is a prerequisite to allocating the new resource in the domain. After this, type-specific initialization is performed.
\begin{itemize}
 \item For \textbf{services}, this verifies that the maximum number of services per Conclave is not exceeded.
 \item  For \textbf{notification resources}, this calls \texttt{exclaves\_notification\_init}, which does only initialize a kernel list field in the resource.
 \item For \textbf{Conclave manager resources}, this calls \texttt{exclaves\_conclave\_init}, which forwards the call to \texttt{exclaves\_conclave\_launcher\_init} parameterized with the discovered resource ID and a connection output parameter. This function call performs what we assume to be Tightbeam connection setup on the discovered resource via \texttt{conclave\_launcher\_conclavecontrol\_\_init}. It stores a \texttt{tb\_client\_connection\_t} in the Conclaves \texttt{r\_control} field. This is the Tightbeam connection that interacts with the Conclave. Tightbeam, as the Exclaves communication framework, will be analyzed in detail in \cref{Tightbeam}.  Based on this type-specific resource initialization, we determine that Conclave manager resources are the only resources to set up and hold a Tightbeam connection.
\end{itemize}

After all resources have been enumerated and type-specific initialization has been performed, \path{populate_conclave_service} is called. This function populates the service bitmaps of all discovered Conclaves for all services belonging to their respective domains. Essentially, this function registers all available resources for each Conclave manager resource, where Conclave manager resources are these Exclave resources that have a designated Conclave. Conclaves will be looked at in more detail in \cref{Conclave}.

\section{Conclaves}
\label{Conclave}
From the \texttt{exclaves\_resource\_t} and \texttt{conclave\_resource\_t} type definitions looked at in \cref{resources}, we know Conclaves to be a specific subtype of resource, optionally contained within an Exclave resource. The type definition of a Conclave are shown in \cref{conclaveTypeDef}.
Conclaves themselves group multiple services in a bitmap, have an assigned task and thread, and store internal state and request information. Additionally, they include a \texttt{tb\_client\_connection\_t}, a \texttt{Tightbeam} connection.
\begin{listing}[H]
    \begin{minted}[linenos, breaklines, bgcolor=LightGray, frame=lines]{c}
typedef struct {
	conclave_state_t       c_state;
	conclave_request_t     c_request;
	bool                   c_active_downcall;
	bool                   c_active_stopcall;
	bool                   c_active_detach;
	tb_client_connection_t c_control;
	task_t                 XNU_PTRAUTH_SIGNED_PTR("conclave.task") c_task;
	thread_t               XNU_PTRAUTH_SIGNED_PTR("conclave.thread") c_downcall_thread;
	bitmap_t               c_service_bitmap[BITMAP_LEN(CONCLAVE_SERVICE_MAX)];
} conclave_resource_t;    
    \end{minted}
    \captionof{lstlisting}[\texttt{conclave\_resource\_t} type definition.]{\texttt{conclave\_resource\_t} type definition from \texttt{osfmk/kern/exclaves\_resource.h}.}
    \label{conclaveTypeDef}
\end{listing}

Conclaves hold an internal state and request. They also have boolean parameters indicating what we assume to be requested transitions. The \texttt{c\_control} field of type \texttt{tb\_client\_connection\_t} holds the Tightbeam communication interface for the Conclave, and is initiated in the previously looked at \texttt{exclaves\_conclave\_init} function. Conclaves furthermore hold a specific task and a downcall thread. Finally, they have a service bitmap field, which is populated with services belonging to the specific Conclave during resource enumeration (\cref{resourceEnumSection}). The services registered in this bitmap are uniquely identified by their own resource ID.
From the previous Exclave resource enumeration, we know Exclave resources holding Conclaves to be of type \path{XNUPROXY_RESOURCE_CONCLAVE_MANAGER}. We denote the Conclave manager resource holding the Conclave as the responsible Conclave manager for that specific Conclave. The \path{osfmk/kern/exclaves_resource.h}
 header reveals that \texttt{conclave\_request\_t} stores a requested state transition. The mapping are depicted in \cref{integerMappingConclaveRequest}.
\begin{table}[h!]
\centering
\begin{tabular}{|l|l|}
\hline
Value & State \\ \hline
0 & \texttt{CONCLAVE\_R\_NONE} \\ \hline
1 & \texttt{CONCLAVE\_R\_LAUNCH\_REQUESTED} \\ \hline
2 & \texttt{CONCLAVE\_R\_SUSPEND\_REQUESTED} \\ \hline
4 & \texttt{CONCLAVE\_R\_STOP\_REQUESTED} \\ \hline
\end{tabular}
\caption[\texttt{conclave\_request\_t} integer to request mapping.]{Integer to request mapping for \texttt{conclave\_request\_t}.}
\label{integerMappingConclaveRequest}
\end{table}
A request may be stored for all transitions into and out of a running Conclave, but not for the attaching and removing. The header also includes information on the Conclave lifecycle, which is discussed in \cref{conclaveLifecycle}.

We conclude that Conclaves are contained within Exclave resources of the Conclave manager type. They scope a variety of different services together, which are in turn also Exclave resources. Services are recognized as belonging to a Conclave via their domain. Every Conclave defines its own domain to which the services are assigned. The services are registered to the Conclave via a bitmap and uniquely identifiable by their own resource ID. We assume communication towards Conclaves to be performed using Tightbeam on the exposed Tightbeam connection field. The responsible Conclave manager appears to handle requests directed to the Conclave and is responsible for its management. We further know that for each Conclave domain, the respective Conclave manager has to be in the \path{com.apple.kernel} domain (\path{osfmk/kern/exclaves_resource.c}, ll. 156f). 
\subsection*{Conclave Lifecyle}
\label{conclaveLifecycle}
\begin{figure}[H]
    \centering
\includegraphics[scale=0.75]{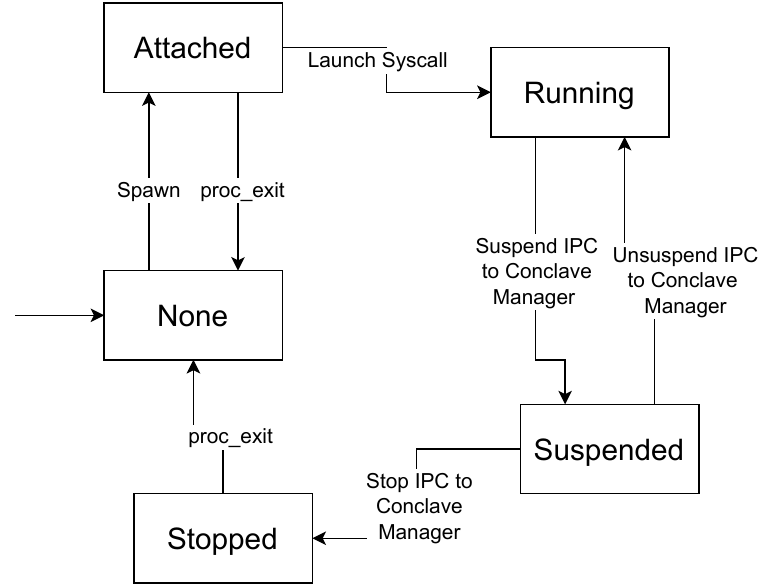}
    \caption[Conclave state machine.]{Conclave state machine adapted from the \path{osfmk/kern/exclaves_resource.h} header file.}
    \label{stateConclave}
\end{figure}

Regarding the Conclave-internal state, we are provided with additional information in the \path{exclaves_resource.h} header file. An included state machine are depicted in \cref{stateConclave}.

Conclaves can be attached to tasks and then started up via a system call. Whilst they are running, the responsible Conclave manager can be called into via the exposed Tightbeam connection. This is not possible whilst they are suspended. Stopping the Conclave also completely terminates \gls{IPC} to the Conclave manager. The respective integer-to-state mapping, as made available in the header file, can be seen below in \cref{integerMappingConclaveState}.

\begin{table}[h!]
\centering
\begin{tabular}{|l|l|}
\hline
Value & State \\ \hline
0 & \texttt{CONCLAVE\_S\_NONE} \\ \hline
1 & \texttt{CONCLAVE\_S\_ATTACHED} \\ \hline
2 & \texttt{CONCLAVE\_S\_RUNNING} \\ \hline
3 & \texttt{CONCLAVE\_S\_STOPPED} \\ \hline
4 & \texttt{CONCLAVE\_S\_SUSPENDED} \\ \hline
\end{tabular}
\caption[\texttt{conclave\_state\_t} integer to state mapping.]{Integer to state mapping for \texttt{conclave\_state\_t}.}
\label{integerMappingConclaveState}
\end{table}

We know that Conclaves can be attached to Mach tasks by the existence of the task field in the Conclave type and the respective Conclave field in the task type. Conclaves are found to be attached to tasks via \path{exclaves_conclave_attach} (\path{exclaves_resource.c}, l. 1347). The function sets the specific Conclave and task fields, and the Conclave state to \texttt{CONCLAVE\_S\_ATTACHED}.

We find it invoked from \texttt{task\_add\_conclave} (\path{osfmk/kern/task.c} l. 9948). The function reveals an entitlement constraint for attaching Conclaves. Only \texttt{launchd} and tasks with the \path{com.apple.private.exclaves.conclave-spawn} entitlement are allowed to attach a Conclave to a task. Furthermore, to have a Conclave attached, a task must have either the \path{com.apple.private.exclaves.conclave-host} or \path{com.apple.private.exclaves.conclave-spawn} entitlement.

\begin{figure}[H]
    \centering
\includegraphics[scale=0.75]{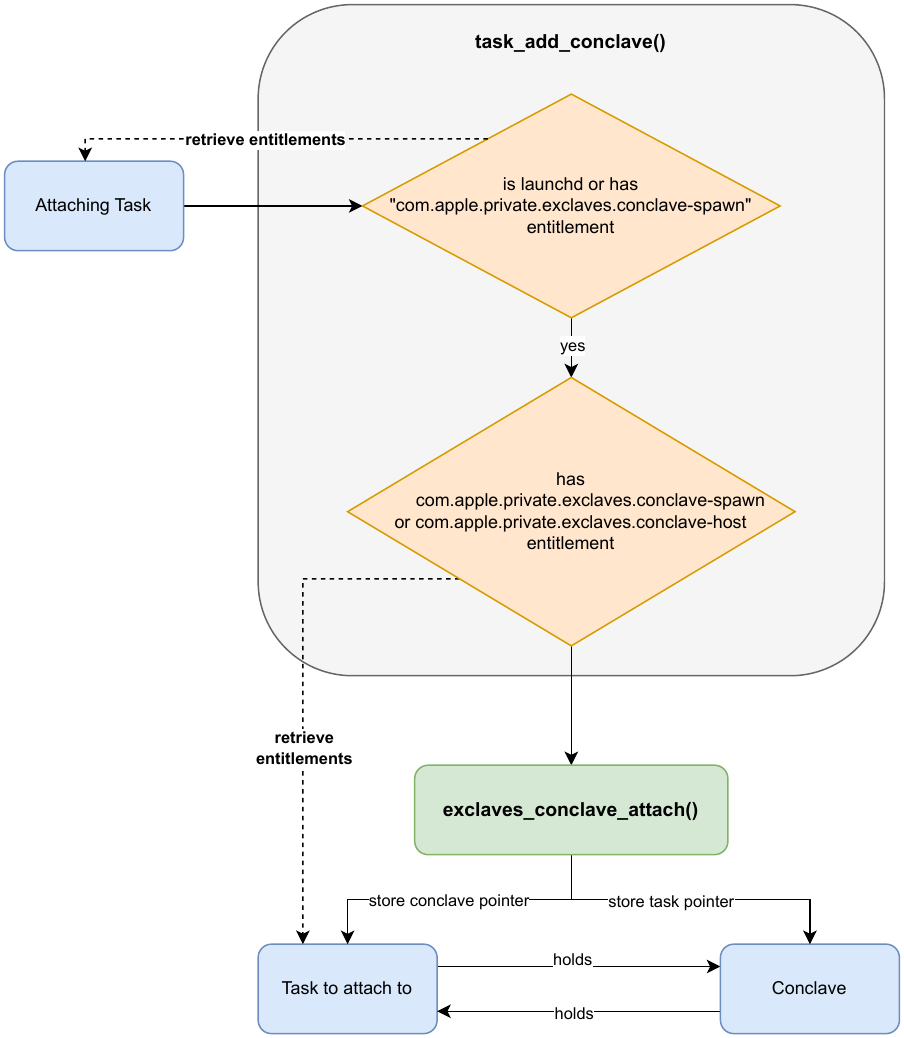}
    \caption[Conclave attachment to task control flow.]{Control flow for attaching a Conclave to a task.}
    \label{conclaveAttach}
\end{figure}

We find \texttt{task\_add\_conclave} to be called from \texttt{posix\_spawn} in \texttt{kern\_exec.c} (l. 4294), which shows that Conclaves are usually attached to a task during task spawn. The process of attaching a Conclave to a task is shown in \cref{conclaveAttach}.

As we have previously found that Conclaves are attached as part of \texttt{posix\_spawn}, it is logical that we find the Conclave ID to be a \gls{POSIX} spawn argument (\path{bsd/sys/spawn_internal.h}, l. 539). We can therefore examine launch property lists for tasks we know to have Conclaves attached (e.g., \texttt{audiomxd}) and, in fact, find a \texttt{\_Conclave} key. The entry in the \texttt{audiomxd} launch property list\footnote{Located at \path{/System/Library/LaunchDaemons/com.apple.audiomxd.plist}.} can be seen in \cref{lst:audiomxdlaunchplist}, and shows the corresponding Conclave identifier.
\begin{listing}[H]
    \begin{minted}[linenos, breaklines, bgcolor=LightGray, frame=lines]{c}
  "_Conclave" => "com.apple.audiomxd.conclave"    
    \end{minted}
    \captionof{lstlisting}{Conclave ID entry in launch property list of \texttt{audiomxd}.}
    \label{lst:audiomxdlaunchplist}
\end{listing}

Theoretically, it is also possible to attach Conclaves to tasks by directly calling \path{task_add_conclave} from an appropriately entitled task. However, we were unable to observe any such invocation. We therefore assume that Conclave attachment is solely done by launchd when spawning a task. Therefore, any task that requires an attached Conclaves needs to have the previously looked at key in its launch property list. %Additionally, for a task to have a conclave attached, %%%% -- incomplete sentence?!

\section{Exclaves Scoped Functionality}
From the retrieved resource list (see \cref{resourceEnumSection}), we can infer that Exclaves mainly scope functionality regarding sensor access and managing audio and video recording. This entails a variety of different resources and domains, ranging from sensor resources specific to audio drivers, and the relevant buffers to allow for copying out sensor data to XNU. We furthermore find that Apple scopes \gls{ANE} related functionality in Exclaves. Resources are subject to change, just like the rest of the Exclaves ecosystem. The amount of resources is increasing with newer firmware versions.

\section{Userspace Exclaves Functions}
\label{UserSpaceExclaveFunctions}
As userspace components need access to Exclaves to invoke operations on them, there exist userspace wrappers to allow for this. We can find these wrapper functions in the open source code XNU files in \path{libsyscall/wrappers/exclaves.c}. In the XNU source code, the libsyscall wrapper published a total of 19 functions. The total list, including function signatures, can be found in \cref{lst:exclave_wrapper}. The following functions are included: 
\begin{itemize}
    \item \texttt{exclaves\_endpoint\_call}
    \item \texttt{exclaves\_outbound\_buffer\_[create/copyout]}
    \item \texttt{exclaves\_inbound\_buffer\_[create/copyin]}
    \item \texttt{exclaves\_named\_buffer\_[create/copyin/copyout]}
    \item \texttt{exclaves\_launch\_conclave}
    \item \texttt{exclaves\_lookup\_service}
    \item \texttt{exclaves\_boot}
    \item \texttt{exclaves\_audio\_buffer\_[create/copyout/copyout]}
    \item \texttt{exclaves\_sensor\_[create/start/stop/status]}
    \item \texttt{exclaves\_notification\_create}    
\end{itemize}

We find most of these functions implemented in the \texttt{libsystem\_kernel.dylib} of the firmware's \gls{DYLD} shared cache. In the analyzed firmware, the \texttt{exclaves\_named\_buffer\_[create/copyin/\-copyout]} functions have not yet been fully implemented, but instead return the error value \texttt{0x46}, which maps to \emph{``(os/kern) service not supported''}. We assume this is, at least with regard to userspace invocation, unimplemented functionality that will be released in the future.

All implemented functions perform the actual invocation via the \texttt{exclaves\_ctl\_trap} Mach trap. Analyzing this function in Ghidra shows the invocation of Mach trap \texttt{-88}. A negative number is expected for a Mach trap, in contrast to standard \gls{POSIX} calls with positive numeration~\cite{kernelSyscall:2025}.

\subsection{Handling of \texttt{exclaves\_ctl\_trap}}
This section focuses on the kernel-side handling of \texttt{exclaves\_ctl\_trap} invocations by user-space components. The \texttt{exclaves\_ctl\_trap} handler is found at \texttt{0xfffffff00813a220} in the iOS 18.4 kernelcache, and in the \path{osfmk/kern/exclaves.c} file in \emph{xnu-11215}.
The handling of the \texttt{exclaves\_boot} is done separately from all other function invocations, as it requires different permissions, because tasks can not yet be associated with Conclaves.

\paragraph{Function Invocation Requirements}
Invoking Exclaves functionality (except for Exclave boot) has a very specific set of requirements. It can only be performed by ``properly entitled tasks which can operate in the kernel domain, or those which have joined conclaves'' (\path{osfmk/kern/exclaves.c}).

Verification on whether a task is allowed to operate in the kernel domain is performed by checking the \path{com.apple.private.exclaves.kernel-domain} entitlement. The kernel itself may also operate in the kernel domain. As another prerequisite for function execution, the execution waits until \texttt{Exclavecore} has successfully booted.

\begin{figure}[H]
  \centering
    \includegraphics[scale=0.75]{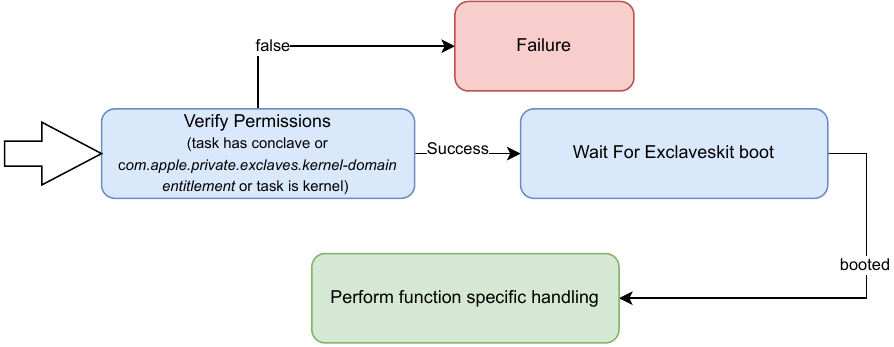}
    \caption[Exclaves user space call validation.]{Process of checking proper caller entitlement for an Exclave functionality user space invocation.}
    \label{fig:a}
\end{figure}

After performing the initial handling verification, the actual operations are performed with further operation-specific parameter verification. As we know, Conclaves are only attached to tasks based on a static set of resources and the correct boot arguments; the list of user-space components able to invoke Exclave functionality is strictly limited to those. 

\paragraph*{Notes on macOS}
While this paper focuses on Apple's mobile operating systems, some interesting findings regarding Exclave functionality in macOS came to light during an attempt to test functionality invocation on an M4 Mac mini. Trying to call any Exclave function from a user-space process context always resulted in a return of \emph{``(os/kern) service not supported''}. An analysis of macOS 15.4.1 revealed that the wrapper functions and the \texttt{exclaves\_ctl\_trap} are indeed implemented. However, they seem unsupported on the kernel side as of now, as the \gls{XNU} Mach control trap handler for the relevant trap is not implemented and instead returns the relevant error code\footnote{\texttt{exclaves\_ctl\_trap} is located at address \texttt{0xfffffe00087bdbf0} in \path{kernelcache.release.Mac16,5_6_7_8_9_11} in macOS 15.4.1 (24E263).}. It appears that Exclaves are only supported on iOS/iPadOS. This might change in the future. 

\subsection{Userspace Exclave Interaction -- Practical Experiment on SRD}
In an attempt to display our ability to interact with Exclaves from userspace, we perform an experiment in which we inject the \texttt{corespeechd} with a \texttt{dylib} that invokes Exclaves functionality. We choose \texttt{corespeechd} as a target for this attempt because we know it to have an attached Conclave, and it remains robust even when tampered with.
Unlike other candidates, \texttt{corespeechd} is stable and not regularly crashes (e.g., \texttt{audiomxd}). We can trace its control flow regarding Exclaves partially. As part of our preparations, we trace invocations of \texttt{exclave\_ctl\_trap} on the \gls{SRD}, and try to recreate them. An excerpt from the logs retrieved during dynamic reverse engineering of the \gls{SRD} can be found in  \cref{audiomxdllbsection} and \ref{sensorCreateSection}, which we provide as the first public look into live logs from Exclave-supported devices.

We successfully create a sensor resource and an audio buffer, receive a return port for both calls, and start the sensor resource; however, copying out of the audio buffer is not successful. Even though the access is permitted and we get successful return codes from the kernel, all bytes returned into the audio buffer are Null.

Furthermore, despite actively accessing an audio buffer, no \acrfull{SIL} appears.
When manually tracing for \gls{SIL} rendering in lldb while triggering the \gls{SIL} via on-board applications on iOS, parts of the rendering appears to be done by the userspace process.
This contradicts our expectation that \gls{SIL} rendering should take place and be monitored in \gls{GL}.

We are unsure why our setup fails, but we assume it is related to the buffer registration for the sensor to write to. Nonetheless, it can still be considered a success that we were able to create Exclaves resources and invoke functions on them from userspace from an appropriately entitled task. The log from the successful creation of the sensor and buffer via our custom \texttt{dylib} and sensor start is contained in \cref{logs}.

\begin{listing}[H]
    \begin{minted}[linenos, breaklines, bgcolor=LightGray, frame=lines]{c}
Jul  1 15:35:05 corespeechd(libExclaves5.A.dylib)[446] <Notice>: Sensor created: port=0x1003
Jul  1 15:35:05 corespeechd(libExclaves5.A.dylib)[446] <Notice>: Buffer created: port=0x1103
Jul  1 15:35:05 corespeechd(libExclaves5.A.dylib)[446] <Notice>: Allocated local memory
Jul  1 15:35:05 corespeechd(libExclaves5.A.dylib)[446] <Notice>: exclaves_sensor_status polled value: 0x1
Jul  1 15:35:06 corespeechd(libExclaves5.A.dylib)[446] <Notice>: Sensor started successfully.    
\end{minted}
\captionof{lstlisting}{Logs written from our custom dylib \texttt{libExclaves5.A.dylib} injected into corespeechd, invoking Exclaves functionalities from userspace.}
\label{logs}
\end{listing}

The source code for the dylib used for this experiment can be found on \href{https://github.com/Moritz209/exclaves_dylib}{GitHub}\footnote{\url{https://github.com/Moritz209/exclaves_dylib}}. It has been based on Apple's guide on creating dynamic libraries~\cite{dynamicLibraries}.

\section{Kernel Entry Point}
In the XNU open source code, we find the Exclave entry point for \gls{IPC} calls from the XNU kernel. The function \texttt{exclaves\_endpoint\_call} is declared in \path{osfmk/kern/exclaves.c}, and shown in \cref{kerneEntryPoint}.
\begin{listing}[H]
    \begin{minted}[linenos, breaklines, breakanywhere, bgcolor=LightGray, frame=lines]{c}
#pragma mark kernel entry points

kern_return_t
exclaves_endpoint_call(ipc_port_t port, exclaves_id_t endpoint_id,
    exclaves_tag_t *tag, exclaves_error_t *error)
{
#if CONFIG_EXCLAVES
	kern_return_t kr = KERN_SUCCESS;
	assert(port == IPC_PORT_NULL);

	Exclaves_L4_IpcBuffer_t *ipcb = Exclaves_L4_IpcBuffer();
	assert(ipcb != NULL);

	exclaves_debug_printf(show_progress,
	    "exclaves: endpoint call:\tendpoint id %lld tag 0x%llx\n",
	    endpoint_id, *tag);

	ipcb->mr[Exclaves_L4_Ipc_Mr_Tag] = *tag;
	kr = exclaves_endpoint_call_internal(port, endpoint_id);
	*tag = ipcb->mr[Exclaves_L4_Ipc_Mr_Tag];
	*error = XNUPROXY_CR_RETVAL(ipcb);

	exclaves_debug_printf(show_progress,
	    "exclaves: endpoint call return:\tendpoint id %lld tag 0x%llx "
	    "error 0x%llx\n", endpoint_id, *tag, *error);

	return kr;
#else /* CONFIG_EXCLAVES */
#pragma unused(port, endpoint_id, tag, error)
	return KERN_NOT_SUPPORTED;
#endif /* CONFIG_EXCLAVES */
}
    \end{minted}
    \captionof{lstlisting}[Kernel \texttt{exclaves\_endpoint\_call} Exclaves entry point function.]{The \texttt{exclaves\_endpoint\_call} function, which is kernel Exclaves entrypoint, from \path{osfmk/kern/exclaves.c}.}
    \label{kerneEntryPoint}
\end{listing}

The function is found to perform a call into Exclaves using \texttt{xnuproxy}. Our analysis further shows that it is called in the context of Tightbeam messaging. We will analyze this in \cref{Tightbeam}.

\section{Calling into Exclaves}
For the XNU open-source code alone, it becomes obvious that calling and scheduling into Exclaves is a notoriously complex task. It appears to be performed in various ways, some of which we will analyze in the following section. Our analysis here has no claim to completeness, and due to the significant parts of the redacted source code that we have to reverse engineer, we also cannot claim correctness.

\label{CallingIntoExclaves}
\subsection{Thread Scheduling}
We find that threads are scheduled into Exclaves when performing endpoint calls. This is confirmed by the inclusion of an interrupt state flag value in \path{thread.h} indicating current thread execution in \gls{SK} or Exclaves userspace. The flag value definition can be seen below in \cref{threadScheduleFlag}.

\begin{listing}[H]
\begin{minted}[linenos, breaklines, breakanywhere, bgcolor=LightGray, frame=lines]{c}
__options_decl(thread_exclaves_intstate_flags_t, uint32_t, {
	/* Thread is currently executing in secure kernel or exclaves userspace
	 * or was interrupted/preempted while doing so. */
	TH_EXCLAVES_EXECUTION                  = 0x1,
});
\end{minted}
\captionof{lstlisting}[Thread interrupt state value definition for execution in Secure Kernel or Exclaves userspace.]{Thread interrupt state value definition for execution in Secure Kernel or Exclaves userspace, in \path{osfmk/kern/thread.h}.}
\label{threadScheduleFlag}
\end{listing}
We find that this value is set before performing \texttt{sk\_enter} and unset afterward. From this, we know that \texttt{sk\_enter} performs the context switch via our known call to SPTM. This appears to be realized via scheduling context IDs, which we find are managed by xnuproxy. Threads get scheduled into the secure world in so-called downcalls, and may perform upcalls back to XNU when requesting XNU services during downcall handling~\cite{OnAppleExclaves:2025}. In this work, we did not focus on the scheduling process itself, but instead on the underlying communication mechanisms. Background on the Exclaves scheduling process can be found in previous work~\cite{DisassemblingAppleExclaves:2025}.

\subsection{Exclave thread execution states}
We can draw information regarding the different calls into Exclaves from \path{osfmk/kern/thread.h}, which lists \path{thread_exclaves_state_flags_t}, the different values it may hold, and provides additional context. We find this value to be consistently altered in \path{osfmk/kern/exclaves.c}, tracking the current execution status of the thread. A listing of the available values, their respective bit in the flag, and context provided in the header file can be seen in \cref{ExclaveCallTable}.

{
\footnotesize
\begin{longtable}{|l|l|p{5cm}|}
\hline
\textbf{Flag Value Name} & \textbf{Flag Value} & \textbf{Context} \\ \hline
\endfirsthead

\multicolumn{3}{c}%
{{\tablename\ \thetable{} -- continued from previous page}} \\
\hline
\textbf{Flag Value Name} & \textbf{Flag Value} & \textbf{Context} \\ \hline
\endhead

\hline \multicolumn{3}{r}{{Continued on next page}} \\ \hline
\endfoot

\hline
\caption[Settable \texttt{thread\_exclaves\_state\_flags\_t} values in a thread.]{Settable \texttt{thread\_exclaves\_state\_flags\_t} values in a thread. The flags indicate the current thread state regarding Exclave execution.}
\label{ExclaveCallTable}
\endlastfoot

\texttt{TH\_EXCLAVES\_RPC} & 0x1 & Thread is currently in RPC handling from either XNU or user space, but it may currently run in XNU whilst performing an upcall. Reentering Exclaves or returning to user space is not permitted. \\ \hline
\texttt{TH\_EXCLAVES\_UPCALL} & 0x2 & During RPC handling, the thread made an upcall RPC request back into XNU. Reentering Exclaves or returning to user space is not permitted. \\ \hline
\texttt{TH\_EXCLAVES\_SCHEDULER\_REQUEST} & 0x4 & Thread made an Exclaves scheduler request (e.g., wait, wake) from the xnu scheduler during RPC handling. Reentering Exclaves or returning to user space is not permitted. \\ \hline
\texttt{TH\_EXCLAVES\_XNUPROXY} & 0x8 & Thread is directly calling into xnuproxy.  Reentering Exclaves or returning to user space is not permitted. \\ \hline
\texttt{TH\_EXCLAVES\_SCHEDULER\_CALL} & 0x10 & Thread is directly calling into the Exclaves scheduler. Reentering Exclaves or returning to user space is not permitted. \\ \hline
\texttt{TH\_EXCLAVES\_STOP\_UPCALL\_PENDING} & 0x20 & Thread has called to stop upcall to XNU. \\ \hline
\end{longtable}
}
We will back-reference these found Exclaves thread states in our analysis of Exclaves communication mechanisms and try to use them to provide context regarding the different options for calling into Exclaves. 

\subsection{Calling into Exclaves -- Mach Ports}

The default mechanism for \gls{IPC} are Mach ports.
As found in \cref{exclaveType}, every Exclave resource holds an \texttt{ipc\_port\_t} field \texttt{r\_port}. We know \gls{IPC} ports to be a named equivalent to Mach ports of type \texttt{mach\_port\_t} (\path{osfmk/ipc/ipc_port.h}, l. 150), which are the core building blocks of \gls{IPC} in Apple operating systems. Mach ports serve as unidirectional communication endpoints for inter-task communication. They are associated with so-called \texttt{port rights} indicating allowed operations on the port (e.g., send or receive) for a specific task~\cite{Mach:2025}. Whilst the exact workings of \gls{IPC} via Mach can be deemed out of scope in this analysis, the existence of a Mach port on an Exclave resource indicates that it can be communicated with via standard \gls{IPC} mechanisms.

We find the Mach port of Exclave resources to get set during the previously looked at \texttt{exclaves\_resource\_init} function via a call to \texttt{exclaves\_resource\_alloc} (\path{osfmk/kern/exclaves_resource.c}, l. 578). The function allocates a kernel object port of the Exclave-specific type \path{IKOT_EXCLAVES_RESOURCE} in the kernel \gls{IPC} space and assigns it to the resource's port. The creation is shown in \cref{lst:portAlloc}.
\begin{listing}[H]
\begin{minted}[linenos, breaklines, bgcolor=LightGray, frame=lines]{c}
  /*
	 * Each resource has an associated kobject of type
	 * IKOT_EXCLAVES_RESOURCE.
	 */
	ipc_port_t port = ipc_kobject_alloc_port((ipc_kobject_t)resource,
	    IKOT_EXCLAVES_RESOURCE, IPC_KOBJECT_ALLOC_NSREQUEST);
	resource->r_port = port;    
\end{minted}
\captionof{lstlisting}{Kernel object port allocation for Exclaves resource in \protect \path{exclaves_resource_alloc}.}
\label{lst:portAlloc}
\end{listing}

This allocation does not yet register any send rights on the \gls{IPC} port; therefore, no tasks can call towards it as of now. We find a function that actually registers a send right on an Exclave resource port. This function is \path{exclaves_resource_create_port_name} in \path{osfmk/kern/exclaves_resource.c}. The function is shown in \cref{lst:createPort}.
\begin{listing}[H]
\begin{minted}[linenos, breaklines, bgcolor=LightGray, frame=lines]{c}
kern_return_t
exclaves_resource_create_port_name(exclaves_resource_t *resource, ipc_space_t space,
    mach_port_name_t *name)
{
	assert3u(os_atomic_load(&resource->r_usecnt, relaxed), >, 0);

	ipc_port_t port = resource->r_port;

	ip_mq_lock(port);

	/* Create an armed send right. */
	kern_return_t ret = ipc_kobject_make_send_nsrequest_locked(port,
	    resource, IKOT_EXCLAVES_RESOURCE);
	if (ret != KERN_SUCCESS &&
	    ret != KERN_ALREADY_WAITING) {
		ip_mq_unlock(port);
		exclaves_resource_release(resource);
		return ret;
	}

	/*
	 * If there was already a send right, then the port already has an
	 * associated use count so drop this one.
	 */
	if (port->ip_srights > 1) {
		assert3u(os_atomic_load(&resource->r_usecnt, relaxed), >, 1);
		exclaves_resource_release(resource);
	}

	ip_mq_unlock(port);

	*name = ipc_port_copyout_send(port, space);
	if (!MACH_PORT_VALID(*name)) {
		/*
		 * ipc_port_copyout_send() releases the send right on failure
		 * (possibly calling exclaves_resource_no_senders() in the
		 * process).
		 */
		return KERN_RESOURCE_SHORTAGE;
	}
	return KERN_SUCCESS;
}    
\end{minted}
\captionof{lstlisting}{The \protect \path{exclaves_resource_create_port_name} in \protect \path{exclaves_resource.c}.}
\label{lst:createPort}
\end{listing}

The function is parameterized with a reference to a resource, an \gls{IPC} space, and a reference to a \texttt{mach\_port\_name\_t}. An \gls{IPC} space is a tasks namespace for its associated ports and defines its \gls{IPC} capabilities such as sending and receiving~\cite{IPCSPACE:2025}. A Mach port name is ``a local identity for a Mach port'', defining rights held by a port in a specific context provided by an implied namespace being task-specific (\path{osfmk/mach/port.h}). Highly simplified, but condensed to the key parts, the function registers a send right on the \gls{IPC} port of the provided resource, and returns a name for it.

Looking for its invokers, we find it referenced from \texttt{\_exclaves\_control\_trap}, the kernel mach trap handler for userspace Exclave functionality invocations. More precisely, the function is called in the handling of \path{EXCLAVES_CTL_OP_SENSOR_CREATE}, \path{EXCLAVES_CTL_OP_AUDIO_BUFFER_CREATE}, \path{EXCLAVES_CTL_OP_NAMED_BUFFER_CREATE}, and \path{EXCLAVES_CTL_OP_NOTIFICATION_RESOURCE_LOOKUP} commands. This indicates that for sensor, audio buffer, named buffer, and notification resources, communication via standard \gls{IPC} is possible.

\begin{figure}[H]
    \centering
\includegraphics[scale=0.75]{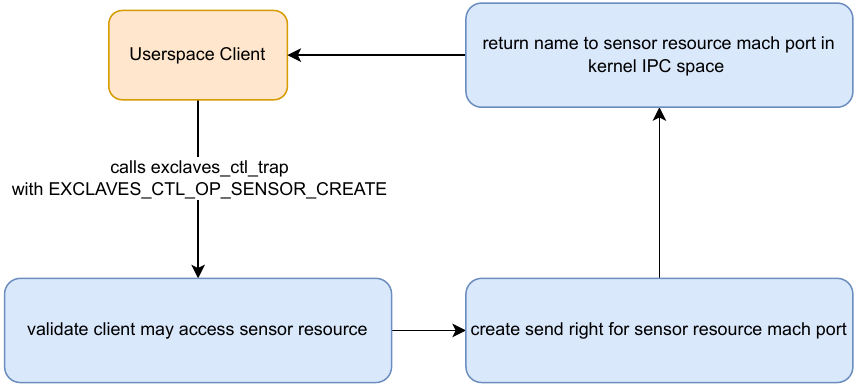}
    \caption{General control flow of \texttt{exclave\_ctl\_trap} call handling for sensor creation invocation.}
    \label{sensorIPCPort}
\end{figure}

Without diving into the details of user space Exclave function invocation, a call to the mach trap with such a creation command does in its essence validate that the client may access the resource (by validating the resource is in the domain defined by the tasks Conclave) and then creates a send right on the specific resources Mach port, and returning a name to it in kernel \gls{IPC} space. The process of send rights creation for a sensor resource Mach port is shown in \cref{sensorIPCPort}.

It is important to note that this process does not provide the calling userspace client with a direct Mach port with send rights into the sensor resource, but instead provides him with the name associated with the port in the kernel \gls{IPC} space. The kernel is now able to send messages to the sensor resource using this port, which is stored in the resources \texttt{r\_port} field, and also associates the created port name with the sensor resource.

This is used when receiving userspace calls for sensor start, as such a call is parameterized with the previously returned name port. The kernel then verifies the caller's entitlement to start the sensor by attempting to retrieve a resource for the provided port name.  It can be noticed that at no point in this shown communication the thread's \texttt{th\_exclaves\_state} field is changed. This confirms that the thread is not actually scheduled into Exclaves, but instead, communication relies on the standard \gls{IPC} mechanism. \todo{??} 

Even though calls into sensors, audio buffers, and named buffers appear to use standard \gls{IPC} mechanisms, they are secured by the usage of Exclave domains. Only properly validated tasks, in terms of the associated Conclave domain, may access resources within that domain.

\subsection{Calling into Exclaves -- seL4-style IPC Buffer}
We find that while specific Exclave resources (sensors, arbitrated buffers, audio buffers) use default \gls{IPC} mechanisms, this does not hold for other resource types, such as services and the Conclave manager. We see no function invocations on these resources assigning their respective port with any rights.
Therefore, a different communication interface must be used for communicating with Conclaves. For this, seL4 microkernel-style \gls{IPC} buffers are employed.

seL4 is an open-source operating system microkernel within the L4 microkernel family. It is supported by the seL4 Foundation. It has a significant focus on security and is provably secure in terms of the security properties of confidentiality, integrity, and availability~\cite{heiser2025sel4}.
Although Apple's implementation differs, it shares core ideas and concepts.

The Exclaves \gls{IPC} buffer definition can be seen in \path{osfmk/mach/exclaves_l4.h}:
\begin{listing}[H]
    \begin{minted}[linenos, breaklines, bgcolor=LightGray, frame=lines]{c}
/* ipc buffer object */
typedef struct __Exclaves_L4_Packed {
	/* message registers */
	Exclaves_L4_Word_t mr[Exclaves_L4_IpcBuffer_Mrs];
	/* source capability registers */
	Exclaves_L4_Word_t scr[Exclaves_L4_IpcBuffer_Crs];
	/* destination capability registers */
	Exclaves_L4_Word_t dcr[Exclaves_L4_IpcBuffer_Crs];
} Exclaves_L4_IpcBuffer_t;  
    \end{minted}
    \captionof{lstlisting}{Type definition for the Exclaves-specific IPC buffer.}
    \label{lst:XNU_exclaves_L4_ipcBuffer}
\end{listing}

In seL4, capabilities refer to ``a unique, unforgeable token that gives the processor permission to access an entity or object in system''~\cite{capabilities:2025}. An instantiation of such a capability can be interpreted as a pointer to an object with specific access rights, and is in that sense comparable to a port right. We can therefore assume that the Exclaves \gls{IPC} buffer can be used to transfer messages from \gls{XNU} to Exclaves and back, as well as references to objects with access rights.

The assumption that XNU employs an \gls{IPC} buffer for Exclaves communication is further supported by the fact that the \texttt{thread\_t} type definition in \path{osfmk/kern/thread.h} lists a \texttt{*th\_exclaves\_ipc\_buffer} field, which is the ``Per-thread IPC buffer for Exclaves communication''. 

\subsubsection{Exclaves IPC Buffer Allocation}
\label{IPCBufferAllocation}
Before proceeding further, we must perform a brief preemption regarding \texttt{xnuproxy}. As noted previously, \texttt{xnuproxy} is no longer a standalone binary in our analyzed firmware, but appears to have been integrated into the \texttt{sharedcache} binary. We will still refer to \texttt{xnuproxy} as a component, as we assume it still holds its central role regarding Exclaves management.

\begin{figure}[H]
%\hspace*{-1cm}% adjust as needed
\centering
\includegraphics[width=\textwidth]{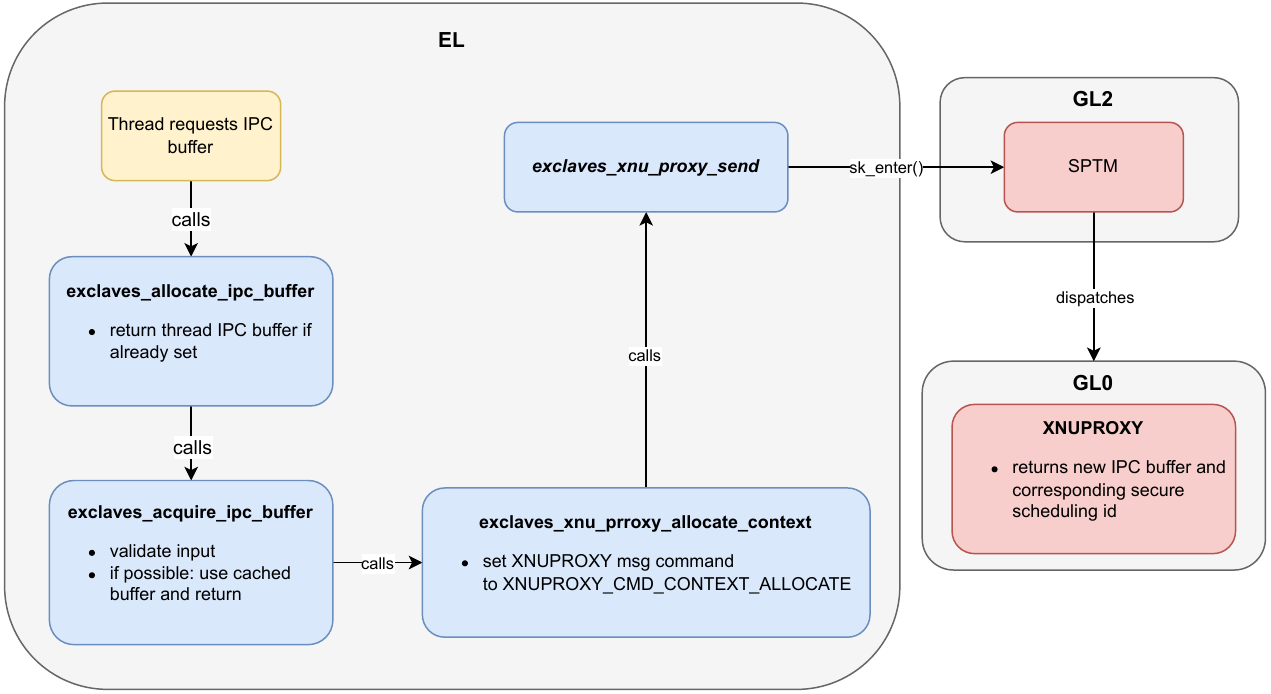}
    \caption[General control flow of an Exclave's IPC buffer request from a thread running in normal mode.]{General control flow of an Exclave's IPC buffer request from a thread running in normal exception levels. The request is forwarded to \texttt{xnuproxy} in the secure world via an \texttt{sk\_enter} call into SPTM.}
    \label{requestIPCBuffer}
\end{figure}

We know threads to be able to hold an \texttt{Exclaves\_L4\_IpcBuffer\_t}. For a corresponding Exclaves \gls{IPC} buffer threads do further hold a scheduling context id \texttt{th\_exclaves\_scheduling\_context\_id}, which is communicated to the exclave scheduling component (\path{osfmk/kern/thread.h}, l. 1007), and is required to correctly schedule requests (and with that, threads) into Exclaves via the \gls{IPC} buffer. The process of a thread requesting and allocating an Exclaves \gls{IPC} buffer is illustrated in \cref{requestIPCBuffer}.

The above figure shows the process of a thread requesting an \gls{IPC} buffer. We will only consider the case for the initial \gls{IPC} buffer request and ignore special cases, such as the reuse of cached buffers. Buffers are requested via an \texttt{xnuproxy} call with message command set to \path{XNUPROXY_CMD_CONTEXT_ALLOCATE}. This request is sent to \gls{SPTM} via \texttt{sk\_enter} and then dispatched to the Exclave GL0 management component, \texttt{xnuproxy}, which serves the request.

A more detailed look into \texttt{xnuproxy}, including how to call into it and request handling, will be provided in \cref{xnuproxySECTION}. As a result of the buffer request, the thread is provided with an \gls{IPC} buffer and a corresponding scheduling context ID, which is known to \texttt{xnuproxy} and can be used to call into Exclaves. 

Kernel threads request Exclaves \gls{IPC} buffers on receiving an Exclaves endpoint call. This indicates that \gls{IPC} buffers are the means of communicating endpoint calls to Exclaves.

\subsubsection{Exclaves IPC Buffer Usage}
Threads use Exclave \gls{IPC} buffers to perform calls into Exclave endpoints. We can infer more detailed usage by examining the process of a user-space endpoint call to Exclaves. A shortened listing of the userspace Exclaves endpoint call wrapper is shown in \cref{lst:exclavesEndpointCallWrapper}.

\begin{listing}[H]
\begin{minted}[linenos, breaklines, bgcolor=LightGray, frame=lines]{c}
kern_return_t
exclaves_endpoint_call(mach_port_t port, exclaves_id_t endpoint_id,
    mach_vm_address_t msg_buffer, mach_vm_size_t size, exclaves_tag_t *tag,
    exclaves_error_t *error)
{
	kern_return_t kr = KERN_SUCCESS;
	if (size != Exclaves_L4_IpcBuffer_Size) {
		return KERN_INVALID_ARGUMENT;
	}
	Exclaves_L4_IpcBuffer_t *ipcb;
	ipcb = Exclaves_L4_IpcBuffer_Ptr((void*)msg_buffer);
	ipcb->mr[Exclaves_L4_Ipc_Mr_Tag] = *tag;
	const uint32_t opf = EXCLAVES_CTL_OP_AND_FLAGS(ENDPOINT_CALL, 0);
	kr = EXCLAVES_CTL_TRAP(port, opf, endpoint_id, msg_buffer, size, 0, 0);
	*tag = ipcb->mr[Exclaves_L4_Ipc_Mr_Tag];
	*error = EXCLAVES_XNU_PROXY_CR_RETVAL(ipcb);
	return kr;
}    
\end{minted}
\captionof{lstlisting}{Simplified Exclave endpoint call userspace wrapper in \protect \path{libsyscall/wrappers/exclaves.c}.}
\label{lst:exclavesEndpointCallWrapper}
\end{listing}

To analyze such calls, we also need to consider the kernel-side handling, which is shown abbreviated for readability in \cref{lst:exclavesEndpointCallKernel}. The full listing can be found in \cref{app:kernelEndpointCallHandling}.
\begin{listing}[H]
\begin{minted}[linenos, breaklines, breakanywhere, bgcolor=LightGray, frame=lines]{c}
case EXCLAVES_CTL_OP_ENDPOINT_CALL: {
    Exclaves_L4_IpcBuffer_t *ipcb;
    if ((error = exclaves_allocate_ipc_buffer((void**)&ipcb))) {
        return error;
    }
    assert(ipcb != NULL);
    if ((error = copyin(ubuffer, ipcb, usize))) {
        return error;
    }
    
    /*
     * Verify that the service actually exists in the current
     * domain (only when the fallbacks are not enabled).
     */
    if (!exclaves_service_fallback &&
        !exclaves_conclave_has_service(task_get_conclave(task),
        identifier)) {
        return KERN_INVALID_ARGUMENT;
    }
    
    kr = exclaves_endpoint_call_internal(IPC_PORT_NULL, identifier);
    error = copyout(ipcb, ubuffer, usize);
   }
}    
\end{minted}
\captionof{lstlisting}{Simplified Exclave endpoint call handling in \protect \path{osfmk/kern/exclaves.c}.}
\label{lst:exclavesEndpointCallKernel}
\end{listing}

From \path{osfmk/kern/exclaves.h} we know the userspace wrapper \path{exclaves_endpoint_call} to be parameterized with a port (which has to be \path{IPC_PORT_NULL}), an endpoint ID being the identifier of the Exclave to send the RPC to, a tag parameter used ``for exclaves IPC tag'', and an error output parameter. We can infer the tag layout from its definition, as shown in \cref{lst:TagDefinition}.

\begin{listing}[H]
    \begin{minted}[linenos, breaklines, bgcolor=LightGray, frame=lines]{c}
/* Exclaves_L4_MessageTag_t
 *
 *  32                             0
 *  llllllllllllllll...nuuucccrrrrrr
 *
 * r: message registers (6 bits)
 * c: capability registers (3 bits)
 * u: unwrapped capabilities (3 bits)
 * n: non-blocking (1 bit)
 * l: label (16 bits)
 */

typedef Exclaves_L4_Word_t Exclaves_L4_MessageTag_t;
    \end{minted}
    \captionof{lstlisting}{\texttt{Exclaves\_L4\_MessageTag\_t} definition in \protect \path{osfmk/mach/exclaves_l4.h}.}
    \label{lst:TagDefinition}
\end{listing}

Whilst the exact usage of the tag has not yet been fully reverse-engineered, we can assume the label of the tag works similarly to standard seL4 \gls{IPC} message tags. We assume the kernel does not process it and instead passes it to the message target. It may hold information on a requested operation~\cite{seL4Manual}.

The \texttt{exclaves\_endpoint\_call} wrapper further receives a userspace buffer as an argument, for which it creates an Exclave's \gls{IPC} buffer pointer, and sets the buffer pointer's tag message register to the user-provided tag. To do so, the user space buffer is interpreted as an \path{Exclaves_L4_IpcBuffer_Ptr}. This buffer is then provided to the kernel via the \path{_exclaves_ctl_trap} mach trap.

The kernel side handling of the request calls the previously analyzed \path{exclaves_allocate_ipc_buffer} to request a new Exclave \gls{IPC} buffer. Then it copies the userspace-provided buffer contents, including the tag, into the newly allocated buffer. We assume that via this, userspace endpoint functionality requests are made available to exclaves.

\begin{figure}[H]
    \centering
\includegraphics[scale=0.75]{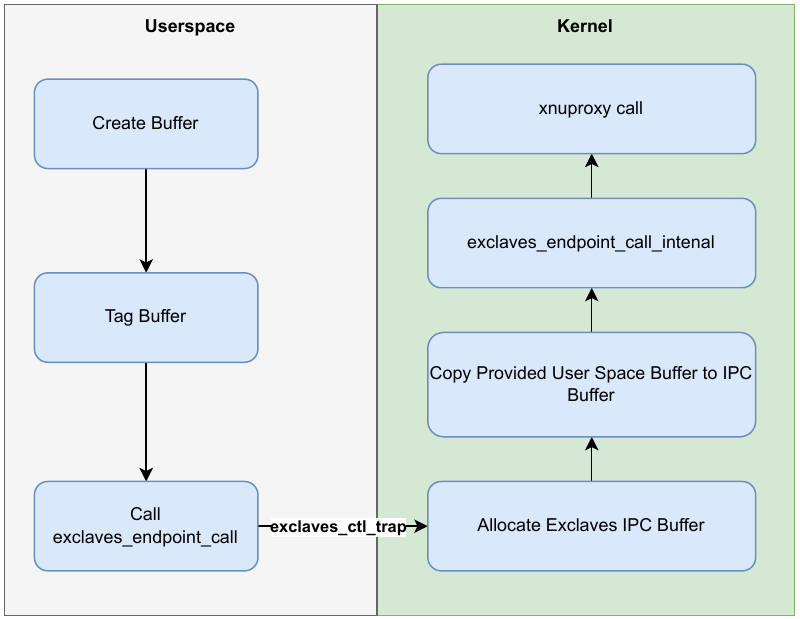}
    \caption{Process of issuing an Exclaves endpoint call from userspace and kernel side handling.}
    \label{endpointCall}
\end{figure}

After further verifying that the requested Exclave service exists, the actual call is performed via \texttt{exclaves\_endpoint\_call\_internal}. This call encodes the requested Exclave resource ID into the \gls{IPC} buffer and performs the actual call via a call to \path{exclaves_xnu_proxy_endpoint_call}. In this call, the thread Exclave state flag for \texttt{TH\_EXCLAVES\_RPC} is set, indicating that Exclave endpoint calls via \gls{IPC} buffers are the Exclaves RPC mechanism. We will look at the actual \texttt{xnuproxy} call execution in \cref{xnuproxySECTION}. The general process of issuing an Exclaves endpoint call from user space and the kernel side handling until the \texttt{xnuproxy} call can be seen in \cref{endpointCall}.

\subsection{Xnuproxy Communication}
\label{xnuproxySECTION}
The open source code parts of Exclaves show a multitude of references to \texttt{xnuproxy}. We initially assumed \texttt{xnuproxy} to be a specific GL0  component of the \texttt{Exclavecore}. We could confirm this in an older firmware version (22D82), where we found an individual \texttt{xnuproxy} binary. Interestingly, in the slightly newer firmware version that we focus on in this paper (22E40), this binary is no longer present. References to \texttt{xnuproxy} still exist. Based on pattern and string matching, we find that \texttt{xnuproxy} was moved into the \texttt{sharedcache} binary of the \texttt{Exclavecore}. This further demonstrates that the Exclaves and entire secure world system are still in active development and subject to constant change. For the sake of better understandability, we will still refer to \texttt{xnuproxy} when discussing such components in the \texttt{sharedcache} binary.

From \cref{ExclaveCallTable}, we find \texttt{TH\_EXCLAVES\_XNUPROXY} to indicate a thread directly calling into xnuproxy. Except for directly calling into \texttt{xnuproxy},
we find it to be responsible for Exclave request forwarding. It receives both downcalls and upcalls and forwards them to the intended targets. Downcalls are calls from XNU into Exclaves, whilst upcalls denote calls back into XNU during Exclave execution, as a result of a required service offered by XNU.
We can infer that \texttt{xnuproxy} is the respective handler for these from string references found in the \texttt{sharedcache} binary. An excerpt of the strings is shown in \cref{lst:xnuproxyString}.

\begin{listing}[H]
\begin{minted}[linenos, breaklines, bgcolor=LightGray, frame=lines]{c}
"[XNUP-D] Forwarding downcall for endpoint id %llu to cap 0x%lx with tag { words: %uh, capabilities: %uh, unwrapped: %uh, non_blocking: %s, label: 0x%04hx }\n"
----------------------------------------------
"[XNUP-D] Forwarded downcall for endpoint id %llu returned %lu with tag { words: %uh, capabilities: %uh, unwrapped: %uh, non_blocking: %s, label: 0x%04hx }\n"
----------------------------------------------
"[XNUP-U] Forwarding upcall to endpoint id %llu with tag 
0x%lx badge %lx\n"
----------------------------------------------
"[XNUP-U] Forwarded upcall to endpoint id %llu returned %lu with tag 0x%lx\n"
 ----------------------------------------------
"[XNUP-U] Invalid upcall endpoint id %llu, not forwarding and returning %lu\n"
\end{minted}
\captionof{lstlisting}[Strings with \texttt{xnuproxy} reference from sharedcache binary.]{Strings with \texttt{xnuproxy} reference from the sharedcache binary, allow for assumption with regards to its responsibilities. \texttt{[XNU-D]} appears to be downcall logging, and \texttt{[XNU-U]} upcall logging respectively.}
\label{lst:xnuproxyString}
\end{listing}

\texttt{xnuproxy} calls are realized via the Exclaves \gls{IPC} buffer, which allocations we looked at in \cref{IPCBufferAllocation}. 

\subsubsection{Direct Calls into \texttt{xnuproxy}}
\begin{figure}[H]
    \centering
\includegraphics[scale=0.75]{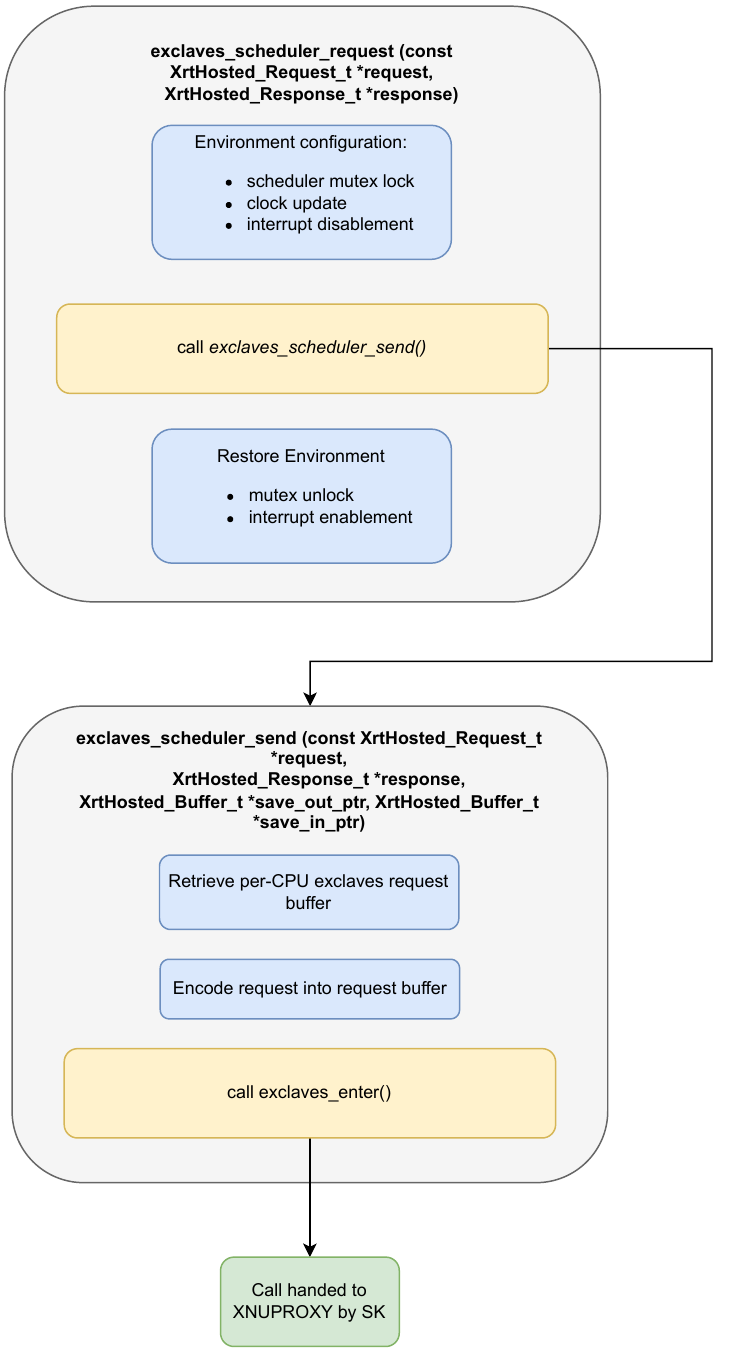}
    \caption[Control flow for direct \texttt{xnuproxy} call.]{Control flow of a call to \texttt{xnuproxy} directly.}
    \label{xnuproxy_direct}
\end{figure}

\cref{xnuproxy_direct}  shows the top-level control flow of a call from an XNU thread to \texttt{xnuproxy} directly as inferred from \gls{XNU} open-source code. It serves as an overview of the more detailed process that we analyze in the following. 
\begin{figure}[H]
    \centering
\includegraphics[scale=0.75]{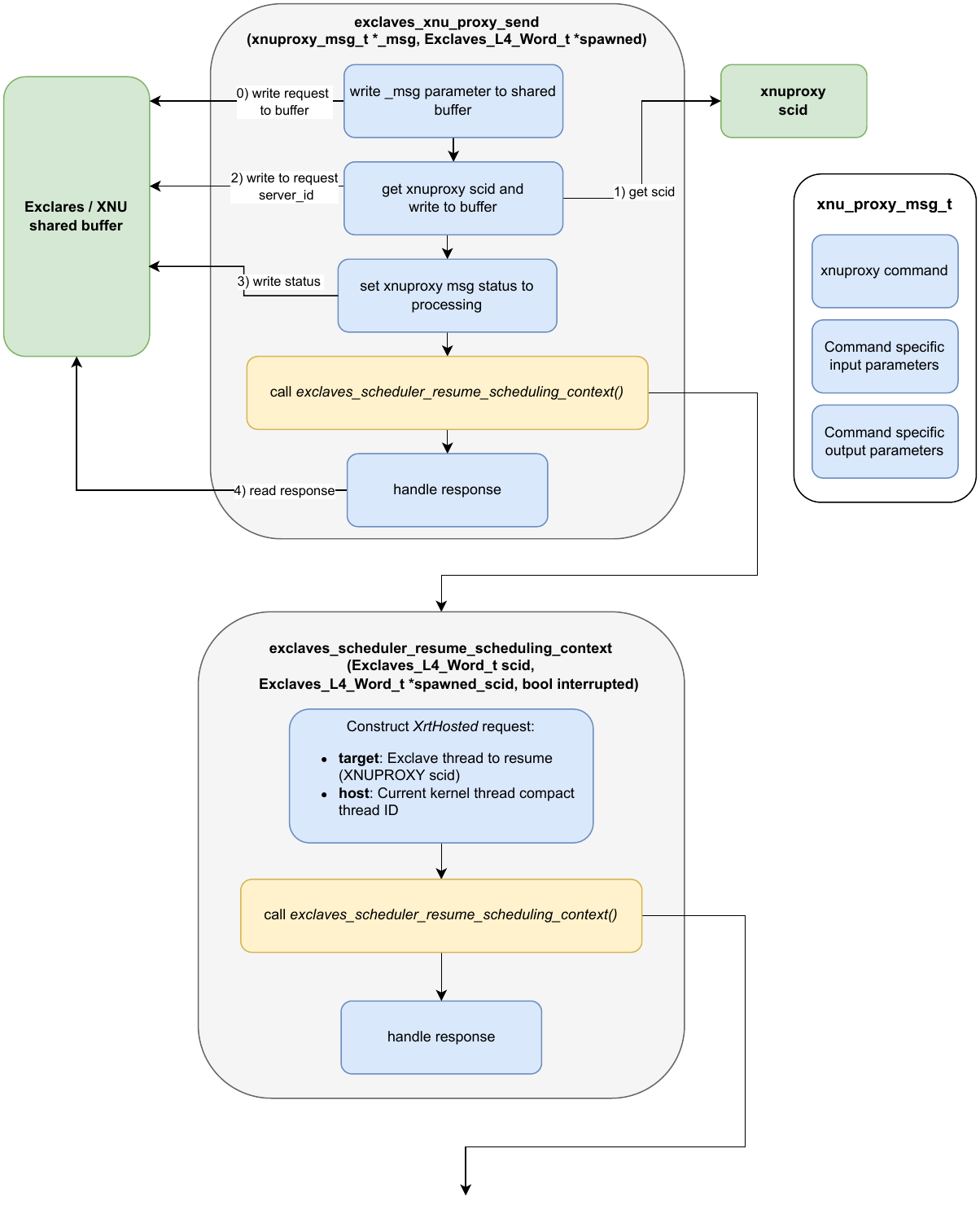}
\end{figure}

Direct calls to \texttt{xnuproxy} are found to be performed for management purposes. An example was seen earlier in \cref{resourceEnumSection}. Such calls are parameterized with an \texttt{xnuproxy\_msg\_t} message, created with a command and what appears to be further command-specific input parameters. The creation of the message for the Exclave resource enumeration is shown in \cref{lst:xnuproxymsgt}.
\begin{listing}[H]
\begin{minted}[linenos, breaklines, bgcolor=LightGray, frame=lines]{c}
xnuproxy_msg_t msg = {
			.cmd = xnuproxy_CMD_RESOURCE_INFO,
			.cmd_resource_info = (xnuproxy_cmd_resource_info_t) {
				.request.index = i,
			},
		};    
\end{minted}
\captionof{lstlisting}[Creation of \texttt{xnuproxy\_msg\_t} value in \protect \path{exclaves_resource_init}.]{Creation of \texttt{xnuproxy\_msg\_t} for Exclave resource enumeration in \path{exclaves_resource_init} (\path{osfmk/kern/exclaves_resources.c}).}
\label{lst:xnuproxymsgt}
\end{listing}

We can enumerate all available \texttt{xnuproxy} management commands from a \texttt{cmd\_to\_str} function in \path{osfmk/kern/exclaves.c} (l. 2311). The full list can be found in \cref{XNUPROXYCommands}. We find a variety of commands similar to those callable from userspace (sensor create/start/stop, buffer copyout/create). Furthermore, there are commands that we have not yet seen before and assume to be relevant for Exclave management (buffer map, buffer layout, context allocate, context free).

\paragraph*{exclaves\_xnu\_proxy\_send}

The previously created \texttt{xnuproxy} message is passed as an input parameter to \texttt{exclaves\_xnu\_proxy\_send} (\texttt{osfmk/kern/exclaves.c} l. 2402). The listing for this core function can be found in \cref{xnuproxysend}. As the process of sending a message to \texttt{xnuproxy} appears to be quite complex, we will attempt to approach it from a high-level perspective.

To send a message directly to \texttt{xnuproxy}, the calling thread writes its request message to the shared Exclaves \gls{IPC} buffer previously allocated (see \cref{IPCBufferAllocation}). It further retrieves the globally stored \texttt{xnuproxy} \gls{SCID}, which is set in the Exclaves setup process, and stores it in the Exclave buffers server ID field. This denotes the target of the call to perform. The call is progressed via \texttt{exclaves\_scheduler\_resume\_scheduling\_context}, which is parameterized with the \texttt{xnuproxy} SCID. Based on the input SCID and the calling threads \gls{CTID}, a so-called \texttt{XrtHosted\_Response\_t} is created, and input into the \texttt{exclaves\_scheduler\_request} call together with an output pointer. We find this function to be responsible for environment configuration in preparation for the call to xnuproxy, after which it forwards the call to \texttt{exclaves\_scheduler\_send}. This function finally writes the request to a per-CPU Exclaves request buffer structure and performs an \texttt{exclaves\_enter} call, which ultimately leads to entering SPTM via a call to \texttt{sk\_enter}. 

Whilst we have not yet fully reverse-engineered the secure world request handling and thread scheduling, our previous analysis of \gls{SK} provides us with a strong base for assumptions.
From the \texttt{sk\_enter} call, we know \gls{SK} to get dispatched from \gls{SPTM}. For an invoking \texttt{exclaves\_enter} call, we know SK to hand control to a GL0 component (see \cref{SK_ERET}. We presume this component to be xnuproxy, which, upon invocation, recognizes that a request has been made and retrieves information via the shared buffer structures. The exact inner workings of this secure world side handling have not yet been reverse-engineered by us and are left as future work. 

\subsubsection{Exclaves Endpoint Calls}
 
Next to explicitly calling into \texttt{xnuproxy} for management purposes, we find that \texttt{xnuproxy} is also responsible for performing endpoint calls to Exclaves. A kernel thread receiving an endpoint call via \texttt{exclaves\_ctl\_trap} serves this call by calling \path{exclaves_endpoint_call_internal} on the provided endpoint identifier to call, which shows to be a wrapper for \path{exclaves_xnu_proxy_endpoint_call}. We find this call to store the provided endpoint ID in the thread's per-thread Exclaves \gls{IPC} buffer endpoint field, and call \path{exclaves_scheduler_resume_scheduling_context} parameterized with the thread \gls{SCID}.
We assume this to be done to correctly schedule the thread once it reaches secure world. At this point, call handling is comparable to calls directly to xnuproxy. We assume that after receiving the call, \texttt{xnuproxy} identifies the call as a proper endpoint call via the set endpoint field in the \gls{IPC} buffer, and performs forwarding for thread scheduling of the requested task into the secure world. 
 
\subsection*{Xnuproxy Conclusions}
We find \texttt{xnuproxy} to be responsible for calling into Exclaves running in the secure world. It appears to interplay with the Exclaves GL0 \texttt{scheduler}, which is yet to be analyzed. It additionally serves as a GL0 management component, being callable with various management commands. A key aspect of this is the provision of shared \gls{IPC} buffer and scheduling context IDs to threads trying to call into exclaves, which appear to be required to perform such calls by scheduling the calling threads into the secure world. Furthermore, it acts in the capacity of a request forwarder in the secure world, dispatching incoming requests to the endpoints of the correct targets.

\section{Tightbeam}
\label{Tightbeam}
Looking into Exclaves communication, we find a variety of references to Tightbeam. Tightbeam is an \texttt{ExclaveKit} private framework. It serves as an Exclave communication framework for secure world components. We find that \gls{XNU} can also utilize it. The underlying transport mechanisms for Tightbeam are found to be \texttt{xnuproxy} for calls from XNU, which we previously found to perform Exclaves endpoint calls via Exclave dedicated \gls{IPC} buffers.

\subsection{Tightbeam Types}
Before diving into a deeper analysis of Tightbeam's inner workings, it is essential to lay the fundamental groundwork by reverse-engineering Tightbeam's specific data types. We are aware of the existence of these types due to their usage in the XNU open-source code. We find the following Tightbeam types directly in the existing source:
\begin{itemize}
    \item \texttt{tb\_message\_t}
    \item \texttt{tb\_endpoint\_t}
    \item \texttt{tb\_client\_connection\_t}
\end{itemize}

We can partially reconstruct these via the Tightbeam binary by looking at getter and setter functions. Following this approach allows us to further define other Tightbeam-specific data types, which will be used in the reverse engineering of Tightbeam. This includes the following types:

\begin{itemize}
    \item \texttt{tb\_transport}
    \item \texttt{tb\_connection}
\end{itemize}

The following will define these types as far as possible without official sources.

\subsubsection{Tightbeam Message Type \texttt{tb\_message\_t}}
\label{section:typedef_tbmessage}
We find multiple setter and getter functions for \texttt{tb\_message\_t} in the Tightbeam binary. The binary provides a large number of exported symbols, and these functions can be found by looking for the characteristic \texttt{\_tb\_message\_set\_<field>} and \texttt{\_tb\_message\_get\_<field>} methods. 
The reconstructed type definitions are shown in \cref{table_typedef_tbmessage}.

\begin{table}[h!]
\centering
\footnotesize
\begin{tabular}{|l|l|l|l|}
\hline
\textbf{Offset} & \textbf{Field} & \textbf{Type} & \textbf{Notes} \\ \hline
0x00 & \texttt{state} & \texttt{uint32\_t} & Message state value \\ \hline
0x04 & \texttt{disposition} & \texttt{uint8\_t} & Message disposition \\ \hline
0x08 & \texttt{connection\_identifier} & \texttt{uint64\_t} & Connection ID \\ \hline
0x10 & \texttt{client\_identifier} & \texttt{uint64\_t} & Client ID \\ \hline
0x18 & \texttt{msg\_identifier} & \texttt{uint64\_t} & Message ID \\ \hline
0x48 & \texttt{num\_caps} & \texttt{uint64\_t} & Unclear \\ \hline
0x50 & \texttt{transport\_buffer} & \texttt{tb\_transport\_message\_buffer\_t} & Transport buffer \\ \hline
\end{tabular}
\caption[\texttt{tb\_message\_t} layout inferred from reverse engineering of Tightbeam.]{Inferred partial layout of \texttt{tb\_message\_t} based on reverse engineering of the Tightbeam framework.}
\label{table_typedef_tbmessage}
\end{table}

We have further inferred on available values for the state and disposition field based on their usage in the Tightbeam binary and corresponding assertion failure strings. The detected values for the message state field are shown in \cref{message_state_tab}.

\begin{table}[h!]
\centering
\begin{tabular}{|l|l|}
\hline
Value & State \\ \hline
0 & \texttt{TB\_MESSAGE\_STATE\_UNINITIALIZED} \\ \hline
1 & \texttt{TB\_MESSAGE\_STATE\_PREPARING} \\ \hline
2 & \texttt{TB\_MESSAGE\_STATE\_READY} \\ \hline
3 & \texttt{TB\_MESSAGE\_STATE\_SENT} \\ \hline
4 & \texttt{TB\_MESSAGE\_STATE\_RECEIVED} \\ \hline
\end{tabular}
\caption[\texttt{tb\_message\_t} state values in Tightbeam binary.]{\texttt{tb\_message\_t} state values detected in the Tightbeam binary.}
\label{message_state_tab}
\end{table}

The corresponding enumeration for the disposition field can be seen in \cref{message_disposition_tab}.
\begin{table}[h!]
\centering
\begin{tabular}{|l|l|}
\hline
Value & State \\ \hline
1 & \texttt{TB\_MESSAGE\_DISPOSITION\_QUERY} \\ \hline
2 & \texttt{TB\_MESSAGE\_DISPOSITION\_REPLY} \\ \hline
\end{tabular}
\caption[\texttt{tb\_message\_t} disposition values detected in Tightbeam.]
{\texttt{tb\_message\_t} disposition values detected in the Tightbeam binary.}
\label{message_disposition_tab}
\end{table}

\subsubsection{Tightbeam Endpoint Type \texttt{tb\_endpoint\_t}}
\label{section:typedef_tbendpoint}
We can perform a similar type reconstruction based on Tightbeam binary getter and setter methods for the \texttt{tb\_endpoint\_t} type. The results can be seen below in \cref{table_typedef_tbendpoint}.
\begin{table}[h!]
\centering
\small
\begin{tabular}{|l|l|l|l|}
\hline
Offset & Field & Type & Notes \\ \hline
0x00 & \texttt{type} & \texttt{uint32\_t} & Endpoint type identifier \\ \hline
0x04 & \texttt{options} & \texttt{uint32\_t} & Endpoints options or flags \\ \hline
0x08 & \texttt{interface\_identifier} & \texttt{uint64\_t} & Interface identifier \\ \hline
0x20 & \texttt{data} / \texttt{value} & \texttt{uint64\_t} & Payload data or value associated \\ \hline
0x28 & \texttt{validity flag }& \texttt{uint8\_t} & Inferred to by usage \\ \hline
\end{tabular}
\caption[\texttt{tb\_endpoint\_t} layout inferred from reverse engineering of Tightbeam.]{Inferred partial layout of \texttt{tb\_endpoint\_t} based on reverse engineering of the Tightbeam framework.}
\label{table_typedef_tbendpoint}.
\end{table}

\subsubsection{Tightbeam Client Connection Type \protect \path{tb_client_connection_t}}
\label{section:typedef_tbclientconnection}
We find no direct setter and getter methods for the \texttt{tb\_client\_connection\_t} type, and assume it to be a wrapper type around the more generic \texttt{tb\_connection\_t}, which will be looked at in \cref{section:typedef_tbconnection}
\subsubsection{Tightbeam Transport Type \texttt{tb\_transport\_t}}
\label{section:typedef_tbtransport}
We find no usage of this type in the XNU open-source code, but we can still infer its layout based on its few getter and setter methods. Despite large parts of the type being unknown, we still list it in \cref{table_typedef_tbtransport} for completeness' sake.

\begin{table}[h!]
\centering
\begin{tabular}{|l|l|l|p{6cm}|}
\hline
Offset & Field & Type & Notes \\ \hline
0x08 & \texttt{endpoint\_data} & \texttt{uint64\_t} & Data field retrieved from endpoint, inferred in \cref{subsub:ClientConnectionCreation} \\ \hline
0x48 & \texttt{vtable} & \texttt{uint64\_t} & Pointer to vtable holding transport type specific function implementations, inferred in \cref{subsub:ClientConnectionCreation} \\ \hline
0x50 & \texttt{message\_handler} & \texttt{uint64\_t} & Message handler \\ \hline
0x68 & \texttt{context} & \texttt{uint64\_t} & Context \\ \hline
\end{tabular}
\caption[\texttt{tb\_transport\_t} layout inferred from reverse engineering of Tightbeam.]{Inferred partial layout of \texttt{tb\_transport\_t} based on reverse engineering of the Tightbeam framework.}
\label{table_typedef_tbtransport}
\end{table}

\subsubsection{Tightbeam Connection Type \texttt{tb\_connection\_type\_t}}
\label{section:typedef_tbconnection}
Similar to the \texttt{tb\_transport\_t} type before, we find some fields via setter and getter methods for the \texttt{tb\_connection\_type\_t}. We include them anyway, and find a \texttt{transport} field in the datatype. We assume this to be of type \texttt{tb\_transport\_t}. The datatype can be seen below in \cref{table_typedef_tbconnection}.
\begin{table}[h!]
\centering
\begin{tabular}{|l|l|l|l|}
\hline
Offset & Field & Type & Notes \\ \hline
0x0 & \texttt{transport} & \texttt{uint64\_t} & Transport, assumed type \texttt{tb\_transport\_t} \\ \hline
0x8 & \texttt{observers} & \texttt{uint64\_t} & Observers \\ \hline
\end{tabular}
\caption[\texttt{tb\_connection\_t} layout inferred by reverse engineering of Tightbeam.]{Inferred partial layout of \texttt{tb\_connection\_t} based on reverse engineering of the Tightbeam framework.}
\label{table_typedef_tbconnection}
\end{table}

\subsubsection{Tightbeam Transport Message Buffer Type \protect \path{tb_transport_message_buffer_t}}

A key type used by Tightbeam is \texttt{tb\_transport\_message\_buffer\_t}. While no fields were directly inferable from getter and setter methods, our reverse engineering revealed large parts of the structure. 

\begin{table}[h!]
\centering
\begin{tabular}{|l|l|l|l|}
\hline
Offset & Field & Type & Notes \\ \hline
0x0 & \texttt{type} & \texttt{uint64\_t} & \begin{tabular}[c]{@{}l@{}}Used to encode function \\ to call at endpoint.\end{tabular} \\ \hline
0x8 & \texttt{wrapping} & \texttt{uint64\_t} &  \\ \hline
0x10 & \texttt{offset} & \texttt{uint64\_t} & Current offset into the buffer. \\ \hline
0x18 & \texttt{size} & \texttt{uint64\_t} & Total buffer size. \\ \hline
\end{tabular}
\caption[\texttt{tb\_transport\_t} layout inferred from reverse engineering of Tightbeam.]{Inferred partial layout of \texttt{tb\_transport\_message\_buffer\_t} based on reverse engineering of the Tightbeam framework.}
\end{table}

The exact usage of the wrapping field remains unclear at this time. Our analysis will further demonstrate that the type field is used to encode a function identifier into the buffer.

\subsection{Tightbeam Communication}
We can reasonably assume that Tightbeam is a mechanism that supports calls towards Exclaves, and specifically Conclaves. This is supported by the fact that, as previously seen, every Conclave resource has a \texttt{tb\_client\_connection\_t}, which we assume to be a Tightbeam communication interface (see \cref{Conclave}). 
Tightbeam usage is largely redacted in the XNU open-source code. Still, we can infer its presence from the general structure of the available sections and create a more complete picture through direct reverse engineering.

We find various references to Tightbeam in \path{osfmk/kern/exclaves_driverkit.c} function \path{hello_driverkit_interrupts}, which appears to be a testing/debug build only test function. Whilst this does mean that the function will not be included in our production build firmware, it also offers more detailed insights into Tightbeam control flow. Communication via Tightbeam can be divided into several phases. An excerpt from the relevant function is shown below in \cref{driverKitTightbeam}. 
\begin{listing}[H]
\begin{minted}[linenos, breaklines, bgcolor=LightGray, frame=lines]{c}
#define EXCLAVES_ID_HELLO_INTERRUPTS_EP              \
    (exclaves_service_lookup(EXCLAVES_DOMAIN_KERNEL, \
    "com.apple.service.HelloDriverInterrupts"))
    
	tb_endpoint_t ep = tb_endpoint_create_with_value(
		TB_TRANSPORT_TYPE_XNU, EXCLAVES_ID_HELLO_INTERRUPTS_EP, 0);

	tb_client_connection_t client =
	    tb_client_connection_create_with_endpoint(ep);

	tb_client_connection_activate(client);
\end{minted}
\captionof{lstlisting}[Simplified initial Tightbeam interaction in \protect \path{hello_driverkit_interrupts}.]{Simplified initial Tightbeam interaction in \texttt{hello\_driverkit\_interrupts} in \path{exclaves_driverkit.c}.}
\label{driverKitTightbeam}
\end{listing}

\subsubsection{Endpoint Creation}
\label{subsub:endpointCreation}
The initial step of Tightbeam communication is the creation of an endpoint. In the above excerpt, this is realized via a call to \texttt{tb\_endpoint\_create\_with\_value}. The call is parameterized with a transport type \footnote{Transport type is \texttt{TB\_TRANSPORT\_TYPE\_XNU} for all \gls{XNU} invocations.}, an identifier, and an endpoint options parameter\footnote{Transport options are \texttt{TB\_ENDPOINT\_OPTIONS\_NONE} or 0 for all XNU invocations.}. 

The identifier appears to be an Exclave resource \texttt{r\_id}\todo{ref to type}, retrievable via a call to \texttt{exclave\_service\_lookup} with resource name and domain. The endpoint creation function has a return type of \texttt{tb\_endpoint\_t}. Its implementation is redacted from the XNU open-source code.

We find the function with a symbolicated name implemented in the Tightbeam binary. The reassembled function is shown in \cref{lst:TB_EP_CREATE}.
\begin{listing}[H]
\begin{minted}[linenos, breaklines, bgcolor=LightGray, frame=lines]{c}
void _tb_endpoint_create_with_value(undefined4 type,long id,undefined4 options)
{
  tb_endpoint_t *puVar1;
  
  puVar1 = (tb_endpoint_t *)0x1;
  func_0x00026258b090(1,0x60,0x1082040faca7f44);
  if (puVar1 != (tb_endpoint_t *)0x0) {
    puVar1->type = type;
    puVar1->options = options;
    puVar1->data = id;
    puVar1->validity_flag = 1;
    return;
  }
  TB_ASSERT_PTR_!=_NULL_FAIL();
}    
\end{minted}
\captionof{lstlisting}[Tightbeam \protect \path{_tb_endpoint_create_with_value} function.]{The \texttt{\_tb\_endpoint\_create\_with\_value} function in Tightbeam, symbolicated by hand. The source code was disassembled by Ghidra.}
\label{lst:TB_EP_CREATE}
\end{listing}

The function appears to show a memory allocation via a function call to a non-resolved function. This function is not resolved in the binary. After the successful allocation, an endpoint of type \texttt{tb\_endpoint\_t} is initialized with the provided parameters. Although not clearly visible in the reassembled code excerpt, based on usage, we assume the call returns the created endpoint.

Based on the implementation in Tightbeam, its control flow, and assumed usage in the kernel, we can find the respective function implementation in the kernel. The reassembled code excerpt can be seen in \cref{lst:TB_EP_CREATE_XNU}.
We reconstructed symbols, in particular types, variable names, and functions.
\begin{listing}[H]
\begin{minted}[linenos, breaklines, bgcolor=LightGray, frame=lines]{c}
tb_endpoint_t * tb_endpoint_create_with_value(undefined4 type,long id,undefined8 ep_options)

{
  tb_endpoint_t *endpoint;
  undefined4 *puVar1;
  undefined8 ep_options_copy;
  undefined4 in_w3;
  undefined1 auVar2 [12];
  
  ep_options_copy = ep_options;
  endpoint = (tb_endpoint_t *)allocate_in_structure(&struct_tb_endpoint_s,4);
  if (endpoint != (tb_endpoint_t *)0x0) {
    endpoint->type = type;
    endpoint->options = (int)ep_options;
    endpoint->data = id;
    endpoint->validity_flag = 1;
    return endpoint;
  }
  auVar2 = TB_ASSERT_PTR_!=_NULL_FAIL();
  puVar1 = auVar2._0_8_;
  *puVar1 = auVar2._8_4_;
  puVar1[1] = in_w3;
  *(undefined1 *)(puVar1 + 10) = 0;
  *(undefined8 *)(puVar1 + 8) = ep_options_copy;
  return (tb_endpoint_t *)0x0;
}    
\end{minted}
\captionof{lstlisting}[Kernel \protect \path{tb_endpoint_create_with_value} function.]{\texttt{tb\_endpoint\_create\_with\_value} in the kernel binary, symbolicated by hand. The source code was disassembled by Ghidra.}
\label{lst:TB_EP_CREATE_XNU}
\end{listing}

\begin{figure}[H]
    \centering
\includegraphics[width=250pt]{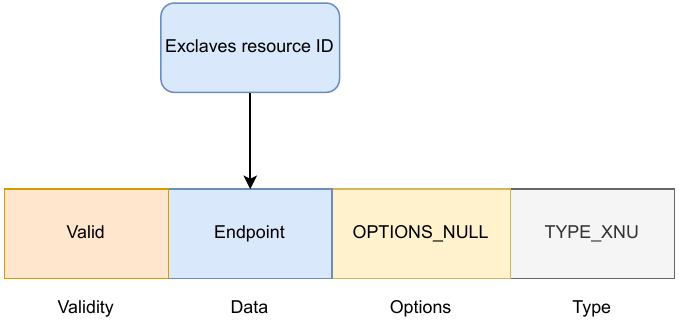}
\caption[Tightbeam \texttt{tb\_endpoint\_t} endpoint created after call to \protect \path{tb_endpoint_create_with_value} in XNU.]{Tightbeam \texttt{tb\_endpoint\_t} endpoint created after call to \path{tb_endpoint_create_with_value} in XNU.}
 \label{fig:endpoint}
 \end{figure}

The previously redacted allocation function in the Tightbeam binary is shown to be an allocator to a passed structure in the kernel implementation. Otherwise, the kernel function appears to be similar to the one found in the Tightbeam binary. Having found the relevant function implementation in the kernel, we can infer that the integer value for \path{TB_TRANSPORT_TYPE_XNU} is 7. This is based on all detectable calls to the function in the kernel. It should be noted that the input identifier, a unique Exclave resource identifier, is written to the endpoint data field. A resulting endpoint structure is shown in \cref{fig:endpoint}.

\subsubsection{Client Connection Creation}
\label{subsub:ClientConnectionCreation}
Recalling \cref{driverKitTightbeam}, the newly created endpoint is now used as an input parameter in a function call to \texttt{tb\_client\_connection\_create\_with\_endpoint}. We find the relevant function in XNU and can label it, as well as its called functions, based on the implementation in Tightbeam. The function implementation in the kernel can be seen in \cref{lst:TB_CONNECTION_CREATE_XNU}.
\begin{listing}[H]
\begin{minted}[linenos, breaklines, bgcolor=LightGray, frame=lines]{c}
tb_connection_t * tb_client_connection_create_with_endpoint(tb_endpoint_t *ep)
{
  tb_connection_t *ptVar1;
  __tb_connection_create_transport_for_endpoint(ep,0);
  ptVar1 = (tb_connection_t *)xnu_tb_connection_create();
  xnu_tb_endpoint_destruct(ep);
  return ptVar1;
}    
\end{minted}
\captionof{lstlisting}[Kernel \texttt{tb\_client\_connection\_create\_with\_endpoint}.]{\texttt{tb\_client\_connection\_create\_with\_endpoint} in the kernel binary, symbolicated by hand. The source code was disassembled by Ghidra.}
\label{lst:TB_CONNECTION_CREATE_XNU}
\end{listing}

The above listing has been symbolized in the kernel binary based on the function implementation in Tightbeam. The \path{__tb_connection_create_transport_for_endpoint} function creates a \texttt{tb\_transport\_t} value based on the input endpoint. In both the kernel and Tightbeam implementations, the function implements a switch on the input endpoint transportation type. We can recall that this was an integer value of 7 for \path{TB_TRANSPORT_TYPE_XNU}. Interestingly, while the kernel implementation supports and implements integer endpoint type value 7, the Tightbeam implementation does not. This indicates that different types of Tightbeam transports can only be created from specific contexts. The reassembled function implementation from Tightbeam can be found in \cref{tb_implementation_transport_create}. It allows us to infer various types of Tightbeam transport.
\begin{table}[h!]
\label{tightbeamTransportTypes}
\centering
\scriptsize
\begin{tabular}{|c|l|c|l|}
\hline
\textbf{Type} & \textbf{Inferred Transport Type Name} & \textbf{Flag} & \textbf{Creation Function} \\
\hline
1   & \texttt{TB\_TRANSPORT\_TYPE\_NULL}           & any  & \texttt{\_tb\_null\_transport\_create} \\
2   & \texttt{TB\_TRANSPORT\_TYPE\_MACH}           & 1    & \texttt{\_tb\_mach\_service\_transport\_create} \\
2   & \texttt{TB\_TRANSPORT\_TYPE\_MACH}           & 0    & \texttt{\_tb\_mach\_client\_transport\_create} \\
4   & \texttt{TB\_TRANSPORT\_TYPE\_EVE}            & 0    & \texttt{\_tb\_eve\_client\_transport\_create} \\
5   & \texttt{TB\_TRANSPORT\_TYPE\_EVE}            & 0    & \texttt{\_tb\_eve\_client\_transport\_create} \\
8   & \texttt{TB\_TRANSPORT\_TYPE\_DARWIN}         & any  & \texttt{\_tb\_darwin\_client\_transport\_create} \\
9   & \texttt{TB\_TRANSPORT\_TYPE\_UNIX}           & 0    & \texttt{\_tb\_unix\_client\_transport\_create\_with\_endpoint} \\
9   & \texttt{TB\_TRANSPORT\_TYPE\_UNIX}           & 1    & \texttt{\_tb\_unix\_service\_transport\_create\_with\_endpoint} \\
10  & \texttt{TB\_TRANSPORT\_TYPE\_DELEGATED}      & 0    & \texttt{\_tb\_delegated\_client\_transport\_create} \\
10  & \texttt{TB\_TRANSPORT\_TYPE\_DELEGATED}      & 1    & \texttt{\_tb\_delegated\_service\_transport\_create} \\
11  & \texttt{TB\_TRANSPORT\_TYPE\_AFK}            & any  & \texttt{\_tb\_afk\_transport\_create} \\
\hline
\end{tabular}
\caption[Endpoint type and endpoint options mapping to transport type name and transport creation function from \protect \path{__tb_connection_create_transport_for_endpoint} in Tightbeam.]{Mapping from endpoint type and endpoint options to inferred transport type name and the relevant transport creation function from \path{__tb_connection_create_transport_for_endpoint} in Tightbeam.}
\label{tab:transport_mapping}
\end{table}

It shows that Tightbeam supports a variety of different transport types. As already noted, \texttt{TB\_TRANSPORT\_TYPE\_XNU} is not implemented in the Tightbeam binary. It is implemented in the kernel in \path{__tb_connection_create_transport_for_endpoint}. The reassembled implementation is shown in \cref{lst:TB_CONNECTION_CREATE_TRANSPORT_FOR_EP}.
\begin{listing}[H]
\begin{minted}[linenos, breaklines, bgcolor=LightGray, frame=lines]{c}
void __tb_connection_create_transport_for_endpoint(tb_endpoint_t *ep)

{
  code *pcVar1;
  int iVar2;
  long lVar3;
  undefined8 extraout_x1;
  undefined8 in_x2;
  undefined4 in_w3;
  
  iVar2 = tb_ep_get_type();
  if (iVar2 == TB_TRANSPORT_TYPE_AFK) {
    lVar3 = _tb_afk_transport_create(ep);
  }
  else {
    if (iVar2 != TB_TRANSPORT_TYPE_XNU) goto LAB_fffffff0088ed670;
    lVar3 = tb_xnu_transport_create(ep,extraout_x1,in_x2,in_w3);
  }    
\end{minted}
\captionof{lstlisting}[Simplified kernel \protect \path{__tb_connection_create_transport_for_endpoint} function.]{Simplified \path{__tb_connection_create_transport_for_endpoint} disassembly of the kernel binary. The source code was disassembled by Ghidra.}
\label{lst:TB_CONNECTION_CREATE_TRANSPORT_FOR_EP}
\end{listing}

\begin{figure}[H]
    \centering
\includegraphics[width=400pt]{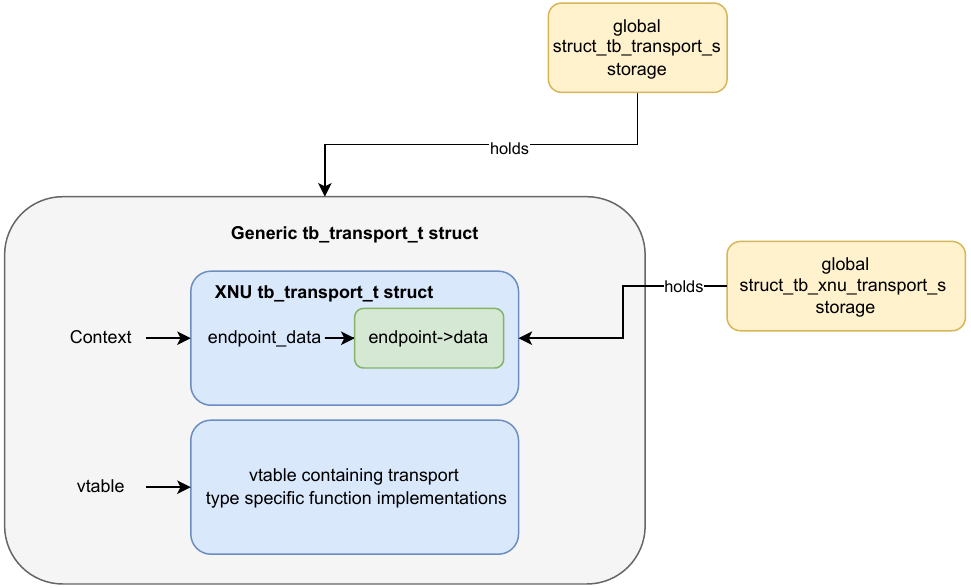}
\caption{Resulting structure of \texttt{tb\_xnu\_transport\_create} in XNU.}
 \label{fig:transportStructure}
 \end{figure}

\gls{XNU} appears to only support \texttt{TB\_TRANSPORT\_TYPE\_XNU} and \path{TB_TRANSPORT_TYPE_AFK}. The transport creation function for the XNU type has been named according to the naming convention used in the Tightbeam binary.
\path{tb_xnu_transport_create} performs a two-step allocation process. The resulting structure of this allocation process can be seen in \cref{fig:transportStructure}

Initially, a generic \texttt{tb\_transport\_t} struct is allocated in a global transport storage. A second XNU type-specific struct is allocated and stored in a further global XNU transport storage. The xnu transport is written to the context field of the generic struct. Furthermore, the data value (recall that this corresponds to the Exclave's resource ID) is written to a field in the XNU struct, which we have denoted \texttt{endpoint\_data}. Furthermore, a vtable pointer is written to a field of the generic struct, which we have denoted \texttt{vtable}. We know this to be a vtable based on a similar implementation in the Tightbeam binary, in which the tables are symbolicated and named. The \gls{XNU} vtable holds a variety of different function pointers. Confirming to the standard, it appears that vtables are used to implement transport type polymorphisms by storing the type-specific functions in the generic transport structure. The XNU Tightbeam transport vtable can be seen below in \cref{XNU_vtable}.
\begin{listing}[H]
\begin{minted}[linenos, breaklines, bgcolor=LightGray, frame=lines]{c}
tb_xnu_client_transport_vtable                  
        tb_xnu_transport_send_message
        LAB_fffffff0088f1090
        __tb_connection_create_transport_for_endpoint
        FUN_fffffff0088f10e0
        __tb_xnu_transport_destruct_message_buffer
        LAB_fffffff0088f1194
        FUN_fffffff0088f11bc
        FUN_fffffff0088f11cc
        FUN_fffffff0088f11d8    
\end{minted}
\captionof{lstlisting}[Function entries in the XNU transport type-specific vtable in kernel.]{Simplified function entries in XNU transport type-specific vtable stored in the generic transport struct in the kernel. Partially symbolized by hand. The source code was disassembled by Ghidra.}
\label{XNU_vtable}
\end{listing}

After the transport has been successfully created, a \texttt{tb\_connection\_t} is created via a call to \texttt{xnu\_tb\_connection\_create}. The function allocates a connection in a global \texttt{struct\_tb\_connection\_s} storage, and stores the previously created transport, and a newly allocated \texttt{tb\_list\_t}. The exact usage of the list is still unknown.

\begin{figure}[H]
    \centering
\includegraphics[width=400pt]{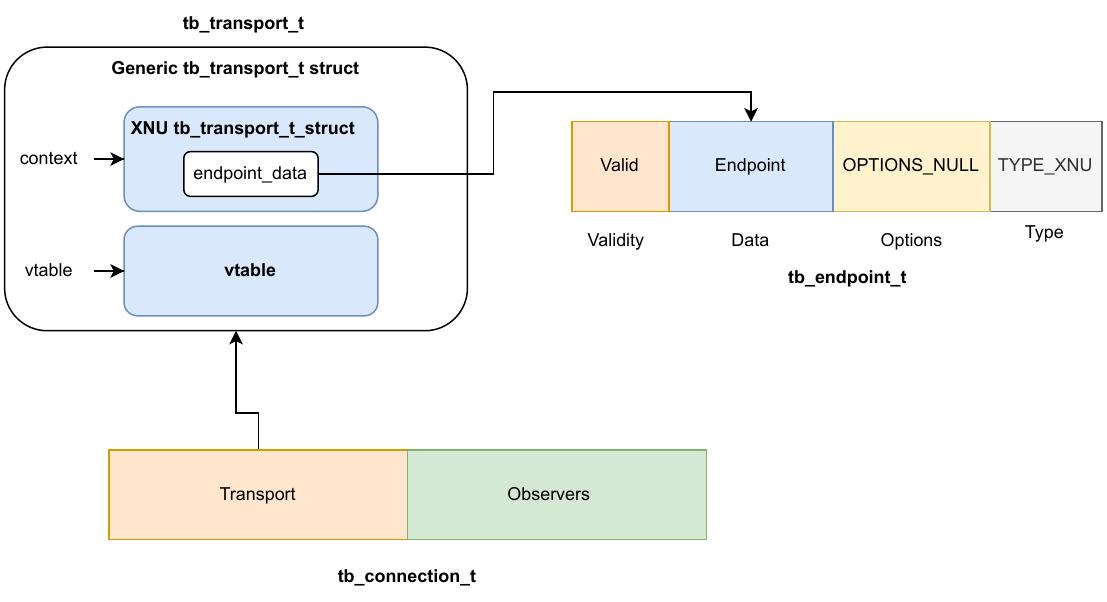}
\caption[Resulting structure of \texttt{tb\_client\_connection\_create\_with\_endpoint} call from XNU.]{Resulting structure of \texttt{tb\_client\_connection\_create\_with\_endpoint} call from XNU. A \texttt{tb\_connection\_t} is returned to the caller.}
 \label{fig:connectionCreated}
 \end{figure}

To finish up the handling of \texttt{tb\_client\_connection\_create\_with\_endpoint}, the previously created endpoint is destroyed, as it has served its purpose and is no longer needed. A \texttt{tb\_client\_connection\_t} is finally returned to the caller. The result of this part of the Tightbeam control flow is shown in \cref{fig:connectionCreated}.

\subsubsection{Client Connection Activation}

Recalling \cref{driverKitTightbeam}, we see that the next step in the Tightbeam communication process is the activation of the previously created connection via a call to \path{tb_client_connection_activate}. The function can be seen in \cref{XNU_activateConnection}.

\begin{listing}[H]
    \begin{minted}[linenos, breaklines, bgcolor=LightGray, frame=lines]{c}
void tb_client_connection_activate(tb_connection_t *client_connection)
{
  long lVar1;
  tb_transport_t *transport;
  
  transport = (tb_transport_t *)client_connection->transport;
  if ((client_connection->observers != 0) &&
     (lVar1 = *(long *)(client_connection->observers + 8), lVar1 != 0)) {
    (**(code **)(lVar1 + 0x10))(lVar1 + 0x10,lVar1,transport);
  }
  activate_connection(transport);
  return;
}    
\end{minted}
\captionof{lstlisting}[Kernel \texttt{tb\_client\_connection\_activate} function.]{\texttt{tb\_client\_connection\_activate} implementation in the kernel binary, symbolicated by hand. The source code was disassembled by Ghidra.}
    \label{XNU_activateConnection}
\end{listing}
Conditional code execution is based on the existence of observers. We have not yet reverse-engineered this part of the Tightbeam framework, but assume this to be an informing call to registered observers that the connection is now active. The actual activation of the connection happens via a function call to what we have denoted \texttt{activate\_connection}, which receives the connection transport as an input. It can be seen in \cref{XNU_activateConnectionProper}.

\begin{listing}[H]
    \begin{minted}[linenos, breaklines, bgcolor=LightGray, frame=lines]{c}
void tb_client_connection_activate(tb_connection_t *client_connection)
{
bool activate_connection(tb_transport_t *param_1)

{
  int iVar1;
  
  if ((tb_transport_t *)param_1->vtable != (tb_transport_t *)0x0) {
    vtable = (tb_transport_t *)param_1->vtable;
  }
  iVar1 = (*(code *)vtable->field_0x8)();
  return iVar1 != 0;
}
}    
\end{minted}
\captionof{lstlisting}[Kernel \texttt{connectin\_activate} function.]{\texttt{connection\_activate} implementation in the kernel binary, symbolicated by hand. The function code has been slightly altered for readability. The source code was disassembled by Ghidra.}
\label{XNU_activateConnectionProper}
\end{listing}

The function accesses the transport vtable and executes the function registered at offset 0x8, which we assume to be the relevant activation function. Interestingly, we find that the specific function is implemented as an empty stub, which returns without taking any action. We assume this to indicate that connections for the XNU transport type do not require activation. In contrast, the specific vtable function in other transport-type vtables, found within the Tightbeam binary, at least partially implements actual functionality.

\subsubsection{Message Preparation}
After the Tightbeam setup for the client connection has been completed, the actual message delivery can be prepared and performed. The relevant code excerpt is listed in \cref{driverkitMessageDelivery}.
\begin{listing}[H]
\begin{minted}[linenos, breaklines, bgcolor=LightGray, frame=lines]{c}
tb_message_t message = NULL;
tb_transport_message_buffer_t tpt_buf = NULL;

message = kalloc_type(struct tb_message_s, Z_WAITOK | Z_ZERO | Z_NOFAIL);
tpt_buf = kalloc_type(struct tb_transport_message_buffer_s,
    Z_WAITOK | Z_ZERO | Z_NOFAIL);

tb_error_t tb_err = TB_ERROR_SUCCESS;
tb_err = tb_client_connection_message_construct(client, message,
    tpt_buf, sizeof(uint8_t), 0);

tb_message_encode_u8(message, (uint8_t) test_type);

tb_message_complete(message);

tb_message_t response = NULL;

tb_err = tb_connection_send_query(client, message, &response,
    TB_CONNECTION_WAIT_FOR_REPLY);
\end{minted}
\captionof{lstlisting}[Tightbeam message preparation and call delivery in \protect \path{hello_driverkit_interrupts}.]{Message preparation and call delivery in \texttt{hello\_driverkit\_interrupts}.}
\label{driverkitMessageDelivery}
\end{listing}

The initial step is the allocation of memory for both the actual message and the \texttt{tpt\_buf}, which we assume to be the transport buffer. After this, the message is constructed via a call to \path{tb_client_connection_message_construct}. This is a wrapper function for \path{__tb_connection_message_construct}, which is listed in \cref{XNU_message_construct}.
\begin{listing}[H]
    \begin{minted}[linenos, breaklines, breakanywhere, bgcolor=LightGray, frame=lines]{c}
undefined8
__tb_connection_message_construct
          (tb_connection_t *connection,int option,tb_message *message,tb_bufffer *tpt_buffer,
          ulong size,undefined8 flag)
{
  int iVar1;
  ulong uVar2;
  undefined8 uVar3;
  undefined1 disposition;
  tb_transport_t *transport;
  
  tb_message_initialize(message);
  transport = (tb_transport_t *)connection->transport;
  iVar1 = tb_transport_supports_multipart_messages(transport);
  if (iVar1 == 0) {
    if (tpt_buffer->wrapping != '\0') goto TPT_BUFFER_WRAPPING_NOT_FALSE;
  }
  else {
    uVar2 = tb_transport_get_tx_buffer_size(transport);
    if (tpt_buffer->wrapping != '\0') {
TPT_BUFFER_WRAPPING_NOT_FALSE:
      __tb_connection_message_construct.cold.1();
      __tb_connection_message_destruct();
      return 0;
    }
    if (uVar2 < size) {
      _tb_connection_alloc_init_owned_transport_message_buffer(tpt_buffer,size);
      goto LAB_fffffff0088ee1e8;
    }
  }
  uVar3 = tb_transport_construct_message_buffer(transport,size,flag,tpt_buffer);
  if ((int)uVar3 != 0) {
    return uVar3;
  }
LAB_fffffff0088ee1e8:
  disposition = TB_MESSAGE_DISPOSITION_QUERY;
  if (option == 1) {
    disposition = TB_MESSAGE_DISPOSITION_REPLY;
  }
  uVar3 = _tb_message_construct(message,tpt_buffer,disposition);
  if ((int)uVar3 == 0) {
    tb_message_set_connection_identifier(message,connection);
    uVar3 = 0;
  }
  return uVar3;
}
    \end{minted}
    \captionof{lstlisting}[Kernel Tightbeam message construction via \protect \path{__tb_connection_message_construct}.]{Message construction via \texttt{\_\_tb\_connection\_message\_construct} in the kernel binary.}
    \label{XNU_message_construct}
\end{listing}

The function initializes an empty Tightbeam message. It further constructs a message buffer for the connection's transport by calling \path{tb_transport_construct_message_buffer}. This invokes a function at offset \texttt{0x18} in the transport vtable. This function initializes the buffer and sets, among others, the size and wrapping field. Based on the provided options, the message disposition is set, differentiating between a query and a reply. Finally, the message is constructed, and the constructed transport buffer and connection identifier fields are set. The message now holds the transport buffer, a disposition determining its communication state, and an identifier for the connection to which it belongs. 

\paragraph*{Message Encoding}

The \texttt{\_tb\_message\_encode\_<Encoding>} function in Tightbeam performs the actual data storage to the message transport buffer. By doing so, they make data available to the underlying call target. Based on invocations in the \gls{XNU} open-source code, we see calls encoding function specifiers (in \path{exclaves_driverkit.c}) or raw byte data derived from a string (in \path{exclaves_test.c}). We can assume it can store arbitrary data, as long as it fits within the buffer's constraints.

\paragraph*{Message Finalization}
The message is further finalized via a call to \texttt{tb\_msg\_complete}. This function sets the message state depending on the current state and the message disposition. For a query message, the state is set to \texttt{TB\_MESSAGE\_STATE\_READY}, indicating its readiness to be sent. 

\subsubsection{Message Delivery}
Finally, the actual message delivery is performed via a call to \path{tb_connection_send_query}. The full function Listing can be found in \cref{tb_connection_send_query_kernel}. The function performs sending validation by confirming the message state, disposition, and connection identifier. There is further setup occurring that has not yet been fully reverse-engineered; however, we know the actual message sending to be invoked by a function call to \path{tb_transport_send_message}, which in turn executes the first function entry in the transport vtable. For XNU, this function is \texttt{tb\_xnu\_transport\_send\_message}. An excerpt from that function is shown in \cref{xnutbtransportmessage}.
\begin{listing}[H]
    \begin{minted}[linenos, breaklines, breakanywhere, bgcolor=LightGray, frame=lines]{c}
undefined8
tb_xnu_transport_send_message
          (tb_transport_t *transport,tb_message *message,undefined8 *param_3,uint param_4)

{
  ulong uVar1;
  int iVar2;
  tb_bufffer *message_buffer;
  long lVar3;
  long error;
  ulong tag;
  
  if (message->msg_state != TB_MESSAGE_STATE_SENT) {
    FUN_fffffff0088f1418();
    return 0;
  }
  message_buffer = (tb_bufffer *)tb_message_get_transport_buffer(message);
  tag =
       (ulong)((int)message_buffer->size + 7U >> 3) & 0x3f |
       (ulong)*(ushort *)&message_buffer->someFlags << 0x10;
  output = 0;
  iVar2 = exclaves_endpoint_call
                    (*(undefined8 *)transport->context,transport->context->endpoint_data,
                     &tag,&error);
    \end{minted}
    \captionof{lstlisting}[Kernel \texttt{tb\_xnu\_transport\_message} excerpt.]{Excerpt from \texttt{tb\_xnu\_transport\_message} in the kernel binary. The source code was disassembled by Ghidra.}
    \label{xnutbtransportmessage}
\end{listing}
We can determine that the called function is \texttt{exclaves\_endpoint\_call}, the kernel's Exclaves entry point, based on its calling context. As known from the open source code, it performs a call to Exclaves via \texttt{exclaves\_xnu\_proxy\_endpoint\_call}.

\begin{figure}[H]
    \centering
\includegraphics[scale=0.75]{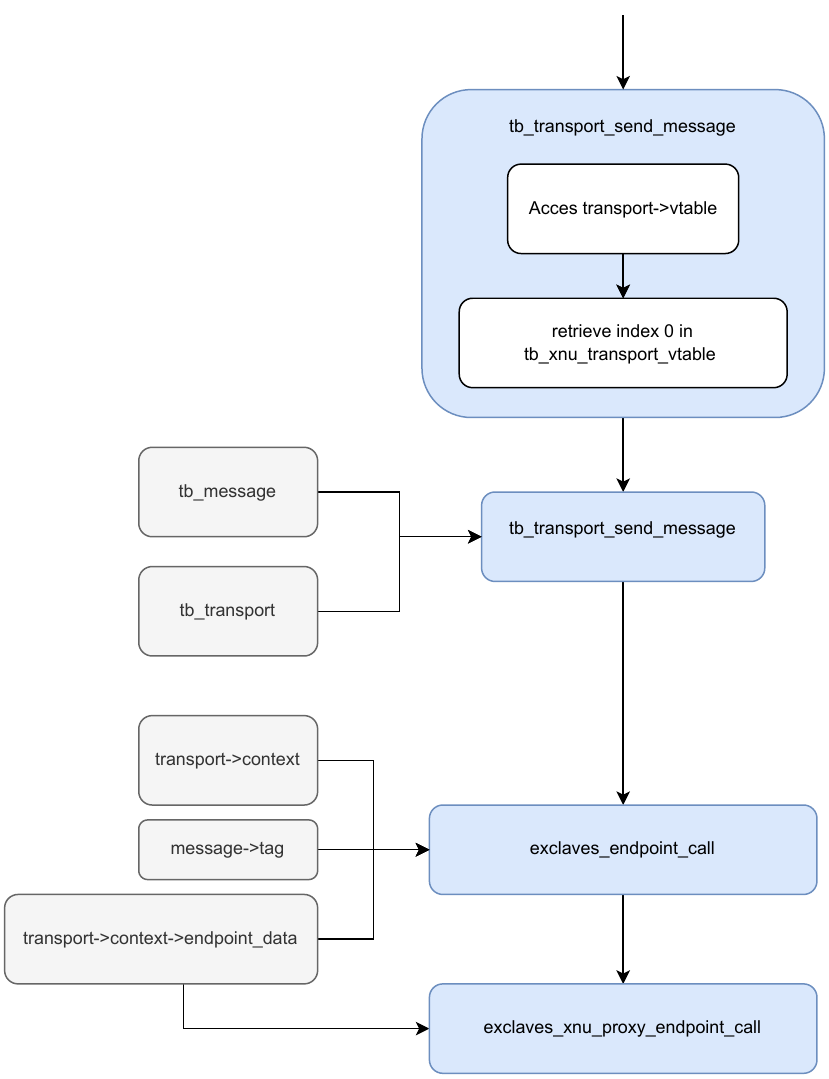}
    \caption[Tightbeam connection setup and call issuing control flow.]{Control flow of setup of Tightbeam connection and performing a call to a specified endpoint.}
    \label{TB_BIG}
\end{figure}

We can infer from this that \texttt{xnuproxy} is the underlying communication mechanism for Tightbeam communication from XNU to Exclaves. We furthermore find that the call from \texttt{tb\_xnu\_transport\_send\_message} is the only invocation of the kernel Exclaves entry point. We can therefore reasonably assert that Tightbeam is in fact the transport interface used by \gls{XNU} to call into Exclaves.
We have provided \cref{TB_BIG} as a full overview of the process of sending a message to an Exclave endpoint via Tightbeam from \gls{XNU}.

\begin{figure}[H]
    \centering
\includegraphics[scale=0.75]{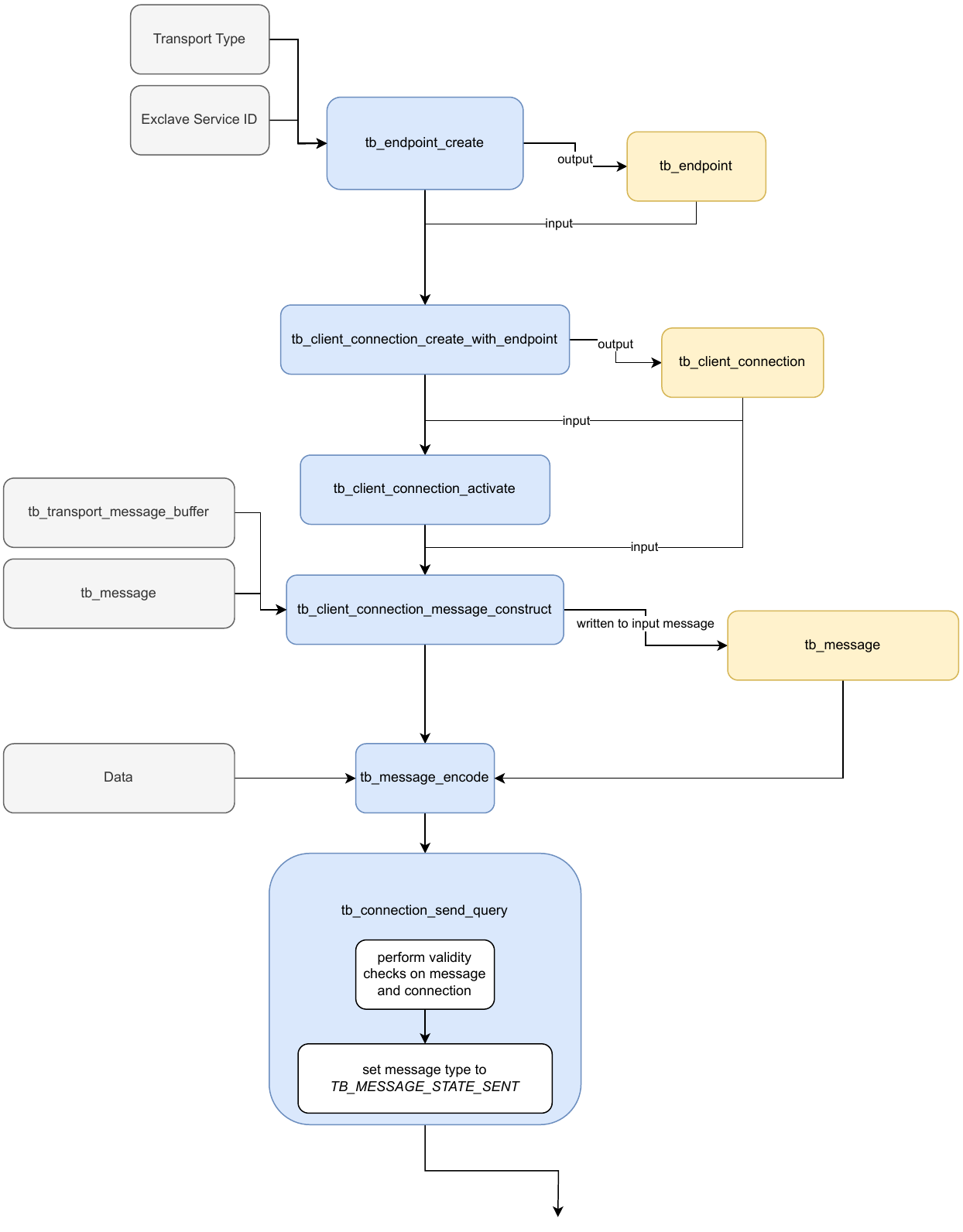}
\end{figure}

\subsection{Tightbeam Usage \& Context}
The reverse-engineered Tightbeam calling process was inferred from the structure provided in XNU. % testing.
\todo[color=red]{was heißt testing hier??}
We do not find any direct usage of Tightbeam communication in the open-source code.
From \cref{Conclave} we know that at Conclave initialization, each Conclave gets set up with a Tightbeam connection.
Additionally, we find a large number of calls to our previously discovered \path{tb_connection_send_query} method call in the kernel binary, which indicates that the kernel does actively use Tightbeam for message delivery.
Tracing back its calling functions, we find that a variety of these uses originate from the aforementioned Conclave setup process.
Further invocations appear on \texttt{exclave\_ctl\_trap} Exclaves boot handling and invocations regarding operations on audio buffers and memory.
Tightbeam is the communication framework used for performing Exclaves calls from the kernel directly, and therefore used whenever Exclaves services are to be invoked. We find that the actual delivery via the underlying Exclaves \gls{IPC} buffers is performed by \texttt{xnuproxy}.

Thightbeam is an Exclave communication framework, which wraps the underlying transport mechanism from the caller. References to Tightbeam are found in various  GL0 Exclavecore components, which support our assumption. We have performed an in-depth analysis of Tightbeam calls executed from kernel; however, recalling \cref{tightbeamTransportTypes}, we found a variety of different transport types supported by Tightbeam. Tightbeam appears to act as an \gls{IPC} mechanism with relatively rigid structuring regarding the underlying transport type. This might make it suitable for communication across security boundaries. 

\section{Real-World Exclave Use Case -- Secure Indicator Light / Privacy Indicator Manipulation} 
With the iOS 14 release in September 2020, Apple introduced a so-called \emph{recording indicator}. This indicator is displayed ``whenever an app has access to the microphone or camera'', aiming to provide increased security against spyware by rendering a highly visible dot on the device screen~\cite{ios14:2024}. With the release of Exclaves, the actual implementation of the indicator light has shifted significantly. In an attempt to highlight security improvements through the implementation and usage of Exclaves, we will examine the initial implementation of the security indicator and explore ways to circumvent it. We compare this with the new implementation using the Exclaves feature, and aim to derive security implications from this. 

\subsection{Recording Indicator}
The \emph{recording indicator} refers to the initial version of the indicator light released in iOS 14. In the following, we will briefly examine its implementation and then demonstrate a way to circumvent it on devices without Exclaves.

\subsubsection{Implementation}
We assume the recording indicator to be realized by the \texttt{mediaserverd}, the daemon responsible for media streaming and sensor access in older iOS versions.

Although the official indicator name after the release was \emph{recording indicator}, a \texttt{frida-trace} call for this was unsuccessful in \texttt{mediaserverd}. However, especially in the macOS context, the name \emph{privacy indicator} circles in official Apple support material\footnote{See \href{https://support.apple.com/en-gb/guide/mac-help/mchlp1446/14.0/mac/14.0}{https://support.apple.com/en-gb/guide/mac-help/mchlp1446/14.0/mac/14.0}, accessed: 15.05.2025.}. 
Tracing for Objective-C method names containing \path{PrivacyIndicator} displays  \texttt{isEligibleForPrivacyIndicator} in \path{STMediaStatusDomainCameraDescriptor} and \path{initWithCameraIdentifier::eligibleForPrivacyIndicator} in \path{STMediaStatusDomainCameraDescriptor}. Unfortunately, overwriting these methods does not lead to a successful deactivation of the indicator, suggesting a more internal embedding and a need to check for its necessity in rendering.

Digging deeper and following the hints provided by the initial trace, tracing for \texttt{*[STMedia* *]} points towards the \texttt{STMutableMediaStatusDomainData}, indicating an alterable parameter set. Tracing for methods of said class leads to \path{-[STMutableMediaStatusDomainData} \path{initWithAttributionCatalog:cameraAttributionListData:]}, which one can assume initializes camera usage with a mutable set of parameters. The following section will show that this is, in fact, the case, and this data set is at least partially responsible for the rendering of the recording indicator. 

\subsubsection{Circumvention}
\label{subsection:RecordingIndicatorCircumvention}
As seen in the previous subsection, the recording indicator is realized in the mediaserverd. 
This is a daemon running in the iOS userspace. We can therefore inject directly into it via FRIDA~\cite{FRIDA:2025}. The following proof of concept was performed on an iPhone 8 running iOS 16.7.10. To inject via FRIDA, the device was jailbroken using the palera1n jailbreak~\cite{palera1n:2025} in rootless mode. 
The following FRIDA script was injected into the \texttt{mediaserverd} process.
\begin{listing}[H]
\begin{minted}[linenos, breaklines, bgcolor=LightGray, frame=lines]{c}
// Usage: frida-trace -U mediaserverd -m '*[STMutableMediaStatusDomainData initWithAttributionCatalog:cameraAttributionListData:]'
// Script Location: Default handler location for STMutableMediaStatusDomainData initWithAttributionCatalog:cameraAttributionListData:
defineHandler({
  onEnter(log, args, state) {
    log(`-[STMutableMediaStatusDomainData initWithAttributionCatalog:${new ObjC.Object(args[2])}  cameraAttributionListData:${new ObjC.Object(args[3])}]`);
    // Empty Attribution Catalog and List
    args[2] = new ObjC.Object(new NativePointer(0));
    args[3] = new ObjC.Object(new NativePointer(0));
  },

  onLeave(log, retval, state) {
  }
});
\end{minted}
\captionof{lstlisting}{FRIDA script for disabling the recording indicator.}
\label{code:disableRecordingIndicator}
\end{listing}

The script in \cref{code:disableRecordingIndicator} targets the attribution catalog, which is used to render the recording indicator. It overwrites the \path{STMutableMediaStatusDomainData} \path{initWithAttributionCatalog:cameraAttributionListData:} method, which was found to be invoked when the camera or microphone is accessed. This disablement of the recording indicator is achieved by nulling the \texttt{AttributionCatalog} catalog passed to the function by overwriting it with a \texttt{nil} object. Using the approach highlighted above, we successfully disabled the recording indicator.

\subsubsection{Implications}
Whilst the recording indicator serves as a functioning indicator indicating sensor (microphone, camera) usage to the user, the initial implementation was easily circumventable. As shown in \cref{subsection:RecordingIndicatorCircumvention}, the iOS component responsible for the recording indicator is the user-space daemon \texttt{mediaserverd}. A capable attacker may be able to reach code execution within \texttt{mediaserverd} to take over the recording indicator, as it has various \gls{XPC} interfaces reachable from sandboxed apps\footnote{Irrespective of Exclaves, Apple limited this risk by splitting \texttt{mediaserverd} into multiple, less-privileged daemons as of iOS 17.}.

In conclusion, the initial recording indicator is only capable of improving user privacy in scenarios where apps trigger microphone and camera recordings over the intended frameworks.
On a compromised device, the recording indicator is part of a highly exposed system daemon, making successful attacks likely.

\subsection{Secure Indicator Light}
With the shift towards Exclaves, a new mechanism for rendering the recording indicator, now \gls{SIL}, emerged.
When enumerating Exclave resources (see \cref{resourceEnumSection}), we find a \texttt{com.apple.service.ExclaveIndicatorController} resource residing in the \texttt{com.apple.kernel} domain. Furthermore, the ExclaveKit \gls{DMG} framework contains a \path{SILManagerComponent.framework}.

We find various references to the \gls{EIC} in \path{osfmk/kern/exclaves_sensors}, where it serves as a management component for sensor control and corresponding memory buffer usage (e.g., it appears to be responsible for starting and stopping sensors). We furthermore find the following function signature with comments depicted in \cref{code:sensorStart} in \path{osfmk/kern/exclaves.h}.
\begin{listing}[H]
\begin{minted}[linenos, breaklines, bgcolor=LightGray, frame=lines]{c}
/*!
 * @function exclaves_sensor_start
 *
 * @abstract
 * Start accessing a sensor and cause any indicators to display.
 *
 * If multiple clients start the same sensor, the sensor will only
 * actually start on the first client.
 *
 * @param sensor_port
 * A sensor buffer port name returned from exclaves_sensor_create()
 * for the sensor.
 *
 * @param flags to pass to the implementation. Must be 0 for now.
 *
 * @param sensor_status
 * Out parameter filled with the sensor status.
 *
 * @result
 * KERN_SUCCESS or mach system call error code.
 */
SPI_AVAILABLE(macos(14.4), ios(17.4), tvos(17.4), watchos(10.4))
kern_return_t
exclaves_sensor_start(mach_port_t sensor_port, uint64_t flags,
    exclaves_sensor_status_t *sensor_status);
\end{minted}
\captionof{lstlisting}[\texttt{exclaves\_sensor\_start} signature and comments.]{The \texttt{exclaves\_sensor\_start} signature and comments, taken from \path{osfmk/kern/exclaves.h}.}
\label{code:sensorStart}
\end{listing}

From the above listing, we see a direct correlation between the call to \path{exclaves_sensor_create}, which is, for example, invoked in the control flow of receiving a sensor start invocation from userspace via \path{exclaves_control_trap}.
Due to the redacted code in the further handling of the call, we are unable to determine the exact mechanisms regarding \gls{SIL} rendering.

\todo[color=red]{Zum DCP nochmal mehr reversen, das ist noch sehr vage und viel angenommen...}

\paragraph*{Assumed Underlying Structure}

We can make assumptions based on our current knowledge. In the \path{com.apple.darwin} component, we find the resource \path{com.apple.service.SecureRTBuddyDCP}. \emph{RTBuddy} is a term coined by the Apple RTKit, its real-time operating system, serving special use cases where standard operating systems would not suffice. Exemplary use cases are the \gls{DCP} and \gls{ANE}~\cite{RTKit}. Based on repeated usage of the \texttt{SecureRT} term in Exclave resources located in the Darwin domain, we assume that Apple introduced an \texttt{RTKit} Exclave equivalent for performing real-time tasks in the guarded world. The suffix \gls{DCP} strongly hints towards a display coprocessor-related resource. Based on these findings, we can speculate that the \texttt{SILManagerComponent} might directly communicate the need to display the \gls{SIL} to the \gls{DCP} via the secure world \texttt{RTKit}. This implies that the decision for this rendering and the communication to the relevant coprocessor occur fully within guarded levels and are therefore significantly harder to tamper with.

 % example
    \chapter{Trusted Execution Monitor}
\label{txm}
\section{TXM Fundamentals}
\gls{TXM} is a guarded level component running in GL0. As previously shown, TXM has its own SPTM domain, featuring a variety of frame types (see \cref{App:SPTMTypesNew}). TXM is responsible for enforcing system policies governing code execution~\cite{appleOSSecurity:2024}. TXM can be called into via SPTM by calling int \texttt{TXM\_DOMAIN}. XNU performs this task to facilitate various code signing and entitlement management operations. TXM itself communicates with \gls{SPTM} via \glspl{SVC}, and sets up the required dispatching structure and for memory frame retyping. We find TXM to be responsible for code signing and entitlement enforcement.

As in this work, we have decided to focus on the inner workings of \gls{SPTM} and the recently released Exclaves. We have chosen not to delve deeply into the workings of TXM, but instead to examine its interaction with SPTM and \gls{XNU}. A more detailed analysis of TXM is left up for future work. We find TXM to be responsible for code signing and entitlement enforcement. 

\section{SPTM Calls}
We have previously discovered SPTM calls from TXM (see \cref{CallsFroMTXM}). We can map them according to the known SPTM dispatch structure for applicable calls. For standard SPTM calls calling into its dispatching logic, we found SVC \#0 to be used. The result can be seen in \cref{TXM_calls_tab}.

\begin{table}[h!]
\footnotesize
\begin{tabular}{|l|l|l|}
\hline
Dispatch Table & Endpoint ID & SPTM Function \\ \hline
\texttt{TXM\_BOOTSTRAP} & 1 & \texttt{type\_TXM-rx\_region\_as\_TXM\_rw\_type} \\ \hline
\texttt{TXM\_BOOTSTRAP} & 2 & \texttt{register\_dispatch\_table} \\ \hline
\texttt{TXM\_BOOTSTRAP} & 3 & \texttt{retype} \\ \hline
\texttt{TXM\_BOOTSTRAP} & 4 & \begin{tabular}[c]{@{}l@{}}Unknown as of now.\\ Thread identifying information read.\end{tabular} \\ \hline
\texttt{RETURN\_TO\_CALLER} & - & Control Function \\ \hline
\texttt{PANIC} & - & Control Function \\ \hline
\texttt{EXCEPTION\_STATE\_SAVED} & - & Control Function \\ \hline
\end{tabular}
\caption{SPTM dispatch calls for TXM.}
\label{TXM_calls_tab}
\end{table}

Comparable to \gls{SK}, we find that TXM registers its own dispatch tables for calls directed towards the \texttt{TXM\_DOMAIN}. It can also employ SPTM to retype memory frames, and further calls into a SPTM function \path{type_TXM-rx_region_as_TXM_rw_type}.
TXM further calls into SPTM using \gls{SVC} \#38 to disable all interrupts and SVC \#37 to enable all interrupts, respectively.

\subsection{Dispatch Table Registration} 
We find that TXM performs dispatch table registration calls in its entry function. The registration process is very comparable to SK's dispatch table registration. TXM also registers two dispatch tables with different dispatch functions into SPTM. We find the first dispatch function \path{TXM_dispatch_function_0} to be registered with permissions allowing XNU to call into it, and the second function \path{TXM_dispatch_function_1} to be registered with permissions allowing SPTM to call into it. The registration can be seen below in \cref{txm_register_tables}. It is performed from the TXM entry.
\begin{listing}[H]
\begin{minted}[linenos, breaklines, bgcolor=LightGray, frame=lines]{c}
register_dispatch_table(0,TXM_dispatch_function_0,2);
register_dispatch_table(1,TXM_disptach_function_1,1);    
\end{minted}
\captionof{lstlisting}[TXM calls to \texttt{sptm\_resgister\_dispatch\_table}.]{Calls to \texttt{sptm\_register\_dispatch\_table} from TXM function \path{FUN_fffffff01701bd60}. TXM registers two dispatching functions that allow calling into TXM when calling the domain \texttt{TXM\_DOMAIN}. The source code was disassembled by Ghidra.}
\label{txm_register_tables}
\end{listing}

\subsection{TXM-\texttt{rx} Region Retyping}
A further action performed in the TXM entry is an SPTM call to \path{type_TXM-rx_region_as_TXM_rw_type}. We found this SPTM function to call \texttt{type\_region}, a function that iterates over a named memory region and retypes all included memory frames to the provided type. The exact reason for this call has not yet been discovered, but we assume it performs TXM lockdown on early boot executable code. 

\section{Calling into TXM}
We know from our previous analysis of SPTM that XNU calls into TXM via \texttt{txm\_enter}, which parameterizes the called domain to \texttt{TXM\_DOMAIN}, sets the target dispatch table to zero, and conditionally sets the endpoint ID based on input parameters. The main XNU call wrapping this actual SPTM call for invoking TXM functionality shows to be \texttt{txm\_kernel\_call}. The call is parameterized with an input structure of type \texttt{txm\_call\_t}. The type definition can be seen \cref{txm_call_type}.

\begin{listing}[H]
\begin{minted}[linenos, breaklines, bgcolor=LightGray, frame=lines]{c}
typedef struct _txm_call {
	/* Input arguments */
	TXMKernelSelector_t selector;
	TXMReturnCode_t failure_code_silent;
	bool failure_fatal;
	bool failure_silent;
	bool skip_logs;
	uint32_t num_input_args;
	uint32_t num_output_args;

	/* Output arguments */
	TXMReturn_t txm_ret;
	uint64_t num_return_words;
	uint64_t return_words[kTXMStackReturnWords];
} txm_call_t;
\end{minted}
\captionof{lstlisting}[\texttt{txm\_call\_t} type definition.]{\texttt{txm\_call\_t} type definition in \path{bsd/sys/trusted_execution.h}.}
\label{txm_call_type}
\end{listing}

The selector field acts as the \gls{SPTM} call endpoint selector. For calls to TXM, a variable number of arguments can be provided, depending on the endpoint called into. Arguments are found to be passed to TXM via thread stacks of type \path{txm_thread_stack_t}. This custom structure is shown in \cref{txm_stack}.

\begin{listing}[H]
\begin{minted}[linenos, breaklines, bgcolor=LightGray, frame=lines]{c}
typedef struct _txm_thread_stack {
	/* Virtual mapping of the thread stack page */
	uintptr_t thread_stack_papt;

	/* Physical page used for the thread stack */
	uintptr_t thread_stack_phys;

	/* Pointer to the thread stack structure on the thread stack page */
	TXMThreadStack_t *thread_stack_data;

	/* Linkage for the singly-linked-list */
	SLIST_ENTRY(_txm_thread_stack) link;
} txm_thread_stack_t;    
\end{minted}
\captionof{lstlisting}[\texttt{txm\_thread\_t} type definition.]{\texttt{txm\_thread\_t} type definition in \path{bsd/sys/trusted_execution.h}.}
\label{txm_stack}
\end{listing}

The actual argument to the \texttt{txm\_enter} call is found to be a pointer to a \path{sptm_call_regs_t} structure, which may hold up to eight arguments. The first argument is always set to the TXM thread stack's physical address, while the other arguments are set based on the TXM function to invoke and the provided input parameters.

\subsection{Selector Enumeration}
We can enumerate known TXM selectors from the XNU open-source code, together with the number of input and output arguments. The full list is shown in \cref{tab:txm_kernel_calls}.

{
\footnotesize
\begin{longtable}{|l|c|c|}
\hline
\textbf{Selector Value} & \textbf{\#Input} & \textbf{\#Output} \\
\hline
\endfirsthead

\hline
\textbf{Selector Value} & \textbf{\#Input} & \textbf{\#Out} \\
\hline
\endhead

\hline
\endfoot

\hline
\caption[TXM kernel call selectors and argument counts.]{TXM kernel call selectors and argument counts extracted from XNU open source.}
\label{tab:txm_kernel_calls} 
\endlastfoot

\texttt{kTXMKernelSelectorGetTrustCacheInfo} & 0 & 4 \\ \hline
\texttt{kTXMKernelSelectorLoadTrustCache} & 7 & 0 \\ \hline
\texttt{kTXMKernelSelectorQueryTrustCache} & 2 & 2 \\ \hline
\texttt{kTXMKernelSelectorCheckTrustCacheRuntimeForUUID} & 1 & 0 \\ \hline
\texttt{kTXMKernelSelectorAddFreeListPage} & 1 & 0 \\ \hline
\texttt{kTXMKernelSelectorGetLogInfo} & 0 & 3 \\ \hline
\texttt{kTXMKernelSelectorGetCodeSigningInfo} & 0 & 6 \\ \hline
\texttt{kTXMKernelSelectorSetSharedRegionBaseAddress} & 2 & 0 \\ \hline
\texttt{kTXMKernelSelectorEnterLockdownMode} & 0 & 0 \\ \hline
\texttt{kTXMKernelSelectorDeveloperModeToggle} & 1 & 0 \\ \hline
\texttt{kTXMKernelSelectorRegisterProvisioningProfile} & 2 & 1 \\ \hline
\texttt{kTXMKernelSelectorUnregisterProvisioningProfile} & 1 & 2 \\ \hline
\texttt{kTXMKernelSelectorAssociateProvisioningProfile} & 2 & 0 \\ \hline
\texttt{kTXMKernelSelectorDisassociateProvisioningProfile} & 1 & 0 \\ \hline
\texttt{kTXMKernelSelectorAuthorizeCompilationServiceCDHash} & 1 & 0 \\ \hline
\texttt{kTXMKernelSelectorMatchCompilationServiceCDHash} & 1 & 1 \\ \hline
\texttt{kTXMKernelSelectorSetLocalSigningPublicKey} & 1 & 0 \\ \hline
\texttt{kTXMKernelSelectorGetLocalSigningPublicKey} & 0 & 1 \\ \hline
\texttt{kTXMKernelSelectorAuthorizeLocalSigningCDHash} & 1 & 0 \\ \hline
\texttt{kTXMKernelSelectorRegisterCodeSignature} & 3 & 2 \\ \hline
\texttt{kTXMKernelSelectorUnregisterCodeSignature} & 1 & 2 \\ \hline
\texttt{kTXMKernelSelectorValidateCodeSignature} & 1 & 0 \\ \hline
\texttt{kTXMKernelSelectorReconstituteCodeSignature} & 1 & 2 \\ \hline
\texttt{kTXMKernelSelectorRegisterAddressSpace} & 2 & 1 \\ \hline
\texttt{kTXMKernelSelectorUnregisterAddressSpace} & 1 & 0 \\ \hline
\texttt{kTXMKernelSelectorAssociateCodeSignature} & 5 & 0 \\ \hline
\texttt{kTXMKernelSelectorAllowJITRegion} & 1 & 0 \\ \hline
\texttt{kTXMKernelSelectorAssociateJITRegion} & 3 & 0 \\ \hline
\texttt{kTXMKernelSelectorAllowInvalidCode} & 1 & 0 \\ \hline
\texttt{kTXMKernelSelectorAcquireSigningIdentifier} & 1 & 1 \\ \hline
\texttt{kTXMKernelSelectorAssociateKernelEntitlements} & 2 & 0 \\ \hline
\texttt{kTXMKernelSelectorResolveKernelEntitlementsAddressSpace} & 1 & 1 \\ \hline
\texttt{kTXMKernelSelectorAccelerateEntitlements} & 1 & 1 \\ \hline
\texttt{kTXMKernelSelectorImage4SetNonce} & 2 & 0 \\ \hline
\texttt{kTXMKernelSelectorImage4RollNonce} & 1 & 0 \\ \hline
\texttt{kTXMKernelSelectorImage4GetNonce} & 1 & 1 \\ \hline
\texttt{kTXMKernelSelectorImage4GetExports} & 0 & 1 \\ \hline
\texttt{kTXMKernelSelectorImage4SetReleaseType} & 1 & 0 \\ \hline
\texttt{kTXMKernelSelectorImage4SetBootNonceShadow} & 1 & 0 \\ \hline
\texttt{kTXMKernelSelectorImage4Dispatch} & 5 & 0 \\ \hline
\end{longtable}

The control flow of calling into TXM from XNU with a specific selector is shown below in \cref{xnuTXMCALL}.
\begin{figure}[h]
\centering
\includegraphics[width=250pt]{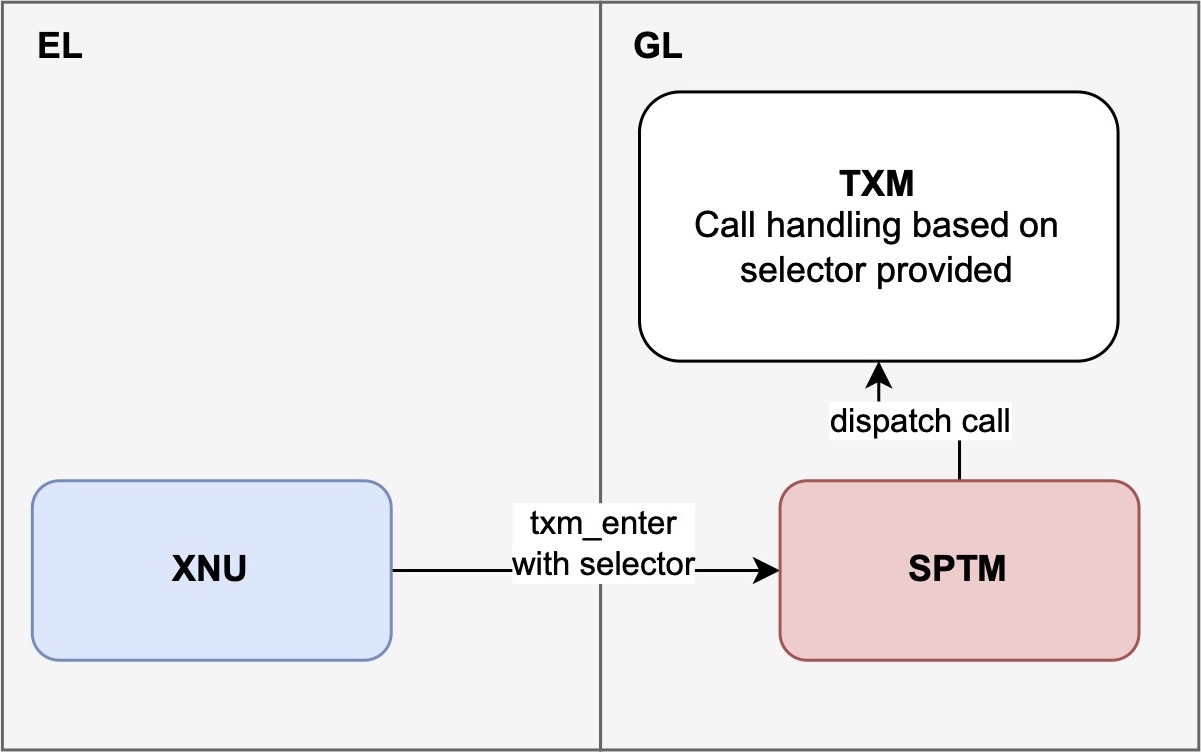}
    \caption[XNU TXM request control flow.]{Control flow of XNU issuing a request to TXM via a call into SPTM, which dispatches the call to TXM.}
    \label{xnuTXMCALL}
\end{figure}
}

\subsection{TXM Responsibilities}
From the selectors available to the \gls{SPTM}-\gls{TXM} call and its integration within the XNU open-source code, we know TXM to be the code-signing component in modern Apple devices that supports SPTM and TXM. In XNU open source \path{bsd/kern/code_signing} directory we find \texttt{ppl.c}, \texttt{xnu.c}, and \texttt{txm.c}. These are the three different code signing monitors XNU employs, with \texttt{xnu.c} indicating XNU itself to handle code signing for devices not supporting \gls{PPL}, \texttt{PPL.c} handling code signing on devices running PPL, and now TXM taking over for supported devices~\cite{df-f-2:2025}.

In such a capability, it is TXM's responsibility to provide signature and entitlement services for XNU. 
For this analysis of guarded level components, we have deemed the complete analysis of the Apple code-signing ecosystem out of scope.

\subsection{TXM Call Handling}
TXM is called from SPTM by invoking its registered dispatching functionality. For calls from XNU, the called TXM dispatch function is \path{TXM_dispatch_function_1}. We find it to confirm the validity of the caller-provided TXM stack, and then call the \texttt{txm\_dispatch\_handler}\todo{table} function. This function performs a switch on the provided selector value, and then invokes the actual handler function. The full listing of \path{txm_dispatch_handler} can be found in \cref{App:txm_dispatch_handler}.

\section{TXM Security}
In our previous analysis of SPTM, we have found TXM to own a variety of different SPTM frame types. Its memory is therefore protected from unwanted tampering from XNU in case of a kernel compromise by the fact that XNU is not allowed to map memory in TXM-owned frames (see \cref{lst:allowedTYpesToMap}). With TXM running in \gls{GXF}, the underlying access and execute permissions directly protect mapped TXM memory from manipulation by normal exception levels, and all alterations of these permissions must be performed through SPTM, as the sole controlling and management instance for page table manipulation. 
For a more precise analysis of the underlying permissions encoded by SPRR, an up-to-date analysis of \gls{SPRR} indices is required to be performed on a system supporting SPTM. Such an analysis was not performed by us and is left open as future work. % example

\chapter{Discussion}

\section{SPTM Security Implications}

The past analysis has shed light on the beginning implementation of a microkernel-like structure in \gls{XNU}. With the introduction of SPTM, Apple has introduced a new root of trust, with the sole authority regarding frame retyping and mapping. We find SPTM to adhere to a strict rule set on allowed and disallowed memory retyping and page mapping operations. Furthermore, SPTM shows how to realize so-called SPTM domains. SPTM domains serve as privilege boundaries, essentially splitting the \gls{XNU} kernel up into subsections with different privileges. By doing so, SPTM significantly limits XNU's ability to interact with security-relevant processes. Whilst we have uncovered the fundamental building blocks within \gls{SPTM} regarding memory security, we have omitted significant parts of the remapping logic. A more detailed analysis would provide a more substantial claim to support the fundamental, underlying security assumptions.

\section{Exclaves Security Implications}

We find Exclaves to isolate a core subset of system functionality from \gls{XNU}, significantly limiting its ability to interact with it. Whilst XNU can invoke services on Exclaves, as it needs to for proper system function, the actual code and memory are guarded via \gls{SPTM} by residing in a different SPTM domain in memory frames of non-shared types. This significantly limits the accessible interface to these components in the event of a kernel compromise, as XNU is no longer able to map memory into them nearly arbitrarily. With more and more parts of the system moving into compartmentalization, it appears the progress towards a true microkernel architecture has been kicked off.  

\section{Limitations}

We were successful in providing an overview of the new Apple system design, particularly in terms of communication across components, but there are obvious limitations.

Despite the XNU open-source code containing partial information regarding Exclaves, we still lack source code for the entire secure world system. As we had to rely entirely on manual reverse engineering of substantial system components, the analysis became very complex. Whilst we were able to infer general information, especially regarding component interaction, the exact inner workings are still unknown. The analysis was further complicated by the nearly complete lack of official sources regarding \gls{SPTM}, which led to us having to operate under non-confirmed assumptions about its workings. 

Furthermore, as both \gls{SPTM} and Exclaves have been released very recently, they are still subject to constant and significant change. While it is possible to perform an analysis on a snapshot of current system specifications, new releases may significantly alter behavior and necessitate new reverse engineering.

\section{Future Work}

The analysis of XNU kernel restructuring is far from done with this work. Even though we were able to shed light on the fundamental underlying concepts, especially concerning SPTM rule sets, the restructuring Apple appears to perform is highly significant. Furthermore, large parts of the architecture have yet to be reverse-engineered. The question of request dispatching with regards to TXM calling into \gls{SPTM} with \glspl{SVC} whilst other GL0 components appear to invoke them to call into SK remains. We assume that underlying thread scheduling and hypervisor configuration are responsible for this handling, but have not yet delved into this. An additional key aspect that has found little focus in our analysis is the analysis of further GL0 \emph{Exclavecore} components. Their interaction and responsibility in the Exclaves ecosystem is entirely unclear as of now, and should be the focus of future research. The same holds for the early boot process of Exclaves, which we have deemed mainly out of scope for this analysis, and the inclusion of what appears to be a seL4-like microkernel operating in the secure world. We have provided an initial look into the secure kernel, yet a complete analysis remains to be performed.

For Exclaves, we shed light on the underlying transport mechanisms, but there is still much to be discovered. The exact Tightbeam implementation and usage from Exclave components are deemed an interesting target concerning system security. As research on SPTM remains sparse, our analysis aimed to provide an architectural overview rather than conducting a comprehensive security analysis. Such an analysis, combined with a well-founded threat model, is key to evaluating SPTM-based security. Finally, as monolithic kernel architectures are often used for their superior performance, the question of performance drawbacks resulting from the implemented compartmentalization arises and should be examined in future research.

\chapter{Conclusion}
In this work, we examined \gls{SPTM}, its callers, request dispatching logic, and provided functionality. We find XNU to call into \gls{SPTM} from normal exception levels, and both \gls{TXM} and \gls{SK} calling into SPTM from guarded levels. \gls{SPTM} calls are parameterized via a call of \gls{SPTM} to be the sole component responsible for page table manipulation and underlying memory frame retyping. It employs a rigid rule set of allowed transitions to limit the caller's ability regarding page table mapping. Via this, \gls{SPTM} introduces the concept of \texttt{SPTM\_DOMAINS}, which are realized as trust domains, scoping different components from each other concerning memory accesses. We assume a gain in system security based on these changes, as kernel compromises no longer impact the entire system.

We have furthermore looked into \gls{SK}, a GL1 component that is part of the \emph{Exclavecore}, and analyzed it in terms of its own request handling capabilities and the requests it performs in turn for \gls{SPTM}. We find it to act as a kernel managing GL0 components. To serve this purpose, \gls{SK} is callable from XNU via SPTM, to allow for the scheduling of threads and requests into secure world components. It acts as the privileged management component for its GL0 clients in \emph{Exclavecore}, and handles issuing frame retyping requests to SPTM on their behalf.

The newest addition to Apple's memory protection mechanisms are Exclaves, which serve as groupings of isolated, sensitive services and resources that require special privileges to call into, and are therefore guarded against unwanted tampering from XNU. Scoped capabilities include sensor accesses and the underlying buffer management and \gls{ANE} services. Calling from XNU into Exclaves revealed a significant amount of complexity surrounding seL4 \gls{IPC} buffers, which suggests that Apple may be implementing the seL4 microkernel with \gls{SK}. We managed to interact with Exclaves from the userspace by invoking the provided functionalities from a suitably privileged task, but were unsuccessful in retrieving actual sensor data from an Exclave.

Finally, we provided a brief overview of \gls{TXM}, which resides in its own \textit{SPTM} domain and is responsible for, among other things, code signing and entitlement verification. These highly sensitive services are therefore scoped away from XNU via SPTM.

In conclusion, we see a clear path towards a more compartmentalized microkernel approach finding its way into Apple's operating systems. The implemented features enhance system security and protect against a total loss of control in the event of a kernel compromise. However, large parts of these mitigations remain unknown, which makes it very difficult to make reliable statements regarding the underlying security assumptions. With Exclaves as a rapidly evolving security enhancement, changes to the system structure in the near future are to be expected and may once again completely alter the system architecture.

	% Bibliographie
	\ifisbook\cleardoubleemptypage\fi
	\phantomsection\addcontentsline{toc}{chapter}{\refname}
	\printbibliography[category=cited]

@online{appleOSSecurity:2024,
	Author = {Apple Platform Security},
	Title = {Operating system integrity},
	Url = {https://support.apple.com/en-gb/guide/security/sec8b776536b/1/web/},
	Urldate = {2025-04-10},
	Year={2024}}

@online{casaDePaPeL:2019,
	Author = {Jonathan Levin},
	Title = {Casa De P(a)P(e)L - Explaining Apple's Page Protections Layer in A12 CPUs},
	Url = {https://newosxbook.com/articles/CasaDePPL.html},
	Urldate = {2025-04-10},
	Year={2019}}

@online{SPRRandGXF:2021,
	Author = {Sven Peter},
	Title = {Apple Silicon Hardware Secrets: SPRR and Guarded Exception Levels (GXF)},
	Url = {https://blog.svenpeter.dev/posts/m1_sprr_gxf/},
	Urldate = {2025-04-10},
	Year={2021}}

@online{ios14:2024,
	Author = {Apple Support},
	Title = {About iOS 14 Updates},
	Url = {https://support.apple.com/en-us/118390},
	Urldate = {2025-05-14},
	Year={2021}}

@online{FRIDA:2025,
	Author = {FRIDA},
	Title = {FRIDA - Dynamic instrumentation toolkit for developers, reverse-engineers, and security researchers},
	Url = {https://frida.re},
	Urldate = {2025-05-14},
	Year={2025}}

@online{palera1n:2025,
	Author = {palera1n team},
	Title = {palera1n - Jailbreak for iPhone, iPad, Macbooks, and AppleTV's for versions 15 and higher},
	Url = {https://palera.in},
	Urldate = {2025-05-14},
	Year={2024}}

@online{AppleSecurityResearchProgram:2025,
	Author = {Apple Inc.},
	Title = {Apple Security Research Device Program},
	Url = {https://security.apple.com/research-device/},
	Urldate = {2025-05-19},
	Year={2025}}

@book{FreeBSD:2015, 
title={The design and implementation of the FreeBSD operating system},
author={Marshall Kirk McKusick, George V. Neville-Neil, Robert N. M. Watson},
publisher={Addison-Wesley},
year={2015},
edition={2nd Edition.},
isbn={978-0-321-96897-5},
pages={45},
}

@online{XNU:2025,
	Author = {Apple Inc.},
	Title = {xnu},
	Url = {https://github.com/apple-oss-distributions/xnu},
	Urldate = {2025-05-19},
	Year={2025}}

@article{roch2004monolithic,
  title={Monolithic kernel vs. Microkernel},
  author={Roch, Benjamin},
  journal={TU Wien},
  volume={1},
  pages={1},
  year={2004}
}

@online{OnAppleExclaves:2025,
	Author = {Random Augustine},
	Title = {{On Apple Exclaves}},
	Url = {https://randomaugustine.medium.com/on-apple-exclaves-d683a2c37194},
	Urldate = {2025-02-18},
	Year={2025}}

@online{RandomAugustine-2:2025,
	Author = {{Random Augustine}},
	Title = {More speculation on exclaves},
	Url = {https://randomaugustine.medium.com/more-speculation-on-exclaves-d4b94182b54b},
	Urldate = {2025-02-26},
	Year={2025}}

@online{DisassemblingAppleExclaves:2025,
	Author = {{Random Augustine}},
	Title = {Disassembling Apple Exclaves},
	Url = {https://randomaugustine.medium.com/disassembling-apple-exclaves-7979bb987f86},
	Urldate = {2025-06-14},
	Year={2025}}

@article{monolithicKernelSecurity,
title = {Improving monolithic kernel security and robustness through intra-kernel sandboxing},
journal = {Computers \& Security},
volume = {127},
pages = {1},
year = {2023},
issn = {0167-4048},
doi = {https://doi.org/10.1016/j.cose.2023.103104},
url = {https://www.sciencedirect.com/science/article/pii/S0167404823000147},
author = {Bojan Novković and Marin Golub},
}

@online{ARM_SVC:2025,
	Author = {ARM},
	Title = {Supervisor Call - SVC},
	Url = {https://developer.arm.com/documentation/107706/0100/System-exceptions/Supervisor-Call---SVC},
	Urldate = {2025-05-21},
	Year={2025}}

@online{df-f-2:2025,
    Author = {{Dataflow Forensics}},
	Title = {{iOS 17: New Version, New Acronyms | Round 2}},
	Url = {https://www.df-f.com/blog/ios-17round2},
	Urldate = {2025-06-12},
	Year={2023}}

@online{df-f-3:2025,
    Author = {{Dataflow Forensics}},
	Title = {{Tracing Back to the Source | SPTM Round 3}},
	Url = {https://www.df-f.com/blog/sptm3},
	Urldate = {2025-02-12},
	Year={2025}}

@online{df-f-1:2023,
    Author = {{Dataflow Forensics}},
	Title = {{iOS 17: New Version, New Acronyms}},
	Url = {https://www.df-f.com/blog/ios17},
	Urldate = {2023-08-08},
	Year={2023}}

@online{disarm:2025,
    Author = {Jonathan Levin},
	Title = {disarm - Quick (\& formerly dirty) CLI instruction lookup for ARM64,
turned full fledged binary analyzer},
	Url = {https://newosxbook.com/tools/disarm.html},
	Urldate = {2025-06-02},
	Year={2025}}

@online{ghidra:2025,
    Author = {National Security Agency (NSA)},
	Title = {Ghidra Software Reverse Engineering Framework},
	Url = {https://github.com/NationalSecurityAgency/ghidra},
	Urldate = {2025-06-05},
	Year={2025}}

@online{appleRegisters:2025,
    Author = {{Asahi Linux}},
	Title = {m1n1: an experimentation playground for Apple Silicon},
	Url = {https://github.com/AsahiLinux/m1n1/blob/main/tools/apple_regs.json},
	Urldate = {2025-06-05},
	Year={2024}}

@online{appleRegistersLeak:2025,
    Author = {plzdonthaxme},
	Title = {accp-h16g-core-sysregs.txt},
	Url = {https://gist.github.com/justtryingthingsout/73bf33903d13a0fba12dbac92ee7cd04/},
	Urldate = {2025-09-20},
	Year={2025}}

@online{VBAR_EL1:2025,
    Author = {ARM},
	Title = {VBAR\_EL1, Vector Base Address Register (EL1)},
	Url = {https://developer.arm.com/documentation/ddi0601/2025-03/AArch64-Registers/VBAR-EL1--Vector-Base-Address-Register--EL1-},
	Urldate = {2025-06-05},
	Year={2025}}

@online{HCR_EL2:2025,
    Author = {ARM},
	Title = {HCR\_EL2, Hypervisor Configuration Register},
	Url = {https://developer.arm.com/documentation/ddi0601/2025-06/AArch64-Registers/HCR-EL2--Hypervisor-Configuration-Register},
	Urldate = {2025-06-05},
	Year={2025}}

@online{TakingAnException:2025,
    Author = {ARM},
	Title = {Taking an exception},
	Url = {https://developer.arm.com/documentation/102412/0103/Handling-exceptions/Taking-an-exception},
	Urldate = {2025-06-05},
	Year={2025}}

@online{ESR_EL1:2025,
    Author = {ARM},
	Title = {ESR\_EL1, Exception Syndrome Register (EL1)},
	Url = {https://developer.arm.com/documentation/ddi0595/2020-12/AArch64-Registers/ESR-EL1--Exception-Syndrome-Register--EL1-},
	Urldate = {2025-06-05},
	Year={2020}}

@online{ARM_Vector:2025,
	Author = {ARM},
	Title = {Registers, vectors, lanes and elements},
	Url = {https://developer.arm.com/documentation/102474/0100/Fundamentals-of-Armv8-Neon-technology/Registers--vectors--lanes-and-elements},
	Urldate = {2025-05-21},
	Year={2025}}

@online{XNU_10063:2024,
	Author = {Apple Inc.},
	Title = {XNU},
	Url = {https://github.com/apple-oss-distributions/xnu/tree/rel/xnu-10063},
	Urldate = {2025-05-21},
	Year={2024}}

@online{ARM-exceptionLevels:2024,
	Author = {ARM},
	Title = {Exception level},
	Url = {https://developer.arm.com/documentation/102412/0103/Privilege-and-Exception-levels/Exception-levels},
	Urldate = {2025-05-21},
	Year={2024}}

@online{ARM-exceptionHandling:2024,
	Author = {ARM},
	Title = {Exception Handling},
	Url = {https://developer.arm.com/documentation/den0013/d/Exception-Handling},
	Urldate = {2025-05-21},
	Year={2025}}

@online{ARM-typesOfPrivilege:2024,
	Author = {ARM},
	Title = {Types of privilege},
	Url = {https://developer.arm.com/documentation/102412/0103/Privilege-and-Exception-levels/Types-of-privilege},
	Urldate = {2025-05-21},
	Year={2025}}

@online{ARM-virtualMemory:2024,
	Author = {ARM},
	Title = {Virtual and physical addresses},
	Url = {https://developer.arm.com/documentation/101811/0104/Virtual-and-physical-addresses},
	Urldate = {2025-05-21},
	Year={2025}}

@online{ARM-multilevelTranslation:2024,
	Author = {ARM},
	Title = {Multilevel translation},
	Url = {https://developer.arm.com/documentation/101811/0104/The-Memory-Management-Unit--MMU-/Multilevel-translation},
	Urldate = {2025-05-21},
	Year={2025}}

@online{ARM-v8AddressTranslations:2024,
	Author = {ARM},
	Title = {Armv8-A Address Translation},
	Url = {https://developer.arm.com/documentation/100940/latest/},
	Urldate = {2025-05-21},
	Year={2025},
    Pages={17}}

@online{ARM-permissions:2024,
	Author = {ARM},
	Title = {Permission attributes},
	Url = {https://developer.arm.com/documentation/102376/0100/Permissions-attributes},
	Urldate = {2025-05-21},
	Year={2025}}

@online{SPRR_Asahi:2024,
	Author = {Asahi Linux Documentation},
	Title = {SPRR and GXF},
	Url = {
https://asahilinux.org/docs/hw/cpu/sprr-gxf/},
	Urldate = {2025-05-21},
	Year={2025}}

@online{P0-PAC:2024,
	Author = {Google Project Zero},
	Title = {Examining Pointer Authentication on the iPhone XS},
	Url = {https://googleprojectzero.blogspot.com/2019/02/examining-pointer-authentication-on.html},
	Urldate = {2025-05-21},
	Year={2025}}

@online{ERET:2025,
    Author = {ARM},
	Title = {ERET},
	Url = {https://developer.arm.com/documentation/ddi0602/2024-03/Base-Instructions/ERET--Exception-Return-},
	Urldate = {2025-06-05},
	Year={2020}}

@online{SPSR:2025,
    Author = {ARM},
	Title = {SPSR - Saved Program Status Register},
	Url = {https://developer.arm.com/documentation/ddi0601/2025-06/AArch32-Registers/SPSR--Saved-Program-Status-Register},
	Urldate = {2025-06-05},
	Year={2020}}

@techreport{heiser2025sel4,
  author      = {Gernot Heiser},
  title       = {The seL4 Microkernel -- An Introduction},
  type        = {White paper},
  institution = {The seL4 Foundation},
  number      = {Revision 1.4},
  year        = {2025},
  month       = jan,
  note        = {Revision 1.4 of 2025-01-08}
}

@online{capabilities:2025,
    Author = {seL4},
	Title = {Capabilities},
	Url = {https://docs.sel4.systems/Tutorials/capabilities.html},
	Urldate = {2025-06-05},
	Year={2020}}

@online{kernelSyscall:2025,
    Author = {The Apple Wiki - Community Forum},
	Title = {Kernel Syscalls},
	Url = {https://theapplewiki.com/wiki/Kernel_Syscalls},
	Urldate = {2025-06-05},
	Year={2020}}

@online{Mach:2025,
    Author = {Apple Documentation Archive},
	Title = {Mach},
	Url = {https://developer.apple.com/library/archive/documentation/Darwin/Conceptual/KernelProgramming/Mach/Mach.html},
	Urldate = {2025-06-05},
	Year={2013}}

@online{IPCSPACE:2025,
    Author = {Amit Singh},
	Title = {Section 9.3. Mach IPC: The Mac OS X Implementation},
	Url = {https://flylib.com/books/en/3.126.1.105/2/},
	Urldate = {2025-06-05},
	Year={2006}}

@manual{seL4Manual,
  title        = {seL4 Reference Manual},
  author       = {{seL4 Foundation}},
  year         = {2024},
  note         = {Accessed: 2025-07-10, page 18},
  url          = {https://sel4.systems/Info/Docs/seL4-manual-latest.pdf}
}

@misc{apple2025sequoia155,
  author       = {{Apple Inc.}},
  title        = {{macOS Sequoia 15.5 Release Notes}},
  howpublished = {\url{https://developer.apple.com/documentation/macos-release-notes/macos-15_5-release-notes}},
  year         = {2025},
  month        = may,
  note         = {Apple Developer Documentation},
}

@misc{wienand_bottomupcs_ch6s04,
  author       = {Ian Wienand},
  title        = {Chapter 6.4: \emph{Physical Memory}},
  howpublished = {\url{https://bottomupcs.com/ch06s04.html}},
  year         = {2004--2022},
  note         = {Accessed: 2025-07-11},
  organization = {Computer Science from the Bottom Up},
  license      = {CC BY‑SA 3.0}
}

@misc{azad2020ppl,
  author       = {Azad, Brandon},
  title        = {The core of Apple is PPL: Breaking the XNU kernel's kernel},
  howpublished = {\url{https://googleprojectzero.blogspot.com/2020/07/}},
  month        = jul,
  year         = {2020},
  note         = {Accessed: 2025-07-11}
}

@misc{applewiki_frameworks_2025,
  title        = {Filesystem:/System/Library/Frameworks},
  author       = {{The Apple Wiki contributors}},
  
  year         = {2025},
  month        = apr,
  day          = {10},
  url          = {https://theapplewiki.com/wiki/Filesystem:/System/Library/Frameworks},
  urldate      = {2025-07-13},
}

@online{ipsw:2025,
    Author = {Blacktop},
	Title = {ipsw},
	Url = {https://github.com/blacktop/ipsw},
	Urldate = {2025-06-05},
	Year={2025}}

@online{symbolicator:2025,
    Author = {Blacktop},
	Title = {ipsw symbolication signatures},
	Url = {https://github.com/blacktop/symbolicator},
	Urldate = {2025-06-05},
	Year={2025}}

@misc{dynamicLibraries,
  title        = {Creating Dynamic Libraries},
  author       = {{Apple Inc. }},
  howpublished = {\url{https://developer.apple.com/library/archive/documentation/DeveloperTools/Conceptual/DynamicLibraries/100-Articles/CreatingDynamicLibraries.html}},
  year         = {2012},
}

@misc{RTKit,
  title        = {RTKit},
  author       = {The Apple Wiki},
  howpublished = {https://theapplewiki.com/wiki/RTKit},
  year         = {2024},
}

@online{LLDB2024,
  author    = {{LLDB Project}},
  title     = {LLDB Debugger},
  year      = {2025},
  url       = {https://lldb.llvm.org},
}

	% ggf. Anhang
	\appendix\chapter{\appendixname}

\section{Symbol to Address Mapping for Analyzed Binaries}
\label{symbolTable}
\begin{longtable}{|l|l|l|}
\caption{Symbols encountered and labeled during the reverse engineering of this work, grouped by binary.}
\\ \hline
\textbf{Binary} & \textbf{Symbol} & \textbf{Address} \\
\hline
\endfirsthead
\hline
\textbf{Binary} & \textbf{Symbol} & \textbf{Address} \\
\hline
\endhead
\hline
\endfoot
\hline
\endlastfoot

% ---------- SPTM ----------
\multicolumn{3}{|c|}{\textbf{Binary: SPTM}} \\
\hline
SPTM & gxf\_setup\_early & fffffff02708b8e8 \\
SPTM & gxf\_setup\_late & fffffff02708b918 \\
SPTM & gxf\_entry\_point & fffffff02708053c \\
SPTM & CORE\_DISPATCH\_STRUCTURE\_POINTER & fffffff027079500 \\
SPTM & Function\_Table\_Structure & fffffff027079508 \\
SPTM & init\_xnu\_ro\_data & fffffff0270987b4 \\
SPTM & IOMMU\_bootstrap & fffffff0270be298 \\
SPTM & register\_iommu & fffffff0270bf164 \\
SPTM & CORE\_SPTM\_FUNCTION & fffffff0270bec68 \\
SPTM & SPECIAL\_DISPATCH\_STRUCTURE & fffffff027078d80 \\
SPTM & low\_level\_logger & fffffff0270a029c \\
SPTM & genter\_dispatch\_entry & fffffff0270bf3f8 \\
SPTM & synchronous\_exception\_handler\_from\_lower & fffffff027081e38 \\
SPTM & retype\_frames & fffffff0270b4118 \\
SPTM & retype & fffffff0270c4ad8 \\
SPTM & SPTM\_Retype\_Global\_Table & fffffff027019120 \\
SPTM & RETYPE\_Typeout\_Structure & fffffff027019150 \\
SPTM & RETYPE\_Flag\_Structure & fffffff027019140 \\
SPTM & PAPT\_permission\_update & fffffff0270b2a38 \\
SPTM & SPRR\_Index\_From\_Type & fffffff027019114 \\
SPTM & sk\_types\_retype\_out & fffffff0270ba318 \\
SPTM & map\_page & fffffff0270c51d4 \\
SPTM & table\_sptm\_type & fffffff0270c51d4 \\
SPTM & REMAP\_STRUCTURE & fffffff027019130 \\
SPTM & NVME\_DISPATCH\_TABLE & fffffff027014710 \\
SPTM & SK\_BOOTSTRAP\_TABLE & fffffff027018c00 \\
SPTM & t8110DART\_DISPATCH\_TABLE & fffffff027014c30 \\
SPTM & CPUTRACE\_DISPATCH\_TABLE & fffffff027014490 \\
SPTM & UAT\_DISPATCH\_TABLE & fffffff0270140c0 \\
SPTM & SART\_DISPATCH\_TABLE & fffffff0270149b0 \\
SPTM & sptm\_dispatch & fffffff0270bf268 \\
SPTM & XNU\_BOOTSTRAP\_TABLE & fffffff0270186d8 \\
SPTM & TXM\_BOOTSTRAP\_TABLE & fffffff027018980 \\
SPTM & AllowedCallerDomains & fffffff027019100 \\
SPTM & tb\_endpoint\_create\_with\_value & fffffff0088ee9ec \\
SPTM & tb\_client\_connection\_create\_with\_endpoint & fffffff0088ed99c \\
SPTM & \_\_tb\_connection\_create\_transport\_for\_endpoint & fffffff0088ed624 \\
SPTM & tb\_xnu\_transport\_create & fffffff0088f0eec \\
SPTM & tb\_xnu\_transport\_send\_message & fffffff0088f0fa4 \\
SPTM & \_\_tb\_connection\_message\_construct & fffffff0088ee154 \\
SPTM & \_tb\_connection\_send\_query & fffffff0088eda7c \\
SPTM & exclaves\_xnu\_proxy\_endpoint\_call & fffffff008148028 \\
\hline

% ---------- XNU ----------
\multicolumn{3}{|c|}{\textbf{Binary: XNU}} \\
\hline
XNU & GENTER\_main\_gate & fffffff00ab3510c \\
XNU & genter\_setup & fffffff0080f933c \\
XNU & txm\_enter & fffffff0088eb384 \\
XNU & txm\_kernel\_call & fffffff00863e048 \\
XNU & exclaves\_enter & fffffff00813b844 \\
XNU & sk\_enter & fffffff0088eb330 \\
XNU & exclaves\_bootinfo & fffffff00813b7c8 \\
XNU & sptm\_retype & fffffff0088ebc0c \\
XNU & sptm\_map\_page & fffffff0088ebc24 \\
XNU & tb\_endpoint\_create\_with\_value & fffffff0088ee9ec \\
XNU & allocate\_in\_structure & fffffff00816b4a4 \\
XNU & xnu\_tb\_connection\_create & fffffff0088ed674 \\
XNU & \_\_tb\_connection\_create\_transport\_for\_endpoint & fffffff0088ed624 \\
XNU & xnu\_tb\_endpoint\_destruct & fffffff0088eea5c \\
XNU & \_\_tb\_connection\_create\_transport\_for\_endpoint & fffffff0088f10a0 \\
XNU & tb\_xnu\_transport\_create & fffffff0088f1228 \\
XNU & tb\_xnu\_client\_transport\_vtable & fffffff007c17938 \\
XNU & xnu\_tb\_connection\_create. & fffffff0088ed674 \\
XNU & tb\_client\_connection\_activate & fffffff0088eda30 \\
XNU & activate\_connection & fffffff0088ee6d4 \\
XNU & \_\_tb\_connection\_message\_construct & fffffff0088ee154 \\
XNU & tb\_transport\_construct\_message\_buffer & fffffff0088ee73c \\
XNU & tb\_connection\_send\_query & fffffff0088eda7c \\
XNU & tb\_xnu\_transport\_send\_message. & fffffff0088f0fa4 \\
XNU & make\_xnuproxy\_endpoint\_call\_from\_tightbeam & fffffff00813af30 \\
\hline

% ---------- SK ----------
\multicolumn{3}{|c|}{\textbf{Binary: SK}} \\
\hline
SK & main\_sptm\_gate & ffffff8000001784 \\
SK & special\_sptm\_gate & ffffff800000178c \\
SK & register\_dispatch\_table & ffffff8000004340 \\
SK & retype & ffffff80000043b8 \\
SK & get\_frame\_type & ffffff80000043a0 \\
SK & sptm\_t8110\_dart\_map\_table & ffffff8000004310 \\
SK & sptm\_t8110\_dart\_unmap\_table & ffffff8000004328 \\
SK & retype\_to\_shared1 & ffffff8000004d44 \\
SK & retype\_to\_shared1 & ffffff8000006ed4 \\
SK & map\_to\_sk\_default1 & ffffff8000004dd4 \\
SK & retype\_to\_shared2 & ffffff8000006ed4 \\
SK & retype\_to\_sk\_default1 & ffffff8000004dd4 \\
SK & retype\_to\_XNU\_default & ffffff8000006f34 \\
SK & retype\_to\_sk\_default2 & ffffff8000007288 \\
SK & SVC\_handler\_from\_lower\_EL & ffffff8000008264 \\
SK & ExceptionHandlerBase & ffffff8000003800 \\
SK & RestoreContextAndEret & ffffff80000040a4 \\
SK & FunctionTable & ffffff8000012480 \\
SK & SPTM\_dispatch\_function\_0 & ffffff8000001858 \\
SK & SPTM\_dispatch\_function\_1 & ffffff80000018b8 \\
SK & SK\_called\_from\_sptm\_dispatch\_function\_0 & ffffff80000066e8 \\
\hline

% ---------- TXM ----------
\multicolumn{3}{|c|}{\textbf{Binary: TXM}} \\
\hline
TXM & core\_SPTM\_gate & fffffff01705c058 \\
TXM & type\_TXM-rx\_region\_as\_TXM\_rw\_type & fffffff01705c070 \\
TXM & register\_dispatch\_table & fffffff017043edc \\
TXM & retype & fffffff017043ec4 \\
TXM & TXM\_dispatch\_function\_0 & fffffff017024c6c \\
TXM & TXM\_disptach\_function\_1 & fffffff017024b00 \\
TXM & txm\_dispatch\_handler & fffffff017022cb0 \\
\end{longtable}

\section{SPTM Domains IDs}
\label{app:domains}
{\footnotesize
\begin{longtable}{|l|r|}
\hline
\textbf{Domain} & \textbf{ID} \\
\hline
\endfirsthead

\hline
\textbf{Domain} & \textbf{ID} \\
\hline
\endhead

\hline
\endfoot

\hline
\caption[SPTM domain IDs.]{SPTM domain IDs retrieved from the \texttt{sptm\_common.h} header file~\cite{apple2025sequoia155}.}
\label{Tab:SPTMDomainIDs} 
\endlastfoot

SPTM\_DOMAIN & 0 \\ \hline
XNU\_DOMAIN & 1 \\ \hline
TXM\_DOMAIN & 2 \\ \hline
SK\_DOMAIN & 3 \\ \hline
XNU\_HIB\_DOMAIN & 4 \\ \hline
MAX\_DOMAINS & 5 \\ \hline

\end{longtable}
}

\section{SPTM Dispatch Table IDs}
\label{app:dispatchTables}
{\footnotesize
\begin{longtable}{|l|r|}
\hline
\textbf{Dispatch Table} & \textbf{ID} \\
\hline
\endfirsthead

\hline
\textbf{Dispatch Table} & \textbf{ID} \\
\hline
\endhead

\hline
\endfoot

\hline
\caption[SPTM dispatch table IDs.]{SPTM dispatch table IDs retrieved from the \texttt{sptm\_common.h} header file~\cite{apple2025sequoia155}.} 
\label{Tab:SPTMDispatchTableIDs} 
\endlastfoot

SPTM\_DISPATCH\_TABLE\_XNU\_BOOTSTRAP & 0 \\ \hline
SPTM\_DISPATCH\_TABLE\_TXM\_BOOTSTRAP & 1 \\ \hline
SPTM\_DISPATCH\_TABLE\_SK\_BOOTSTRAP & 2 \\ \hline
SPTM\_DISPATCH\_TABLE\_T8110\_DART\_XNU & 3 \\ \hline
SPTM\_DISPATCH\_TABLE\_T8110\_DART\_SK & 4 \\ \hline
SPTM\_DISPATCH\_TABLE\_SART & 5 \\ \hline
SPTM\_DISPATCH\_TABLE\_NVME & 6 \\ \hline
SPTM\_DISPATCH\_TABLE\_UAT & 7 \\ \hline
SPTM\_DISPATCH\_TABLE\_SHART & 8 \\ \hline
SPTM\_DISPATCH\_TABLE\_RESERVED & 9 \\ \hline
SPTM\_DISPATCH\_TABLE\_HIB & 10 \\ \hline
SPTM\_DISPATCH\_TABLE\_INVALID & 11 \\ \hline

\end{longtable}
}

\section{SPTM Endpoint IDs}
\label{app:functionIDs}
{\footnotesize
\begin{longtable}{|l|r|}
\hline
\textbf{Function} & \textbf{ID} \\
\hline
\endfirsthead

\hline
\textbf{Function} & \textbf{ID} \\
\hline
\endhead

\hline
\endfoot

\hline
\caption[SPTM endpoint IDs.]{SPTM endpoint IDs retrieved from the \texttt{sptm\_xnu.h} header~\cite{apple2025sequoia155}.} 
\label{Tab:SPTMFunctionIDs} 
\endlastfoot

SPTM\_FUNCTIONID\_LOCKDOWN & 0 \\ \hline
SPTM\_FUNCTIONID\_RETYPE & 1 \\ \hline
SPTM\_FUNCTIONID\_MAP\_PAGE & 2 \\ \hline
SPTM\_FUNCTIONID\_MAP\_TABLE & 3 \\ \hline
SPTM\_FUNCTIONID\_UNMAP\_TABLE & 4 \\ \hline
SPTM\_FUNCTIONID\_UPDATE\_REGION & 5 \\ \hline
SPTM\_FUNCTIONID\_UPDATE\_DISJOINT & 6 \\ \hline
SPTM\_FUNCTIONID\_UNMAP\_REGION & 7 \\ \hline
SPTM\_FUNCTIONID\_UNMAP\_DISJOINT & 8 \\ \hline
SPTM\_FUNCTIONID\_CONFIGURE\_SHAREDREGION & 9 \\ \hline
SPTM\_FUNCTIONID\_NEST\_REGION & 10 \\ \hline
SPTM\_FUNCTIONID\_UNNEST\_REGION & 11 \\ \hline
SPTM\_FUNCTIONID\_CONFIGURE\_ROOT & 12 \\ \hline
SPTM\_FUNCTIONID\_SWITCH\_ROOT & 13 \\ \hline
SPTM\_FUNCTIONID\_REGISTER\_CPU & 14 \\ \hline
SPTM\_FUNCTIONID\_FIXUPS\_COMPLETE & 15 \\ \hline
SPTM\_FUNCTIONID\_SIGN\_USER\_POINTER & 16 \\ \hline
SPTM\_FUNCTIONID\_AUTH\_USER\_POINTER & 17 \\ \hline
SPTM\_FUNCTIONID\_REGISTER\_EXC\_RETURN & 18 \\ \hline
SPTM\_FUNCTIONID\_CPU\_ID & 19 \\ \hline
SPTM\_FUNCTIONID\_SLIDE\_REGION & 20 \\ \hline
SPTM\_FUNCTIONID\_UPDATE\_DISJOINT\_MULTIPAGE & 21 \\ \hline
SPTM\_FUNCTIONID\_REG\_READ & 22 \\ \hline
SPTM\_FUNCTIONID\_REG\_WRITE & 23 \\ \hline
SPTM\_FUNCTIONID\_GUEST\_VA\_TO\_IPA & 24 \\ \hline
SPTM\_FUNCTIONID\_GUEST\_STAGE1\_TLBOP & 25 \\ \hline
SPTM\_FUNCTIONID\_GUEST\_STAGE2\_TLBOP & 26 \\ \hline
SPTM\_FUNCTIONID\_GUEST\_DISPATCH & 27 \\ \hline
SPTM\_FUNCTIONID\_GUEST\_EXIT & 28 \\ \hline
SPTM\_FUNCTIONID\_MAP\_SK\_DOMAIN & 29 \\ \hline
SPTM\_FUNCTIONID\_HIB\_BEGIN & 30 \\ \hline
SPTM\_FUNCTIONID\_HIB\_VERIFY\_HASH\_NON\_WIRED & 31 \\ \hline
SPTM\_FUNCTIONID\_HIB\_FINALIZE\_NON\_WIRED & 32 \\ \hline
SPTM\_FUNCTIONID\_IOFILTER\_PROTECTED\_WRITE & 33 \\ \hline

\end{longtable}
}

\section{IOMMU IDs}
\label{app:iommuIDs}
{\footnotesize
\begin{longtable}{|l|c|}
\hline
\textbf{IOMMU} & \textbf{ID} \\
\hline
\endfirsthead

\hline
\textbf{IOMMU} & \textbf{ID} \\
\hline
\endhead

\hline
\endfoot

\hline
\caption[IOMMU IDs.]{IOMMU IDs retrieved from the \texttt{sptm\_xnu.h} header~\cite{apple2025sequoia155}.} 
\label{Tab:IOMMUIdentifiers} 
\endlastfoot

\texttt{IOMMU\_ID\_SHART} & 0 \\ \hline
\texttt{IOMMU\_ID\_SART} & 1 \\ \hline
\texttt{IOMMU\_ID\_NVME} & 2 \\ \hline
\texttt{IOMMU\_ID\_UAT} & 3 \\ \hline
\texttt{IOMMU\_ID\_DART\_T8020} & 4 \\ \hline
\texttt{IOMMU\_ID\_DART\_T8110} & 5 \\ \hline
\texttt{IOMMU\_ID\_DART\_T6000} & 6 \\ \hline

\end{longtable}
}

\section{\texttt{IOMMU\_bootstrap} Function in SPTM}
\label{IOMMU_bootstrap}

\begin{minted}[linenos, breaklines, breakanywhere, bgcolor=LightGray, frame=lines]{c}
void IOMMU_bootstrap(ulong iommu_id?)

{
  uint bootstrap_stage;
  byte bVar1;
  int iVar2;
  undefined8 PAPT_Address;
  ulong iommu_id;
  ulong permissions;
  IOMMU_struct *IOMMU_Struct;
  long current_GL;
  TPIDR *tpidr_e/gL2;
  TPIDR *tpidr_el2/gl2;
  long gxf_status;
  long iommu_id_offset;
  byte *valid_flag;
  
  if ((bootstrap_stage >> 7 & 1) == 0) {
                    /* does the iommu id check out compared with the current state? */
    iommu_id_offset = (iommu_id? & 0xffffffff) * 0x20;
    valid_flag = &LAB_fffffff027078c50 + iommu_id_offset;
                    /* if a functions resided there call it and store value */
    if (*(code **)(*(long *)(&IOMMUStruct + iommu_id_offset) + 8) == (code *)0x0) {
      *valid_flag = 1;
    }
    else {
      bVar1 = (**(code **)(*(long *)(&IOMMUStruct + iommu_id_offset) + 8))();
      *valid_flag = bVar1;
      if ((bVar1 & 1) == 0) {
        return;
      }
    }
                    /* if we are in the expected boot stage, check if the iommu is valid */
    if ((bootstrap_stage >> 7 & 1) == 0) {
      if ((*valid_flag & 1) == 0) {
                    /* WARNING: Subroutine does not return */
        cleanup_and_panic("IOMMU with id %d not supported");
      }
      IOMMU_Struct = *(IOMMU_struct **)(&IOMMUStruct + iommu_id_offset);
      if ((bootstrap_stage >> 7 & 1) == 0) {
                    /* GXF Status */
        gxf_status = UnkSytemRegRead(3,6,0xf,8,0);
        if (gxf_status == 0) {
          tpidr_el2/gl2 = (TPIDR *)tpidr_el2;
        }
        else {
          tpidr_el2/gl2 = (TPIDR *)UnkSytemRegRead(3,6,0xf,0xb,1);
        }
        *(ulong *)&tpidr_el2/gl2->dispatchID = (iommu_id? & 0xffffffff) + 1;
        if ((bootstrap_stage >> 7 & 1) == 0) {
          PAPT_Address = reserve_pages_for_IOMMU?
                                   (SPTM_IOMMU_BOOTSTRAP,
                                    (int)IOMMU_Struct->size? * (uint)IOMMU_Struct->pages_address? +
                                    0x3fff >> 0xe);
          *(undefined8 *)(&IOMMU_Struct_8c60 + iommu_id_offset) = PAPT_Address;
          *(undefined2 *)(&IOMMU_Struct_8c68 + iommu_id_offset) = 0;
          iVar2 = (*(code *)IOMMU_Struct->field16_0x10)();
                    /* verify we have a dispatch table here */
          if (IOMMU_Struct->dispatchTable == 0) {
                    /* WARNING: Subroutine does not return */
            cleanup_and_panic("IOMMU \"%s\" does not provide an XNU-facing dispatch table");
          }
                    /* if NVME? */
          if ((int)iommu_id? == 2) {
            iommu_id = 2;
            permissions = 0x12;
          }
          else {
            permissions = 2;
            iommu_id = iommu_id?;
          }
                    /* then we call register dispatch table with the dispatch_table pointer, and
                       some others
                       +0x38 has the table */
          register_iommu(iommu_id,IOMMU_Struct->dispatchTable_id,IOMMU_Struct->dispatchTable,
                         permissions);
          if (IOMMU_Struct->dispatch_table != 0) {
                    /* is this kinda like SK dart table specific? */
            register_iommu(iommu_id?,IOMMU_Struct->dispatchTable_id + 1,IOMMU_Struct->dispatch_table
                           ,8);
          }
          if ((bootstrap_stage >> 7 & 1) == 0) {
            current_GL = UnkSytemRegRead(3,6,0xf,8,0);
            if (current_GL == 0) {
              tpidr_e/gL2 = (TPIDR *)tpidr_el2;
            }
            else {
              tpidr_e/gL2 = (TPIDR *)UnkSytemRegRead(3,6,0xf,0xb,1);
            }
            *(undefined8 *)&tpidr_e/gL2->dispatchID = 9;
            if (iVar2 == 0) {
              return;
            }
                    /* WARNING: Subroutine does not return */
            cleanup_and_panic("IOMMU \"%s\" bootstrap failure");
          }
        }
      }
    }
  }
                    /* WARNING: Subroutine does not return */
  panic?("%s: Unexpected bootstrap stages reached. Expected: %p, Actual: %p");
}
\end{minted}
\captionof{lstlisting}[\texttt{IOMMU\_bootstrap} function in SPTM.]{\texttt{IOMMU\_bootstrap} function in SPTM.}
\label{lst:IOMMU_bootstrap}

\section{SPTM State Transition Actions}
\label{App:StateTransitionTable}
\label{app:stateTransitions}
{
\footnotesize
\begin{longtable}{|l|l|l|l|l|l|l|}
\hline
\textbf{State} & \textbf{Event} & \textbf{Transition Action Label} & \textbf{Next State} & \textbf{Domain} & \textbf{Flag} & \textbf{Domain Name} \\
\hline
\endfirsthead

\hline
\textbf{State} & \textbf{Event} & \textbf{Transition Action Label} & \textbf{Next State} & \textbf{Domain} & \textbf{Flag} & \textbf{Domain Name} \\
\hline
\endhead

0x0 & 0x0 & FUN\_fffffff0270bf150 & 0x2 &  &  &  \\ \hline
0x0 & 0x1 & FUN\_fffffff0270bf13c & 0x1 &  &  &  \\ \hline
0x0 & 0x9 & FUN\_fffffff0270816e4 & 0x13 &  &  &  \\ \hline
0x1 & 0x2 & FUN\_fffffff027080248 & 0x4 & 0x3 & 0x1 & SK\_DOMAIN \\ \hline
0x1 & 0x5 & FUN\_fffffff0270812ac & 0x0 &  &  &  \\ \hline
0x1 & 0x9 & FUN\_fffffff02708173c & 0x13 & 0x3 &  & SK\_DOMAIN \\ \hline
0x2 & 0x2 & FUN\_fffffff027015fc0 & 0x3 & 0x2 & 0x1 & TXM\_DOMAIN \\ \hline
0x2 & 0x5 & FUN\_fffffff027081258 & 0x0 &  &  &  \\ \hline
0x2 & 0x9 & FUN\_fffffff0270816e4 & 0x13 & 0x2 &  & TXM\_DOMAIN \\ \hline
0x3 & 0x5 & FUN\_fffffff027080e1c & 0x2 &  &  &  \\ \hline
0x3 & 0x9 & FUN\_fffffff0270816e4 & 0x13 &  &  &  \\ \hline
0x4 & 0x5 & FUN\_fffffff027080eb4 & 0x1 &  &  &  \\ \hline
0x4 & 0x9 & FUN\_fffffff0270816e4 & 0x13 &  &  &  \\ \hline
0x5 & 0x2 & FUN\_fffffff027080248 & 0xb & 0x1 & 0x1 & XNU\_DOMAIN \\ \hline
0x5 & 0x3 & FUN\_fffffff027080000 & 0x8 & 0x1 & 0x3 & XNU\_DOMAIN \\ \hline
0x5 & 0x4 & FUN\_fffffff027080100 & 0x10 & 0x1 & 0x1 & XNU\_DOMAIN \\ \hline
0x5 & 0xb & FUN\_fffffff027081798 & 0x15 & 0x1 &  & XNU\_DOMAIN \\ \hline
0x5 & 0xc & FUN\_fffffff027080248 & 0x6 & 0x1 & 0x1 & XNU\_DOMAIN \\ \hline
0x5 & 0xe & FUN\_fffffff027080248 & 0xc & 0x1 & 0x1 & XNU\_DOMAIN \\ \hline
0x6 & 0x5 & FUN\_fffffff02708048c & 0x6 &  &  &  \\ \hline
0x6 & 0x6 & FUN\_fffffff0270bf098 & 0x6 &  &  &  \\ \hline
0x6 & 0x9 & FUN\_fffffff027081570 & 0x13 &  &  &  \\ \hline
0x6 & 0xb & FUN\_fffffff027081798 & 0x15 &  &  &  \\ \hline
0x6 & 0xd & FUN\_fffffff027080248 & 0xb & 0x1 & 0x1 & XNU\_DOMAIN \\ \hline
0x7 & 0x2 & FUN\_fffffff027080248 & 0xc & 0x4 & 0x1 & XNU\_HIB\_DOMAIN \\ \hline
0x7 & 0xb & FUN\_fffffff027081798 & 0x15 & 0x4 &  & XNU\_HIB\_DOMAIN \\ \hline
0x8 & 0x2 & FUN\_fffffff027080248 & 0xe & 0x2 & 0x1 & TXM\_DOMAIN \\ \hline
0x8 & 0x5 & FUN\_fffffff02708048c & 0x5 & 0x2 &  & TXM\_DOMAIN \\ \hline
0x8 & 0x6 & FUN\_fffffff027080164 & 0x9 & 0x2 &  & TXM\_DOMAIN \\ \hline
0x8 & 0x9 & FUN\_fffffff027081570 & 0x13 & 0x2 &  & TXM\_DOMAIN \\ \hline
0x8 & 0xb & FUN\_fffffff027081798 & 0x15 & 0x2 &  & TXM\_DOMAIN \\ \hline
0x9 & 0x7 & FUN\_fffffff0270bf098 & 0x5 & 0x2 &  & TXM\_DOMAIN \\ \hline
0xa & 0x5 & FUN\_fffffff027081014 & 0xb & 0x1 &  & XNU\_DOMAIN \\ \hline
0xa & 0x9 & FUN\_fffffff027081570 & 0x13 & 0x2 &  & TXM\_DOMAIN \\ \hline
0xa & 0xb & FUN\_fffffff027081798 & 0x2 &  &  &  \\ \hline
0xb & 0x3 & FUN\_fffffff027080000 & 0xa &  & 0x1 &  \\ \hline
0xb & 0x5 & FUN\_fffffff02708048c & 0x5 &  &  &  \\ \hline
0xb & 0x6 & FUN\_fffffff0270bf098 & 0xd &  &  &  \\ \hline
0xb & 0x9 & FUN\_fffffff027081570 & 0x13 &  &  &  \\ \hline
0xb & 0xb & FUN\_fffffff027081798 & 0x15 &  &  &  \\ \hline
0xc & 0x4 & FUN\_fffffff027080100 & 0x11 &  & 0x1 &  \\ \hline
0xc & 0x5 & FUN\_fffffff02708048c & 0x7 &  &  &  \\ \hline
0xc & 0x9 & FUN\_fffffff027081570 & 0x13 &  &  &  \\ \hline
0xc & 0xb & FUN\_fffffff027081798 & 0x15 &  &  &  \\ \hline
0xd & 0x2 & FUN\_fffffff027080308 & 0x16 & 0x1 & 0x1 & XNU\_DOMAIN \\ \hline
0xd & 0x8 & FUN\_fffffff027080930 & 0xb & 0x1 &  & XNU\_DOMAIN \\ \hline
0xd & 0xb & FUN\_fffffff027081798 & 0x15 & 0x1 &  & XNU\_DOMAIN \\ \hline
0xe & 0x5 & FUN\_fffffff027080e1c & 0x8 & 0x1 &  & XNU\_DOMAIN \\ \hline
0xe & 0x9 & FUN\_fffffff027081488 & 0x13 &  &  &  \\ \hline
0xe & 0xb & FUN\_fffffff027081798 & 0x15 &  &  &  \\ \hline
0xf & 0x5 & FUN\_fffffff027080eb4 & 0x10 & 0x1 &  & XNU\_DOMAIN \\ \hline
0xf & 0x9 & FUN\_fffffff0270814d0 & 0x13 &  &  &  \\ \hline
0xf & 0xb & FUN\_fffffff027081798 & 0x15 &  &  &  \\ \hline
0x10 & 0x2 & FUN\_fffffff027080248 & 0xf & 0x3 & 0x1 & SK\_DOMAIN \\ \hline
0x10 & 0x5 & FUN\_fffffff02708048c & 0x5 & 0x3 &  & SK\_DOMAIN \\ \hline
0x10 & 0x7 & FUN\_fffffff0270bf098 & 0x5 & 0x3 &  & SK\_DOMAIN \\ \hline
0x10 & 0x9 & FUN\_fffffff0270814f4 & 0x13 & 0x3 &  & SK\_DOMAIN \\ \hline
0x10 & 0xb & FUN\_fffffff027081798 & 0x15 & 0x3 &  & SK\_DOMAIN \\ \hline
0x11 & 0x5 & FUN\_fffffff027081014 & 0xc & 0x3 &  & SK\_DOMAIN \\ \hline
0x11 & 0xb & FUN\_fffffff027081798 & 0x15 & 0x3 &  & SK\_DOMAIN \\ \hline
0x16 & 0x5 & FUN\_fffffff02708048c & 0xd &  &  &  \\ \hline
0x16 & 0x9 & FUN\_fffffff027081570 & 0x13 &  &  &  \\ \hline
0x16 & 0xb & FUN\_fffffff027081798 & 0x15 &  &  &  \\ \hline

\caption[SPTM state transition entries.]{The table shows SPTM-internal state transitions, based on a current state and an event. For valid state transitions, a transition action is defined and executed. The predetermined structure further defines the next state to transition into, an SPTM domain field, and a flag, which are used for validity checking.}

\end{longtable}
}

\section{\texttt{genter\_dispatch\_entry} Function}
\label{app:genterDispatchEntry}

\begin{minted}[linenos, breaklines, bgcolor=LightGray, frame=lines]{c}
void genter_dispatch_entry(ulong sptm_dispatch_target,undefined8 param_2)
{
  ushort domain_id;
  undefined4 internal_event_id;
  uint endpoint_id;
  
  domain_id = (ushort)(sptm_dispatch_target >> 0x30) & 0xff;
  if ((sptm_dispatch_target & 0xff000000000000) == 0) {
                    // Domain: SPTM_DOMAIN
    if ((sptm_dispatch_target & 0xff00000000) == 0) {
                    // Dispatch Table: 
                    // SPTM_DISPATCH_TABLE_XNU_BOOTSTRAP      
      endpoint_id = (uint)sptm_dispatch_target & 0xff;
      if (endpoint_id == 0x1b) {
        internal_event_id = 0xc;
      }
      else {
        if (endpoint_id != 0x1c) {
          internal_event_id = 0xe;
          if (endpoint_id != 0x1e) {
            internal_event_id = 2;
          }
          goto CORE_FUNCTION_CALLER;
        }
        internal_event_id = 0xd;
      }
      if ((global_flag & 1) == 0) {
        debug?(0x56,param_2,"%s(%s:%d) - %s(%#llx), %s(%#llx), %s(%#llx), %s(%#llx), %s(%#llx)\n");
      }
    }
    else {
      internal_event_id = 2;
    }
  }
  else {
        // Domain called not SPTM_DOMAIN
    if (dispatch_table_id == 2) {
        // Domain: TXM_DOMAIN
      internal_event_id = 3;
    }
    else {
      if (dispatch_table_id != 3) {
        debug?(0x25,param_2,"%s(%s:%d) - %s(%#llx), %s(%#llx), %s(%#llx)\n");
      }
      // Domain: SK_DOMAIN
      internal_event_id = 4;
    }
  }
CORE_FUNCTION_CALLER:
  CORE_SPTM_FUNCTION(internal_event_id,sptm_dispatch_target);
  return;
}
\end{minted}
\captionof{lstlisting}[SPTM \texttt{genter\_dispatch\_entry} function.]{The \texttt{genter\_dispatch\_entry} is function responsible for dispatching \texttt{GENTER} calls into SPTM.}
\label{lst:genter_dispatch_entry}

\section{\texttt{synchronous\_handler\_from\_lower} SPTM Function}
\label{app_syncrhonous_from_lower}

\begin{minted}[linenos, breaklines, bgcolor=LightGray, frame=lines]{c}
void synchronous_exception_handler_from_lower
   (undefined8 param_1,undefined1 param_2 [16],undefined1 param_3 [16],
   undefined1 param_4 [16],undefined1 param_5 [16],undefined1 param_6 [16],
   undefined1 param_7 [16],undefined1 param_8 [16],undefined1 param_9 [16],
   undefined8 param_10,ulong param_11,undefined8 param_12,undefined8 param_13,
   undefined8 param_14,undefined8 param_15,undefined8 param_16,undefined8 param_17)

{
  undefined8 uVar1;
  undefined8 uVar2;
  undefined8 uVar3;
  char *pcVar4;
  ulong EC;
  ulong SVC_Imm;
  undefined8 in_x9;
  ulong dispatch_table;
  ulong dispatch_table_id;
  long lVar5;
  undefined8 *puVar6;
  undefined8 in_x10;
  ulong some_flag_from_TPIDR_GL2;
  undefined8 in_x11;
  undefined8 in_x12;
  undefined8 in_x13;
  undefined8 in_x14;
  undefined8 in_x15;
  ulong in_x16;
  undefined8 in_x17;
  undefined8 in_x18;
  undefined8 unaff_x19;
  undefined8 unaff_x20;
  undefined8 unaff_x21;
  undefined8 unaff_x22;
  undefined8 unaff_x23;
  undefined8 unaff_x24;
  undefined8 unaff_x25;
  undefined8 unaff_x26;
  undefined8 unaff_x27;
  undefined8 unaff_x28;
  undefined8 unaff_x29;
  undefined8 unaff_x30;
  undefined8 unaff_d8;
  undefined8 in_register_00005108;
  undefined8 unaff_d9;
  undefined8 in_register_00005128;
  undefined8 unaff_d10;
  undefined8 in_register_00005148;
  undefined8 unaff_d11;
  undefined8 in_register_00005168;
  undefined8 unaff_d12;
  undefined8 in_register_00005188;
  undefined8 unaff_d13;
  undefined8 in_register_000051a8;
  undefined8 unaff_d14;
  undefined8 in_register_000051c8;
  undefined8 unaff_d15;
  undefined8 in_register_000051e8;
  undefined8 in_d16;
  undefined8 in_register_00005208;
  undefined8 in_d17;
  undefined8 in_register_00005228;
  undefined8 in_d18;
  undefined8 in_register_00005248;
  undefined8 in_d19;
  undefined8 in_register_00005268;
  undefined8 in_d20;
  undefined8 in_register_00005288;
  undefined8 in_d21;
  undefined8 in_register_000052a8;
  undefined8 in_d22;
  undefined8 in_register_000052c8;
  undefined8 in_d23;
  undefined8 in_register_000052e8;
  undefined8 in_d24;
  undefined8 in_register_00005308;
  undefined8 in_d25;
  undefined8 in_register_00005328;
  undefined8 in_d26;
  undefined8 in_register_00005348;
  undefined8 in_d27;
  undefined8 in_register_00005368;
  undefined8 in_d28;
  undefined8 in_register_00005388;
  undefined8 in_d29;
  undefined8 in_register_000053a8;
  undefined8 in_d30;
  undefined8 in_register_000053c8;
  undefined8 in_d31;
  undefined8 in_register_000053e8;
  ulong hcr_el2;
  ulong ESR_GL1;
  ulong ELR_GL1;
  ulong FAR_GL1;
  ulong ESR_GL1__;
  TPIDR *tmp_TPIDR;
  TPIDR *TPIDR;
  ulong ?daif?;
  ulong ??daif?;
  ulong SPSR_GL1;
  undefined8 SPSR_GL1_store;
  long TPIDR_GL2;
  
  pan = 0;
  tmp_TPIDR = (TPIDR *)UnkSytemRegRead(3,6,0xf,0xb,1);
  do {
  } while (&stack0x00000000 != *(undefined1 **)&tmp_TPIDR->field_0x470);
  hcr_el2 = hcr_el2;
  if ((hcr_el2 & 0x8000001) == 1) {
    pan = 0;
    TPIDR = (TPIDR *)UnkSytemRegRead(3,6,0xf,0xb,1);
    *(undefined8 *)&TPIDR->field_0x650 = param_12;
    *(undefined8 *)&TPIDR->field_0x658 = param_13;
    *(undefined8 *)&TPIDR->field_0x660 = param_14;
    *(undefined8 *)&TPIDR->field_0x668 = param_15;
    *(undefined8 *)&TPIDR->field_0x670 = param_16;
    *(undefined8 *)&TPIDR->field_0x678 = param_17;
    *(undefined8 *)&TPIDR->field_0x680 = param_1;
    *(undefined8 *)&TPIDR->field_0x688 = in_x9;
    *(undefined8 *)&TPIDR->field_0x690 = in_x10;
    *(undefined8 *)&TPIDR->field_0x698 = in_x11;
    *(undefined8 *)&TPIDR->field_0x6a0 = in_x12;
    *(undefined8 *)&TPIDR->field_0x6a8 = in_x13;
    *(undefined8 *)&TPIDR->field_0x6b0 = in_x14;
    *(undefined8 *)&TPIDR->field_0x6b8 = in_x15;
    *(ulong *)&TPIDR->field_0x6c0 = in_x16;
    *(undefined8 *)&TPIDR->field_0x6c8 = in_x17;
    *(undefined8 *)&TPIDR->field_0x6d0 = in_x18;
    *(undefined8 *)&TPIDR->field_0x6d8 = unaff_x19;
    *(undefined8 *)&TPIDR->field_0x6e0 = unaff_x20;
    *(undefined8 *)&TPIDR->field_0x6e8 = unaff_x21;
    *(undefined8 *)&TPIDR->field_0x6f0 = unaff_x22;
    *(undefined8 *)&TPIDR->field_0x6f8 = unaff_x23;
    *(undefined8 *)&TPIDR->field_0x700 = unaff_x24;
    *(undefined8 *)&TPIDR->field_0x708 = unaff_x25;
    *(undefined8 *)&TPIDR->field_0x710 = unaff_x26;
    *(undefined8 *)&TPIDR->field_0x718 = unaff_x27;
    *(undefined8 *)&TPIDR->field_0x720 = unaff_x28;
    *(undefined8 *)&TPIDR->field_0x728 = unaff_x29;
    *(undefined8 *)&TPIDR->field_0x730 = unaff_x30;
    *(long *)&TPIDR->field_0x768 = param_2._8_8_;
    *(long *)&TPIDR->field_0x760 = param_2._0_8_;
    *(long *)&TPIDR->field_0x778 = param_3._8_8_;
    *(long *)&TPIDR->field_0x770 = param_3._0_8_;
    *(long *)&TPIDR->field_0x788 = param_4._8_8_;
    *(long *)&TPIDR->field_0x780 = param_4._0_8_;
    *(long *)&TPIDR->field_0x798 = param_5._8_8_;
    *(long *)&TPIDR->field_0x790 = param_5._0_8_;
    *(long *)&TPIDR->field_0x7a8 = param_6._8_8_;
    *(long *)&TPIDR->field_0x7a0 = param_6._0_8_;
    *(long *)&TPIDR->field_0x7b8 = param_7._8_8_;
    *(long *)&TPIDR->field_0x7b0 = param_7._0_8_;
    *(long *)&TPIDR->field_0x7c8 = param_8._8_8_;
    *(long *)&TPIDR->field_0x7c0 = param_8._0_8_;
    *(long *)&TPIDR->field_0x7d8 = param_9._8_8_;
    *(long *)&TPIDR->field_0x7d0 = param_9._0_8_;
    *(undefined8 *)&TPIDR->field_0x7e8 = in_register_00005108;
    *(undefined8 *)&TPIDR->field_0x7e0 = unaff_d8;
    *(undefined8 *)&TPIDR->field_0x7f8 = in_register_00005128;
    *(undefined8 *)&TPIDR->field_0x7f0 = unaff_d9;
    *(undefined8 *)&TPIDR->field_0x808 = in_register_00005148;
    *(undefined8 *)&TPIDR->field_0x800 = unaff_d10;
    *(undefined8 *)&TPIDR->field_0x818 = in_register_00005168;
    *(undefined8 *)&TPIDR->field_0x810 = unaff_d11;
    *(undefined8 *)&TPIDR->field_0x828 = in_register_00005188;
    *(undefined8 *)&TPIDR->field_0x820 = unaff_d12;
    *(undefined8 *)&TPIDR->field_0x838 = in_register_000051a8;
    *(undefined8 *)&TPIDR->field_0x830 = unaff_d13;
    *(undefined8 *)&TPIDR->field_0x848 = in_register_000051c8;
    *(undefined8 *)&TPIDR->field_0x840 = unaff_d14;
    *(undefined8 *)&TPIDR->field_0x858 = in_register_000051e8;
    *(undefined8 *)&TPIDR->field_0x850 = unaff_d15;
    *(undefined8 *)&TPIDR->field_0x868 = in_register_00005208;
    *(undefined8 *)&TPIDR->field_0x860 = in_d16;
    *(undefined8 *)&TPIDR->field_0x878 = in_register_00005228;
    *(undefined8 *)&TPIDR->field_0x870 = in_d17;
    *(undefined8 *)&TPIDR->field_0x888 = in_register_00005248;
    *(undefined8 *)&TPIDR->field_0x880 = in_d18;
    *(undefined8 *)&TPIDR->field_0x898 = in_register_00005268;
    *(undefined8 *)&TPIDR->field_0x890 = in_d19;
    *(undefined8 *)&TPIDR->field_0x8a8 = in_register_00005288;
    *(undefined8 *)&TPIDR->field_0x8a0 = in_d20;
    *(undefined8 *)&TPIDR->field_0x8b8 = in_register_000052a8;
    *(undefined8 *)&TPIDR->field_0x8b0 = in_d21;
    *(undefined8 *)&TPIDR->field_0x8c8 = in_register_000052c8;
    *(undefined8 *)&TPIDR->field_0x8c0 = in_d22;
    *(undefined8 *)&TPIDR->field_0x8d8 = in_register_000052e8;
    *(undefined8 *)&TPIDR->field_0x8d0 = in_d23;
    *(undefined8 *)&TPIDR->field_0x8e8 = in_register_00005308;
    *(undefined8 *)&TPIDR->field_0x8e0 = in_d24;
    *(undefined8 *)&TPIDR->field_0x8f8 = in_register_00005328;
    *(undefined8 *)&TPIDR->field_0x8f0 = in_d25;
    *(undefined8 *)&TPIDR->field_0x908 = in_register_00005348;
    *(undefined8 *)&TPIDR->field_0x900 = in_d26;
    *(undefined8 *)&TPIDR->field_0x918 = in_register_00005368;
    *(undefined8 *)&TPIDR->field_0x910 = in_d27;
    *(undefined8 *)&TPIDR->field_0x928 = in_register_00005388;
    *(undefined8 *)&TPIDR->field_0x920 = in_d28;
    *(undefined8 *)&TPIDR->field_0x938 = in_register_000053a8;
    *(undefined8 *)&TPIDR->field_0x930 = in_d29;
    *(undefined8 *)&TPIDR->field_0x948 = in_register_000053c8;
    *(undefined8 *)&TPIDR->field_0x940 = in_d30;
    *(undefined8 *)&TPIDR->field_0x950 = in_d31;
    *(undefined8 *)&TPIDR->field_0x958 = in_register_000053e8;
    ELR_GL1 = UnkSytemRegRead(3,6,0xf,10,6);
    SPSR_GL1_store = UnkSytemRegRead(3,6,0xf,10,3);
    uVar2 = fpsr;
    uVar1 = fpcr;
    *(ulong *)&TPIDR->ELR_store = ELR_GL1;
    *(int *)&TPIDR->SPSR_store = (int)SPSR_GL1_store;
    *(int *)&TPIDR->floating_point_status_register = (int)uVar2;
    *(int *)&TPIDR->floating_point_control_register = (int)uVar1;
    FAR_GL1 = UnkSytemRegRead(3,6,0xf,10,7);
    ESR_GL1__ = UnkSytemRegRead(3,6,0xf,10,5);
    *(ulong *)&TPIDR->FAR_store = FAR_GL1;
    *(ulong *)&TPIDR->ESR_store = ESR_GL1__;
    *(undefined8 *)&TPIDR->field_0x640 = param_10;
    *(ulong *)&TPIDR->field_0x648 = param_11;
    SPSR_GL1_store = sp_el0;
    *(undefined8 *)&TPIDR->field_0x738 = SPSR_GL1_store;
    calls_CSPTM_event6_x16_param2(3,3);
    return;
  }
                    /* GL1 Exception Syndrome Register
                        */
  ESR_GL1 = UnkSytemRegRead(3,6,0xf,10,5);
  EC = ESR_GL1 >> 0x1a & 0x3f;
  if (EC == 0b00010101) {
    SPSR_GL1 = UnkSytemRegRead(3,6,0xf,10,5);
    SVC_Imm = SPSR_GL1 & 0xffff;
                    /* we have SVC 0 */
    if (SVC_Imm == 0) {
      dispatch_table = in_x16 >> 0x20 & 0xff;
      if (dispatch_table == 0xfd) {
        SVC_0_is_0xfd/RETURN_TO_CALLER();
        return;
      }
      if (dispatch_table != 0xfe) {
        if (dispatch_table != 0xff) {
          TPIDR_GL2 = UnkSytemRegRead(3,6,0xf,0xb,1);
          lVar5 = *(long *)(TPIDR_GL2 + 0x470);
          *(undefined8 *)(TPIDR_GL2 + 0x490) = param_10;
          *(ulong *)(TPIDR_GL2 + 0x498) = param_11;
          *(undefined8 *)(TPIDR_GL2 + 0x4a0) = param_12;
          *(undefined8 *)(TPIDR_GL2 + 0x4a8) = param_13;
          *(undefined8 *)(TPIDR_GL2 + 0x4b0) = param_14;
          *(undefined8 *)(TPIDR_GL2 + 0x4b8) = param_15;
          *(undefined8 *)(TPIDR_GL2 + 0x4c0) = param_16;
          *(undefined8 *)(TPIDR_GL2 + 0x4c8) = param_17;
          some_flag_from_TPIDR_GL2 = *(ulong *)(TPIDR_GL2 + 0x488);
          do {
          } while (1 < some_flag_from_TPIDR_GL2);
          do {
          } while (1 < some_flag_from_TPIDR_GL2);
          if (some_flag_from_TPIDR_GL2 == 1) {
            puVar6 = (undefined8 *)(TPIDR_GL2 + 0x588);
          }
          else {
            puVar6 = (undefined8 *)(TPIDR_GL2 + 0x4d0);
          }
          *puVar6 = unaff_x19;
          puVar6[1] = unaff_x20;
                    /* we basically put TPIDR_GL2 on the stack? */
          puVar6[2] = unaff_x21;
          puVar6[3] = unaff_x22;
          puVar6[4] = unaff_x23;
          puVar6[5] = unaff_x24;
          puVar6[6] = unaff_x25;
          puVar6[7] = unaff_x26;
          puVar6[8] = unaff_x27;
          puVar6[9] = unaff_x28;
          puVar6[0x12] = unaff_x29;
          puVar6[0x13] = unaff_x30;
          puVar6[10] = unaff_d8;
          puVar6[0xb] = unaff_d9;
          puVar6[0xc] = unaff_d10;
          puVar6[0xd] = unaff_d11;
          puVar6[0xe] = unaff_d12;
          puVar6[0xf] = unaff_d13;
          puVar6[0x10] = unaff_d14;
          puVar6[0x11] = unaff_d15;
          *(ulong *)(TPIDR_GL2 + 0x488) = some_flag_from_TPIDR_GL2 + 1;
                    /* SPSR_GL1 - holds the saved process stat when an exception is taken to EL1 */
          SPSR_GL1_store = UnkSytemRegRead(3,6,0xf,10,3);
          puVar6[0x16] = SPSR_GL1_store;
          SPSR_GL1_store = sp_el0;
          puVar6[0x14] = SPSR_GL1_store;
                    /* ELR_GL1 - when taking an exception to EL1, holds the address to return to
                        */
          SPSR_GL1_store = UnkSytemRegRead(3,6,0xf,10,6);
          puVar6[0x15] = SPSR_GL1_store;
          *(undefined8 *)(lVar5 + -0x10) = unaff_x29;
          *(undefined8 *)(lVar5 + -8) = unaff_x30;
          CORE_SPTM_FUNCTION(2);
          return;
        }
        SVC_0_is_0xff();
        return;
      }
                    /* TPIDR_GL0 */
      TPIDR_GL2 = UnkSytemRegRead(3,6,0xf,0xb,1);
                    /* SPRR_UPERM_EL0 - SPRR User Permissions Configuration Register
                       we set this to a value from TPIDR_GL2 */
      UnkSytemRegWrite(3,6,0xf,1,5,*(undefined8 *)(TPIDR_GL2 + 0x9a0));
      mdscr_el1 = *(undefined8 *)(TPIDR_GL2 + 0x9a8);
                    /* AGTCNTRDIR_EL1 */
      UnkSytemRegWrite(3,1,0xf,1,5,*(undefined8 *)(TPIDR_GL2 + 0x9b0));
                    /* 3_4_c15_c0_4 */
      UnkSytemRegWrite(3,4,0xf,0,4,*(undefined8 *)(TPIDR_GL2 + 0x9b8));
      sctlr_el1 = *(undefined8 *)(TPIDR_GL2 + 0x9c0);
      tcr_el1 = *(undefined8 *)(TPIDR_GL2 + 0x9c8);
      spsel = 0;
      calls_CSPTM_both_variable(param_10,0,1);
      return;
    }
    if (SVC_Imm == 0x25) {
      ?daif? = UnkSytemRegRead(3,6,0xf,10,3);
      UnkSytemRegWrite(3,6,0xf,10,3,?daif? & 0xfffffffffffffe3f);
      ExceptionReturn();
      return;
    }
    if (SVC_Imm == 0x26) {
      ??daif? = UnkSytemRegRead(3,6,0xf,10,3);
      UnkSytemRegWrite(3,6,0xf,10,3,??daif? | 0x1c0);
      ExceptionReturn();
      return;
    }
    do {
      WaitForEvent();
    } while( true );
  }
  if (EC != 0b00010110) {
    TPIDR_GL2 = UnkSytemRegRead(3,6,0xf,0xb,1);
    *(undefined8 *)(TPIDR_GL2 + 0x640) = param_10;
    *(ulong *)(TPIDR_GL2 + 0x648) = param_11;
    SPSR_GL1_store = sp_el0;
    *(undefined8 *)(TPIDR_GL2 + 0x738) = SPSR_GL1_store;
    *(undefined8 *)(TPIDR_GL2 + 0x650) = param_12;
    *(undefined8 *)(TPIDR_GL2 + 0x658) = param_13;
    *(undefined8 *)(TPIDR_GL2 + 0x660) = param_14;
    *(undefined8 *)(TPIDR_GL2 + 0x668) = param_15;
    *(undefined8 *)(TPIDR_GL2 + 0x670) = param_16;
    *(undefined8 *)(TPIDR_GL2 + 0x678) = param_17;
    *(undefined8 *)(TPIDR_GL2 + 0x680) = param_1;
    *(undefined8 *)(TPIDR_GL2 + 0x688) = in_x9;
    *(undefined8 *)(TPIDR_GL2 + 0x690) = in_x10;
    *(undefined8 *)(TPIDR_GL2 + 0x698) = in_x11;
    *(undefined8 *)(TPIDR_GL2 + 0x6a0) = in_x12;
    *(undefined8 *)(TPIDR_GL2 + 0x6a8) = in_x13;
    *(undefined8 *)(TPIDR_GL2 + 0x6b0) = in_x14;
    *(undefined8 *)(TPIDR_GL2 + 0x6b8) = in_x15;
    *(ulong *)(TPIDR_GL2 + 0x6c0) = in_x16;
    *(undefined8 *)(TPIDR_GL2 + 0x6c8) = in_x17;
    *(undefined8 *)(TPIDR_GL2 + 0x6d0) = in_x18;
    *(undefined8 *)(TPIDR_GL2 + 0x6d8) = unaff_x19;
    *(undefined8 *)(TPIDR_GL2 + 0x6e0) = unaff_x20;
    *(undefined8 *)(TPIDR_GL2 + 0x6e8) = unaff_x21;
    *(undefined8 *)(TPIDR_GL2 + 0x6f0) = unaff_x22;
    *(undefined8 *)(TPIDR_GL2 + 0x6f8) = unaff_x23;
    *(undefined8 *)(TPIDR_GL2 + 0x700) = unaff_x24;
    *(undefined8 *)(TPIDR_GL2 + 0x708) = unaff_x25;
    *(undefined8 *)(TPIDR_GL2 + 0x710) = unaff_x26;
    *(undefined8 *)(TPIDR_GL2 + 0x718) = unaff_x27;
    *(undefined8 *)(TPIDR_GL2 + 0x720) = unaff_x28;
    *(undefined8 *)(TPIDR_GL2 + 0x728) = unaff_x29;
    *(undefined8 *)(TPIDR_GL2 + 0x730) = unaff_x30;
    *(long *)(TPIDR_GL2 + 0x768) = param_2._8_8_;
    *(long *)(TPIDR_GL2 + 0x760) = param_2._0_8_;
    *(long *)(TPIDR_GL2 + 0x778) = param_3._8_8_;
    *(long *)(TPIDR_GL2 + 0x770) = param_3._0_8_;
    *(long *)(TPIDR_GL2 + 0x788) = param_4._8_8_;
    *(long *)(TPIDR_GL2 + 0x780) = param_4._0_8_;
    *(long *)(TPIDR_GL2 + 0x798) = param_5._8_8_;
    *(long *)(TPIDR_GL2 + 0x790) = param_5._0_8_;
    *(long *)(TPIDR_GL2 + 0x7a8) = param_6._8_8_;
    *(long *)(TPIDR_GL2 + 0x7a0) = param_6._0_8_;
    *(long *)(TPIDR_GL2 + 0x7b8) = param_7._8_8_;
    *(long *)(TPIDR_GL2 + 0x7b0) = param_7._0_8_;
    *(long *)(TPIDR_GL2 + 0x7c8) = param_8._8_8_;
    *(long *)(TPIDR_GL2 + 0x7c0) = param_8._0_8_;
    *(long *)(TPIDR_GL2 + 0x7d8) = param_9._8_8_;
    *(long *)(TPIDR_GL2 + 2000) = param_9._0_8_;
    *(undefined8 *)(TPIDR_GL2 + 0x7e8) = in_register_00005108;
    *(undefined8 *)(TPIDR_GL2 + 0x7e0) = unaff_d8;
    *(undefined8 *)(TPIDR_GL2 + 0x7f8) = in_register_00005128;
    *(undefined8 *)(TPIDR_GL2 + 0x7f0) = unaff_d9;
    *(undefined8 *)(TPIDR_GL2 + 0x808) = in_register_00005148;
    *(undefined8 *)(TPIDR_GL2 + 0x800) = unaff_d10;
    *(undefined8 *)(TPIDR_GL2 + 0x818) = in_register_00005168;
    *(undefined8 *)(TPIDR_GL2 + 0x810) = unaff_d11;
    *(undefined8 *)(TPIDR_GL2 + 0x828) = in_register_00005188;
    *(undefined8 *)(TPIDR_GL2 + 0x820) = unaff_d12;
    *(undefined8 *)(TPIDR_GL2 + 0x838) = in_register_000051a8;
    *(undefined8 *)(TPIDR_GL2 + 0x830) = unaff_d13;
    *(undefined8 *)(TPIDR_GL2 + 0x848) = in_register_000051c8;
    *(undefined8 *)(TPIDR_GL2 + 0x840) = unaff_d14;
    *(undefined8 *)(TPIDR_GL2 + 0x858) = in_register_000051e8;
    *(undefined8 *)(TPIDR_GL2 + 0x850) = unaff_d15;
    *(undefined8 *)(TPIDR_GL2 + 0x868) = in_register_00005208;
    *(undefined8 *)(TPIDR_GL2 + 0x860) = in_d16;
    *(undefined8 *)(TPIDR_GL2 + 0x878) = in_register_00005228;
    *(undefined8 *)(TPIDR_GL2 + 0x870) = in_d17;
    *(undefined8 *)(TPIDR_GL2 + 0x888) = in_register_00005248;
    *(undefined8 *)(TPIDR_GL2 + 0x880) = in_d18;
    *(undefined8 *)(TPIDR_GL2 + 0x898) = in_register_00005268;
    *(undefined8 *)(TPIDR_GL2 + 0x890) = in_d19;
    *(undefined8 *)(TPIDR_GL2 + 0x8a8) = in_register_00005288;
    *(undefined8 *)(TPIDR_GL2 + 0x8a0) = in_d20;
    *(undefined8 *)(TPIDR_GL2 + 0x8b8) = in_register_000052a8;
    *(undefined8 *)(TPIDR_GL2 + 0x8b0) = in_d21;
    *(undefined8 *)(TPIDR_GL2 + 0x8c8) = in_register_000052c8;
    *(undefined8 *)(TPIDR_GL2 + 0x8c0) = in_d22;
    *(undefined8 *)(TPIDR_GL2 + 0x8d8) = in_register_000052e8;
    *(undefined8 *)(TPIDR_GL2 + 0x8d0) = in_d23;
    *(undefined8 *)(TPIDR_GL2 + 0x8e8) = in_register_00005308;
    *(undefined8 *)(TPIDR_GL2 + 0x8e0) = in_d24;
    *(undefined8 *)(TPIDR_GL2 + 0x8f8) = in_register_00005328;
    *(undefined8 *)(TPIDR_GL2 + 0x8f0) = in_d25;
    *(undefined8 *)(TPIDR_GL2 + 0x908) = in_register_00005348;
    *(undefined8 *)(TPIDR_GL2 + 0x900) = in_d26;
    *(undefined8 *)(TPIDR_GL2 + 0x918) = in_register_00005368;
    *(undefined8 *)(TPIDR_GL2 + 0x910) = in_d27;
    *(undefined8 *)(TPIDR_GL2 + 0x928) = in_register_00005388;
    *(undefined8 *)(TPIDR_GL2 + 0x920) = in_d28;
    *(undefined8 *)(TPIDR_GL2 + 0x938) = in_register_000053a8;
    *(undefined8 *)(TPIDR_GL2 + 0x930) = in_d29;
    *(undefined8 *)(TPIDR_GL2 + 0x948) = in_register_000053c8;
    *(undefined8 *)(TPIDR_GL2 + 0x940) = in_d30;
    *(undefined8 *)(TPIDR_GL2 + 0x950) = in_d31;
    *(undefined8 *)(TPIDR_GL2 + 0x958) = in_register_000053e8;
    SPSR_GL1_store = UnkSytemRegRead(3,6,0xf,10,6);
    uVar1 = UnkSytemRegRead(3,6,0xf,10,3);
    uVar3 = fpsr;
    uVar2 = fpcr;
    *(undefined8 *)(TPIDR_GL2 + 0x740) = SPSR_GL1_store;
    *(int *)(TPIDR_GL2 + 0x748) = (int)uVar1;
    *(int *)(TPIDR_GL2 + 0x960) = (int)uVar3;
    *(int *)(TPIDR_GL2 + 0x964) = (int)uVar2;
    SPSR_GL1_store = UnkSytemRegRead(3,6,0xf,10,7);
    uVar1 = UnkSytemRegRead(3,6,0xf,10,5);
    *(undefined8 *)(TPIDR_GL2 + 0x750) = SPSR_GL1_store;
    *(undefined8 *)(TPIDR_GL2 + 0x758) = uVar1;
    lVar5 = UnkSytemRegRead(3,6,0xf,0xb,1);
    lVar5 = *(long *)(lVar5 + 0x470);
    SPSR_GL1_store = UnkSytemRegRead(3,6,0xf,10,6);
    *(undefined8 *)(lVar5 + -0x10) = unaff_x29;
    *(undefined8 *)(lVar5 + -8) = SPSR_GL1_store;
    lVar5 = hcr_el2;
    if (lVar5 == 0x30480000000) {
      SPSR_GL1_store = 2;
      SPSR_GL1 = UnkSytemRegRead(3,6,0xf,10,3);
      if ((SPSR_GL1 & 0xc) == 0) {
        pcVar4 = "[SK] Unhandled synchronous exception taken from GL0";
      }
      else {
        pcVar4 = "[SK] Unhandled synchronous exception taken from GL1";
      }
    }
    else {
      SPSR_GL1_store = 1;
      pcVar4 = "[TXM] Unhandled synchronous exception taken from GL0";
    }
    if (DAT_fffffff027085028 != 0) {
      *(undefined8 *)(DAT_fffffff027085028 + 0x40) = SPSR_GL1_store;
    }
    FUN_fffffff0270cc638((undefined8 *)(TPIDR_GL2 + 0x640),pcVar4);
    return;
  }
  dispatch_table_id = in_x16 >> 0x20 & 0xff;
  if (dispatch_table_id == 0xfd) {
    FUN_fffffff027082530();
    return;
  }
  if (dispatch_table_id == 0xfe) {
    FUN_fffffff027082218();
    return;
  }
  if (dispatch_table_id == 0xff) {
    TPIDR_GL2 = UnkSytemRegRead(3,6,0xf,0xb,1);
    lVar5 = *(long *)(TPIDR_GL2 + 0x470);
    *(undefined8 *)(lVar5 + -0x10) = 0;
    *(undefined1 **)(lVar5 + -8) = &LAB_fffffff0270825b0;
    *(undefined8 *)(TPIDR_GL2 + 0x998) = param_10;
    SPSR_GL1 = *(long *)(TPIDR_GL2 + 0x488) - 1;
    do {
    } while (1 < SPSR_GL1);
    do {
    } while (1 < SPSR_GL1);
    if (SPSR_GL1 == 1) {
      lVar5 = TPIDR_GL2 + 0x588;
    }
    else {
      lVar5 = TPIDR_GL2 + 0x4d0;
    }
    SPSR_GL1_store = *(undefined8 *)(lVar5 + 0x90);
    uVar1 = *(undefined8 *)(lVar5 + 0x98);
    *(undefined1 **)(TPIDR_GL2 + 0x740) = &LAB_fffffff0270825b0;
    *(undefined8 *)(TPIDR_GL2 + 0x728) = SPSR_GL1_store;
    *(undefined8 *)(TPIDR_GL2 + 0x730) = uVar1;
    CORE_SPTM_FUNCTION(7,(param_11 & 1) << 2 | 2);
    return;
  }
  leads_to_CORE_SPTM_FUNCTION();
  return;
}

\end{minted}
\captionof{lstlisting}[SPTM \texttt{synchronous\_from\_lower} function.]{\texttt{synchronous\_handler\_from\_lower} function in SPTM responsible for handling incoming SVC and HVC exceptions.}
\label{lst:svcHandlerSynchronous}

\section{SPTM SVC Handling Excerpt from \protect \path{synchronous_exception_handler_from_lower}}
\label{lst:sptm_SVC-handler}
\begin{listing}[H]
    \begin{minted}[linenos, breaklines, bgcolor=LightGray, frame=lines]{asm}
                             SVC_HANDLER                                     
     ff027081f88 a8 fa 3e d5     mrs        EC,sreg(0x3, 0x6, c0xf, c0xa, 0x5)
     ff027081f8c 08 3d 40 92     and        SVC_Imm,SVC_Imm,#0xffff
                             we have SVC 0
     ff027081f90 1f 01 00 f1     cmp        SVC_Imm,#0x0
     ff027081f94 e0 02 00 54     b.eq       SVC_0_HANDLER
     ff027081f98 1f 95 00 f1     cmp        SVC_Imm,#0x25
     ff027081f9c c0 00 00 54     b.eq       SVC_37_HANDLER_enableInterrupts
     ff027081fa0 1f 99 00 f1     cmp        SVC_Imm,#0x26
     ff027081fa4 80 01 00 54     b.eq       SVC_38_HANDLER_disableInterrupts
     ff027081fa8 a0 d5 9b d2     mov        x0,#0xdead
                             LAB_fffffff027081fac                           
     ff027081fac 5f 20 03 d5     wfe
     ff027081fb0 ff ff ff 17     b          LAB_fffffff027081fac

    \end{minted}
    \captionof{lstlisting}{SPTM synchronous\_exception\_handler\_from\_lower function.}
\end{listing}

\section{SPTM Frame Types}
\label{App:SPTMTypesNew} 
{\footnotesize
\begin{longtable}{|l|r|}

\hline
\textbf{Type} & \textbf{Value} \\
\hline
\endfirsthead

\hline
\textbf{Type} & \textbf{Value} \\
\hline
\endhead

\hline
\endfoot

\hline
\caption[SPTM frame type to integer mapping.]{SPTM frame types and their corresponding integer values~\cite{apple2025sequoia155}.}
\endlastfoot

SPTM\_UNTYPED & 0 \\ \hline
SPTM\_UNUSED & 1 \\ \hline
SPTM\_DEFAULT & 2 \\ \hline
SPTM\_RO & 3 \\ \hline
SPTM\_CODE & 4 \\ \hline
SPTM\_TXM\_CODE & 5 \\ \hline
SPTM\_XNU\_CODE & 6 \\ \hline
SPTM\_XNU\_CODE\_DBG\_RW & 7 \\ \hline
SPTM\_KERNEL\_ROOT\_TABLE & 8 \\ \hline
SPTM\_PAGE\_TABLE & 9 \\ \hline
SPTM\_IOMMU\_BOOTSTRAP & 10 \\ \hline

XNU\_DEFAULT & 11 \\ \hline
XNU\_RO & 12 \\ \hline
XNU\_RO\_DBG\_RW & 13 \\ \hline
XNU\_USER\_EXEC & 14 \\ \hline
XNU\_USER\_DEBUG & 15 \\ \hline
XNU\_USER\_JIT & 16 \\ \hline
XNU\_USER\_ROOT\_TABLE & 17 \\ \hline
XNU\_SHARED\_ROOT\_TABLE & 18 \\ \hline
XNU\_PAGE\_TABLE & 19 \\ \hline
XNU\_PAGE\_TABLE\_SHARED & 20 \\ \hline
XNU\_PAGE\_TABLE\_ROZONE & 21 \\ \hline
XNU\_PAGE\_TABLE\_COMMPAGE & 22 \\ \hline
XNU\_IOMMU & 23 \\ \hline
XNU\_ROZONE & 24 \\ \hline
XNU\_IO & 25 \\ \hline
XNU\_PROTECTED\_IO & 26 \\ \hline
XNU\_COMMPAGE\_RW & 27 \\ \hline
XNU\_COMMPAGE\_RO & 28 \\ \hline
XNU\_COMMPAGE\_RX & 29 \\ \hline
XNU\_TAG\_STORAGE & 30 \\ \hline
XNU\_STAGE2\_ROOT\_TABLE & 31 \\ \hline
XNU\_STAGE2\_PAGE\_TABLE & 32 \\ \hline
XNU\_KERNEL\_RESTRICTED & 33 \\ \hline
XNU\_RESERVED\_1 & 34 \\ \hline
XNU\_RESERVED\_2 & 35 \\ \hline
XNU\_RESTRICTED\_IO & 36 \\ \hline
XNU\_RESTRICTED\_IO\_TELEMETRY & 37 \\ \hline

TXM\_DEFAULT & 38 \\ \hline
TXM\_RO & 39 \\ \hline
TXM\_RW & 40 \\ \hline
TXM\_CPU\_STACK & 41 \\ \hline
TXM\_THREAD\_STACK & 42 \\ \hline
TXM\_ADDRESS\_SPACE\_TABLE & 43 \\ \hline
TXM\_MALLOC\_PAGE & 44 \\ \hline
TXM\_FREE\_LIST & 45 \\ \hline
TXM\_SLAB\_TRUST\_CACHE & 46 \\ \hline
TXM\_SLAB\_PROFILE & 47 \\ \hline
TXM\_SLAB\_CODE\_SIGNATURE & 48 \\ \hline
TXM\_SLAB\_CODE\_REGION & 49 \\ \hline
TXM\_SLAB\_ADDRESS\_SPACE & 50 \\ \hline
TXM\_BUCKET\_1024 & 51 \\ \hline
TXM\_BUCKET\_2048 & 52 \\ \hline
TXM\_BUCKET\_4096 & 53 \\ \hline
TXM\_BUCKET\_8192 & 54 \\ \hline
TXM\_BULK\_DATA & 55 \\ \hline
TXM\_BULK\_DATA\_READ\_ONLY & 56 \\ \hline
TXM\_LOG & 57 \\ \hline
TXM\_SEP\_SECURE\_CHANNEL & 58 \\ \hline

SK\_DEFAULT & 59 \\ \hline
SK\_SHARED\_RO & 60 \\ \hline
SK\_SHARED\_RW & 61 \\ \hline
SK\_IO & 62 \\ \hline

\end{longtable}
}

\section{Retyping Allowed Caller Domains}
\label{app:AllowedCallerDomain}
{
\small
\begin{longtable}{|r|l|r|l|r|}
\hline
\textbf{Type} & \textbf{Type Name} & \textbf{Domain} & \textbf{Domain Name} & \textbf{Retype Flag} \\ \hline
\endfirsthead
\hline
\textbf{Type} & \textbf{Type Name} & \textbf{Domain} & \textbf{Domain Name} & \textbf{Retype Flag} \\ \hline
\endhead
\hline
\endfoot
\hline
\caption[Allowed caller domains for SPTM \texttt{retype} based on the frame type.]{The table shows valid caller domains for memory frame retyping of frames of the specified type via the SPTM \texttt{retype} function. The domain ID is resolved to the corresponding domain. The table also contains the retyping flag that is used for conditional handling in the process.}
\endlastfoot
0 & SPTM\_UNTYPED & 0 & SPTM\_DOMAIN & 0 \\ \hline
1 & SPTM\_UNUSED & 0 & SPTM\_DOMAIN & 0 \\ \hline
2 & SPTM\_DEFAULT & 0 & SPTM\_DOMAIN & 0 \\ \hline
3 & SPTM\_RO & 0 & SPTM\_DOMAIN & 0 \\ \hline
4 & SPTM\_CODE & 0 & SPTM\_DOMAIN & 0 \\ \hline
5 & SPTM\_TXM\_CODE & 0 & SPTM\_DOMAIN & 0 \\ \hline
6 & SPTM\_XNU\_CODE & 0 & SPTM\_DOMAIN & 0 \\ \hline
7 & SPTM\_XNU\_CODE\_DBG\_RW & 0 & SPTM\_DOMAIN & 0 \\ \hline
8 & SPTM\_KERNEL\_ROOT\_TABLE & 0 & SPTM\_DOMAIN & 0 \\ \hline
9 & SPTM\_PAGE\_TABLE & 0 & SPTM\_DOMAIN & 255 \\ \hline
10 & SPTM\_IOMMU\_BOOTSTRAP & 0 & SPTM\_DOMAIN & 0 \\ \hline
11 & XNU\_DEFAULT & 1 & XNU\_DOMAIN & 3840 \\ \hline
12 & XNU\_RO & 1 & XNU\_DOMAIN & 3840 \\ \hline
13 & XNU\_RO\_DBG\_RW & 1 & XNU\_DOMAIN & 3840 \\ \hline
14 & XNU\_USER\_EXEC & 1 & XNU\_DOMAIN & 3840 \\ \hline
15 & XNU\_USER\_DEBUG & 1 & XNU\_DOMAIN & 3840 \\ \hline
16 & XNU\_USER\_JIT & 1 & XNU\_DOMAIN & 3840 \\ \hline
17 & XNU\_USER\_ROOT\_TABLE & 1 & XNU\_DOMAIN & 48 \\ \hline
18 & XNU\_SHARED\_ROOT\_TABLE & 1 & XNU\_DOMAIN & 0 \\ \hline
19 & XNU\_PAGE\_TABLE & 1 & XNU\_DOMAIN & 0 \\ \hline
20 & XNU\_PAGE\_TABLE\_SHARED & 1 & XNU\_DOMAIN & 0 \\ \hline
21 & XNU\_PAGE\_TABLE\_ROZONE & 1 & XNU\_DOMAIN & 0 \\ \hline
22 & XNU\_PAGE\_TABLE\_COMMPAGE & 1 & XNU\_DOMAIN & 0 \\ \hline
23 & XNU\_IOMMU & 1 & XNU\_DOMAIN & 0 \\ \hline
24 & XNU\_ROZONE & 1 & XNU\_DOMAIN & 3840 \\ \hline
25 & XNU\_IO & 1 & XNU\_DOMAIN & 0 \\ \hline
26 & XNU\_PROTECTED\_IO & 1 & XNU\_DOMAIN & 0 \\ \hline
27 & XNU\_COMMPAGE\_RW & 1 & XNU\_DOMAIN & 0 \\ \hline
28 & XNU\_COMMPAGE\_RO & 1 & XNU\_DOMAIN & 0 \\ \hline
29 & XNU\_COMMPAGE\_RX & 1 & XNU\_DOMAIN & 0 \\ \hline
30 & XNU\_TAG\_STORAGE & 0 & SPTM\_DOMAIN & 0 \\ \hline
31 & XNU\_STAGE2\_ROOT\_TABLE & 1 & XNU\_DOMAIN & 0 \\ \hline
32 & XNU\_STAGE2\_PAGE\_TABLE & 1 & XNU\_DOMAIN & 0 \\ \hline
33 & XNU\_KERNEL\_RESTRICTED & 1 & XNU\_DOMAIN & 3840 \\ \hline
34 & XNU\_RESERVED\_1 & 1 & XNU\_DOMAIN & 0 \\ \hline
35 & XNU\_RESERVED\_2 & 1 & XNU\_DOMAIN & 0 \\ \hline
36 & XNU\_RESTRICTED\_IO & 1 & XNU\_DOMAIN & 0 \\ \hline
37 & XNU\_RESTRICTED\_IO\_TELEMETRY & 1 & XNU\_DOMAIN & 0 \\ \hline
38 & TXM\_DEFAULT & 2 & TXM\_DOMAIN & 3840 \\ \hline
39 & TXM\_RO & 2 & TXM\_DOMAIN & 0 \\ \hline
40 & TXM\_RW & 2 & TXM\_DOMAIN & 0 \\ \hline
41 & TXM\_CPU\_STACK & 2 & TXM\_DOMAIN & 0 \\ \hline
42 & TXM\_THREAD\_STACK & 2 & TXM\_DOMAIN & 0 \\ \hline
43 & TXM\_ADDRESS\_SPACE\_TABLE & 2 & TXM\_DOMAIN & 0 \\ \hline
44 & TXM\_MALLOC\_PAGE & 2 & TXM\_DOMAIN & 0 \\ \hline
45 & TXM\_FREE\_LIST & 2 & TXM\_DOMAIN & 3840 \\ \hline
46 & TXM\_SLAB\_TRUST\_CACHE & 2 & TXM\_DOMAIN & 0 \\ \hline
47 & TXM\_SLAB\_PROFILE & 2 & TXM\_DOMAIN & 0 \\ \hline
48 & TXM\_SLAB\_CODE\_SIGNATURE & 2 & TXM\_DOMAIN & 0 \\ \hline
49 & TXM\_SLAB\_CODE\_REGION & 2 & TXM\_DOMAIN & 0 \\ \hline
50 & TXM\_SLAB\_ADDRESS\_SPACE & 2 & TXM\_DOMAIN & 0 \\ \hline
51 & TXM\_BUCKET\_1024 & 2 & TXM\_DOMAIN & 0 \\ \hline
52 & TXM\_BUCKET\_2048 & 2 & TXM\_DOMAIN & 0 \\ \hline
53 & TXM\_BUCKET\_4096 & 2 & TXM\_DOMAIN & 0 \\ \hline
54 & TXM\_BUCKET\_8192 & 2 & TXM\_DOMAIN & 0 \\ \hline
55 & TXM\_BULK\_DATA & 2 & TXM\_DOMAIN & 3840 \\ \hline
56 & TXM\_BULK\_DATA\_READ\_ONLY & 2 & TXM\_DOMAIN & 3840 \\ \hline
57 & TXM\_LOG & 2 & TXM\_DOMAIN & 0 \\ \hline
58 & TXM\_SEP\_SECURE\_CHANNEL & 2 & TXM\_DOMAIN & 0 \\ \hline
59 & SK\_DEFAULT & 3 & SK\_DOMAIN & 0 \\ \hline
60 & SK\_SHARED\_RO & 3 & SK\_DOMAIN & 3840 \\ \hline
61 & SK\_SHARED\_RW & 3 & SK\_DOMAIN & 3840 \\ \hline
62 & SK\_IO & 3 & SK\_DOMAIN & 0 \\ \hline
\end{longtable}
}

\section{SPTM Type to SPRR Index Mapping}
\label{SPTM_to_SPRR}
\begin{longtable}{|r|l|l|}
\hline
\textbf{Index} & \textbf{Name} & \textbf{Value} \\ \hline
\endfirsthead

\hline
\textbf{Index} & \textbf{Name} & \textbf{Value} \\ \hline
\endhead

\hline

\endfoot

\hline
\caption{SPRR index retrieved for the specified SPTM frame type in SPTM \texttt{retype}.}
\endlastfoot

0 & SPTM\_UNTYPED & 0x1 \\ \hline
1 & SPTM\_UNUSED & 0xff \\ \hline
2 & SPTM\_DEFAULT & 0x1 \\ \hline
3 & SPTM\_RO & 0xb \\ \hline
4 & SPTM\_CODE & 0x0 \\ \hline
5 & SPTM\_TXM\_CODE & 0x8 \\ \hline
6 & SPTM\_XNU\_CODE & 0xa \\ \hline
7 & SPTM\_XNU\_CODE\_DBG\_RW & 0xa \\ \hline
8 & SPTM\_KERNEL\_ROOT\_TABLE & 0x1 \\ \hline
9 & SPTM\_PAGE\_TABLE & 0x1 \\ \hline
10 & SPTM\_IOMMU\_BOOTSTRAP & 0x1 \\ \hline
11 & XNU\_DEFAULT & 0x3 \\ \hline
12 & XNU\_RO & 0xb \\ \hline
13 & XNU\_RO\_DBG\_RW & 0xb \\ \hline
14 & XNU\_USER\_EXEC & 0xb \\ \hline
15 & XNU\_USER\_DEBUG & 0xb \\ \hline
16 & XNU\_USER\_JIT & 0xb \\ \hline
17 & XNU\_USER\_ROOT\_TABLE & 0x1 \\ \hline
18 & XNU\_SHARED\_ROOT\_TABLE & 0x1 \\ \hline
19 & XNU\_PAGE\_TABLE & 0x1 \\ \hline
20 & XNU\_PAGE\_TABLE\_SHARED & 0x1 \\ \hline
21 & XNU\_PAGE\_TABLE\_ROZONE & 0x1 \\ \hline
22 & XNU\_PAGE\_TABLE\_COMMPAGE & 0x1 \\ \hline
23 & XNU\_IOMMU & 0x1 \\ \hline
24 & XNU\_ROZONE & 0x2 \\ \hline
25 & XNU\_IO & 0xff \\ \hline
26 & XNU\_PROTECTED\_IO & 0x1 \\ \hline
27 & XNU\_COMMPAGE\_RW & 0x3 \\ \hline
28 & XNU\_COMMPAGE\_RO & 0xb \\ \hline
29 & XNU\_COMMPAGE\_RX & 0xb \\ \hline
30 & XNU\_TAG\_STORAGE & 0x0 \\ \hline
31 & XNU\_STAGE2\_ROOT\_TABLE & 0x1 \\ \hline
32 & XNU\_STAGE2\_PAGE\_TABLE & 0x1 \\ \hline
33 & XNU\_KERNEL\_RESTRICTED & 0x3 \\ \hline
34 & XNU\_RESERVED\_1 & 0xb \\ \hline
35 & XNU\_RESERVED\_2 & 0xb \\ \hline
36 & XNU\_RESTRICTED\_IO & 0xff \\ \hline
37 & XNU\_RESTRICTED\_IO\_TELEMETRY & 0xff \\ \hline
38 & TXM\_DEFAULT & 0x4 \\ \hline
39 & TXM\_RO & 0xb \\ \hline
40 & TXM\_RW & 0x4 \\ \hline
41 & TXM\_CPU\_STACK & 0x4 \\ \hline
42 & TXM\_THREAD\_STACK & 0x4 \\ \hline
43 & TXM\_ADDRESS\_SPACE\_TABLE & 0x4 \\ \hline
44 & TXM\_MALLOC\_PAGE & 0x4 \\ \hline
45 & TXM\_FREE\_LIST & 0x4 \\ \hline
46 & TXM\_SLAB\_TRUST\_CACHE & 0x4 \\ \hline
47 & TXM\_SLAB\_PROFILE & 0x4 \\ \hline
48 & TXM\_SLAB\_CODE\_SIGNATURE & 0x4 \\ \hline
49 & TXM\_SLAB\_CODE\_REGION & 0x4 \\ \hline
50 & TXM\_SLAB\_ADDRESS\_SPACE & 0x4 \\ \hline
51 & TXM\_BUCKET\_1024 & 0x4 \\ \hline
52 & TXM\_BUCKET\_2048 & 0x4 \\ \hline
53 & TXM\_BUCKET\_4096 & 0x4 \\ \hline
54 & TXM\_BUCKET\_8192 & 0x4 \\ \hline
55 & TXM\_BULK\_DATA & 0x4 \\ \hline
56 & TXM\_BULK\_DATA\_READ\_ONLY & 0xb \\ \hline
57 & TXM\_LOG & 0x4 \\ \hline
58 & TXM\_SEP\_SECURE\_CHANNEL & 0x4 \\ \hline
59 & SK\_DEFAULT & 0xff \\ \hline
60 & SK\_SHARED\_RO & 0xb \\ \hline
61 & SK\_SHARED\_RW & 0x3 \\ \hline
62 & SK\_IO & 0xff \\ \hline
\end{longtable}
\newpage

\section{Allowed SPTM Frame Types to Map Into a Table of the Specified Type}
\label{appendixAllowedTableMaps}
{
\footnotesize
\begin{longtable}{|p{1.5cm}|p{5cm}|p{2.5cm}|p{5cm}|}
\hline
\textbf{Table Type ID} & \textbf{Table Type Name} & \textbf{Entry Value} & \textbf{Allowed Frame Type to Insert} \\
\hline
\endfirsthead

\hline
\textbf{Table Type ID} & \textbf{Table Type Name} & \textbf{Entry Value} & \textbf{Allowed Frame Type to Insert} \\
\hline
\endhead

\hline
\endfoot

\hline
\caption[Allowed SPTM frame types to map into a table of the specified type.]{Allowed SPTM frame types to map into a table of the specified type.}
\endlastfoot

8 & SPTM\_KERNEL\_ROOT\_TABLE & 0x280000 & XNU\_PAGE\_TABLE, XNU\_PAGE\_TABLE\_ROZONE \\
\hline

9 & SPTM\_PAGE\_TABLE & 0xFFFFFFFF & all types \\
\hline

17 & XNU\_USER\_ROOT\_TABLE & 0x48000 & XNU\_PAGE\_TABLE, XNU\_PAGE\_TABLE\_COMMPAGE \\
\hline

18 & XNU\_SHARED\_ROOT\_TABLE & 0x10000 & XNU\_PAGE\_TABLE\_SHARED \\
\hline

19 & XNU\_PAGE\_TABLE & 0x679E881 & SPTM\_RO, SPTM\_XNU\_CODE\_DBG\_RW, SPTM\_PAGE\_TABLE, SPTM\_IOMMU\_BOOTSTRAP, XNU\_DEFAULT, XNU\_RO, XNU\_USER\_DEBUG, XNU\_USER\_JIT, XNU\_USER\_ROOT\_TABLE, XNU\_SHARED\_ROOT\_TABLE, XNU\_PAGE\_TABLE\_ROZONE, XNU\_PAGE\_TABLE\_COMMPAGE \\
\hline

20 & XNU\_PAGE\_TABLE\_SHARED & 0x2104801 & SPTM\_UNTYPED, XNU\_DEFAULT, XNU\_USER\_EXEC, XNU\_PAGE\_TABLE\_SHARED, XNU\_IO \\
\hline

21 & XNU\_PAGE\_TABLE\_ROZONE & 0x1200000 & XNU\_PAGE\_TABLE\_ROZONE, XNU\_ROZONE \\
\hline

22 & XNU\_PAGE\_TABLE\_COMMPAGE & 0x00004038 & SPTM\_RO, SPTM\_CODE, SPTM\_TXM\_CODE, XNU\_USER\_EXEC \\
\hline

31 & XNU\_STAGE2\_ROOT\_TABLE & 0x0 & None \\
\hline

32 & XNU\_STAGE2\_PAGE\_TABLE & 0x800 & XNU\_DEFAULT \\
\hline

\end{longtable}
}
\todo{einrückungen fixen, außerdem: continues on next page string}

\section{\texttt{Frameworks} and \texttt{PrivateFrameworks} Found in the \texttt{ExclaveKit} DMG}
\label{frameworkList}
\subsection{Frameworks}
\begin{itemize}
  \item AOPVoiceTriggerSecure.framework
  \item AVFAudio.framework
  \item Accelerate.framework
  \item AppleProxExclaveComponent.framework
  \item Common\_ISP\_EK\_TBModule.framework
  \item CoreAudioTypes.framework
  \item CoreFoundation.framework
  \item DeviceTreeKit.framework
  \item EXDataLoader.framework
  \item EXMobileAssetLoader.framework
  \item EXSimpleFileIO.framework
  \item ExclaveFDRDecodeRawDataStoreKitComponent.framework
  \item ExclavesAudioDrivers.framework
  \item Foundation.framework
  \item IR\_ISP\_EK\_TBModule.framework
  \item IsolatedAudioMeterClientExclaveComponent.framework
  \item IsolatedCoreAudioClientComponent.framework
  \item MobileAssetExclaveComponent.framework
  \item MobileGestalt.framework
  \item OSLogExclaves.framework
  \item RGB\_ISP\_EK\_TBModule.framework
  \item SILManagerComponent.framework
  \item SharedMemory.framework
  \item SoundAnalysis.framework
  \item StackshotDelegateComponent.framework
  \item T8140\_IR\_ISP\_EK\_Component.framework
  \item T8140\_RGB\_ISP\_EK\_Component.framework
  \item TransitKit.framework
  \item Vision.framework
  \item VoiceTriggerCommon.framework
  \item VoiceTriggerSecure.framework
\end{itemize}

\subsection{PrivateFrameworks}
\begin{itemize}
  \item ANEExclaveServices.framework
  \item ANSTCommon.framework
  \item AppleCVA.framework
  \item AppleSauce.framework
  \item AudioCaptureServer.framework
  \item AudioCaptureServerComponent.framework
  \item AudioDSP.framework
  \item AudioDSPControllerComponent.framework
  \item AudioDSPGraph.framework
  \item AudioDSPProcessorComponent.framework
  \item AudioToolboxCore.framework
  \item CollectionsInternal.framework
  \item CoreANSTKit.framework
  \item CoreAudioResources.framework
  \item CoreERClientKit.framework
  \item CoreMDClientKit.framework
  \item CoreMedia.framework
  \item CoreSpeechExclave.framework
  \item CoreSpeechExclaveComponent.framework
  \item CoreSpeechUtils.framework
  \item CoreVideo.framework
  \item DebugExfiltration.framework
  \item EXDisplayPipeSwapClient.framework
  \item EXSurface.framework
  \item Espresso.framework
  \item ExclaveCredentialManager.framework
  \item ExclaveFDRDecode.framework
  \item MIL.framework
  \item MLCompilerRuntime.framework
  \item MLCompilerServices.framework
  \item SISP\_EK\_AlgoModels.framework
  \item SecureVoiceTriggerAssets\_exclavekit.framework
  \item ShareLibCommon\_EK.framework
  \item ShazamKit.framework
  \item SpeakerRecognitionKit.framework
  \item SphinxProxTrustedFDR.framework
  \item T8140\_CoreAAClientKit.framework
  \item T8140\_ExclaveISPSharedLib\_exclavekit.framework
  \item Tightbeam.framework
  \item VoiceTriggerEventProviderExclave.framework
  \item VoiceTriggerExclave.framework
  \item caulk.framework
\end{itemize}

\section{Exclave Resource Type Specific Data Types}
\label{typeDefinitionsExclaves}

\begin{listing}[H]
    \begin{minted}[linenos, breaklines, bgcolor=LightGray, frame=lines]{c}
#define CONCLAVE_SERVICE_MAX 192

typedef struct {
	conclave_state_t       c_state;
	conclave_request_t     c_request;
	bool                   c_active_downcall;
	bool                   c_active_stopcall;
	bool                   c_active_detach;
	tb_client_connection_t c_control;
	task_t XNU_PTRAUTH_SIGNED_PTR("conclave.task") c_task;
	thread_t XNU_PTRAUTH_SIGNED_PTR("conclave.thread") c_downcall_thread;
	bitmap_t               c_service_bitmap[BITMAP_LEN(CONCLAVE_SERVICE_MAX)];
} conclave_resource_t;

typedef struct {
	size_t sm_size;
	exclaves_buffer_perm_t sm_perm;
	char *sm_addr;
	sharedmemorybase_mapping_s sm_mapping;
	sharedmemorybase_segxnuaccess_s sm_client;
} shared_memory_resource_t;

typedef struct {
	/* how many times *this* sensor resource handle has been
	 * used to call sensor_start */
	uint64_t s_startcount;
} sensor_resource_t;

typedef struct {
	struct klist notification_klist;
} exclaves_notification_t;
    \end{minted}
    \captionof{lstlisting}[Exclave resource type-specific data types.]{Exclave resource type-specific data types as made available in \path{osfmk/kern/exclaves_resource.h}~\cite{XNU_10063:2024}.}
\end{listing}

\section{Exclaves Resources Grouped by Domain}
\label{exclaveResources}

\begingroup
\small

\paragraph{Domain: \texttt{com.apple.audiomxd.conclave}}
\begin{itemize}
  \item \texttt{com.apple.audio.lpmic}
  \item \texttt{com.apple.audio.lpmic}
  \item \texttt{com.apple.audio.mic}
  \item \texttt{com.apple.audio.mic}
  \item \texttt{com.apple.audiomxd.AudioCaptureServer}
  \item \texttt{com.apple.audiomxd.MTDAudioDSPControl}
  \item \texttt{com.apple.audiomxd.PerceptionAudioDSPControl}
  \item \texttt{com.apple.audiomxd.SharedAudioDSPControl}
  \item \texttt{com.apple.audiomxd.SiriAudioDSPControl}
  \item \texttt{com.apple.audiomxd.SoundAnalysisAudioDSPControl}
  \item \texttt{com.apple.inbound\_buffer.hpmic16\_injection}
  \item \texttt{com.apple.inbound\_buffer.hpmic\_injection}
  \item \texttt{com.apple.inbound\_buffer.muted\_talker\_detection\_ref\_stream}
  \item \texttt{com.apple.inbound\_buffer.muted\_talker\_detection\_ref\_stream}
  \item \texttt{com.apple.inbound\_buffer.perception\_ref\_stream}
  \item \texttt{com.apple.inbound\_buffer.shared\_audiodsp\_ref\_stream}
  \item \texttt{com.apple.inbound\_buffer.siri\_ref\_stream}
  \item \texttt{com.apple.inbound\_buffer.siri\_ref\_stream}
  \item \texttt{com.apple.inbound\_buffer.sound\_analysis\_ref\_stream}
  \item \texttt{com.apple.inbound\_buffer.sound\_analysis\_ref\_stream}
  \item \texttt{com.apple.notification.audiocapture}
  \item \texttt{com.apple.notification.audiodsp}
  \item \texttt{com.apple.notification.audiodsp.analysis}
  \item \texttt{com.apple.sensors.mic}
\end{itemize}

\paragraph{Domain: \texttt{com.apple.backboardd.conclave}}
\begin{itemize}
  \item \texttt{com.apple.backboardd.AppleProxExclaveService}
  \item \texttt{com.apple.backboardd.ExclaveFDRDecodeRawDataStoreKitService}
  \item \texttt{com.apple.backboardd.SILManager}
\end{itemize}

\paragraph{Domain: \texttt{com.apple.cameracaptured.conclave}}
\begin{itemize}
  \item \texttt{com.apple.cameracaptured.ISPIRManager}
  \item \texttt{com.apple.cameracaptured.ISPIRService}
  \item \texttt{com.apple.cameracaptured.ISPRGBManager}
  \item \texttt{com.apple.cameracaptured.ISPRGBService}
\end{itemize}

\paragraph{Domain: \texttt{com.apple.conclave.mediaserverd}}
\begin{itemize}
  \item \texttt{com.apple.mediaserverd.ISPIRManager}
  \item \texttt{com.apple.mediaserverd.ISPIRService}
  \item \texttt{com.apple.mediaserverd.ISPRGBManager}
  \item \texttt{com.apple.mediaserverd.ISPRGBService}
\end{itemize}

\paragraph{Domain: \texttt{com.apple.conclave.test1}}
\begin{itemize}
  \item \texttt{com.apple.service.ExclavesCHelloServer}
\end{itemize}

\paragraph{Domain: \texttt{com.apple.corespeechd.conclave}}
\begin{itemize}
  \item \texttt{com.apple.audio.siri\_history}
  \item \texttt{com.apple.audio.siri\_history}
  \item \texttt{com.apple.corespeechd.SiriVoiceTriggerService}
  \item \texttt{com.apple.isolatedcoreaudioclient.service}
  \item \texttt{com.apple.sensors.mic}
\end{itemize}

\paragraph{Domain: \texttt{com.apple.darwin}}
\begin{itemize}
  \item \texttt{com.apple.service.ANEEngine}
  \item \texttt{com.apple.service.ANEExclave}
  \item \texttt{com.apple.service.ANEExclaveTestClient}
  \item \texttt{com.apple.service.ANEExclave\_EDK}
  \item \texttt{com.apple.service.AudioDriver}
  \item \texttt{com.apple.service.AudioDriverLPMic}
  \item \texttt{com.apple.service.AudioDriverLPMic\_EDK}
  \item \texttt{com.apple.service.AudioDriverVT}
  \item \texttt{com.apple.service.AudioDriverVT\_EDK}
  \item \texttt{com.apple.service.AudioDriver\_EDK}
  \item \texttt{com.apple.service.AudioSharedDARTMapper}
  \item \texttt{com.apple.service.AudioSharedDARTMapper\_EDK}
  \item \texttt{com.apple.service.EXBrightCore}
  \item \texttt{com.apple.service.EXBrightCore\_EDK}
  \item \texttt{com.apple.service.EXDisplayPipeDriver\_EDK\_xnuproxy}
  \item \texttt{com.apple.service.EXDisplayPipeDriver\_xnuproxy}
  \item \texttt{com.apple.service.ExclaveDriverKit}
  \item \texttt{com.apple.service.ExclaveFDRDecodeTestClientExclave}
  \item \texttt{com.apple.service.ExclaveSEPManager}
  \item \texttt{com.apple.service.ExclaveSEPManager\_EDK}
  \item \texttt{com.apple.service.ExclaveSEPTestClient}
  \item \texttt{com.apple.service.ExclaveTestService}
  \item \texttt{com.apple.service.HPMic16DMA}
  \item \texttt{com.apple.service.HPMic16DMA\_EDK}
  \item \texttt{com.apple.service.HPMic16Device}
  \item \texttt{com.apple.service.HPMic16Device\_EDK}
  \item \texttt{com.apple.service.HPMicAudioDriver}
  \item \texttt{com.apple.service.HPMicAudioDriver\_EDK}
  \item \texttt{com.apple.service.HelloDrivers}
  \item \texttt{com.apple.service.HelloExclaveCxx}
  \item \texttt{com.apple.service.HelloExclaveKit}
  \item \texttt{com.apple.service.HelloStorage}
  \item \texttt{com.apple.service.SecureRTBuddyAOP}
  \item \texttt{com.apple.service.SecureRTBuddyAOP\_EDK}
  \item \texttt{com.apple.service.SecureRTBuddyAOP\_IOReporting}
  \item \texttt{com.apple.service.SecureRTBuddyDCP}
  \item \texttt{com.apple.service.SecureRTBuddyDCP\_EDK}
  \item \texttt{com.apple.service.SecureRTBuddyDCP\_IOReporting}
  \item \texttt{com.apple.service.TightbeamTestsServer}
  \item \texttt{com.apple.service.user\_app}
  \item \texttt{com.apple.service.user\_app2}
  \item \texttt{com.apple.service.user\_app3}
\end{itemize}
\paragraph{Domain: \texttt{com.apple.exclavetestrunnerd.conclave}}
\begin{itemize}
  \item \texttt{com.apple.exclavetestrunnerd.service}
\end{itemize}

\paragraph{Domain: \texttt{com.apple.facekittestd.conclave}}
\begin{itemize}
  \item \texttt{com.apple.facekittestd.service}
\end{itemize}

\paragraph{Domain: \texttt{com.apple.isolatedaudiometerclientd.conclave}}
\begin{itemize}
  \item \texttt{com.apple.isolatedaudiometerclientd.service}
  \item \texttt{com.apple.isolatedcoreaudioclient.service}
\end{itemize}

\paragraph{Domain: \texttt{com.apple.kernel}}
\begin{itemize}
  \item \texttt{com.apple.audio.lpmic}
  \item \texttt{com.apple.audio.lpmic}
  \item \texttt{com.apple.audio.mic}
  \item \texttt{com.apple.audio.mic}
  \item \texttt{com.apple.audio.siri\_history}
  \item \texttt{com.apple.audio.siri\_history}
  \item \texttt{com.apple.audio.test}
  \item \texttt{com.apple.audio.test}
  \item \texttt{com.apple.audiomxd.conclave}
  \item \texttt{com.apple.backboardd.conclave}
  \item \texttt{com.apple.cameracaptured.conclave}
  \item \texttt{com.apple.conclave.mediaserverd}
  \item \texttt{com.apple.conclave.test1}
  \item \texttt{com.apple.corespeechd.conclave}
  \item \texttt{com.apple.exclavetestrunnerd.conclave}
  \item \texttt{com.apple.facekittestd.conclave}
  \item \texttt{com.apple.inbound\_buffer.exfiltration}
  \item \texttt{com.apple.inbound\_buffer.hpmic16\_injection}
  \item \texttt{com.apple.inbound\_buffer.hpmic\_injection}
  \item \texttt{com.apple.inbound\_buffer.muted\_talker\_detection\_ref\_stream}
  \item \texttt{com.apple.inbound\_buffer.siri\_ref\_stream}
  \item \texttt{com.apple.inbound\_buffer.sound\_analysis\_ref\_stream}
  \item \texttt{com.apple.isolatedaudiometerclientd.conclave}
  \item \texttt{com.apple.mobileassetd.conclave}
  \item \texttt{com.apple.named\_buffer.1}
  \item \texttt{com.apple.named\_buffer.1}
  \item \texttt{com.apple.named\_buffer.10}
  \item \texttt{com.apple.named\_buffer.112}
  \item \texttt{com.apple.named\_buffer.113}
  \item \texttt{com.apple.named\_buffer.16}
  \item \texttt{com.apple.named\_buffer.2}
  \item \texttt{com.apple.named\_buffer.2}
  \item \texttt{com.apple.named\_buffer.3}
  \item \texttt{com.apple.named\_buffer.32}
  \item \texttt{com.apple.named\_buffer.36}
  \item \texttt{com.apple.named\_buffer.4}
  \item \texttt{com.apple.named\_buffer.41}
  \item \texttt{com.apple.named\_buffer.6}
  \item \texttt{com.apple.named\_buffer.7}
  \item \texttt{com.apple.notification.audiocapture}
  \item \texttt{com.apple.notification.audiodsp}
  \item \texttt{com.apple.notification.audiodsp.analysis}
  \item \texttt{com.apple.notification.hello}
  \item \texttt{com.apple.outbound\_buffer.exfiltration}
  \item \texttt{com.apple.perceptiond.conclave}
  \item \texttt{com.apple.securdled.conclave}
  \item \texttt{com.apple.sensors.cam}
  \item \texttt{com.apple.sensors.cam\_alt\_faceid}
  \item \texttt{com.apple.sensors.cam\_alt\_faceid\_delayed}
  \item \texttt{com.apple.sensors.mic}
  \item \texttt{com.apple.service.ConclaveLauncherControl}
  \item \texttt{com.apple.service.ConclaveLauncherDebug}
  \item \texttt{com.apple.service.ExclaveIndicatorController}
  \item \texttt{com.apple.service.ExclaveKeyStoreTestClient}
  \item \texttt{com.apple.service.ExclavesCHelloServer}
  \item \texttt{com.apple.service.Exfiltration\_xnuproxy}
  \item \texttt{com.apple.service.FrameMint}
  \item \texttt{com.apple.service.HelloDriverInterrupts}
  \item \texttt{com.apple.service.ISPCamEngine}
  \item \texttt{com.apple.service.ISPCamEngine\_EDK}
  \item \texttt{com.apple.service.ISPEngine}
  \item \texttt{com.apple.service.ISPEngine\_EDK}
  \item \texttt{com.apple.service.LogServer\_xnuproxy}
  \item \texttt{com.apple.service.PanicInit}
  \item \texttt{com.apple.service.Stackshot}
  \item \texttt{com.apple.shareddspd.conclave}
  \item \texttt{com.apple.sharedmem.stackshotserver}
  \item \texttt{com.apple.soundanalysisd.conclave}
  \item \texttt{com.apple.storage.backend}
\end{itemize}
\paragraph{Domain: \texttt{com.apple.mobileassetd.conclave}}
\begin{itemize}
  \item \texttt{com.apple.mobileassetd.service}
\end{itemize}

\paragraph{Domain: \texttt{com.apple.perceptiond.conclave}}
\begin{itemize}
  \item \texttt{com.apple.isolatedcoreaudioclient.service}
  \item \texttt{com.apple.perceptiond.services}
\end{itemize}

\paragraph{Domain: \texttt{com.apple.securdled.conclave}}
\begin{itemize}
  \item \texttt{com.apple.securdled.service}
\end{itemize}

\paragraph{Domain: \texttt{com.apple.shareddspd.conclave}}
\begin{itemize}
  \item \texttt{com.apple.isolatedcoreaudioclient.service}
\end{itemize}

\paragraph{Domain: \texttt{com.apple.soundanalysisd.conclave}}
\begin{itemize}
  \item \texttt{com.apple.isolatedcoreaudioclient.service}
  \item \texttt{com.apple.soundanalysisd.service}
\end{itemize}

\endgroup

\section{Kernel Handling of Userspace Exclaves Endpoint Call}
\label{app:kernelEndpointCallHandling}
\begin{minted}[linenos, breaklines, breakanywhere, bgcolor=LightGray, frame=lines]{c}
	switch (operation) {
	case EXCLAVES_CTL_OP_ENDPOINT_CALL: {
		if (name != MACH_PORT_NULL) {
			/* Only accept MACH_PORT_NULL for now */
			return KERN_INVALID_CAPABILITY;
		}
		if (ubuffer == USER_ADDR_NULL || usize == 0 ||
		    usize != Exclaves_L4_IpcBuffer_Size) {
			return KERN_INVALID_ARGUMENT;
		}

		Exclaves_L4_IpcBuffer_t *ipcb;
		if ((error = exclaves_allocate_ipc_buffer((void**)&ipcb))) {
			return error;
		}
		assert(ipcb != NULL);
		if ((error = copyin(ubuffer, ipcb, usize))) {
			return error;
		}

		if (identifier >= CONCLAVE_SERVICE_MAX) {
			return KERN_INVALID_ARGUMENT;
		}

		/*
		 * Verify that the service actually exists in the current
		 * domain (only when the fallbacks are not enabled).
		 */
		if (!exclaves_service_fallback &&
		    !exclaves_conclave_has_service(task_get_conclave(task),
		    identifier)) {
			return KERN_INVALID_ARGUMENT;
		}

		kr = exclaves_endpoint_call_internal(IPC_PORT_NULL, identifier);
		error = copyout(ipcb, ubuffer, usize);
		/*
		 * Endpoint call to conclave may have trigger a stop upcall,
		 * check if stop upcall completion handler needs to run.
		 */
		task_stop_conclave_upcall_complete();
		if (error) {
			return error;
		}
		break;
	}
\end{minted}
\captionof{lstlisting}[Exclaves endpoint call handling in kernel.]{Kernel handling of Exclave endpoint call invocation performed via \texttt{\_exclaves\_ctl\_trap} with operation \texttt{EXCLAVES\_CTL\_OP\_ENDPOINT\_CALL}.}
\label{lst:appendixListingEndpointCall}

\section{Xnuproxy available commands, from \texttt{cmd\_to\_string}}
Also see \texttt{osfmk/kern/exclavs.c} l. 2311~\cite{XNU_10063:2024}.
\label{XNUPROXYCommands}
\begin{itemize}
    \item XNUPROXY\_CMD\_UNDEFINED
    \item XNUPROXY\_CMD\_SETUP
    \item XNUPROXY\_CMD\_CONTEXT\_ALLOCATE
    \item XNUPROXY\_CMD\_CONTEXT\_FREE
    \item XNUPROXY\_CMD\_NAMED\_BUFFER\_CREATE
    \item XNUPROXY\_CMD\_NAMED\_BUFFER\_DELETE
    \item XNUPROXY\_CMD\_RESOURCE\_INFO
    \item XNUPROXY\_CMD\_AUDIO\_BUFFER\_CREATE
    \item XNUPROXY\_CMD\_AUDIO\_BUFFER\_COPYOUT
    \item XNUPROXY\_CMD\_AUDIO\_BUFFER\_DELETE
    \item XNUPROXY\_CMD\_SENSOR\_START
    \item XNUPROXY\_CMD\_SENSOR\_STOP
    \item XNUPROXY\_CMD\_SENSOR\_STATUS
    \item XNUPROXY\_CMD\_DISPLAY\_HEALTHCHECK\_RATE
    \item XNUPROXY\_CMD\_NAMED\_BUFFER\_MAP
    \item XNUPROXY\_CMD\_NAMED\_BUFFER\_LAYOUT
    \item XNUPROXY\_CMD\_AUDIO\_BUFFER\_MAP
    \item XNUPROXY\_CMD\_AUDIO\_BUFFER\_LAYOUT
    \item XNUPROXY\_CMD\_REPORT\_MEMORY\_USAGE
    \item XNUPROXY\_CMD\_UPCALL\_READY
\end{itemize}

\section{Kernel \texttt{exclaves\_xnu\_proxy\_send} Function}
\label{xnuproxysend}

\begin{minted}[linenos, breaklines, breakanywhere, bgcolor=LightGray, frame=lines]{c}
kern_return_t
exclaves_xnu_proxy_send(xnuproxy_msg_t *_msg, Exclaves_L4_Word_t *spawned)
{
    assert3p(_msg, !=, NULL);

    thread_t thread = current_thread();

    if (exclaves_xnu_proxy_msg_buffer == NULL) {
	return KERN_FAILURE;
    }

    kern_return_t kr = KERN_SUCCESS;
    xnuproxy_msg_t *msg = exclaves_xnu_proxy_msg_buffer;
    bool interrupted = false;

    lck_mtx_lock(&exclaves_xnu_proxy_lock);

    assert3u(thread->th_exclaves_state & TH_EXCLAVES_STATE_ANY, ==, 0);
    thread->th_exclaves_state |= TH_EXCLAVES_XNUPROXY;

    KDBG_RELEASE(MACHDBG_CODE(DBG_MACH_EXCLAVES, MACH_EXCLAVES_XNUPROXY)
        | DBG_FUNC_START, exclaves_xnu_proxy_scid, _msg->cmd);

    *msg = *_msg;
    msg->server_id = exclaves_xnu_proxy_scid;

    os_atomic_store(&msg->status, XNUPROXY_MSG_STATUS_PROCESSING,
	release);

    while (os_atomic_load(&msg->status, relaxed) ==
	XNUPROXY_MSG_STATUS_PROCESSING) {
	exclaves_xnu_proxy_show_progress(in progress, msg);
	kr = exclaves_scheduler_resume_scheduling_context(msg->server_id,
		spawned, interrupted);
	assert(kr == KERN_SUCCESS || kr == KERN_ABORTED);

	/* A wait was interrupted. */
	interrupted = kr == KERN_ABORTED;

	if (NULL != exclaves_xnu_proxy_upcall_ipcb) {
		if (XNUPROXY_MSG_STATUS_UPCALL == XNUPROXY_CR_STATUS(exclaves_xnu_proxy_upcall_ipcb)) {
				xnuproxy_msg_status_t status = (xnuproxy_msg_status_t)
				    XNUPROXY_CR_STATUS(exclaves_xnu_proxy_upcall_ipcb);
				(void) exclaves_handle_upcall(thread, exclaves_xnu_proxy_upcall_ipcb,
				    exclaves_xnu_proxy_scid, status);
		}
	}
    }

    if (os_atomic_load(&msg->status, acquire) ==
	XNUPROXY_MSG_STATUS_NONE) {
		exclaves_xnu_proxy_show_progress(complete, msg);
    } else {
		kr = KERN_FAILURE;
		exclaves_xnu_proxy_show_error(msg);
    }

    *_msg = *msg;

    KDBG_RELEASE(MACHDBG_CODE(DBG_MACH_EXCLAVES, MACH_EXCLAVES_XNUPROXY)
	| DBG_FUNC_END);

    thread->th_exclaves_state &= ~TH_EXCLAVES_XNUPROXY;
    lck_mtx_unlock(&exclaves_xnu_proxy_lock);

    return kr;
}
\end{minted}
\captionof{lstlisting}[\texttt{exclaves\_xnu\_proxy\_send} function used for xnuproxy command invocation.]{\texttt{exclaves\_xnu\_proxy\_send} function used for xnuproxy command invocation, from \texttt{osfmk/kern/exclaves.c}~\cite{XNU_10063:2024}.}

\section{Tightbeam \protect \path{__tb_connection_create_transport_for_endpoint} Function}
\label{tb_implementation_transport_create}
\begin{minted}[linenos, breaklines, bgcolor=LightGray, frame=lines]{c}
void __tb_connection_create_transport_for_endpoint(tb_endpoint_t *tb_endpoint,int some_option)

{
  code *pcVar1;
  uint type;
  tb_endpoint_t *ptVar2;
  
  ptVar2 = tb_endpoint;
  _tb_endpoint_get_type(tb_endpoint);
  type = (uint)ptVar2;
  if (type == 1) {
    _tb_null_transport_create();
  }
  else if (type == 2) {
    if (some_option == 1) {
      _tb_mach_service_transport_create();
    }
    else {
      if (some_option != 0) goto LAB_25db24dac;
      _tb_mach_client_transport_create();
    }
  }
  else {
    if (type == 0) goto LAB_25db24dac;
    if ((some_option == 0) && ((type & 0b11111111111111111111111111111110) == 4)) {
      _tb_eve_client_transport_create();
    }
    else if ((int)type < 10) {
      if (type == 8) {
        _tb_darwin_client_transport_create();
      }
      else {
        if (type != 9) goto LAB_25db24dac;
        if (some_option == 0) {
          _tb_unix_client_transport_create_with_endpoint();
        }
        else {
          if (some_option != 1) goto LAB_25db24dac;
          _tb_unix_service_transport_create_with_endpoint();
        }
      }
    }
    else if (type == 10) {
      if (some_option == 0) {
        _tb_delegated_client_transport_create();
      }
      else {
        if (some_option != 1) goto LAB_25db24dac;
        _tb_delegated_service_transport_create();
      }
    }
    else {
      if (type != 0xb) goto LAB_25db24dac;
      _tb_afk_transport_create();
    }
  }
  if (tb_endpoint != (tb_endpoint_t *)0x0) {
    return;
  }
LAB_25db24dac:
                    /* WARNING: Does not return */
  pcVar1 = (code *)SoftwareBreakpoint(1,0x25db24db0);
  (*pcVar1)();
}
\end{minted}
\captionof{lstlisting}{Tightbeam \protect \path{__tb_connection_create_transport_for_endpoint} function.}

\section{\texttt{txm\_dispatch\_handler} Function in TXM}
\label{App:txm_dispatch_handler}

\begin{minted}[linenos, breaklines, breakanywhere, bgcolor=LightGray, frame=lines]{c}
void txm_dispatch_handler
               (int selector,ulong param_2,undefined8 param_3,undefined8 param_4,undefined8 param_5,
               undefined8 param_6)

{
  ulong uVar1;
  long lVar2;
  undefined8 uVar3;
  long lVar4;
  uint uVar5;
  ulong local_70 [4];
  
  lVar2 = FUN_fffffff017024e44();
  local_70[3] = 1;
  if (DAT_fffffff017060380 != '\0') {
    uVar3 = 0xa0;
LAB_fffffff017022d1c:
                    /* WARNING: Subroutine does not return */
    FUN_fffffff017021bd8(uVar3,0);
  }
  uVar1 = (ulong)(local_70 + 3) & 0xffffffffffffc000;
  if (uVar1 == 0) {
    uVar3 = 0x40;
    goto LAB_fffffff017022d1c;
  }
  if ((ulong *)0xffffffffffffbfff < local_70 + 3) {
    uVar3 = 0x42;
    goto LAB_fffffff017022d1c;
  }
  local_70[2] = 0x4000;
  local_70[1] = 0x4000;
  local_70[0] = uVar1;
  FUN_fffffff017024790(local_70,0x2a);
  *(undefined8 *)(lVar2 + 0x18) = 0;
  if (0x32 < selector - 1U) {
    uVar3 = 0xa1;
    goto LAB_fffffff017022d1c;
  }
  uVar5 = (uint)param_2;
  switch(selector) {
  default:
    lVar4 = FUN_fffffff017024e44();
    *(undefined8 *)(lVar4 + 0x18) = 3;
    *(undefined8 *)(lVar4 + 0x20) = DAT_fffffff017010500;
    *(undefined4 **)(lVar4 + 0x28) = &DAT_fffffff017060384;
    *(undefined4 **)(lVar4 + 0x30) = &DAT_fffffff017060388;
  case 0x1f:
code_r0xfffffff017023048:
    local_70[3] = 0;
    break;
  case 2:
    FUN_fffffff017023258();
    goto code_r0xfffffff017023048;
  case 3:
    lVar4 = FUN_fffffff017024e44();
    *(undefined8 *)(lVar4 + 0x18) = 4;
    *(undefined8 **)(lVar4 + 0x20) = &DAT_fffffff017010590;
    *(ulong *)(lVar4 + 0x28) = (ulong)DAT_fffffff0170105b8;
    *(undefined8 *)(lVar4 + 0x30) = 0;
    *(undefined8 *)(lVar4 + 0x38) = 0;
    goto code_r0xfffffff017023048;
  case 4:
    lVar4 = FUN_fffffff017024e44();
    *(undefined8 *)(lVar4 + 0x18) = 1;
    uVar3 = get_build_string();
    *(undefined8 *)(lVar4 + 0x20) = uVar3;
    goto code_r0xfffffff017023048;
  case 5:
    FUN_fffffff01701be44();
    goto code_r0xfffffff017023048;
  case 6:
    local_70[3] = FUN_fffffff0170233d0();
    break;
  case 7:
    local_70[3] = FUN_fffffff0170232c0();
    break;
  case 8:
    local_70[3] = FUN_fffffff01701ba98();
    break;
  case 9:
    FUN_fffffff017023348(param_2);
    goto code_r0xfffffff017023048;
  case 10:
    FUN_fffffff017023464(param_2);
    goto code_r0xfffffff017023048;
  case 0xb:
    local_70[3] = FUN_fffffff0170234c0();
    break;
  case 0xc:
    local_70[3] = FUN_fffffff01702351c(uVar5 & 0xff,param_3,param_4,param_5,param_6);
    break;
  case 0xd:
    if (param_2 + 0x10 < param_2) goto LAB_fffffff01702317c;
    local_70[3] = FUN_fffffff01701e3f4(param_2);
    break;
  case 0xe:
    local_70[3] = FUN_fffffff017023638(uVar5 & 0xff,param_3);
    break;
  case 0xf:
    local_70[3] = FUN_fffffff017023720(param_2);
    break;
  case 0x10:
    local_70[3] = FUN_fffffff0170237f4(param_2);
    break;
  case 0x11:
    local_70[3] = FUN_fffffff017023898(param_2,param_3);
    break;
  case 0x12:
    local_70[3] = FUN_fffffff017023958(param_2,param_3,param_4);
    break;
  case 0x13:
    local_70[3] = FUN_fffffff017023a2c(param_2);
    break;
  case 0x14:
    local_70[3] = FUN_fffffff017023a88(param_2,param_3);
    break;
  case 0x15:
    local_70[3] = FUN_fffffff017023ad0(param_2);
    break;
  case 0x16:
    local_70[3] = FUN_fffffff017023b04(param_2,param_3,param_4);
    break;
  case 0x17:
    local_70[3] = FUN_fffffff017023bdc(param_2);
    break;
  case 0x18:
    FUN_fffffff01701f684(param_2);
    local_70[3] = FUN_fffffff01701c728();
    break;
  case 0x19:
    local_70[3] = FUN_fffffff017023c38(param_2);
    break;
  case 0x1a:
    if (param_2 + 0x61 < param_2) goto LAB_fffffff01702317c;
    local_70[3] = FUN_fffffff01701d398(param_2);
    break;
  case 0x1b:
    local_70[3] = FUN_fffffff017023c98();
    break;
  case 0x1c:
    if (param_2 + 0x14 < param_2) goto LAB_fffffff01702317c;
    local_70[3] = FUN_fffffff01701d430(param_2);
    break;
  case 0x1d:
    if (param_2 + 0x14 < param_2) goto LAB_fffffff01702317c;
    local_70[3] = FUN_fffffff01701d5b4(param_2);
    break;
  case 0x1e:
    local_70[3] = FUN_fffffff017023ce0(param_2);
    break;
  case 0x20:
    local_70[3] = FUN_fffffff017023d4c(param_2);
    break;
  case 0x21:
    uVar3 = FUN_fffffff01701f684(param_2);
    local_70[3] = FUN_fffffff01701cd70(uVar3,param_3);
    break;
  case 0x22:
    local_70[3] = FUN_fffffff017023da4(param_2);
    break;
  case 0x23:
    local_70[3] = FUN_fffffff01701a268(param_2,param_3);
    break;
  case 0x24:
    local_70[3] = FUN_fffffff017023e34(uVar5 & 0xffff,param_3);
    break;
  case 0x25:
    FUN_fffffff017020280(param_2);
    local_70[3] = FUN_fffffff01701a3b0();
    break;
  case 0x26:
    local_70[3] = associate_code_signature(param_2,param_3,param_4,param_5,param_6);
    break;
  case 0x27:
    local_70[3] = FUN_fffffff017023f08(param_2);
    break;
  case 0x28:
    uVar3 = FUN_fffffff017020280(param_2);
    local_70[3] = FUN_fffffff01701b1fc(uVar3,param_3,param_4);
    break;
  case 0x29:
    FUN_fffffff017020280(param_2);
    local_70[3] = allow_invalid_code?();
    break;
  case 0x2a:
    uVar3 = FUN_fffffff017020280(param_2);
    local_70[3] = FUN_fffffff01701b328(uVar3,param_3,param_4);
    break;
  case 0x2b:
    local_70[3] = FUN_fffffff017023f54(param_2);
    break;
  case 0x2c:
    local_70[3] = FUN_fffffff017023fe0(param_2);
    break;
  case 0x2d:
    local_70[3] = FUN_fffffff017024038(param_2,param_3,param_4);
    break;
  case 0x2e:
  case 0x2f:
  case 0x30:
  case 0x31:
  case 0x32:
  case 0x33:
    local_70[3] = 0x26;
  }
  *(ulong *)(lVar2 + 8) = local_70[3];
  FUN_fffffff017022a30();
LAB_fffffff01702317c:
                    /* WARNING: Subroutine does not return */
  FUN_fffffff017021d2c(0x19);
\end{minted}
\captionof{lstlisting}[TXM \texttt{txm\_dispatch\_handler} function.]{\texttt{txm\_dispatch\_handler} found in the TXM binary, invoked on TXM calls from XNU to perform actual function handling based on a provided selector.}
\label{lst:txm_dispatch_handler}
\newpage
\section{Allowed SPTM Frame Type Transitions}

\begin{longtable}{|p{6cm}|P{8cm}|}
\hline
\textbf{Type} & \textbf{Allowed retypes} \\
\hline
\endfirsthead

\hline
\textbf{Type} & \textbf{Allowed retypes} \\
\hline
\endhead

\hline
\endfoot

\hline
\caption[SPTM allowed frame retype operations.]{The table lists allowed frame retypings extracted from the SPTM binary based on conditional checks in its \texttt{retype} operation. The type column shows the current frame type, while allowed retypes shows valid new frame types retype to.}
\label{App:allowedTransition}
\endlastfoot

 \texttt{SPTM\_UNTYPED} & \texttt{SPTM\_UNTYPED}, \texttt{SPTM\_UNUSED}, \texttt{SPTM\_DEFAULT}, \texttt{SPTM\_RO}, \texttt{SPTM\_CODE}, \texttt{SPTM\_TXM\_CODE}, \texttt{SPTM\_XNU\_CODE}, \texttt{SPTM\_XNU\_CODE\_DBG\_RW}, \texttt{SPTM\_KERNEL\_ROOT\_TABLE}, \texttt{SPTM\_PAGE\_TABLE}, \texttt{SPTM\_IOMMU\_BOOTSTRAP}, \texttt{XNU\_DEFAULT}, \texttt{XNU\_RO}, \texttt{XNU\_RO\_DBG\_RW}, \texttt{XNU\_USER\_EXEC}, \texttt{XNU\_USER\_DEBUG}, \texttt{XNU\_USER\_JIT}, \texttt{XNU\_USER\_ROOT\_TABLE}, \texttt{XNU\_SHARED\_ROOT\_TABLE}, \texttt{XNU\_PAGE\_TABLE}, \texttt{XNU\_PAGE\_TABLE\_SHARED}, \texttt{XNU\_PAGE\_TABLE\_ROZONE}, \texttt{XNU\_PAGE\_TABLE\_COMMPAGE}, \texttt{XNU\_IOMMU}, \texttt{XNU\_ROZONE}, \texttt{XNU\_IO}, \texttt{XNU\_PROTECTED\_IO}, \texttt{XNU\_COMMPAGE\_RW}, \texttt{XNU\_COMMPAGE\_RO}, \texttt{XNU\_COMMPAGE\_RX}, \texttt{XNU\_TAG\_STORAGE}, \texttt{XNU\_STAGE2\_ROOT\_TABLE}, \texttt{XNU\_STAGE2\_PAGE\_TABLE}, \texttt{XNU\_KERNEL\_RESTRICTED}, \texttt{XNU\_RESERVED\_1}, \texttt{XNU\_RESERVED\_2}, \texttt{XNU\_RESTRICTED\_IO}, \texttt{XNU\_RESTRICTED\_IO\_TELEMETRY}, \texttt{TXM\_DEFAULT}, \texttt{TXM\_RO}, \texttt{TXM\_RW}, \texttt{TXM\_CPU\_STACK}, \texttt{TXM\_THREAD\_STACK}, \texttt{TXM\_ADDRESS\_SPACE\_TABLE}, \texttt{TXM\_MALLOC\_PAGE}, \texttt{TXM\_FREE\_LIST}, \texttt{TXM\_SLAB\_TRUST\_CACHE}, \texttt{TXM\_SLAB\_PROFILE}, \texttt{TXM\_SLAB\_CODE\_SIGNATURE}, \texttt{TXM\_SLAB\_CODE\_REGION}, \texttt{TXM\_SLAB\_ADDRESS\_SPACE}, \texttt{TXM\_BUCKET\_1024}, \texttt{TXM\_BUCKET\_2048}, \texttt{TXM\_BUCKET\_4096}, \texttt{TXM\_BUCKET\_8192}, \texttt{TXM\_BULK\_DATA}, \texttt{TXM\_BULK\_DATA\_READ\_ONLY}, \texttt{TXM\_LOG}, \texttt{TXM\_SEP\_SECURE\_CHANNEL}, \texttt{SK\_DEFAULT}, \texttt{SK\_SHARED\_RO}, \texttt{SK\_SHARED\_RW}, \texttt{SK\_IO} \\
\hline
\texttt{SPTM\_UNUSED} &  \\
\hline
\texttt{SPTM\_DEFAULT} &  \\
\hline
\texttt{SPTM\_RO} &  \\
\hline
\texttt{SPTM\_CODE} &  \\
\hline
\texttt{SPTM\_TXM\_CODE} &  \\
\hline
\texttt{SPTM\_XNU\_CODE} &  \\
\hline
\texttt{SPTM\_XNU\_CODE\_DBG\_RW} &  \\
\hline
\texttt{SPTM\_KERNEL\_ROOT\_TABLE} &  \\
\hline
\texttt{SPTM\_PAGE\_TABLE} &  \\
\hline
\texttt{SPTM\_IOMMU\_BOOTSTRAP} &  \\
\hline
\texttt{XNU\_DEFAULT} & \texttt{SPTM\_UNUSED}, \texttt{XNU\_DEFAULT}, \texttt{XNU\_USER\_EXEC}, \texttt{XNU\_USER\_DEBUG}, \texttt{XNU\_USER\_JIT}, \texttt{XNU\_USER\_ROOT\_TABLE}, \texttt{XNU\_PAGE\_TABLE}, \texttt{XNU\_PAGE\_TABLE\_SHARED}, \texttt{XNU\_PAGE\_TABLE\_ROZONE}, \texttt{XNU\_PAGE\_TABLE\_COMMPAGE}, \texttt{XNU\_IOMMU}, \texttt{XNU\_ROZONE}, \texttt{XNU\_COMMPAGE\_RW}, \texttt{XNU\_COMMPAGE\_RO}, \texttt{XNU\_COMMPAGE\_RX}, \texttt{XNU\_STAGE2\_ROOT\_TABLE}, \texttt{XNU\_STAGE2\_PAGE\_TABLE}, \texttt{XNU\_KERNEL\_RESTRICTED}, 
\texttt{XNU\_RESERVED\_2},
\texttt{TXM\_DEFAULT}
\texttt{SK\_DEFAULT}\\
\hline
\texttt{XNU\_RO} &  \\
\hline
\texttt{XNU\_RO\_DBG\_RW} &  \\
\hline
\texttt{XNU\_USER\_EXEC} & \texttt{XNU\_DEFAULT} \\
\hline
\texttt{XNU\_USER\_DEBUG} & \texttt{XNU\_DEFAULT} \\
\hline
\texttt{XNU\_USER\_JIT} & \texttt{XNU\_DEFAULT} \\
\hline
\texttt{XNU\_USER\_ROOT\_TABLE} & \texttt{XNU\_USER\_DEFAULT}, \texttt{XNU\_SHARED\_ROOT\_TABLE} \\
\hline
\texttt{XNU\_SHARED\_ROOT\_TABLE} & \texttt{XNU\_DEFAULT} \\
\hline
\texttt{XNU\_PAGE\_TABLE} & \texttt{XNU\_DEFAULT} \\
\hline
\texttt{XNU\_PAGE\_TABLE\_SHARED} & \texttt{XNU\_DEFAULT} \\
\hline
\texttt{XNU\_PAGE\_TABLE\_ROZONE} &  \\
\hline
\texttt{XNU\_PAGE\_TABLE\_COMMPAGE} &  \\
\hline
\texttt{XNU\_IOMMU} & \texttt{XNU\_DEFAULT} \\
\hline
\texttt{XNU\_ROZONE} & \texttt{XNU\_DEFAULT} \\
\hline
\texttt{XNU\_IO} &  \\
\hline
\texttt{XNU\_PROTECTED\_IO} &  \\
\hline
\texttt{XNU\_COMMPAGE\_RW} &  \\
\hline
\texttt{XNU\_COMMPAGE\_RO} &  \\
\hline
\texttt{XNU\_COMMPAGE\_RX} &  \\
\hline
\texttt{XNU\_TAG\_STORAGE} &  \\
\hline
\texttt{XNU\_STAGE2\_ROOT\_TABLE} & \texttt{XNU\_DEFAULT} \\
\hline
\texttt{XNU\_STAGE2\_PAGE\_TABLE} & \texttt{XNU\_DEFAULT} \\
\hline
\texttt{XNU\_KERNEL\_RESTRICTED} & \texttt{XNU\_DEFAULT},
\texttt{XNU\_KERNEL\_RESTRICTED}\\
\hline
\texttt{XNU\_RESERVED\_1} &  \\
\hline
\texttt{XNU\_RESERVED\_2} & \texttt{XNU\_DEFAULT} \\
\hline
\texttt{XNU\_RESTRICTED\_IO} &  \\
\hline
\texttt{XNU\_RESTRICTED\_IO\_TELEMETRY} &  \\
\hline
\texttt{TXM\_DEFAULT} & \texttt{TXM\_FREE\_LIST}, \texttt{TXM\_BULK\_DATA}, \texttt{TXM\_BULK\_DATA\_READ\_ONLY}
\\
\hline
\texttt{TXM\_RO} &  \\
\hline
\texttt{TXM\_RW} &  \\
\hline
\texttt{TXM\_CPU\_STACK} &  \\
\hline
\texttt{TXM\_THREAD\_STACK} &  \\
\hline
\texttt{TXM\_ADDRESS\_SPACE\_TABLE} &  \\
\hline
\texttt{TXM\_MALLOC\_PAGE} &  \\
\hline
\texttt{TXM\_FREE\_LIST} & \texttt{XNU\_DEFAULT},  \texttt{TXM\_SLAB\_TRUST\_CACHE}, \texttt{TXM\_SLAB\_PROFILE}, \texttt{TXM\_SLAB\_CODE\_SIGNATURE}, \texttt{TXM\_SLAB\_CODE\_REGION}, \texttt{TXM\_SLAB\_ADDRESS\_SPACE}, \texttt{TXM\_BUCKET\_1024}, \texttt{TXM\_BUCKET\_2048}, \texttt{TXM\_BUCKET\_4096}, \texttt{TXM\_BUCKET\_8192}
\\
\hline
\texttt{TXM\_SLAB\_TRUST\_CACHE} &  \\
\hline
\texttt{TXM\_SLAB\_PROFILE} &  \\
\hline
\texttt{TXM\_SLAB\_CODE\_SIGNATURE} &  \\
\hline
\texttt{TXM\_SLAB\_CODE\_REGION} &  \\
\hline
\texttt{TXM\_SLAB\_ADDRESS\_SPACE} &  \\
\hline
\texttt{TXM\_BUCKET\_1024} &  \\
\hline
\texttt{TXM\_BUCKET\_2048} &  \\
\hline
\texttt{TXM\_BUCKET\_4096} &  \\
\hline
\texttt{TXM\_BUCKET\_8192} &  \\
\hline
\texttt{TXM\_BULK\_DATA} & \texttt{XNU\_DEFAULT}, \texttt{TXM\_BULK\_DATA\_READ\_ONLY}\\
\hline
\texttt{TXM\_BULK\_DATA\_READ\_ONLY} & \texttt{XNU\_DEFAULT} \\
\hline
\texttt{TXM\_LOG} &  \\
\hline
\texttt{TXM\_SEP\_SECURE\_CHANNEL} &  \\
\hline
\texttt{SK\_DEFAULT} & \texttt{XNU\_DEFAULT}, \texttt{SK\_SHARED\_RO}, \texttt{SK\_SHARED\_RW} \\
\hline
\texttt{SK\_SHARED\_RO} & \texttt{SK\_DEFAULT}, \texttt{SK\_SHARED\_RW} \\
\hline
\texttt{SK\_SHARED\_RW} & \texttt{SK\_DEFAULT}, \texttt{SK\_SHARED\_RO}  \\
\hline
\texttt{SK\_IO} &  \\
\hline

\end{longtable}

\section{Type-Specific Retype Functions}
\begin{longtable}{|l|l|l|l|}

\hline
Type & Type Name & type\_out\_function & type\_in\_function \\ 
\hline
\endfirsthead

\hline
Type & Type Name & type\_out\_function & type\_in\_function \\ 
\hline
\endhead

\hline
\multicolumn{4}{r}{\texttt{Continued on next page}} \\
\endfoot

\hline
\caption[SPTM \texttt{retype\_out} and \texttt{retype\_in} functions.]{Type-specific \texttt{retype\_in} and \texttt{retype\_out} functions called during an SPTM \texttt{retype} operation.} 
\label{tab:type-function-table} 
\endlastfoot

8 & SPTM\_KERNEL\_ROOT\_TABLE & FUN\_fffffff0270bacd0 & FUN\_fffffff0270baffc \\ \hline
11 & XNU\_DEFAULT & FUN\_fffffff0270ba9e8 & FUN\_fffffff0270bacb8 \\ \hline
14 & XNU\_USER\_EXEC & FUN\_fffffff0270ba918 & - \\ \hline
15 & XNU\_USER\_DEBUG & FUN\_fffffff0270ba918 & - \\ \hline
16 & XNU\_USER\_JIT & FUN\_fffffff0270ba918 & - \\ \hline
17 & XNU\_USER\_ROOT\_TABLE & FUN\_fffffff0270bacd0 & FUN\_fffffff0270baffc \\ \hline
18 & XNU\_SHARED\_ROOT\_TABLE & FUN\_fffffff0270bacd0 & FUN\_fffffff0270baffc \\ \hline
19 & XNU\_PAGE\_TABLE & FUN\_fffffff0270ba7c4 & FUN\_fffffff0270ba8a8 \\ \hline
20 & XNU\_PAGE\_TABLE\_SHARED & FUN\_fffffff0270ba7c4 & FUN\_fffffff0270ba8a8 \\ \hline
21 & XNU\_PAGE\_TABLE\_ROZONE & - & FUN\_fffffff0270ba8a8 \\ \hline
22 & XNU\_PAGE\_TABLE\_COMMPAGE & - & FUN\_fffffff0270ba8a8 \\ \hline
23 & XNU\_IOMMU & FUN\_fffffff0270ba540 & FUN\_fffffff0270ba66c \\ \hline
24 & XNU\_ROZONE & FUN\_fffffff0270ba45c & - \\ \hline
31 & XNU\_STAGE2\_ROOT\_TABLE & FUN\_fffffff0270bacd0 & FUN\_fffffff0270baffc \\ \hline
32 & XNU\_STAGE2\_PAGE\_TABLE & FUN\_fffffff0270ba8a8 & FUN\_fffffff0270ba7c4 \\ \hline
33 & XNU\_KERNEL\_RESTRICTED & FUN\_fffffff0270ba9e8 & - \\ \hline
35 & XNU\_RESERVED\_2 & FUN\_fffffff0270ba9e8 & FUN\_fffffff0270bacb8 \\ \hline
59 & SK\_DEFAULT & FUN\_fffffff0270ba318 & - \\ \hline
60 & SK\_SHARED\_RO & FUN\_fffffff0270ba318 & - \\ \hline
61 & SK\_SHARED\_RW & FUN\_fffffff0270ba318 &  \\ \hline

\end{longtable}

\section{SPTM Exception Handler Function \protect \path{synchronous_exception_handler_from_lower}}
\begin{listing}[H]
    \begin{minted}[linenos, breaklines, bgcolor=LightGray, frame=lines]{asm}
**************************************************************
*                          FUNCTION                          *
**************************************************************
            undefined synchronous_exception_handler_from_lower()
fffffff027081e38 9f 40 00 d5     msr        PState.PAN,#0x0
fffffff027081e3c e8 27 3f a9     stp        x8,x9,[sp, #local_10]
fffffff027081e40 28 fb 3e d5     mrs        x8,sreg(0x3, 0x6, c0xf, c0xb, 0x1)
fffffff027081e44 08 01 08 aa     orr        x8,x8,x8
fffffff027081e48 08 41 11 91     add        x8,x8,#0x450
fffffff027081e4c 08 11 40 f9     ldr        x8,[x8, #0x20]
fffffff027081e50 ff 63 28 eb     cmp        sp,x8
                    LAB_fffffff027081e54                              
fffffff027081e54 01 00 00 54     b.ne       LAB_fffffff027081e54
fffffff027081e58 08 11 3c d5     mrs        x8,hcr_el2
fffffff027081e5c 08 6d 40 92     and        x8,x8,#0xfffffff
fffffff027081e60 08 95 65 92     and        x8,x8,#-0x7ffffff
fffffff027081e64 1f 05 00 f1     cmp        x8,#0x1
fffffff027081e68 00 01 00 54     b.eq       LAB_fffffff027081e88
                    GL1 Exception Syndrome Register
fffffff027081e6c a8 fa 3e d5     mrs        x8,sreg(0x3, 0x6, c0xf, c0xa, 0x5)
fffffff027081e70 08 7d 5a d3     ubfx       x8,x8,#0x1a,#0x6
fffffff027081e74 1f 55 00 f1     cmp        x8,#0x15
fffffff027081e78 80 08 00 54     b.eq       SVC_HANDLER
fffffff027081e7c 1f 59 00 f1     cmp        x8,#0x16
fffffff027081e80 e0 0f 00 54     b.eq       LAB_fffffff02708207c
fffffff027081e84 8b 00 00 14     b          LAB_fffffff0270820b0    
    \end{minted}
    \captionof{lstlisting}[SPTM synchronous exception handler.]{SPTM synchronous exception handler function \texttt{synchronous\_exception\_handler\_from\_lower}.}
    \label{lst:sptm_synchronous-handler}
\end{listing}

\section{SPTM SVC \#0 Handler}

\begin{minted}[linenos, breaklines, bgcolor=LightGray, frame=lines]{asm}
                    SVC_0_HANDLER                                   
fffffff027081ff0 09 9e 60 d3     ubfx       x9,x16,#0x20,#0x8
fffffff027081ff4 aa 1f 80 d2     mov        x10,#0xfd
fffffff027081ff8 3f 01 0a eb     cmp        x9,x10
fffffff027081ffc 41 00 00 54     b.ne       SVC_0_not_0xfd
fffffff027082000 18 01 00 14     b          SVC_0_is_0xfd                                    undefined SVC_0_is_0xfd()

          
                    SVC_0_not_0xfd                               
fffffff027082004 ca 1f 80 d2     mov        x10,#0xfe
fffffff027082008 3f 01 0a eb     cmp        x9,x10
fffffff02708200c 01 03 00 54     b.ne       SVC_0_not_0xfe
fffffff027082010 28 fb 3e d5     mrs        x8,sreg(0x3, 0x6, c0xf, c0xb, 0x1)
fffffff027082014 08 01 08 aa     orr        x8,x8,x8
fffffff027082018 08 41 11 91     add        x8,x8,#0x450
fffffff02708201c 0a 41 15 91     add        x10,x8,#0x550
fffffff027082020 49 01 40 f9     ldr        x9,[x10]
fffffff027082024 a9 f1 1e d5     msr        sreg(0x3, 0x6, c0xf, c0x1, 0x5),x9
fffffff027082028 49 05 40 f9     ldr        x9,[x10, #0x8]
fffffff02708202c 49 02 10 d5     msr        mdscr_el1,x9
fffffff027082030 49 09 40 f9     ldr        x9,[x10, #0x10]
fffffff027082034 a9 f1 19 d5     msr        sreg(0x3, 0x1, c0xf, c0x1, 0x5),x9
fffffff027082038 49 0d 40 f9     ldr        x9,[x10, #0x18]
fffffff02708203c 89 f0 1c d5     msr        sreg(0x3, 0x4, c0xf, c0x0, 0x4),x9
fffffff027082040 49 11 40 f9     ldr        x9,[x10, #0x20]
fffffff027082044 09 10 18 d5     msr        sctlr_el1,x9
fffffff027082048 49 15 40 f9     ldr        x9,[x10, #0x28]
fffffff02708204c 49 20 18 d5     msr        tcr_el1,x9
fffffff027082050 bf 40 00 d5     msr        PState.SP,#0x0
fffffff027082054 01 00 80 d2     mov        x1,#0x0
fffffff027082058 22 00 80 d2     mov        x2,#0x1
fffffff02708205c c1 fc ff 17     b          FUN_fffffff027081360  

...               
                    SVC_0_is_0xfd                                   
fffffff027082460 28 fb 3e d5     mrs        x8,sreg(0x3, 0x6, c0xf, c0xb, 0x1)
fffffff027082464 08 01 08 aa     orr        x8,x8,x8
fffffff027082468 08 41 11 91     add        x8,x8,#0x450
fffffff02708246c 09 11 40 f9     ldr        x9,[x8, #0x20]
fffffff027082470 3f 01 00 91     mov        sp,x9
fffffff027082474 1e 00 00 90     adrp       x30,-0xfd8f7e000
fffffff027082478 de 93 16 91     add        x30,x30,#0x5a4
fffffff02708247c 1d 00 80 d2     mov        x29,#0x0

...

                    SVC_0_not_0xfe  
fffffff02708206c ea 1f 80 d2     mov        x10,#0xff
fffffff027082070 3f 01 0a eb     cmp        x9,x10
fffffff027082074 a1 19 00 54     b.ne       SVC_0_not_0xff
fffffff027082078 1d 01 00 14     b          SVC_0_is_0xff     

...

                    SVC_0_not_0xff                                   
fffffff0270823a8 28 fb 3e d5     mrs        x8,sreg(0x3, 0x6, c0xf, c0xb, 0x1)
fffffff0270823ac 08 01 08 aa     orr        x8,x8,x8
fffffff0270823b0 08 41 11 91     add        x8,x8,#0x450
fffffff0270823b4 09 11 40 f9     ldr        x9,[x8, #0x20]
fffffff0270823b8 3f 01 00 91     mov        sp,x9
fffffff0270823bc 09 01 01 91     add        x9,x8,#0x40
fffffff0270823c0 20 05 00 a9     stp        x0,x1,[x9]
fffffff0270823c4 22 0d 01 a9     stp        x2,x3,[x9, #0x10]
fffffff0270823c8 24 15 02 a9     stp        x4,x5,[x9, #0x20]
fffffff0270823cc 26 1d 03 a9     stp        x6,x7,[x9, #0x30]
fffffff0270823d0 0a 1d 40 f9     ldr        x10,[x8, #0x38]
fffffff0270823d4 5f 05 00 f1     cmp        x10,#0x1

...

                    SVC_0_is_0xff                                   
fffffff0270824ec 28 fb 3e d5     mrs        x8,sreg(0x3, 0x6, c0xf, c0xb, 0x1)
fffffff0270824f0 08 01 08 aa     orr        x8,x8,x8
fffffff0270824f4 08 41 11 91     add        x8,x8,#0x450
fffffff0270824f8 09 11 40 f9     ldr        x9,[x8, #0x20]
fffffff0270824fc 3f 01 00 91     mov        sp,x9
fffffff027082500 1e 00 00 90     adrp       x30,-0xfd8f7e000
fffffff027082504 de 93 16 91     add        x30,x30,#0x5a4
fffffff027082508 1d 00 80 d2     mov        x29,#0x0    
\end{minted}
\captionof{lstlisting}{SPTM SVC \#0 handler.}
\label{lst:sptm_SVC0-handler}

\section{Function Signatures for User-Space Available Exclaves Functions} 
\label{lst:exclave_wrapper}
Taken from \texttt{libsyscall/wrappers/exclaves.c}.

\begin{minted}[linenos, breaklines, bgcolor=LightGray, frame=lines]{c}
kern_return_t exclaves_endpoint_call(mach_port_t port, exclaves_id_t endpoint_id,
    mach_vm_address_t msg_buffer, mach_vm_size_t size, exclaves_tag_t *tag,
    exclaves_error_t *error)

kern_return_t exclaves_outbound_buffer_create(mach_port_t port, const char *buffer_name,
    mach_vm_size_t size, mach_port_t *out_outbound_buffer_port)

kern_return_t exclaves_outbound_buffer_copyout(mach_port_t outbound_buffer_port,
    mach_vm_address_t dst_buffer, mach_vm_size_t size1, mach_vm_size_t offset1,
    mach_vm_size_t size2, mach_vm_size_t offset2)

kern_return_t exclaves_inbound_buffer_create(mach_port_t port, const char *buffer_name,
    mach_vm_size_t size, mach_port_t *out_inbound_buffer_port)

kern_return_t exclaves_inbound_buffer_copyin(mach_port_t inbound_buffer_port,
    mach_vm_address_t src_buffer, mach_vm_size_t size1, mach_vm_size_t offset1,
    mach_vm_size_t size2, mach_vm_size_t offset2)

kern_return_t exclaves_named_buffer_create(mach_port_t port, exclaves_id_t buffer_id,
    mach_vm_size_t size, mach_port_t *out_named_buffer_port)

kern_return_t exclaves_named_buffer_copyin(mach_port_t named_buffer_port,
    mach_vm_address_t src_buffer, mach_vm_size_t size, mach_vm_size_t offset)

kern_return_t exclaves_named_buffer_copyout(mach_port_t named_buffer_port,
    mach_vm_address_t dst_buffer, mach_vm_size_t size, mach_vm_size_t offset)

kern_return_t exclaves_launch_conclave(mach_port_t port, void *arg1,
    uint64_t arg2)

kern_return_t exclaves_lookup_service(mach_port_t port, const char *name,
    exclaves_id_t *resource_id)

kern_return_t exclaves_boot(mach_port_t port, exclaves_boot_stage_t stage)

kern_return_t exclaves_audio_buffer_create(mach_port_t port, const char *buffer_name,
    mach_vm_size_t size, mach_port_t* out_audio_buffer_port)

kern_return_t exclaves_audio_buffer_copyout(mach_port_t audio_buffer_port,
    mach_vm_address_t dst_buffer,
    mach_vm_size_t size1, mach_vm_size_t offset1,
    mach_vm_size_t size2, mach_vm_size_t offset2)

kern_return_t exclaves_sensor_create(mach_port_t port, const char *sensor_name,
    mach_port_t *sensor_port)

kern_return_t exclaves_sensor_start(mach_port_t sensor_port, uint64_t flags,
    exclaves_sensor_status_t *sensor_status)
    
kern_return_t exclaves_sensor_stop(mach_port_t sensor_port, uint64_t flags,
    exclaves_sensor_status_t *sensor_status)

kern_return_t exclaves_sensor_status(mach_port_t sensor_port, uint64_t flags,
    exclaves_sensor_status_t *sensor_status)

kern_return_t exclaves_notification_create(__unused mach_port_t port, const char *name,
    uint64_t *notification_id) 
\end{minted}
\captionof{lstlisting}{User-space Exclave function signatures.}
\label{lst:exclave_wrapper}

\section{Ghidra Jython Script Extracting Exclave Resources and Corresponding Domains}

This script extracts Exclave resources and domains from the \texttt{sharedcache} binary, grouping them by domains.

\begin{minted}[linenos, breaklines, bgcolor=LightGray, frame=lines]{c}
# -*- coding: utf-8 -*-
#@author Moritz Steffin
#@category Analysis

"""
Ghidra script to extract and group resource entries by domain.

Each entry:
+0x00 : resource_name (C string)
+0x80 : domain_name (C string)
Size  : 0x110 bytes
Count : 0xa3 entries

Output: Grouped text file
"""

ENTRY_COUNT = 0xa3
ENTRY_SIZE = 0x110
RESOURCE_NAME_OFFSET = 0x0
DOMAIN_NAME_OFFSET = 0x80
STRING_READ_MAXLEN = 256

def askBaseAddress():
    input_str = askString("Base Address", "Enter the resource table base address (hex):")
    if input_str is None:
        exit(0)
    input_str = input_str.strip()
    if input_str.startswith("0x") or input_str.startswith("0X"):
        input_str = input_str[2:]
    try:
        return toAddr(int(input_str, 16))
    except:
        popup("Invalid address: {}".format(input_str))
        exit(1)

def read_c_string(addr, maxlen=256):
    result = []
    for i in range(maxlen):
        b = getByte(addr.add(i)) & 0xFF
        if b == 0:
            break
        if 32 <= b <= 126:
            result.append(chr(b))
        else:
            result.append('.')
    return ''.join(result)

def main():
    base_addr = askBaseAddress()
    print("[*] Reading resource table at {}".format(base_addr))

    # Build domain -> [resources] map
    domain_map = {}

    for i in range(ENTRY_COUNT):
        entry_addr = base_addr.add(i * ENTRY_SIZE)
        resource_addr = entry_addr.add(RESOURCE_NAME_OFFSET)
        domain_addr = entry_addr.add(DOMAIN_NAME_OFFSET)

        resource_name = read_c_string(resource_addr, STRING_READ_MAXLEN)
        domain_name = read_c_string(domain_addr, STRING_READ_MAXLEN)

        if not resource_name:
            resource_name = "<undefined>"
        if not domain_name:
            domain_name = "<undefined>"

        if domain_name not in domain_map:
            domain_map[domain_name] = []
        domain_map[domain_name].append(resource_name)

    # Ask where to save
    output_file = askFile("Save Text File", "Save").absolutePath

    with open(output_file, "w") as f:
        for domain in sorted(domain_map.keys()):
            f.write("Domain: {}\n".format(domain))
            for res in sorted(domain_map[domain]):
                f.write("  - {}\n".format(res))
            f.write("\n")

    popup("Done!\n\nExtracted and grouped {} domains.\nSaved to:\n{}".format(len(domain_map), output_file))

if __name__ == "__main__":
    main()


\end{minted}
\captionof{lstlisting}{Ghidra Script for Extracting Exclave Resources and Grouping them by Corresponding Domains from the sharedcache Binary.}
\label{lst:resourceEnumScript}

\section{Kernel \texttt{tb\_connection\_send\_query} Function}

\begin{minted}[linenos, breaklines, breakanywhere, bgcolor=LightGray, frame=lines]{c}
ulong _tb_connection_send_query
                (tb_connection_t *connection,tb_message *message,undefined8 *response,
                undefined8 some_sort_of_state)

{
  int iVar1;
  int splitRequired?;
  long tb_message_tpt_buffer;
  ulong send_return;
  long oberserver;
  undefined8 *puVar2;
  tb_message *message_00;
  tb_bufffer *tpt_buffer;
  undefined *puVar3;
  tb_transport_t *transport;
  undefined8 *puVar4;
  ulong uVar5;
  long local_70;
  
  if (message->msg_state == TB_MESSAGE_STATE_READY) {
    if (message->disposition != TB_MESSAGE_DISPOSITION_QUERY) goto LAB_fffffff0088ede10;
    iVar1 = tb_message_check_connection_identifier(message,connection);
    if (iVar1 == 0) {
      return TB_ERROR;
    }
    tb_message_set_state(message,TB_MESSAGE_STATE_SENT);
    transport = (tb_transport_t *)connection->transport;
    tb_message_tpt_buffer = tb_message_get_transport_buffer(message);
    if (((uint)some_sort_of_state >> 1 & 1) == 0) {
      *(ushort *)(tb_message_tpt_buffer + 0x2a) = *(ushort *)(tb_message_tpt_buffer + 0x2a) | 0x10;
    }
    splitRequired? =
         _tb_message_splitter_split_required
                   (transport,*(undefined8 *)(tb_message_tpt_buffer + 0x18));
    if (splitRequired? == 0) {
      if (((long *)connection->observers != (long *)0x0) &&
         (oberserver = *(long *)connection->observers, oberserver != 0)) {
        (**(code **)(oberserver + 0x10))
        (oberserver + 0x10,oberserver,transport,message,response,some_sort_of_state);
      }
      send_return = tb_transport_send_message(transport);
    }
    else {
      send_return = _tb_message_splitter_send
            (connection,transport,message,response,some_sort_of_state);
    }
    if ((int)send_return != 0) {
      return send_return;
    }
    uVar5 = (ulong)((*(ushort *)(tb_message_tpt_buffer + 0x2a) & 8) >> 1);
    if (((uint)some_sort_of_state >> 1 & 1) == 0) {
      return uVar5;
    }
    if ((*(ushort *)(tb_message_tpt_buffer + 0x2a) & 8) != 0) {
      return uVar5;
    }
    if ((response == (undefined8 *)0x0) || ((tb_message *)*response == (tb_message *)0x0)) {
      return 4;
    }
    tb_message_set_state((tb_message *)*response,TB_MESSAGE_STATE_RECEIVED);
    set_tb_msg_disposition((tb_message *)*response,2);
    tb_message_tpt_buffer = tb_message_get_transport_buffer((tb_message *)*response);
    if ((*(ushort *)(tb_message_tpt_buffer + 0x2a) & 1) == 0) {
                    /* Failure */
      return 0;
    }
    local_70 = FUN_fffffff0088eeef4((tb_message *)*response);
    if (local_70 == 0) {
      local_70 = FUN_fffffff0082bab90();
      FUN_fffffff0088eef00(*response);
    }
    puVar4 = connection->tb_list;
    if (puVar4 == (undefined8 *)0x0) goto LAB_fffffff0088ede14;
    puVar2 = (undefined8 *)_tb_message_accumulator_accumulate(puVar4,*response);
    message_00 = (tb_message *)allocate_in_structure(&DAT_fffffff007c2a670,4);
    if (message_00 != (tb_message *)0x0) {
      tpt_buffer = (tb_bufffer *)allocate_in_structure(&DAT_fffffff007c2a6b0,4);
      if (tpt_buffer != (tb_bufffer *)0x0) {
        uVar5 = __tb_connection_message_construct(connection,0,message_00,tpt_buffer,0,0);
        if ((int)uVar5 != 0) {
          FUN_fffffff00816b9e8(&DAT_fffffff007c2a6f0,tpt_buffer);
          puVar3 = &DAT_fffffff007c2a730;
LAB_fffffff0088edc98:
          FUN_fffffff00816b9e8(puVar3,message_00);
          return uVar5;
        }
        if (puVar2 == (undefined8 *)0x0) {
          while( true ) {
            message_complete(message_00);
            *(ushort *)&tpt_buffer->someFlags = *(ushort *)&tpt_buffer->someFlags | 4;
            tb_message_set_state(message_00,3);
            if (((long *)connection->observers != (long *)0x0) &&
               (tb_message_tpt_buffer = *(long *)connection->observers, tb_message_tpt_buffer != 0))
            {
              (**(code **)(tb_message_tpt_buffer + 0x10))
                        (tb_message_tpt_buffer + 0x10,tb_message_tpt_buffer,transport,message,
                         response,some_sort_of_state);
            }
            uVar5 = tb_transport_send_message(transport);
            if ((int)uVar5 != 0) {
              __tb_connection_message_destruct(connection);
              FUN_fffffff00816b9e8(&DAT_fffffff007c2a770,tpt_buffer);
              puVar3 = &DAT_fffffff007c2a7b0;
              goto LAB_fffffff0088edc98;
            }
            FUN_fffffff0088eef00(message_00,local_70);
            puVar2 = (undefined8 *)_tb_message_accumulator_accumulate(puVar4,message_00);
            if (puVar2 != (undefined8 *)0x0) break;
            FUN_fffffff0088ee078(connection,message_00,0,0,0);
          }
        }
        __tb_connection_message_destruct(connection,message_00);
        FUN_fffffff00816b9e8(&DAT_fffffff007c2a7f0,tpt_buffer);
        FUN_fffffff00816b9e8(&DAT_fffffff007c2a830,message_00);
        puVar4 = (undefined8 *)tb_message_get_transport_buffer((tb_message *)*response);
        FUN_fffffff0088ee764(transport,puVar4);
        FUN_fffffff0088ee80c(puVar4);
        *puVar4 = *puVar2;
        puVar4[3] = puVar2[3];
        *(undefined1 *)(puVar4 + 5) = 1;
        *(undefined2 *)((long)puVar4 + 0x2a) = *(undefined2 *)((long)puVar2 + 0x2a);
        FUN_fffffff00816b9e8(&DAT_fffffff007c2a870,puVar2);
        return 0;
      }
      goto LAB_fffffff0088ede1c;
    }
  }
  else {
    ASSERT_FAIL_QUERY_STATE==TB_MESSAGE_STATE_READY();
LAB_fffffff0088ede10:
    TB_ASSERT_FAIL_QUERY_DISPOSITION==TB_MESSAGE_DISPOSITON_QUERY();
LAB_fffffff0088ede14:
    FUN_fffffff0088ee4b4();
  }
  FUN_fffffff0088ee488();
LAB_fffffff0088ede1c:
  uVar5 = FUN_fffffff0088ee45c();
  if ((*(ulong **)(uVar5 + 8) != (ulong *)0x0) && (uVar5 = **(ulong **)(uVar5 + 8), uVar5 != 0)) {
                    /* WARNING: Could not recover jumptable at 0xfffffff0088ede3c. Too many branches
                        */
                    /* WARNING: Treating indirect jump as call */
    uVar5 = (**(code **)(uVar5 + 0x10))();
    return uVar5;
  }
  return uVar5;
}
\end{minted}
\captionof{lstlisting}[Kernel \texttt{tb\_connection\_send\_query} function for sending a Tightbeam message.]{Kernel \texttt{tb\_connection\_send\_query} function for sending a Tightbeam message.}
\label{tb_connection_send_query_kernel}

\section{lldb Trace on \texttt{audiomxd} on Startup with Audio Capture}
The breakpoint is set to \texttt{exclaves\_*}, reduced excerpt.

\label{audiomxdllbsection}
\begin{listing}[H]
\begin{minted}[fontsize=\footnotesize, breaklines, breakanywhere, bgcolor=LightGray, frame=lines]{text}
lldb)  bt
  thread #49, queue = 'AudioControl', stop reason = breakpoint 1.18
    frame #0: 0x00000001f18d0c0c libsystem_kernel.dylib`exclaves_sensor_start
    frame #1: 0x00000001b2a4a7a8 MediaExperience`-[MXExclaves updateSensorStatus:reason:]
(lldb)  bt
  thread #62, name = 'audio IO: VAD [vdef] AggDev 11', queue = 'com.apple.AudioServerDriverTransports.HPMic.serial', stop reason = breakpoint 1.17
    frame #0: 0x00000001f18d0bd8 libsystem_kernel.dylib`exclaves_sensor_create
    frame #1: 0x0000000264839a7c AudioServerDriverTransports_Base'
(lldb)  bt
  thread #62, name = 'audio IO: VAD [vdef] AggDev 11', queue = 'com.apple.AudioServerDriverTransports.HPMic.serial', stop reason = breakpoint 1.3
    frame #0: 0x00000001f18d0b38 libsystem_kernel.dylib`exclaves_audio_buffer_create
    frame #1: 0x000000026484b78c AudioServerDriverTransports_Base
  thread #62, name = 'audio IO: VAD [vdef] AggDev 11', queue = 'com.apple.AudioServerDriverTransports.HPMic.serial', stop reason = breakpoint 1.19
    frame #0: 0x00000001f18d0c74 libsystem_kernel.dylib`exclaves_sensor_status
    frame #1: 0x0000000264839e5c AudioServerDriverTransports_Base
(lldb)  bt
  thread #62, name = 'audio IO: VAD [vdef] AggDev 11', queue = 'com.apple.AudioServerDriverTransports.IOA2.device.HPMic.ioReference', stop reason = breakpoint 1.19
    frame #0: 0x00000001f18d0c74 libsystem_kernel.dylib`exclaves_sensor_status
    frame #1: 0x0000000264839e5c AudioServerDriverTransports_Base
(lldb)  bt
  thread #62, name = 'audio IO: VAD [vdef] AggDev 11', queue = 'com.apple.AudioServerDriverTransports.IOA2.device.HPMic.ioReference', stop reason = breakpoint 1.19
    frame #0: 0x00000001f18d0c74 libsystem_kernel.dylib`exclaves_sensor_status
    frame #1: 0x0000000264839e5c AudioServerDriverTransports_Base`
(lldb)  bt
  thread #62, name = 'audio IO: VAD [vdef] AggDev 11', stop reason = breakpoint 1.2
    frame #0: 0x00000001f18d0ba0 libsystem_kernel.dylib`exclaves_audio_buffer_copyout_with_status
    frame #1: 0x000000026484bb10 AudioServerDriverTransports_Base
(lldb)  bt
  thread #62, name = 'audio IO: VAD [vdef] AggDev 11', stop reason = breakpoint 1.2
    frame #0: 0x00000001f18d0ba0 libsystem_kernel.dylib`exclaves_audio_buffer_copyout_with_status
    frame #1: 0x000000026484bb10 AudioServerDriverTransports_Base

    \end{minted}
    \captionof{lstlisting}{lldb trace on \texttt{audiomxd} on startup with audio capture, breakpoint set to \texttt{exclaves\_*}, reduced excerpt.}
    \label{lldb_startup_audiomxd}
\end{listing}

\section{lldb Trace Excerpt on \texttt{exclaves\_sensor\_start}}

Trace excerpt taken from \texttt{corespeechd}.
\label{sensorCreateSection}
\begin{minted}[fontsize=\footnotesize, breaklines, breakanywhere, bgcolor=LightGray, frame=lines]{text}
[ Legend: Modified register | Code | Heap | Stack | String ]
-----------------registers-----------------
x0     : 0x0           
x1     : 0x1d0cb7436     -> ("com.apple.sensors.mic"?)
x2     : 0x10069d340   
x3     : 0x16fdb9cf0   
x4     : 0x20140b098    (libsystem_trace.dylib 0x1ebb8b098)  -><_os_log_current_test_callback+0>
x5     : 0x3b34dcdd02040004
x6     : 0x0           
x7     : 0x0           
x8     : 0xf52979e45a5f00a3
x9     : 0xf52979e45a5f00a3
x10    : 0x100000100000000
x11    : 0x64d0        
x12    : 0x0           
x13    : 0x200000110000e40
x14    : 0x20000001000 
x15    : 0x200000010000df0
x16    : 0x1ea190bd8    (libsystem_kernel.dylib 0x1d4910bd8)  -><exclaves_sensor_create+0>
x17    : 0x204d4fbc0   
x18    : 0x0           
x19    : 0x10069d330   
x20    : 0x100695b00   
x21    : 0x1fcba6000   
x22    : 0x20384a000    (CoreSpeechFoundation 0x1edfca000)  -><CSAudioFileManager+24>
x23    : 0x0           
x24    : 0x0           
x25    : 0x0           
x26    : 0x0           
x27    : 0x0           
x28    : 0x0           
fp     : 0x16fdbb0d0   
lr     : 0x1d0c09234    (CoreSpeechFoundation 0x1bb389234)  -><-[CSExclaveIndicatorController init]+180>
sp     : 0x16fdbb070   
pc     : 0x1ea190bd8    (libsystem_kernel.dylib 0x1d4910bd8)  -><exclaves_sensor_create+0>
cpsr   : [n Z C v q ssbs pan dit ge e a i f m]
-----------------stack-----------------
0x16fdbb070|+0000: 0x0000000000000000  <- $sp
0x16fdbb078|+0008: 0x0000000000000000
0x16fdbb080|+0010: 0x000000010069d330
0x16fdbb088|+0018: 0x0000000203846f40 (CoreSpeechFoundation 0x1edfc6f40)  -> <CSExclaveIndicatorController+0>
0x16fdbb090|+0020: 0xd0cb741108200102
0x16fdbb098|+0028: 0x0000000000000001
0x16fdbb0a0|+0030: 0x01338001a8e7e978
0x16fdbb0a8|+0038: 0xf52979e45a5f00a3
-----------------code-----------------
libsystem_kernel.dylib`exclaves_sensor_create:
0x1ea190bd8 <+0>:  pacibsp 
0x1ea190bdc <+4>:  stp    x29, x30, [sp, #-0x10]!
0x1ea190be0 <+8>:  mov    x29, sp
0x1ea190be4 <+12>: mov    x3, x2
0x1ea190be8 <+16>: mov    x2, x1
0x1ea190bec <+20>: mov    w1, #0xa000000 ; =167772160 
0x1ea190bf0 <+24>: mov    x4, #0x0 ; =0 
0x1ea190bf4 <+28>: mov    x5, #0x0 ; =0 
0x1ea190bf8 <+32>: mov    x6, #0x0 ; =0 
-----------------threads-----------------
thread #1: tid = 0x64d0, 0x00000001ea190bd8 libsystem_kernel.dylib`exclaves_sensor_create, queue = 'com.apple.main-thread', stop reason = breakpoint 1.21
thread #2: tid = 0x64d1, 0x00000001ea179a90 libsystem_kernel.dylib`__workq_kernreturn + 8
thread #3: tid = 0x64d3, 0x00000002234e4e68 libxpc.dylib`xpc_int64_get_value, queue = 'com.apple.xpc.activity.com.apple.cs.postinstall'
-----------------trace-----------------
[#0] 0x1ea190bd8 (libsystem_kernel.dylib 0x1d4910bd8)   -> exclaves_sensor_create()
[#1] 0x1d0c09234 (CoreSpeechFoundation 0x1bb389234)   -> -[CSExclaveIndicatorController init]()
[#2] 0x1d0c28e50 (CoreSpeechFoundation 0x1bb3a8e50)   -> -[CSExclaveRecordClient init]()
[#3] 0x1d0c28f68 (CoreSpeechFoundation 0x1bb3a8f68)   -> __37+[CSExclaveRecordClient sharedClient]_block_invoke()
[#4] 0x1a049f584 (libdispatch.dylib 0x18ac1f584)   -> _dispatch_client_callout()
[#5] 0x1a04889a8 (libdispatch.dylib 0x18ac089a8)   -> _dispatch_once_callout()
[#6] 0x1d0c28f44 (CoreSpeechFoundation 0x1bb3a8f44)   -> +[CSExclaveRecordClient sharedClient]()
[#7] 0x1d0c2ca04 (CoreSpeechFoundation 0x1bb3aca04)   -> +[CSExclaveMessageHandlingFactory commonExclaveMessageHandler]()
[#8] 0x1d0c25d18 (CoreSpeechFoundation 0x1bb3a5d18)   -> -[CSExclaveAssetManagerProxy init]()
[#9] 0x1d0c25dc0 (CoreSpeechFoundation 0x1bb3a5dc0)   -> __43+[CSExclaveAssetManagerProxy sharedManager]_block_invoke()
[#10] 0x1a049f584 (libdispatch.dylib 0x18ac1f584)   -> _dispatch_client_callout()
[#11] 0x1a04889a8 (libdispatch.dylib 0x18ac089a8)   -> _dispatch_once_callout()
[#12] 0x1d0c25d9c (CoreSpeechFoundation 0x1bb3a5d9c)   -> +[CSExclaveAssetManagerProxy sharedManager]()
[#13] 0x1d0c80520 (CoreSpeechFoundation 0x1bb400520)   -> -[CSUAFAssetManagerBase init]()
[#14] 0x1d0c80408 (CoreSpeechFoundation 0x1bb400408)   -> -[CSUAFAssetManagerBase initWithForceSetIsExclave:exclaveManagerProxy:]()
[#15] 0x1d0c4520c (CoreSpeechFoundation 0x1bb3c520c)   -> __35+[CSUAFAssetManager sharedInstance]_block_invoke()
[#16] 0x1a049f584 (libdispatch.dylib 0x18ac1f584)   -> _dispatch_client_callout()
[#17] 0x1a04889a8 (libdispatch.dylib 0x18ac089a8)   -> _dispatch_once_callout()
[#18] 0x1d0c451cc (CoreSpeechFoundation 0x1bb3c51cc)   -> +[CSUAFAssetManager sharedInstance]()
[#19] 0x1000957f4 (corespeechd 0x1001517f4)   -> ___lldb_unnamed_symbol3044()
\end{minted}
\captionof{lstlisting}{lldb trace excerpt on \texttt{exclaves\_sensor\_start} from \texttt{corespeechd}.}
\label{lldb_sensor_create_off}

 % example

	% ggf. bei englischen Arbeiten den deutschen Abstract nach hinten verschieben
	% \ifisbook\pagestyle{plain}\cleardoubleemptypage\include{content/abstract_deu}\fi

	% Eigenständigkeitserklärung
	\ifisbook\pagestyle{plain}\cleardoubleemptypage% => Laut Aussage des Studienreferats braucht es - auch wenn die Arbeit in englischer Sprache verfasst ist - KEINE separate Version der Eigenständigkeitserklärung auf Englisch. Sowohl für Arbeiten in deutscher Sprache als auch für Arbeiten in englischer Sprache genügt EINE EINZIGE Eigenständigkeitserklärung auf DEUTSCH.
\begin{otherlanguage}{ngerman}

\begin{center}\textsf{\textbf{Eidesstattliche Erklärung}}\end{center}
Hiermit versichere ich, dass meine {\thesistype} mit dem Titel \enquote{\thesistitle} (\enquote{\thesistitleother}) selbständig verfasst wurde und dass keine anderen Quellen und Hilfsmittel als die angegebenen benutzt wurden. Diese Aussage trifft auch für alle Implementierungen und Dokumentationen im Rahmen dieses Projektes zu.\\

\noindent
Potsdam, den \thesishandindate
\vspace{2cm}

\begin{center}
\begin{tabular}{C{6cm}}
\hline
{\small({\thesisauthor})}
\end{tabular}
\end{center}

\end{otherlanguage}\fi

\end{document}